\documentclass[%
 aip,
 amsmath,amssymb,
 reprint,%
]{revtex4-1}

\usepackage{CJKutf8}

\usepackage{amssymb}
\usepackage{subfigure}

\usepackage{graphicx}
\usepackage{dcolumn}
\usepackage{bm}
\usepackage{amsmath}

\usepackage[utf8]{inputenc}
\usepackage[T1]{fontenc}
\usepackage{mathptmx}
\usepackage{etoolbox}

\makeatletter
\def\@email#1#2{%
 \endgroup
 \patchcmd{\titleblock@produce}
  {\frontmatter@RRAPformat}
  {\frontmatter@RRAPformat{\produce@RRAP{*#1\href{mailto:#2}{#2}}}\frontmatter@RRAPformat}
  {}{}
}%
\makeatother
\begin{document}
\begin{CJK*}{UTF8}{gbsn}

\preprint{AIP/123-QED}

\title{Data-driven modeling of unsteady flow based on deep operator network}
\author{Heming Bai(白鹤鸣)}
 \affiliation{ 
 State Key Laboratory of Fluid Power and Mechatronic Systems, Department of Engineering Mechanics, Zhejiang University, Hangzhou 310027, P.R.China
 }%

\author{Zhicheng Wang(王志成)}%
 \affiliation{ 
 School of Energy and Power Engineering, Dalian University of Technology, Dalian 116081, P.R.China 
 }%

\author{Xuesen Chu(褚学森)}
 \affiliation{ 
 China Ship Scientific Research Center, Wuxi 214082, P.R.China
 }

\author{Jian Deng(邓见)}
 \affiliation{ 
 State Key Laboratory of Fluid Power and Mechatronic Systems, Department of Engineering Mechanics, Zhejiang University, Hangzhou 310027, P.R.China
 }%

\author{Xin Bian*(边鑫)}
 \affiliation{ 
 State Key Laboratory of Fluid Power and Mechatronic Systems, Department of Engineering Mechanics, Zhejiang University, Hangzhou 310027, P.R.China
 }%
 \email{bianx@zju.edu.cn}

\date{\today}

\begin{abstract}
Time-dependent flow fields are typically generated by a computational fluid dynamics~(CFD) method, which is an extremely time-consuming process.
However, the latent relationship between the flow fields is governed by the Navier-Stokes equations and can be described by an operator. We therefore train a deep operator network, or simply DeepONet, to learn the temporal evolution between flow snapshots. Once properly trained, given a few consecutive snapshots as input, the network has a great potential to generate the next snapshot accurately and quickly. Using the output as a new input, the network iterates the process,
generating a series of successive snapshots with little wall time.
Specifically, we consider 2D flow around a circular cylinder at Reynolds number $1000$, and prepare a set of high-fidelity data using a high-order spectral/{\em hp} element method as ground truth. Although the flow fields are periodic, there are many small-scale features in the wake flow that are difficult to generate accurately. Furthermore, any discrepancy between the prediction and the ground truth for the first snapshots can easily accumulate during the iterative process, which eventually amplifies the overall deviations. Therefore, we propose two alternative techniques to improve the training of DeepONet. The first one enhances the feature extraction of the network by harnessing the "multi-head non-local block". The second one refines the network parameters by leveraging the local smooth optimization technique. Both techniques prove to be highly effective in reducing the cumulative errors and our results outperform those of the dynamic mode decomposition method. 
\end{abstract}

\maketitle
\end{CJK*}

\section{INTRODUCTION}
\label{introduction}

Computational fluid dynamics (CFD) methods require significant resources and time when dealing with complex unsteady flows. 
This can limit their applicability in several engineering applications, such as structural shape optimization~\cite{zhang2021shapeoptir1, structuralshapeoptimizationr2} and active flow control,~\cite{afcr30,zheng2021activer31, Wang2023, mo2023} where accurate solutions are time sensitive. It is therefore crucial to develop surrogate models that can accurately predict flows with low computational cost. 
To this end, several reduced-order models have been proposed as low-dimensional representations of a high-dimensional dynamical system.~\cite{ROMr40,ROMr41} This allows effects of different parameters to be explored in a cost effective manner and thus provides optimal results quickly. Two influential methods are proper orthogonal decomposition~(POD)~\cite{r5POD} and dynamic mode decomposition~(DMD).~\cite{r6DMD} 
POD takes a set of orthogonal modes containing most of the energy to represent the fluid dynamics; DMD constructs an optimally fitted linear dynamical system via singular value decomposition to approximate the time evolution of the nonlinear fluid system. 
Despite their success, Kutz~\cite{drawbackPODDMDr7} identified two limitations inherent to these methods: the inability to capture intermittent, transient or multiscale phenomena; the inability to capture invariance due to rotation, translation or scaling. 

Deep neural networks~(DNNs) can effectively capture the non-linear features in data, as confirmed by their remarkable success in the fields of natural language processing~\cite{NLPr80,NLPr81} and computer vision.~\cite{CVr90,CVr91}
Therefore, they seem to be well suited for modeling fluid dynamics and may avoid the drawbacks of the traditional surrogate models.~\cite{drawbackPODDMDr7}
To facilitate the discussion, we classify the flow surrogate models based on DNNs into three categories. The first type is called the "unsupervised model", which does not rely on flow data for training. Instead, it uses DNNs to learn the fluid dynamics by minimizing the residuals of the governing partial differential equations~(PDEs) for given boundary and initial conditions. Successful examples include physics-informed neural networks~(PINNs), deep Ritz, deep Galerkin and ODIL methods.~\cite{2019PINNsr10, zhu2024physics,song2024vwpinns, deepritzr13, Sirignano2018, karnakov2022ODILr161}  The second category is referred to as the  “weakly supervised model”. It combines a limited amount of data and PDEs to construct the loss function and generate the flow fields,~\cite{jin2021nsfnetsr18,mao2020pinnsr20,cai2021flowr21}
which has the advantage of using all available information. The third category is the “supervised model”, where the DNNs are trained by a large amount of high-fidelity data.~\cite{Brunton2018}
Advances in high performance scientific computing and experimental techniques in recent years have led to a significant accumulation of flow data.
This provides a solid basis for the construction of a supervised flow model,
which is the focus of this study.

Unsteady flows are characterized by the temporal sequence and causality of flow snapshots, which must be recognized by any surrogate model.
In natural language processing, recurrent neural networks~(RNNs) are widely used for time series forecasting. Among the most commonly used methods are the long short-term memory~(LSTM) networks,~\cite{LSTM1997r22} which have been applied to predict the evolution of flow fields. Han et al.~\cite{han2019pofr23} developed a hybrid of convolutional neural network~(CNN) and convolutional LSTM networks for the flow generation around a cylinder and forcasted the flow evolution in the near future. 
Nakamura et al.~\cite{nakamura2021r26} constructed a surrogate model based on LSTM and CNN autoencoders to capture the temporal evolution of turbulent channel flow. 
Wu et al.~\cite{wu2021pofr27} employed a combination of dilated convolutional encoder, LSTM, deconvolution decoder and "non-local block" based on self-attention mechanism to model the 2D flow around a cylinder and accurately extrapolate $400$ snapshots of the flow fields. Zhang et al.~\cite{zhang2022CMAMEr28} reconstructed a complete flow field using sparse data, where they established a mapping between the latter and the mode coefficients of "sparsity promoting DMD" through a feed-forward neural network~(FNN). They then used the LSTM to capture the temporal evolution of the mode coefficients of the DMD. Although the LSTM-based neural networks have shown remarkable achievements in predicting the spatio-temporal evolution of fluid dynamics, their complex architecture and suppressed parallel computation are considered as critical limitations. 
This has motivated researchers to explore alternative architectures of neural networks, and the design of operator networks is a notable endeavor of these efforts.

An operator represents a mapping between different function spaces and can therefore describe the latent relationship between flow fields at different time instants.
"Operator learning" refers to the utilization of DNNs 
to model an operator based on a set of high fidelity data. 
Deep operator network~(DeepONet),~\cite{lu2021DNOr30} Fourier neural operator~(FNO)~\cite{li2020FNOr31} and Transformer operator~\cite{li2022transformerr32,xu2024pof} are three representative frameworks.
In particular, DeepONet was developed by Lu et al.~\cite{lu2021DNOr30} based on the "universal operator approximation theorem" proposed by Chen and Chen.~\cite{chenandchen1995r33} It provides a succinct and universal framework for approximating different operators, while facilitating the encoding of spatio-temporal coordinates. Extensive work has been carried out based on DeepONet. Lin et al.~\cite{lin2021JFMr35} developed a surrogate based on DeepONet to accurately model the evolution of a bubble growth at different spatio-temporal scales. Oommen et al.~\cite{oommen2022operatorr36} applied the DeepONet to generate the evolution of a two-phase mixture using an image-based representation for the flow field and a CNN-Encoder to extract the flow characteristics. Mao et al.~\cite{mao2021deepmr37} developed the DeepMMNet framework that employs a combination of DeepONet structures to accurately predict the velocity field, temperature field and five gas compositions in the downstream of a shock wave. Wang et al.~\cite{wang2023longphysicDNO} solved time-dependent PDEs with long-time integration based on the physics-informed DeepONet. Michalowska et al.~\cite{michalowska2023r38} integrated DeepONet and FNO with RNNs to improve the prediction accuracy of PDEs to certain extent.

\begin{figure*}[!htb] 
 \centering  
  \subfigure[]{
  \label{originUMSE}
  \includegraphics[scale=0.255]{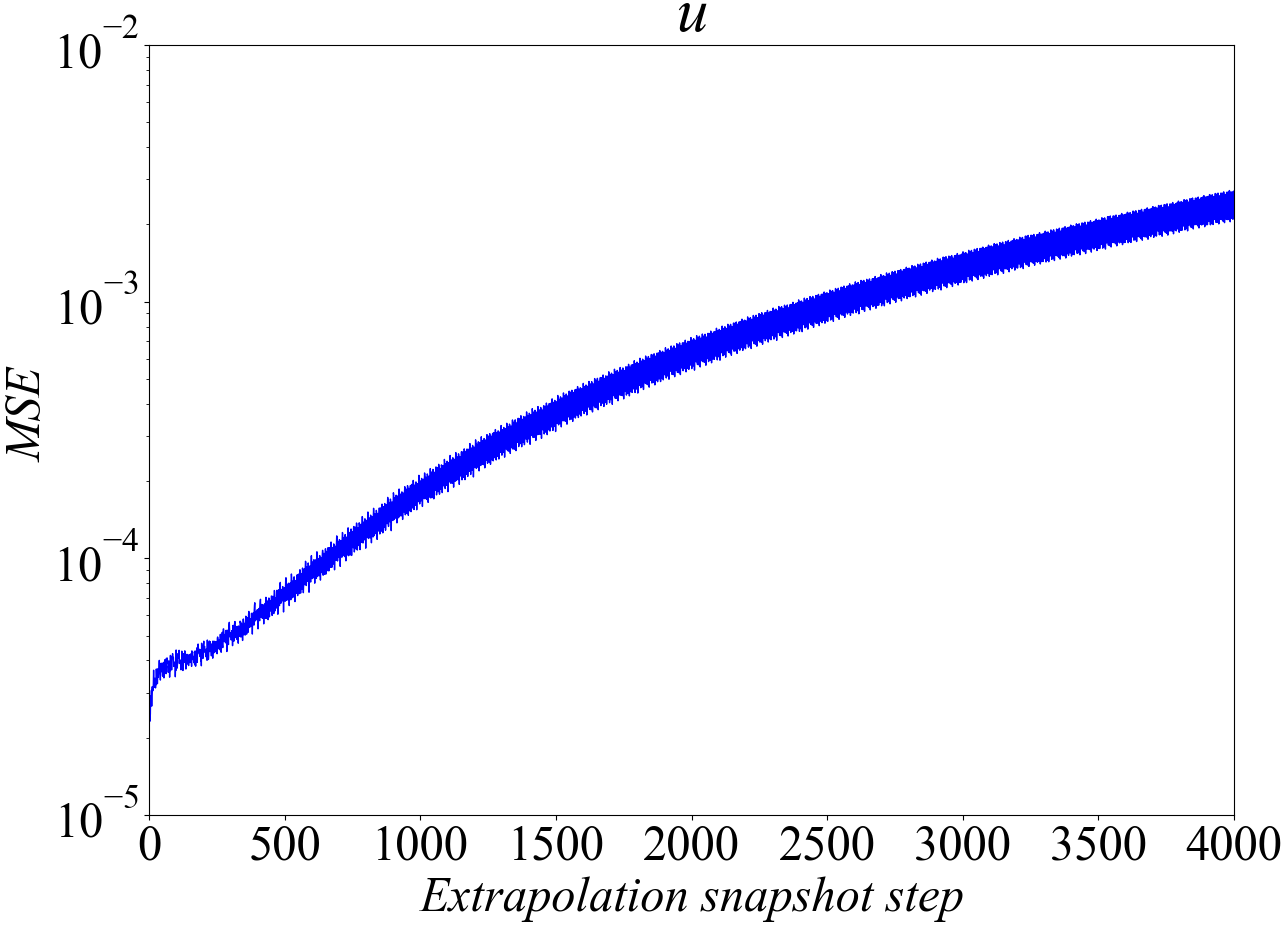}}
  \subfigure[]{
  \label{originVMSE}
  \includegraphics[scale=0.255]{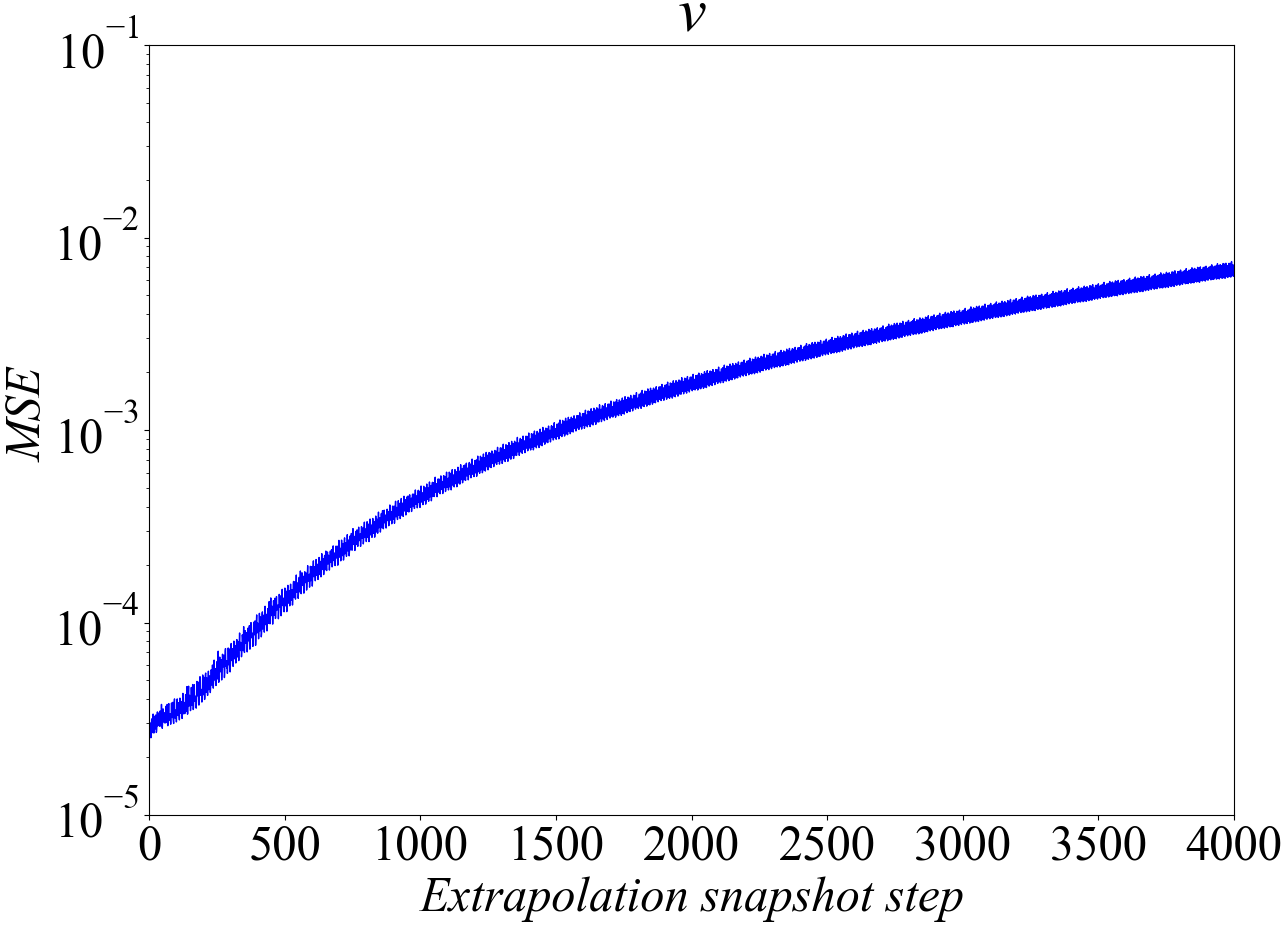}}
  \subfigure[]{
  \label{originPMSE}
  \includegraphics[scale=0.255]{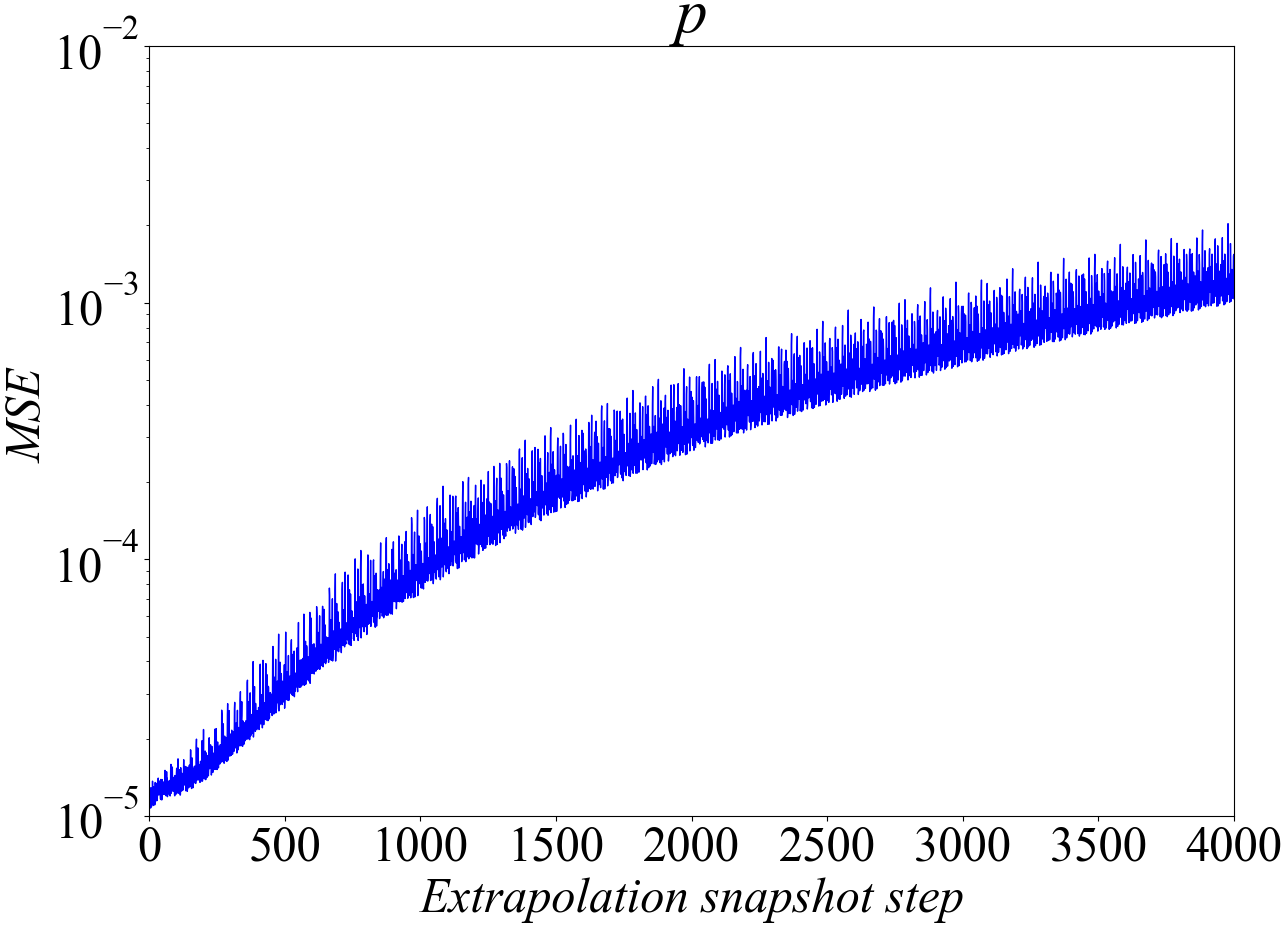}}
 \caption{The mean squared errors~(MSEs) between the DNS data and the predictions of a plain DeepONet over $4000$ extrapolation snapshots.
 Time interval between adjacent snapshots is $\Delta t/\delta t=200$ times of the CFD time step.
 (a) $u_{MSE}$: error for $x$-component of the velocity. (b) $v_{MSE}$: error for $y$-component of the velocity. (c) $p_{MSE}$: error for the pressure.  Given the first $10$ consecutive snapshots, the DeepOnet generates the $11th$ snapshot. When the $2-11$ snapshots are available, the DeepOnet generates the $12th$ one. This procedure iterates itself until $4000$ generated snapshots are available.}
 \label{MSE of origin DeepONet}
\end{figure*}

The accurate prediction of long-term sequence presents a critical challenge in the application of DNN-based models to fluid dynamics.
We take a 2D unsteady flow around a circular cylinder at Reynolds number $1000$ as demonstration.
We present the mean squared errors~(MSEs) between the direct numerical simulation~(DNS) data and the predictions of a plain DeepONet over $4000$ extrapolation snapshots on Fig.~\ref{MSE of origin DeepONet}.
The technical details will be described in section~\ref{method}. 
It is evident that the errors are acceptable for a few hundreds extrapolation snapshots, but increases quickly as the prediction further advances. 
These results examplify that the plain DeepONet or other operator networks
may fail to accurately predict the evolution of an unsteady flow over a long time, even though the flow is periodic. 
The previous LSTM-based works did not account for extrapolation of snapshots over a few hundreds either.~\cite{han2019pofr23, wu2021pofr27} 
It is the aim of this study to explore the possibility of accurate prediction of flow fields over a long time.
To achieve this goal, we propose two different routes for improving the training of DeepONet: enhancing the extraction capability of flow characteristics and refining optimization of neural network parameters.

This work is organized as follows: Section \ref{dataset} presents the procedure of generating the dataset by DNS, 
which is followed by an overview of the surrogate model based on DeepONet including the novel neural network architectures and training algorithms in Section \ref{method}. In Section \ref{result}, extensive results and comprehensive discussions are provided. Finally, a summary of this work along with prospects are presented in Section \ref{conclusion}.

\section{Flow simulation configuration}
\label{dataset}

\begin{figure*}[!htb] 
 \centering  
  \subfigure[]{
  \label{domainmesh}
  \includegraphics[scale=0.37]{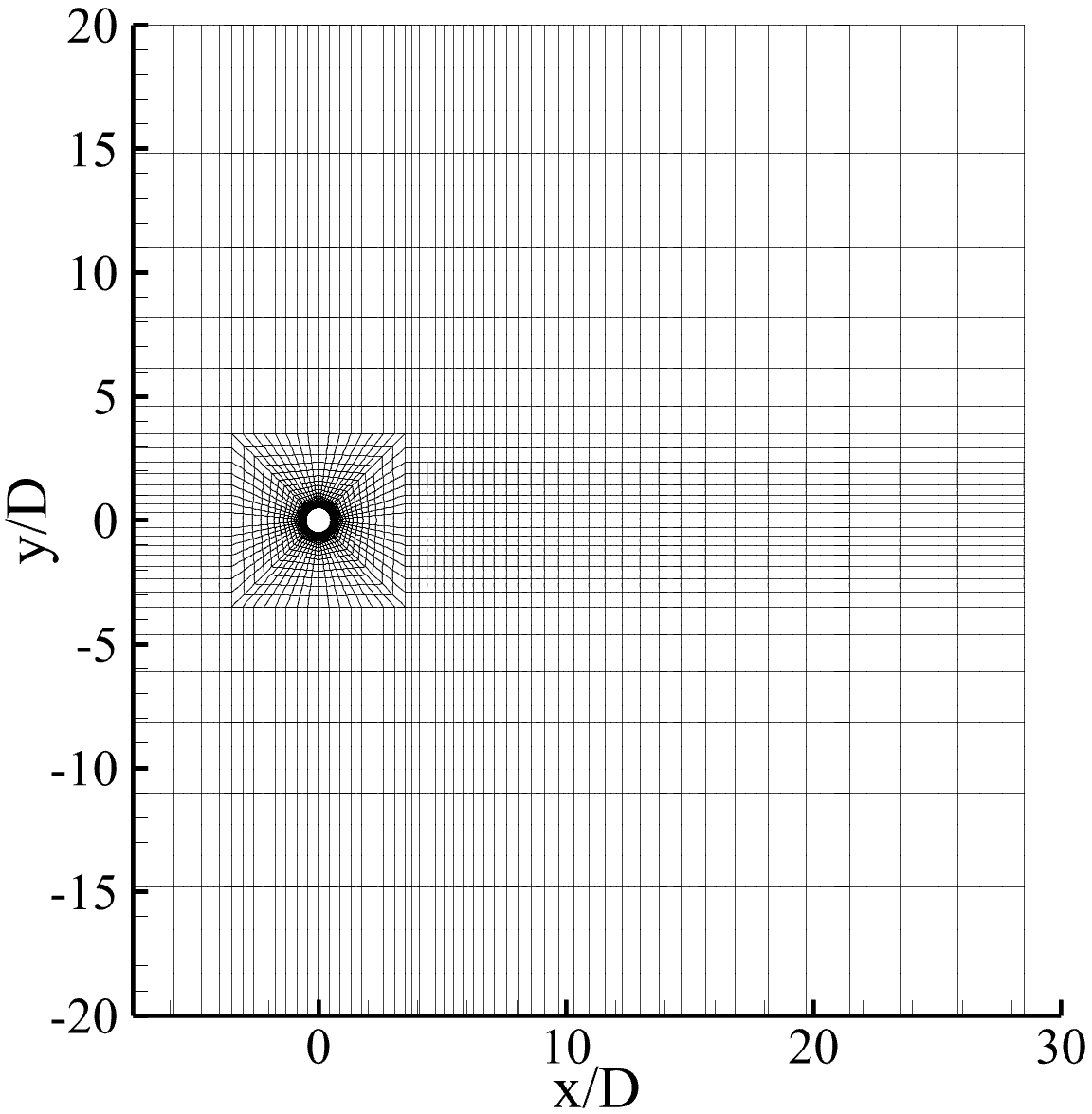}}
  \subfigure[]{
  \label{domainmeshpart}
  \includegraphics[scale=0.375]{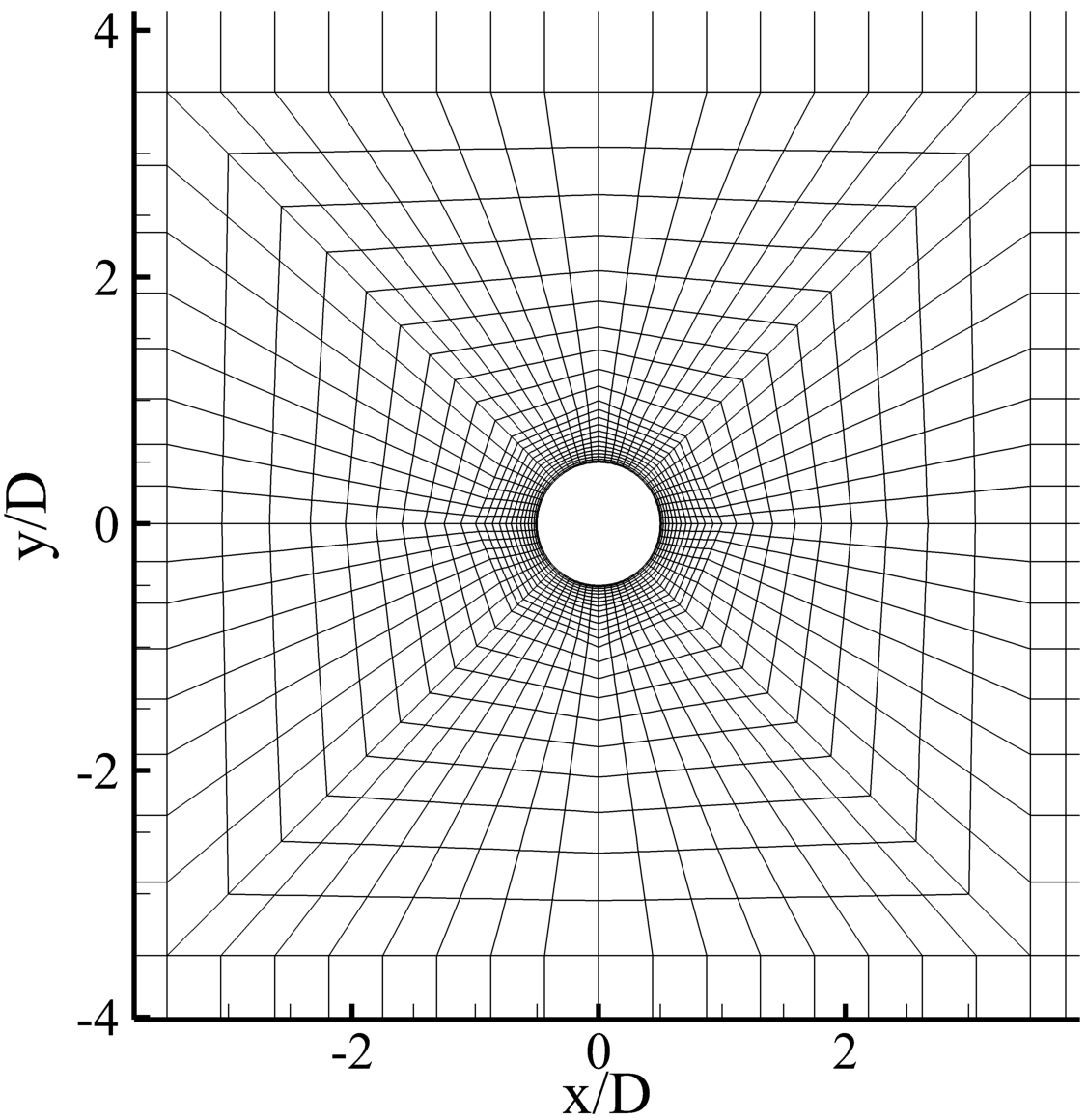}}
 \caption{Computational mesh in the x-y plane with $D$ as the cylinder's diameter. (a) Complete view. (b) Local view.}
 \label{Computational mesh}
\end{figure*}

A 2D flow around a circular cylinder is simulated by a spectral/{\em hp} element method.~\cite{GK_CFDbook} The computational domain is   $[-7.5\,D,28.5\,D]\times[-20\,D,20\, D]$ consisting of $2616$ quadrilateral elements. The cylinder has diameter $D = 1$ and is located at $(0,0)$. 
The computational mesh is shown in Fig.~\ref{Computational mesh}.

On the cylinder surface the flow has no-slip and cannot penetrate the solid. A uniform velocity ($u = U_{\infty}, v=0$) is imposed along the x-direction on the inflow boundary where $x/D=-7.5$, periodic boundary condition is applied in the y-direction at $y/D=\pm 20$, and zero-pressure boundary is prescribed on the outflow boundary at $x/D = 28.5$. The thickness of the mesh adjacent to the cylinder is $0.01\,D$. 

The governing equations are the incompressible Navier-Stokes equations
and the Reynolds number $\text{Re}=U_{\infty}D/\nu = 1\,000$, where $U_{\infty} = 1$ and kinematic viscosity $\nu=0.001$. The spectral element mode $M=3$ and there are $5$ quadrature points along each spatial direction. 
The simulation runs from scratch with time step $\delta t=0.002$.
Verification of the numerical simulation is briefly presented in Appendix~\ref{appendixA}.

The data is collected after $t\cdot U_{\infty}/D=300$ when the flow is fully developed. Thereafter, the flow fields in the focus window $[-2\,D,6\,D]\times[-2\,D,2\,D]$ at every $N_s$ time steps are cropped and linearly interpolated onto a uniform grid with $125 \times 250$ cells,
which form as the training and testing datasets.
The interpolation is necessary due to the requirement of the CNN
as discussed later.

\section{Methodology}
\label{method}
The principle and architecture of DeepONet are presented in section~\ref{DeepONetdetail},
while the procedure of modeling flow fields based on DeepONet is explained in
section~\ref{workflow}.
As demonstrated in Fig.~\ref{MSE of origin DeepONet}, 
the prediction error increases quickly as the number of extrapolation snapshots increases. Thus, we propose two alternative techniques to improve the training of DeepONet: enhancing the extraction capability of flow characteristics and refining optimization of neural network parameters. 
The corresponding algorithms are elaborated in section~\ref{MHnl} and~\ref{SAMGSAMsection}, respectively. 

\subsection{DeepONet}
\label{DeepONetdetail}

According to the universal operator approximation theorem,~\cite{chenandchen1995r33} a neural network has the capability to approximate nonlinear continuous operators. Lu et al.~\cite{lu2021DNOr30} proposed the DeepONet architecture based on this theorem and applied it to learn mappings between different function spaces in systems involving ordinary differential equations~(ODEs), PDEs and stochastic PDEs. The architecture of DeepONet consists of two components: the branch network and the trunk network. The branch network is responsible for encoding the characteristics of the input function space, which can be approximated using discrete function values at a fixed number of points ${\bf x}_i, i=1,2,3, \cdots,m$. Meanwhile, the trunk network is responsible for encoding the domains of the output function space. The outputs of the branch network and the trunk network are combined by a dot product operation to produce the final output as summarized in Eq.(~\ref{operator}):
\begin{eqnarray}
\label{operator}
 G(U)(Y) \approx \sum_{i=1}^{p}b_it_i.
\end{eqnarray}

\begin{figure}[!htb] 
 \centering  
 \includegraphics[scale=0.115]{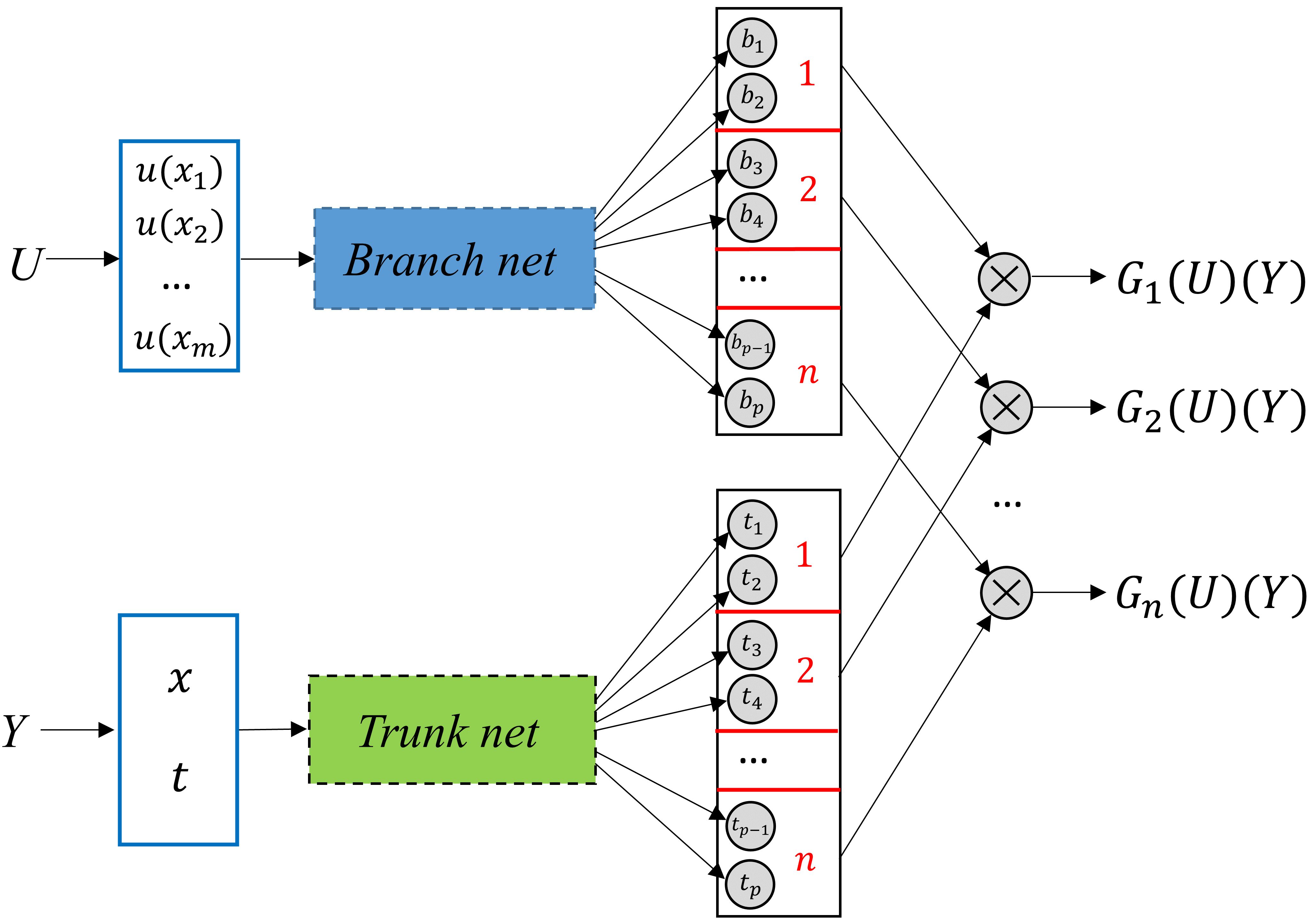}
 \caption{The schematic of unstacked DeepONet with multiple outputs. $\otimes$ represents dot product.}
 \label{unstackDeepONetmulti}
\end{figure}
We adopt the unstacked verison of DeepONet due to its superior performance and more streamlined architecture.
However, in practical applications, multiple output functions may be required. Therefore, we employ the approach suggested by Lu et al.,~\cite{lu2022comprehensiver43}
where both outputs are split into $n$ segments, with each segment generating a corresponding prediction, as shown by a schematic on Fig.~\ref{unstackDeepONetmulti}.

\begin{figure*}[htbp]
\centering
  \includegraphics[scale=0.12]{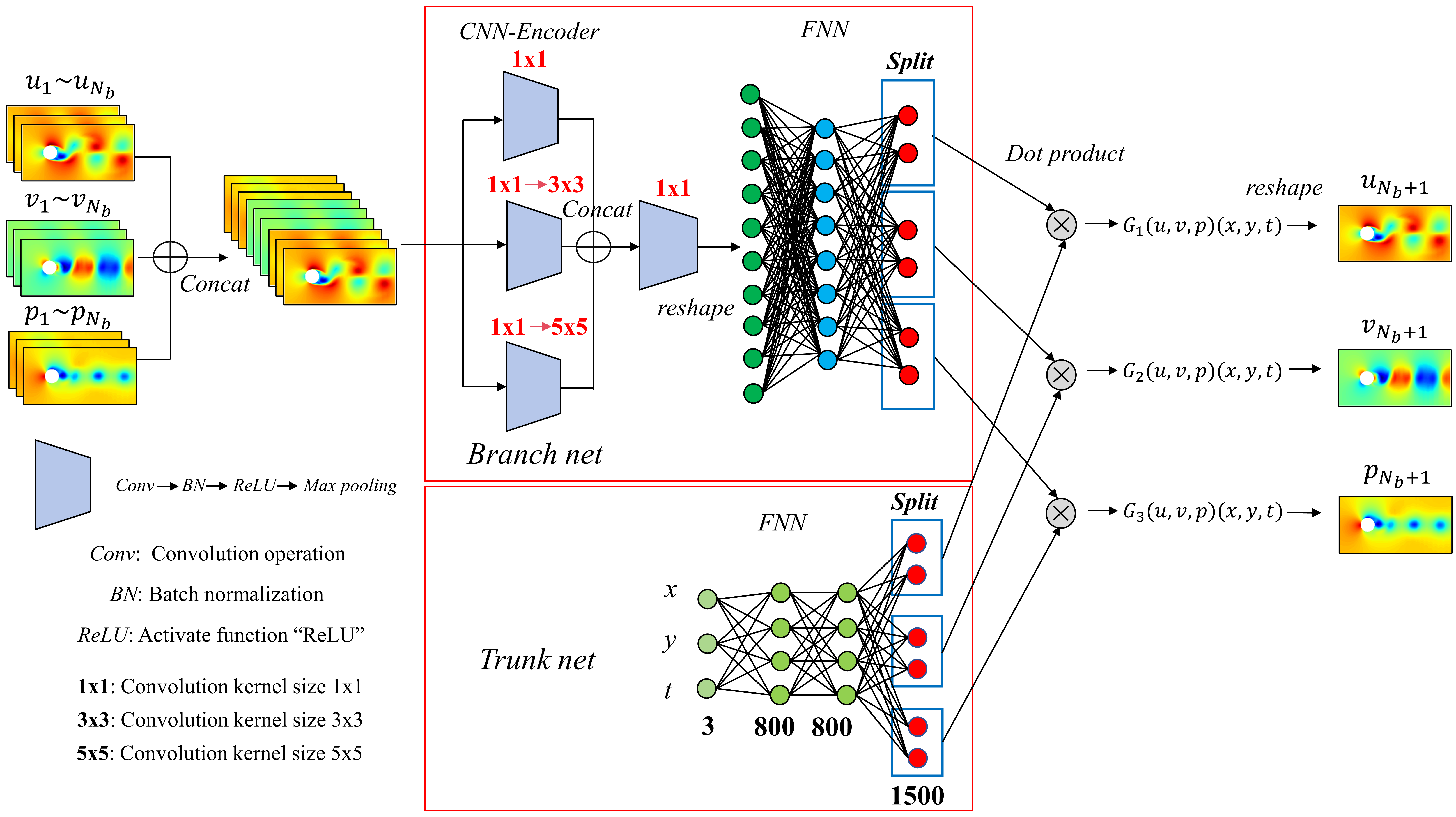}\\
  \caption{The overall architecture of DeepONet for modeling flow fields.}\label{DeepONet s model}
\end{figure*}

\begin{figure*}[htbp]
\centering
  \includegraphics[scale=0.14]{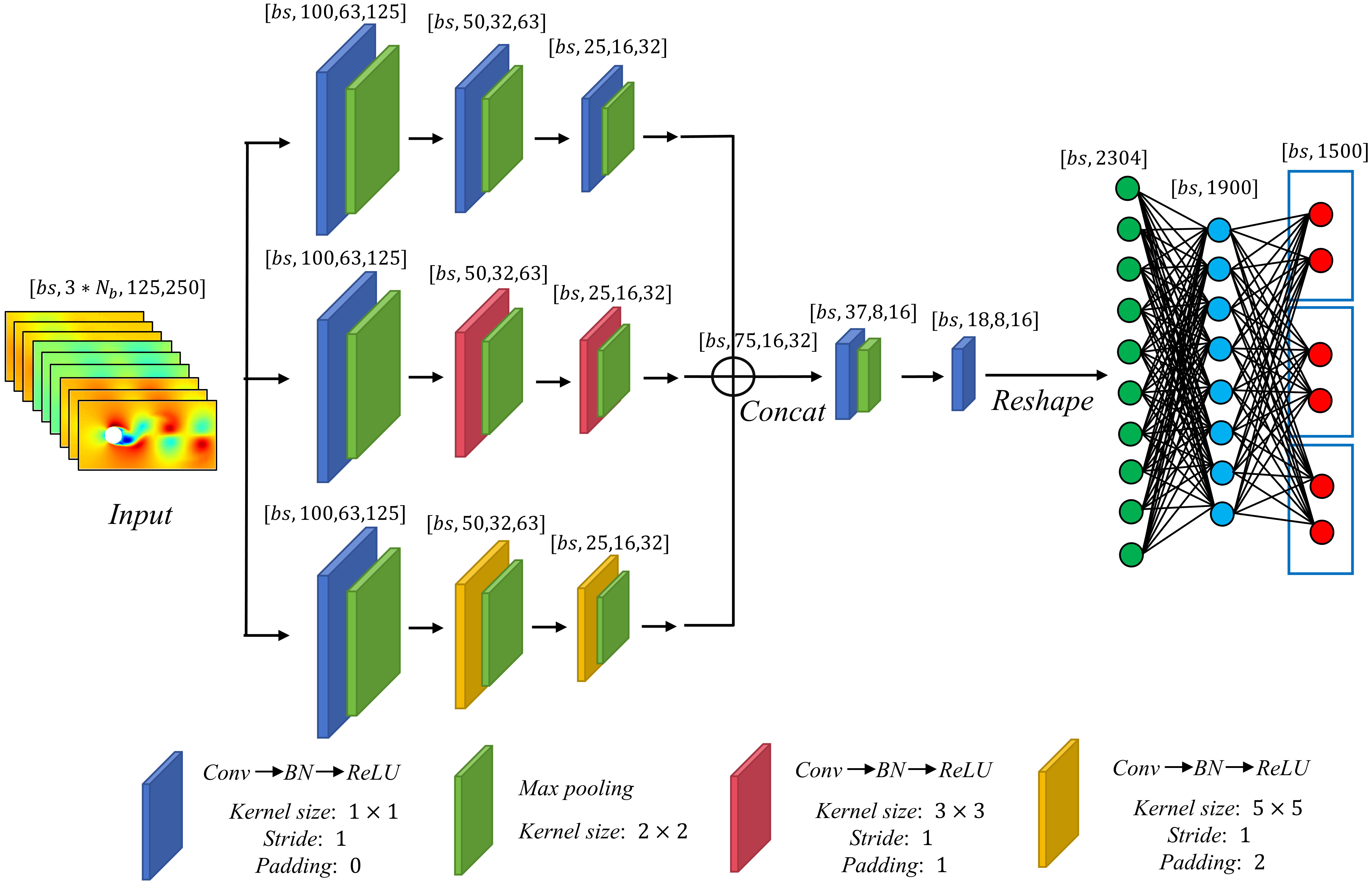}\\
  \caption{The detailed architecture and parameters of the branch network. $[bs,c,h,w]$ represents the feature map, where $bs$ is $batchsize$, $c$ is the number of channels, $h$ and $w$ are the size of the feature map in the two spatial directions.}\label{Detail DNO Origin}
\end{figure*}

In this work, CNN and FNN are baseline models for the branch network and trunk network, respectively.
Therefore, the overall architecture of DeepONet utilized for modeling flow fields is depicted in Fig.~\ref{DeepONet s model}. We refer to the inception module from GoogleNet~\cite{googlenet2015r42} as basis to construct the branch network. The flow data is initially fed into three CNN-Encoders  equipped with convolution kernels of size $1 \times 1$, $3 \times 3$ and $5 \times 5$. This design facilitates an efficient extraction of the flow features across multiple scales. The last two CNN-Encoders begin with a single convolution operation utilizing a convolution kernel of size $1 \times 1$, which aims to reduce network parameters. The outputs of three CNN-Encoders are concatenated, and then the resulting feature map is fused through a convolution operation using a $1 \times 1$ convolution kernel. 
The details of the branch network are further illustrated in Fig.~\ref{Detail DNO Origin},
where each parameter explains itself in the caption and on the figure.

\subsection{Procedure of modeling flow fields}
\label{workflow}

\begin{figure*}[htbp]
\centering
  \includegraphics[scale=0.20]{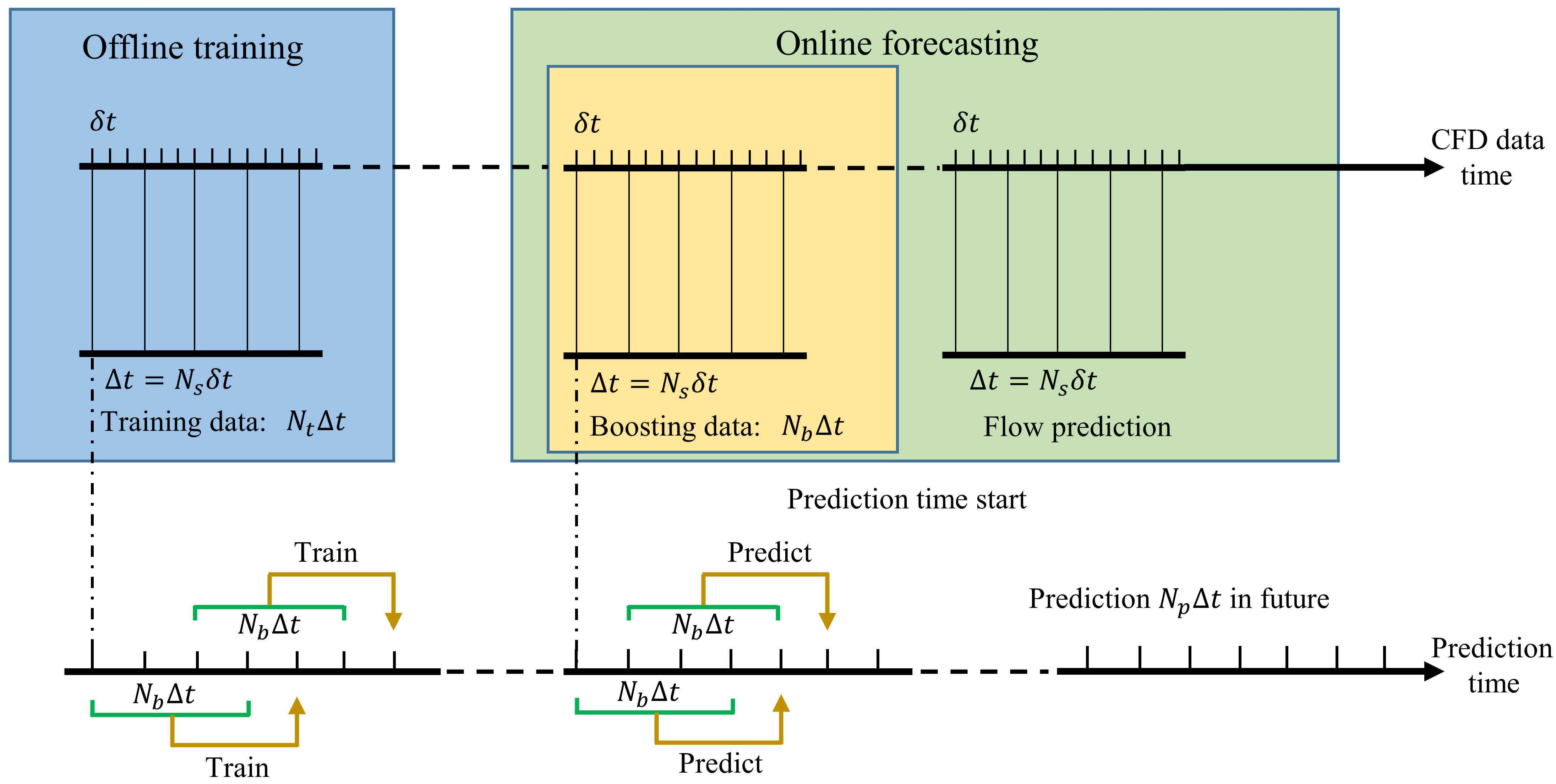}\\
  \caption{Schematic of the procedure for the offline training and online forecasting.}\label{procedure}
\end{figure*}

The procedure of modeling flow fields based on DeepONet consists of
offline training and online forecasting, as summarized in Fig.~\ref{procedure}.
Firstly, we generate a sequence of high-fidelity CFD data,
where $\delta t$ is the time step.
To reduce the total amount of data,
we select every $N_{s}$ steps so that
the time interval between selected snapshots is $\Delta t=N_{s}\delta t$.
$N_{t}=2000$ is the total number of snapshots in the training dataset. 
$N_b$ is called the boosting number,
denoting the number of the snapshots as input for DeepONet,
which always generates one subsequent snapshot.

During the offline training stage,
DeepONet employs the flow fields of streamwise velocity~($u$), transverse velocity~$(v)$ and pressure~$(p)$ from snapshot $1$ to $N_{b}$,
which are concatenated in the channel dimension and fed into the branch network. 
The trunk network takes the coordinates of the flow field and new time~$(N_{b}+1) \cdot \Delta t$ as input.
Moreover, the output tensor of DeepONet are reshaped to obtain the snapshot of the flow field at time $(N_{b}+1) \cdot \Delta t$.
The generated flow field at snapshot $(N_{b}+1)$
may be different from the true $(N_b+1)th$ snapshot of the CFD data,
and the difference contributes to the loss function.
Furthermore, DeepONet employs the $2 -(N_b+1)$ true snapshots to generate the $(N_b+2)th$ snapshot,
which may be different from the true $(N_b+2)th$ snapshot.
This difference further adds to the loss function. 
This process goes on until it uses all $N_t$ snapshots of the training dataset.

According to the aforementioned procedure, 
we employ the concept of “random mini-batch”~\cite{wandel2022spliner16} and
deploy DeepONet to absorb consecutive true snapshots of length $N_b$ of $batchsize$~($bs$) groups from the training dataset to generate the corresponding output snapshots. Subsequently, the total loss function is formulated by summarizing the disparity between these output snapshots and their respective true counterparts.

During the online forecasting phase, the flow fields of any $1-N_b$ consecutive snapshots in the testing dataset are fed into the trained DeepONet as boosting data. DeepONet then generates the flow fields at snapshot $N_b+1$.
Thereafter, the input data shift to be the $2-(N_b+1)$ snapshots to generate the $(N_b+2)th$ snapshot. This process iterates and enables the generation of flow fields for a required time horizon of $N_{p}$ snapshots.

We will not explore all the possibilities of $N_t$,
but will present prediction errors over different $N_p$ and
examine effects of different values of $N_s$ and $N_b$ on the results. 

\subsection{Multi-head non-local block}
\label{MHnl}

Due to the limited size of the convolution kernel, the convolution operation can only capture partial information. Although increasing the size of the convolution kernel can expand the receptive field of convolutional neurons, this approach leads to a substantial increase in the computational cost and still fails to capture long-range dependencies. Wang et al.~\cite{wang2018nonlocalr45} proposed a "non-local block" architecture inspired by the non-local means method~\cite{buades2005nonlocalalgorithmr46} in computer vision to effectively capture long-range dependencies in images. Wu et al.~\cite{wu2021pofr27} integrated the non-local block into the development of a flow surrogate model, leading to enhanced prediction performance. 

The generic non-local operation in DNNs is defined as follows:
\begin{eqnarray}
\label{non-local operation}
 y_{m} = \frac{1}{C(x)} \sum_{\substack{\forall n}}g(x_m,x_n)h(x_n),
\end{eqnarray}
where $x$ is the network input signal, and $y$ is the output signal of the same size as $x$. $m$ represents the specific location of the response to be computed, which can be in time, space or spacetime, and $n$ represents all possible locations that can be enumerated. The function $g$ represents the correlation between $m$ and all $n$. The function $h$ is the representation of the input signal at position $n$. $C(x)$ is the normalized function.

There are various options for the function $g$, and we take the “embedded gaussian version”:
\begin{eqnarray}
\label{Embedden Guassian}
 g(x_m,x_n) = e^{\theta(x_m)^\mathrm{T}\phi(x_n)},
\end{eqnarray}
where $\theta(x_{m})=W_{\theta} \ast x_{m}$ and $\phi(x_{n}) = W_{\phi} \ast x_{n}$ are doing convolution operation on the input signal with a convolution kernel of size $1 \times 1$.
The normalized function is defined as: 
\begin{eqnarray}
\label{normalized function}
  C(x) = \sum_{\substack{\forall n}}g(x_m,x_n).
\end{eqnarray}

It is worth mentioning that the output signal of the non-local operation based on embedded gaussian version precisely corresponds to the self-attention form in Vaswani et al.:~\cite{vaswani2017attentionr47}
\begin{eqnarray}
\label{attentionform}
 y = softmax[(W_{\theta} \ast x)^\mathrm{T}(W_{\phi} \ast x)]h(x).
\end{eqnarray}
The function $h$ can be defined as a convolution operation with a convolution kernel size of $1 \times 1$. The architecture of the non-local block is illustrated in Fig.~\ref{origin NLblock}:

\begin{figure}[htbp]
\centering
  \includegraphics[scale=0.13]{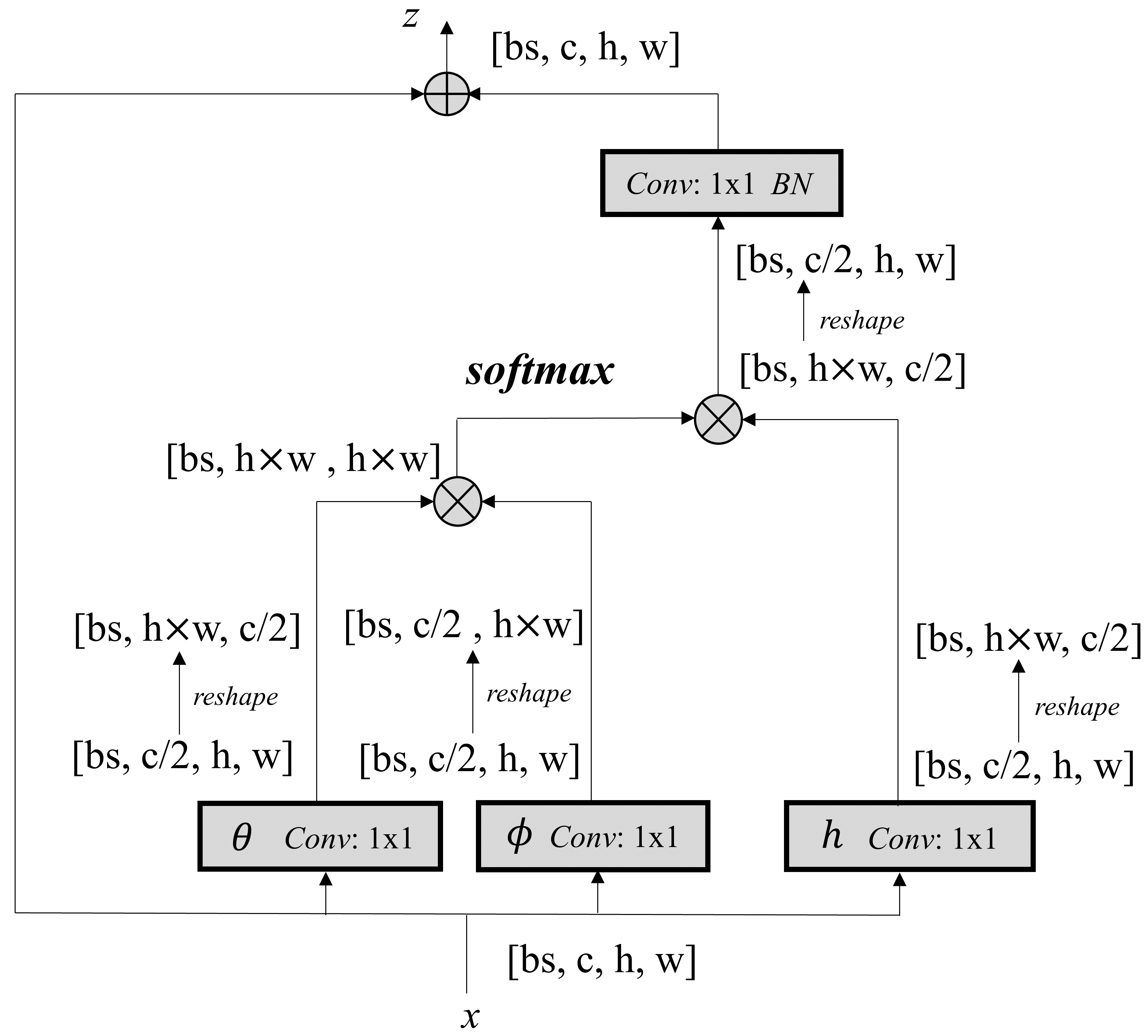}\\
  \caption{The architecture of non-local block. $\oplus$ represents element-wise sum, $\otimes$ represents matrix multiplication.}\label{origin NLblock}
\end{figure}

The output signal of the non-local block can be formulated as follows
\begin{eqnarray}
\label{E1}
 z = W_z \ast y + x,
\end{eqnarray}
\noindent where $y$ can be computed by Eq.~(\ref{attentionform}), $"\ast"$ represents a convolution operation on $y$ with a convolution kernel of size $1 \times 1$, and $"+x"$ represents a residual connection.~\cite{he2016residualconnection}

\begin{figure*}[htbp]
\centering
  \includegraphics[scale=0.125]{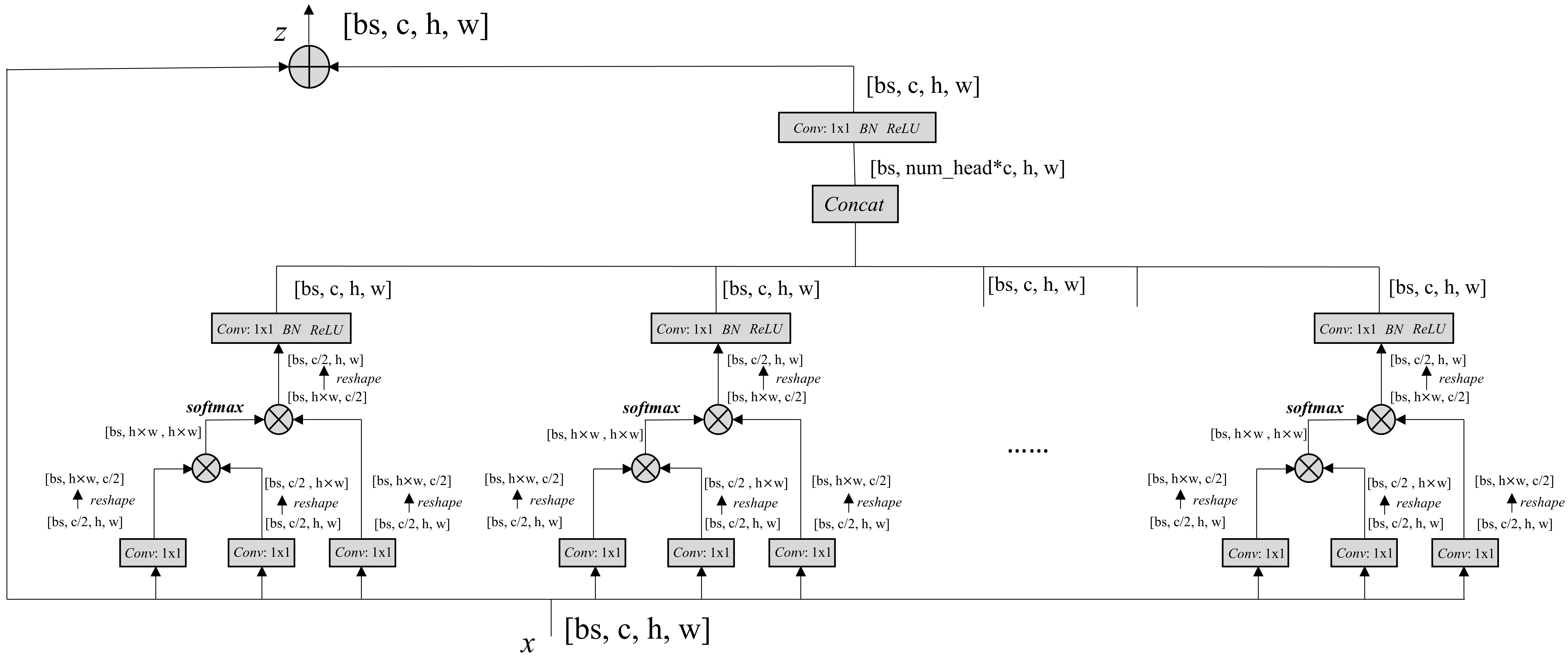}\\
  \caption{The architecture of a multi-head non-local block.}\label{MH NLblock}
\end{figure*}

The "multi-head attention mechanism" is a fundamental component of the transformer architecture.~\cite{vaswani2017attentionr47} Park et al.~\cite{park2022visiontfworkr48} pointed out that the multi-head self-attention mechanism employed in the “vision transformer”~\cite{visiontransformerr49}  architecture for computer vision enhances the generalization ability and performance of the neural network when sufficient training data is available. 
Inspired by this concept, we extend the classical non-local block to 
a "multi-head" non-local block. The fundamental idea is to apply multiple non-local operations to the input signal $x$ for attention computation, followed by concatenating the outputs of each non-local operation and conducting information fusion. The architecture of the multi-head non-local block is depicted in Fig.~\ref{MH NLblock}, where the number of non-local operations corresponds to the number of heads.

\begin{figure*}[htbp]
\centering
  \center{}
  \includegraphics[scale=0.12]{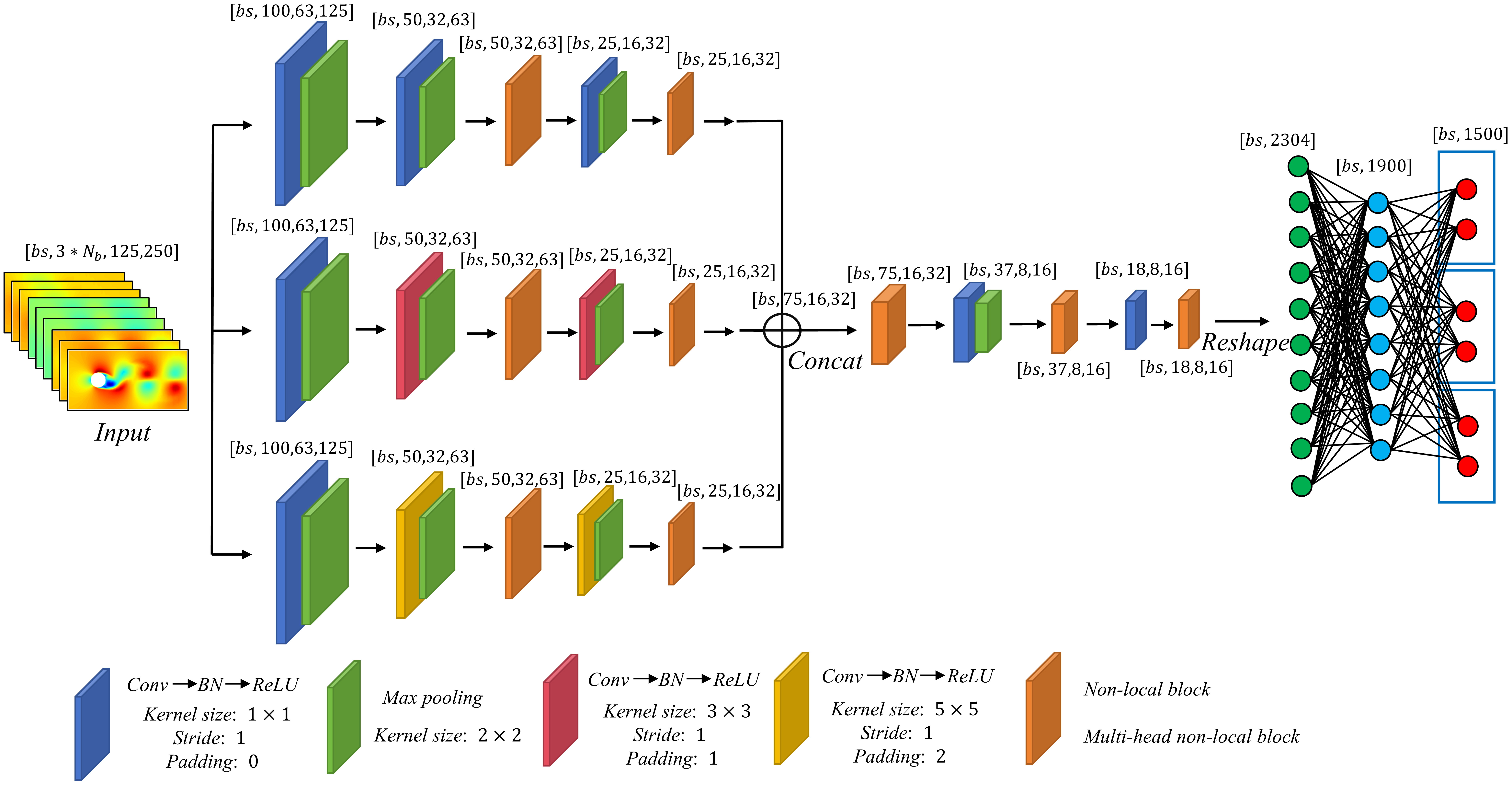}\\
  \caption{Details of the branch network integrating non-local/multi-head non-local block.}\label{Detail of MH branch}
\end{figure*}

The multi-head non-local block can be seamlessly incorporated into the CNN-based neural network architecture as an auxiliary module, requiring no modifications to the original network. Based on the neural network architecture shown in Fig.~\ref{DeepONet s model}, the multi-head non-local block is incorporated into the deeper layer of the branch network. Details of the enhanced branch network are depicted in Fig.~\ref{Detail of MH branch}.

\subsection{Local smooth optimization technique of the loss landscape}
\label{SAMGSAMsection}

Complex neural network frequently demonstrates intricate loss landscape, which can make the network more prone to poor generalization.~\cite{keskar2016larger530,dziugaite2017computingr531,jiang2019fantasticr532} A technique to enhance the smoothness of the loss landscape of the neural network, referred to as “sharpness-aware minimization~(SAM)”,~\cite{foret2020SAMr50} is introduced here first.
Its fundamental concept is to address the following optimization problem

\begin{eqnarray}
\begin{aligned}
\label{SAM method}
 &\min_{\omega}\{[\max_{\Vert \epsilon \Vert_2 \leq \rho} L_S(\omega + \epsilon) - L_S(\omega)] + L_S(\omega)\}\\
 &=\min_{\omega}[\max_{\Vert \epsilon \Vert_2 \leq \rho} L_S(\omega + \epsilon)] \triangleq \min_{\omega}L_S^{SAM}(\omega),
\end{aligned}
\end{eqnarray}
where $L_s$ represents the loss function and $S$ is the dataset.
$[\max_{\Vert \epsilon \Vert_2 \leq \rho} L_S(\omega + \epsilon) - L_S(\omega)]$ represents the "sharpness" of the loss function with neural network parameter $\omega$, meaning the maximum change of loss function when a perturbation $\epsilon$ is added to $\omega$. 
The maximum perturbation magnitude is $\rho$,
which is a hyper-parameter.
By solving Eq.~(\ref{SAM method}), the loss landscape of the neural network is smoothed, resulting in an enhancement of its generalization capability. The SAM algorithm can be seamlessly integrated with any neural network parameter optimizer. In this work, automatic differentiation and Adam algorithm~\cite{kingma2014adamr44} are used to solve the $\nabla_{\omega}L_S^{SAM}(\omega) $ and update the network parameters, respectively.

Following the above procedure, the initial step involves finding an approximate solution for the inner maximization problem:
\begin{eqnarray}
\begin{aligned}
\label{inner max}
 \hat{\epsilon}(\omega) &=\mathop{\mathrm{arg\,max}}\limits_{\Vert \epsilon \Vert_{2} \leq \rho}{L_{S}(\omega + \epsilon)}\approx\mathop{\mathrm{arg\,max}}\limits_{\Vert \epsilon \Vert_{2} \leq \rho}[L_S(\omega) + \epsilon^\mathrm{T}\nabla_{\omega}L_S(\omega)]\\
 &=\mathop{\mathrm{arg\,max}}\limits_{\Vert \epsilon \Vert_{2} \leq \rho}\epsilon^\mathrm{T}\nabla_{\omega}L_S(\omega).
\end{aligned}
\end{eqnarray}
By solving a classical dual norm problem, the value of  $\epsilon(\omega)$ can be obtained:
\begin{eqnarray}
\label{ew1}
 \hat{\epsilon}(\omega) = \rho\, sign(\nabla_{\omega}L_S(\omega)) \frac{\vert \nabla_{\omega}L_S(\omega) \vert^{q-1}}{(\Vert \nabla_{\omega}L_S(\omega) \Vert_q^{q})^{\frac{1}{p}}},
\end{eqnarray}
where $1/p+1/q=1$, and $p$ is taken to be $2$ as in Foret et al~\cite{foret2020SAMr50}. So $\hat{\epsilon}(\omega)$ is simplified as follows
\begin{eqnarray}
\label{ew2}
 \hat{\epsilon}(\omega) = \rho\,\frac{\nabla_{\omega}L_S(\omega)}{\Vert \nabla_{\omega}L_S(\omega) \Vert_2}.
\end{eqnarray}

\begin{figure}[htbp]
\centering
  \center{}
  \includegraphics[scale=0.15]{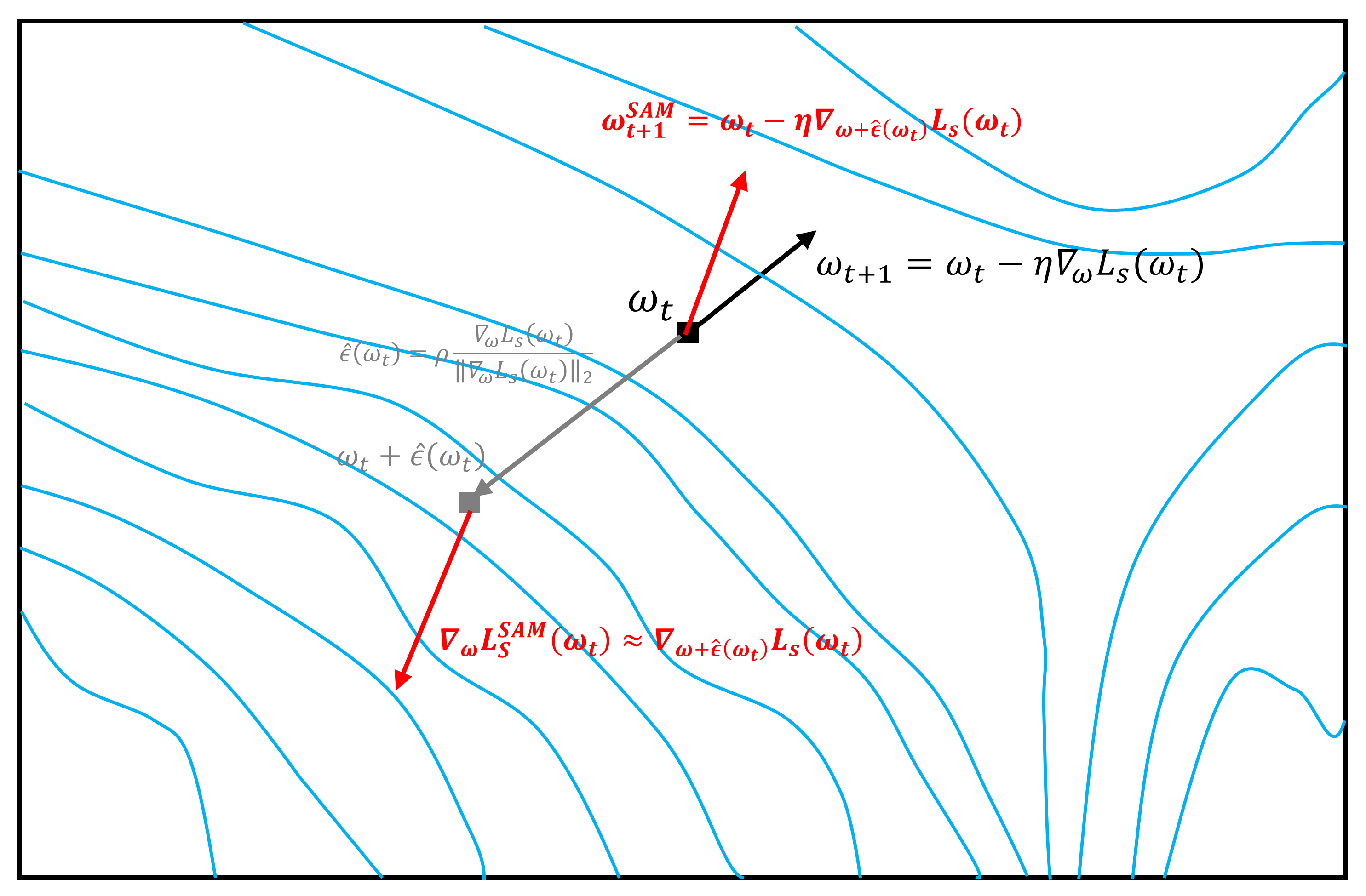}\\
  \caption{Procedure of parameter update in the SAM algorithm.}\label{SAM para update}
\end{figure}

The gradient approximation and neural network parameter updating process are shown below:
\begin{eqnarray}
\label{finallygradappro}
 \nabla_{\omega}L_S^{SAM}(\omega) &\approx& \nabla_{\omega+\hat{\epsilon}(\omega)}L_S(\omega),
\\
\label{finallyparaupdateSAM}
 \omega_{t+1}^{SAM} &=& \omega_{t} -\eta\nabla_{\omega+\hat{\epsilon}(\omega_t)}L_S(\omega_t).
\end{eqnarray}
The procedure of the SAM algorithm is depicted in Fig.~\ref{SAM para update}.

However, the SAM algorithm has an inherent limitation, as it cannot guarantee a low level of sharpness. To mitigate the limitation, Zhuang et al.~\cite{zhuang2022GSAMr51} proposed the “surrogate gap guided sharpness-aware minimization~(GSAM)” algorithm. 

\begin{figure*}[htbp]
\centering
  \center{}
  \includegraphics[scale=0.2]{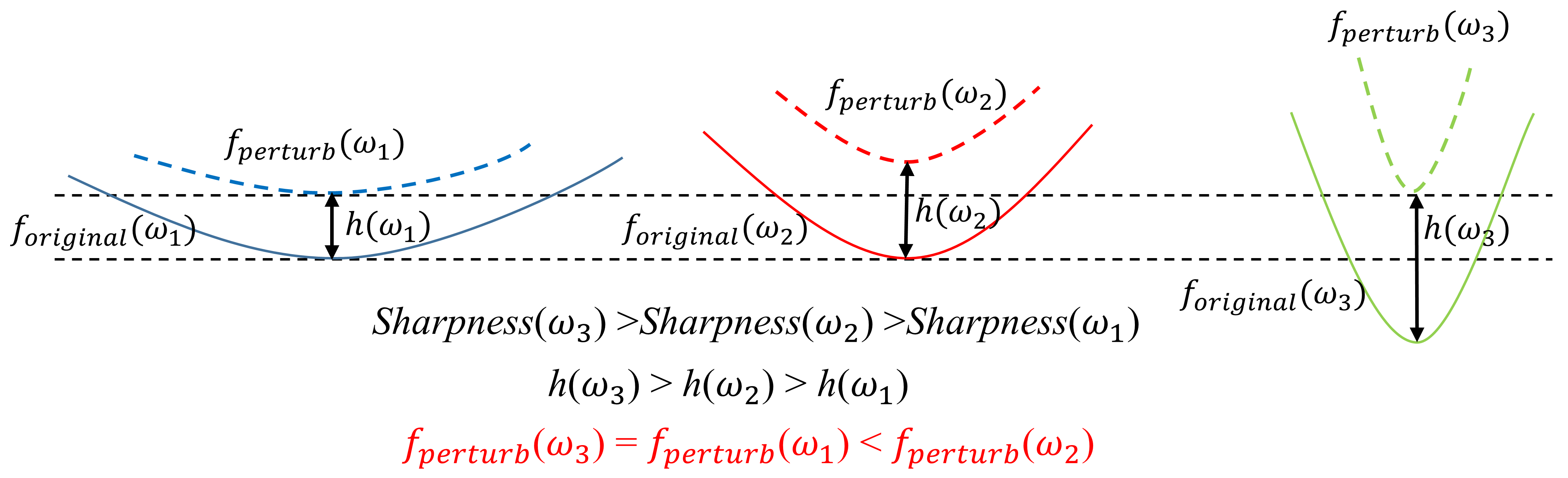}\\
  \caption{Drawbacks of the SAM algorithm.}\label{SAM drawback}
\end{figure*}
The inadequacy of the SAM algorithm is illustrated in  Fig.~\ref{SAM drawback}. The solid line depicts the value of the original loss function $f_{original}=L_S(\omega)$, while the dashed line depicts the maximum perturb loss function  $f_{perturb}=\max_{\Vert \epsilon \Vert_2 \leq \rho} L_S(\omega + \epsilon)$ within a Euclidean sphere centered on the network parameter $\omega$ with a radius of $\rho$. The term $h=f_{perturb}-f_{original}$ is referred to as the "surrogate gap", ~\cite{zhuang2022GSAMr51} which essentially represents the sharpness. The $f_{perturb}$ can be minimized by optimizing the neural network parameter $\omega$ using Eq.~(\ref{SAM method}). However, a comparison of $f_{perturb}$ between $\omega_{2}$ and $\omega_{3}$, as shown in Fig.~\ref{SAM drawback}, reveals that although $f_{perturb}$ is smaller for $\omega_{3}$, its sharpness is significantly greater than that of $\omega_{2}$. Therefore, the GSAM algorithm incorporates the surrogate gap term into Eq.~(\ref{SAM method}) to establish a novel loss function, as depicted in Eq.~(\ref{GSAMbasic}):
\begin{eqnarray}
\label{GSAMbasic}
 \min_{\omega} \{f_{perturb}+\alpha \times h\}, 
\end{eqnarray}
where $\alpha \geq 0$ is a hyperparameter. When $\alpha = 0$, GSAM degrades to SAM.

Eq.~(\ref{GSAMbasic}) guarantees the optimization of both perturb loss and sharpness towards smaller values. It is worth noting that optimizing the neural network parameters in the GSAM algorithm should not increase $f_{perturb}$. Therefore, Zhuang et al.~\cite{zhuang2022GSAMr51} employed "gradient decomposition" to separate $\nabla h$ and $\nabla f_{original}$ into two components: one parallel and one perpendicular to $\nabla f_{perturb}$:
\begin{eqnarray}
\label{hgrad}
 \nabla h &=& \nabla f_{perturb}-\nabla f_{original},
\\
\label{hdecomp}
 \nabla h &=& \nabla h_{\|} +\nabla h_{\bot}, 
\\
\label{fdecomp}
 \nabla f_{original} &=& \nabla f_{original \|} +\nabla f_{original \bot},
\\
\label{h=-f}
 \nabla h_{\bot}  &=& -\nabla f_{original \bot}.
\end{eqnarray}

\begin{figure}[htbp]
\centering
  \center{}
  \includegraphics[scale=0.15]{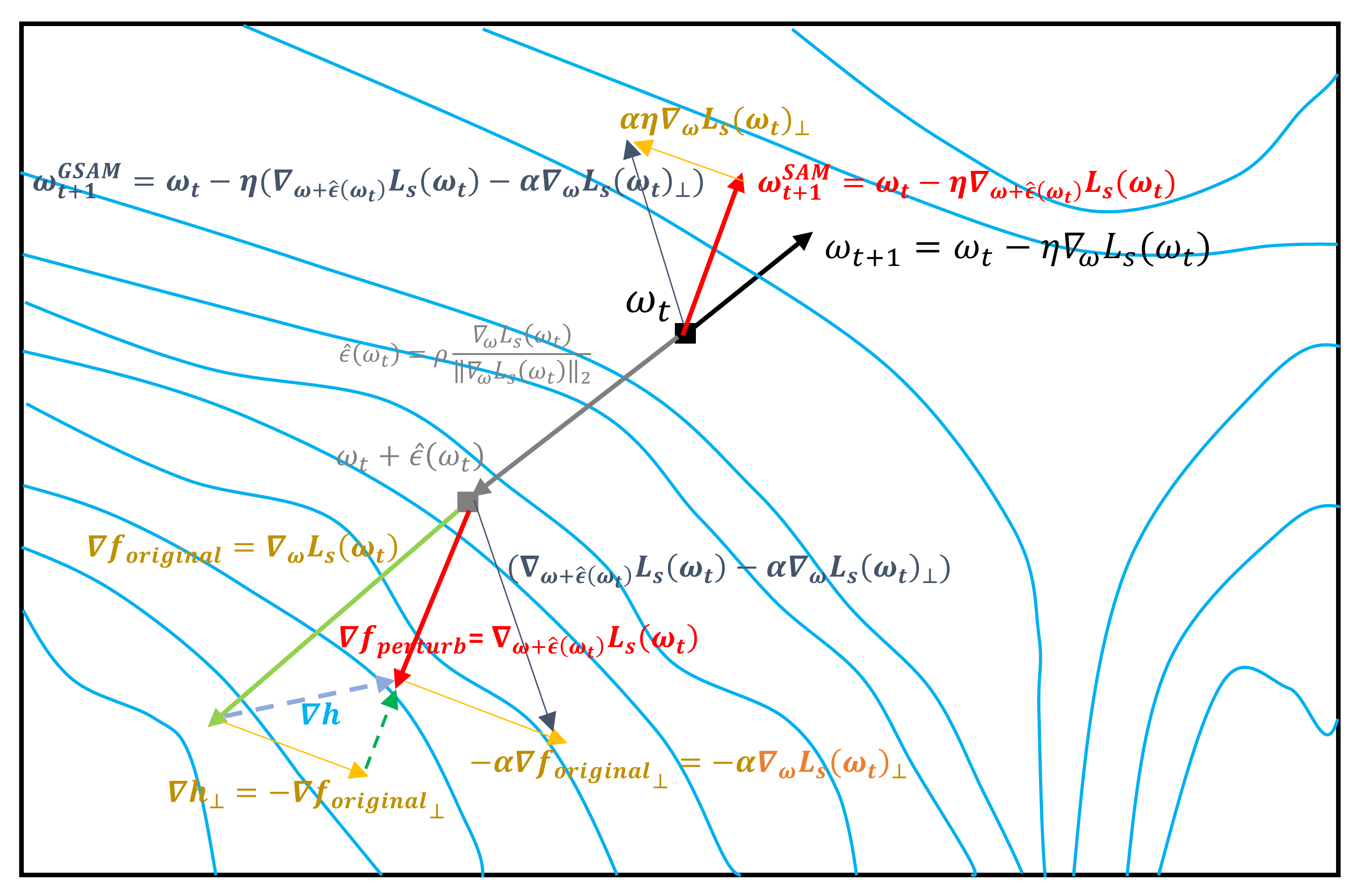}\\
  \caption{Procedure of parameter update in the GSAM algorithm.}\label{GSAM para update}
\end{figure}

Ultimately, the updating directions of the neural network parameters are determined by $\nabla h_{\bot}$ and $\nabla f_{perturb}$:
\begin{eqnarray}
\label{finallyparaupdate0GSAM}
 \nabla f_{perturb} &=& \nabla_{\omega+\hat{\epsilon}(\omega)}L_S(\omega),
\\
\label{finallyparaupdate1GSAM}
 \nabla h_{\bot} &=& -\nabla_{\omega}L_{S}(\omega)_{\bot},
\\
\label{finallyparaupdateGSAM}
 \omega_{t+1}^{GSAM} &=& \omega_{t} -\eta(\nabla_{\omega+\hat{\epsilon}(\omega_t)}L_S(\omega_t)-\alpha\nabla_{\omega}L_{S}(\omega_t)_{\bot}).
\end{eqnarray}
The procedure of GSAM algorithm is depicted in Fig.~\ref{GSAM para update}.

\section{Results and discussions}
\label{result}

In this section, the proposed DeepONet-based surrogate model is utilized for predicting the flow fields of the 2D flow around a circular cylinder at $Re = 1000$. We provide a comprehensive analysis and discussion on the results from the plain version of DeepONet, 
as well as the enhanced versions by one-head/multi-head non-local block and SAM/GSAM algorithms.

The MSE is utilized to define the loss function as
\begin{eqnarray}
\label{loss_total}
\begin{aligned}
 loss_{total}=& \frac{1}{N}\sum_{i=1}^{N}[(u_i^{True} - u_i^{Predict})^2 + (v_i^{True} - v_i^{Predict})^2 \\
 &+ (p_i^{True} - p_i^{Predict})^2], 
\end{aligned}
\end{eqnarray}
where $N$ represents the total number of grid points. 
The concept of random mini-batch is employed, with a $batchsize$~($bs$) of $20$ and each epoch comprises $1000$ iterations. The neural network parameters are updated using the Adam algorithm or a combination of the Adam algorithm and local smooth optimization technique~(when SAM/GSAM is activated). The initial learning rate~($lr$) is set to $0.0001$, and a $cosine$ $lr$ schedule is utilized. The neural network is trained for $1000$ epochs. The proposed neural network architecture is implemented using the deep learning framework Pytorch and trained on an NVIDIA GeForce RTX 4090.

In this work, $N_{s} = 100~(\Delta t=0.2), 200~(\Delta t=0.4)$ and $300~(\Delta t=0.6)$ are selected. 
The detailed discussion of the results  is about $N_{s} = 200$, 
while results for $N_{s} = 100$ and $300$ are presented in Appendix~\ref{appendixC} and Appendix~\ref{appendixD}, respectively.

\subsection{Vanilla DeepONet}
\label{originDNO}

A plain version of DeepONet as illustrated in Fig.~\ref{DeepONet s model} is trained using the high-fidelity DNS data. The specific configuration of the network, as well as the procedure for offline training and online forecasting are provided in Section \ref{DeepONetdetail} and \ref{workflow}. The network parameters are updated exclusively using the Adam algorithm.

\begin{figure*}[htbp] 
 \centering  
  \subfigure[]{
  \label{UDNSorgin}
  \includegraphics[scale=0.178]{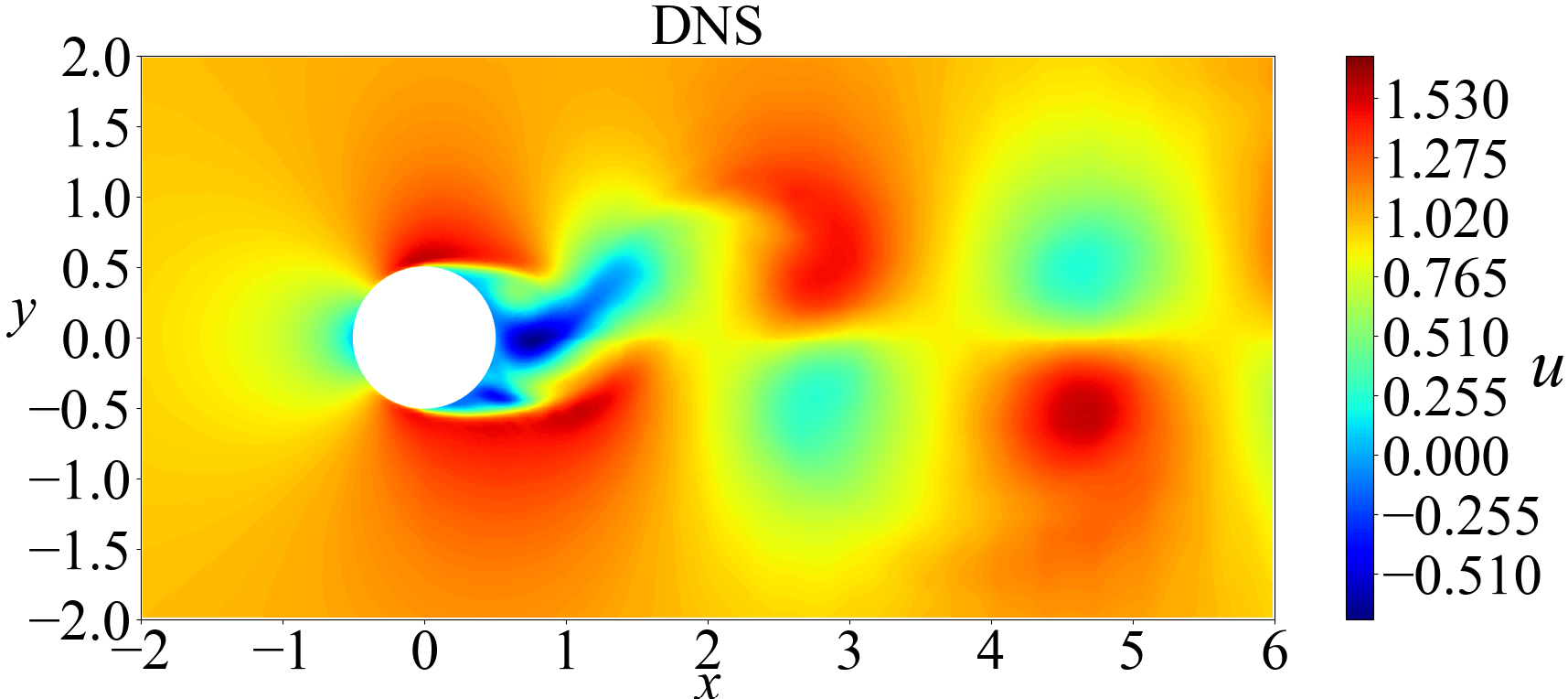}}
  \subfigure[]{
  \label{Upreorigin}
  \includegraphics[scale=0.178]{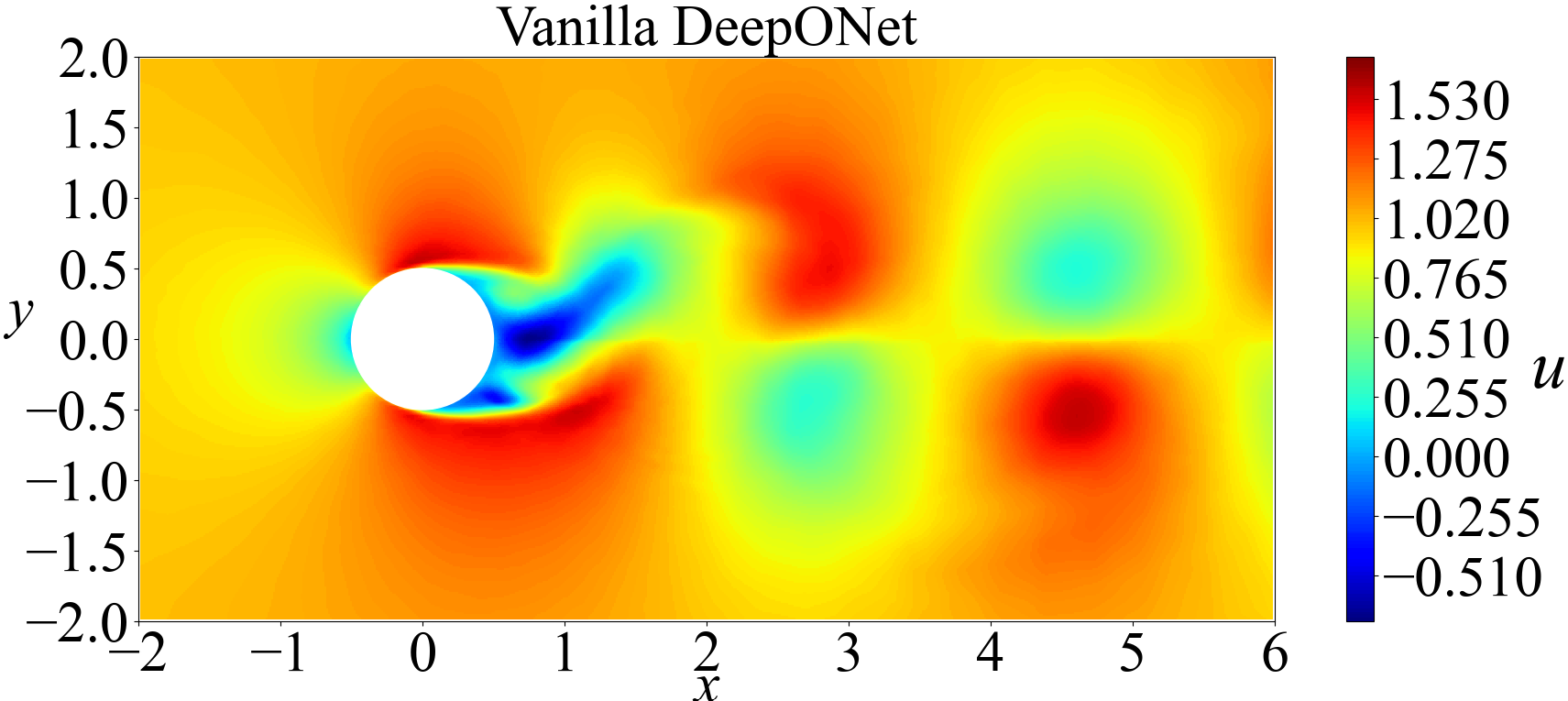}}
  \subfigure[]{
  \label{Udeltaorigin}
  \includegraphics[scale=0.178]{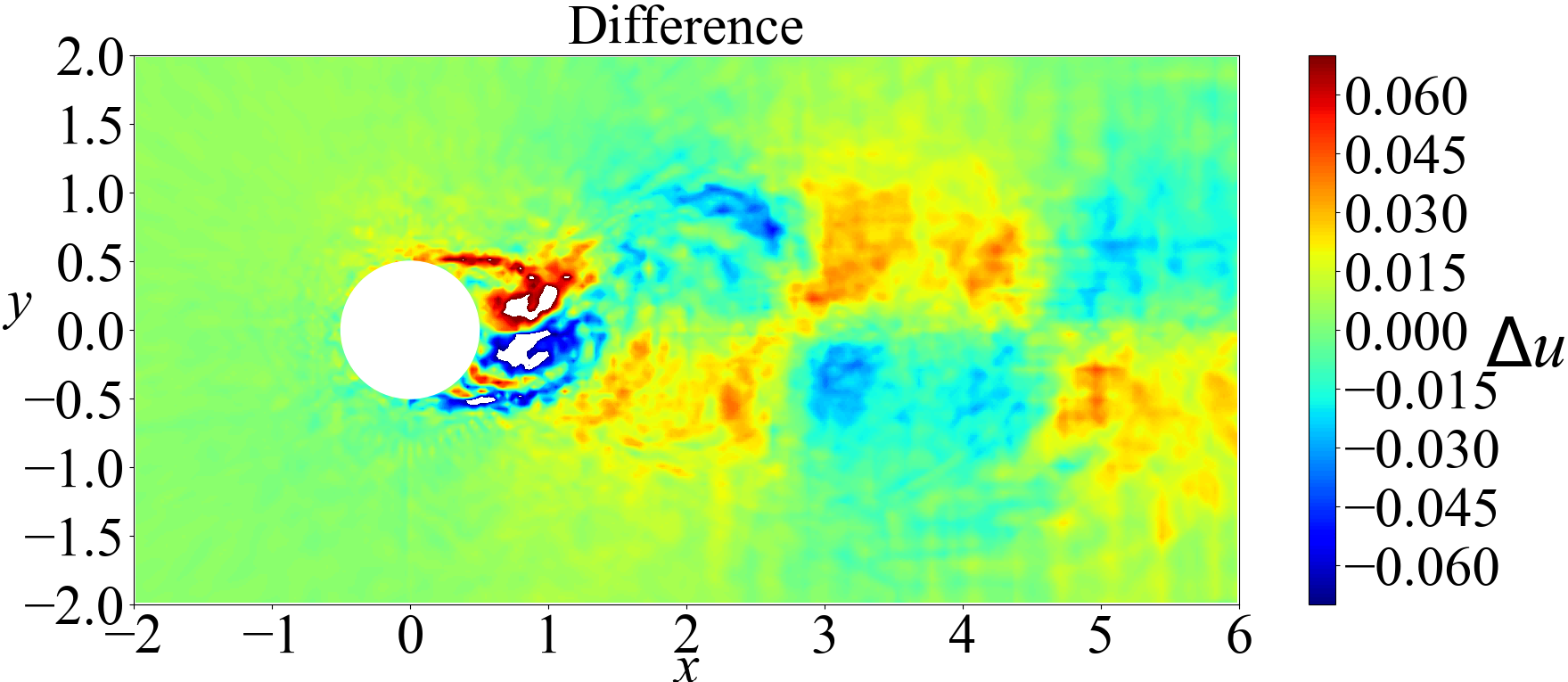}}
  \subfigure[]{
  \label{VDNSorigin}
  \includegraphics[scale=0.178]{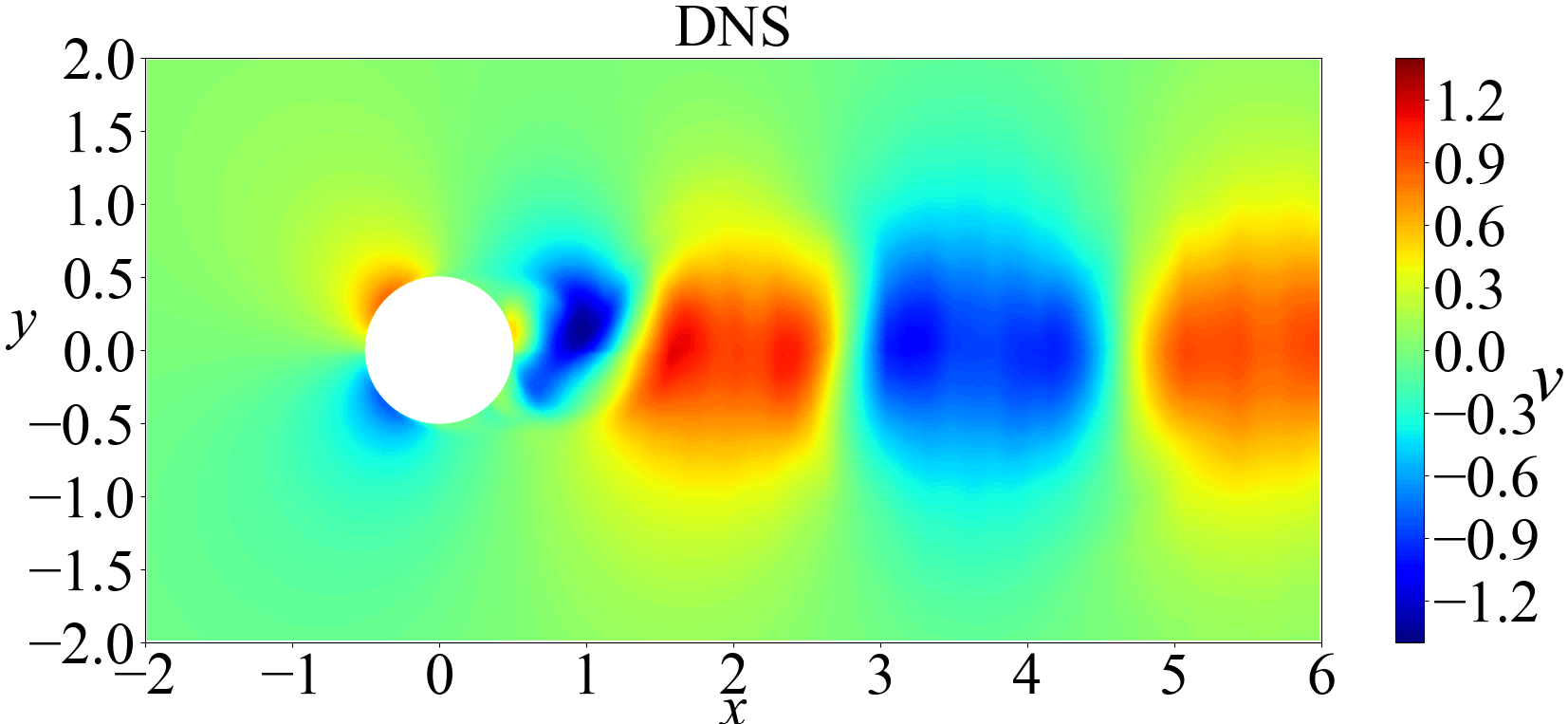}}
  \subfigure[]{
  \label{Vpreorigin}
  \includegraphics[scale=0.178]{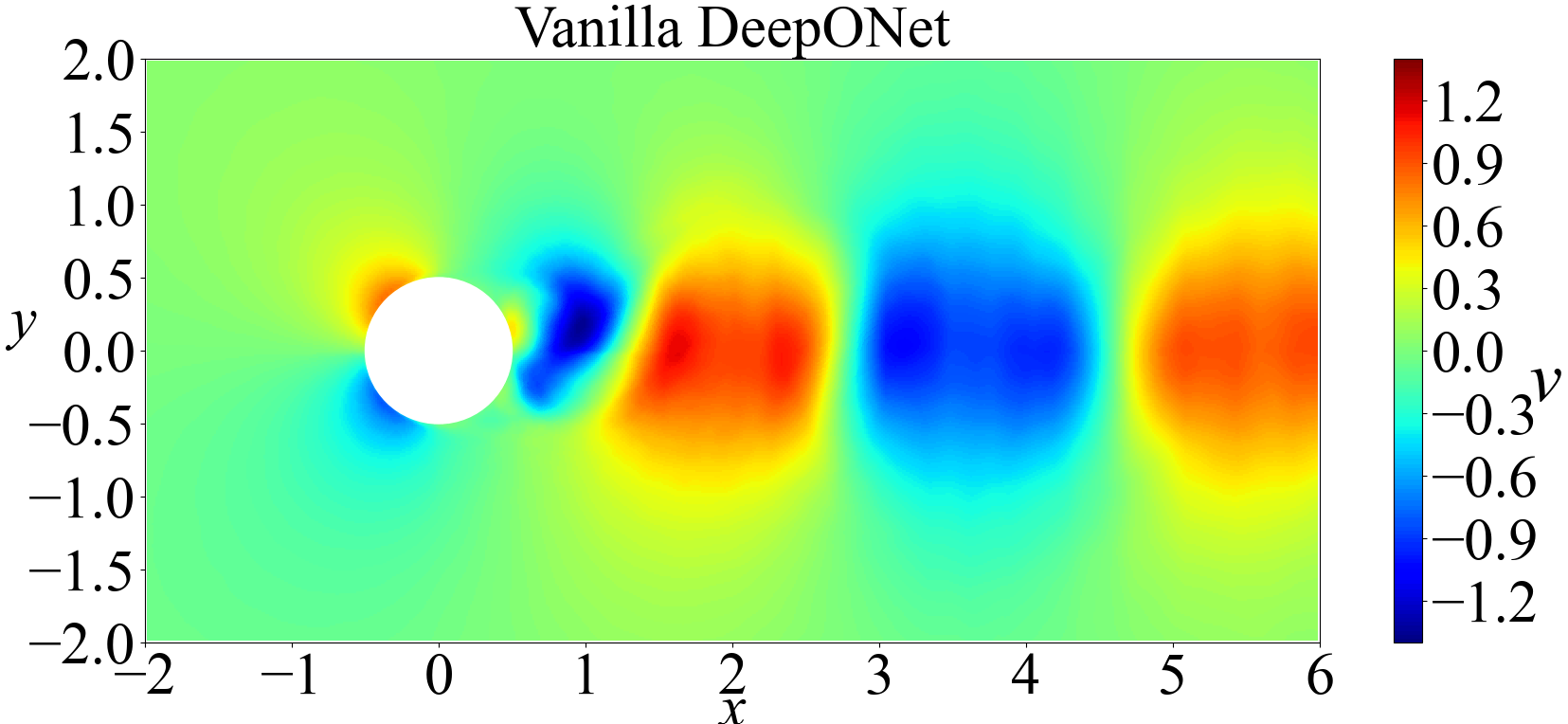}}
  \subfigure[]{
  \label{Vdeltaorigin}
  \includegraphics[scale=0.178]{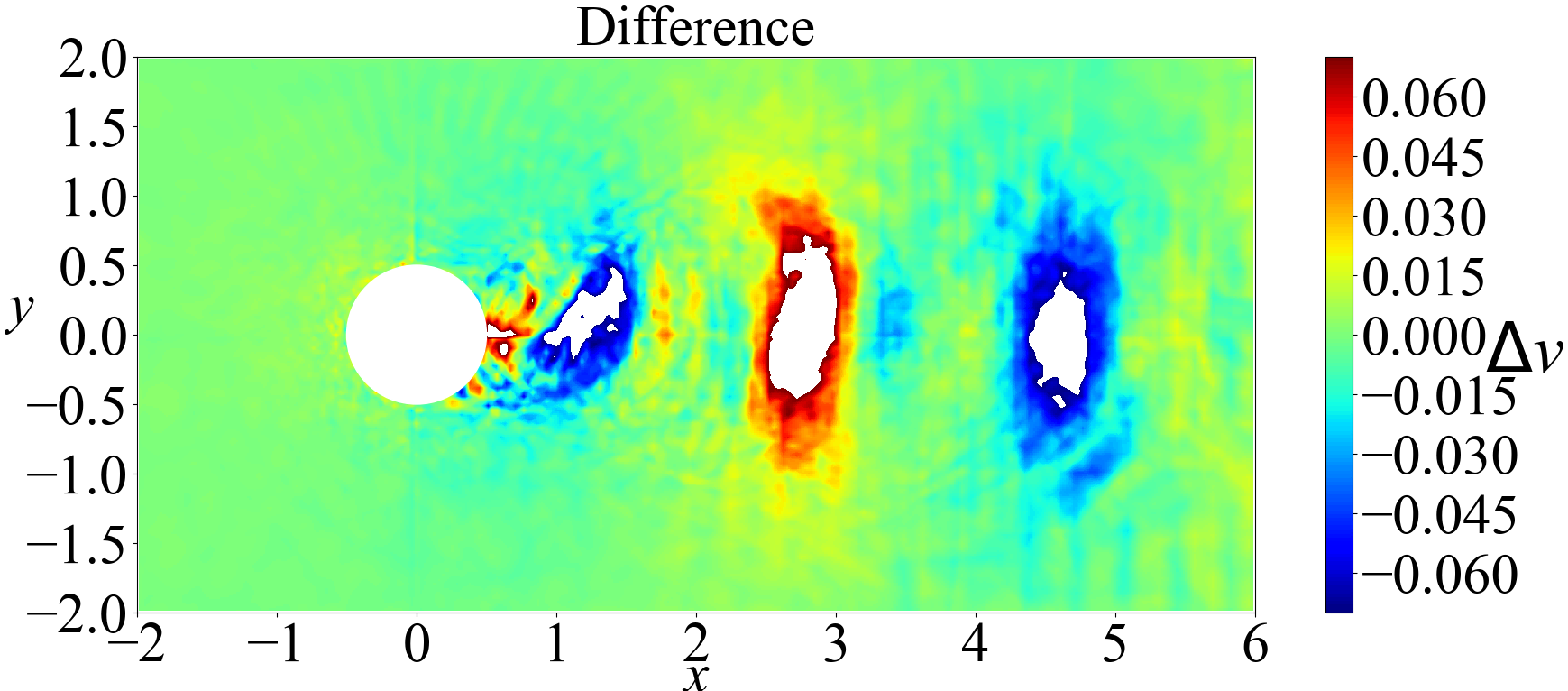}}
  \subfigure[]{
  \label{PDNSorigin}
  \includegraphics[scale=0.178]{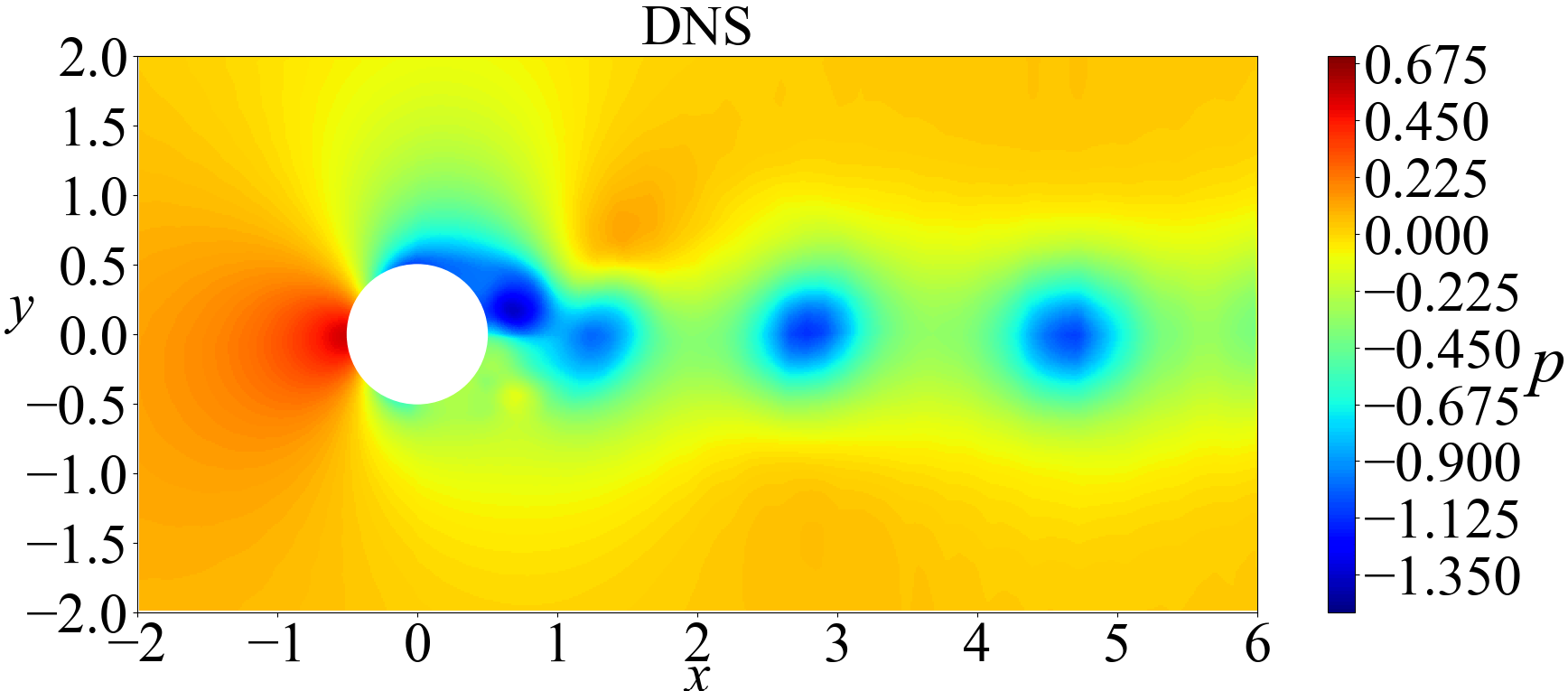}}
  \subfigure[]{
  \label{Ppreorigin}
  \includegraphics[scale=0.178]{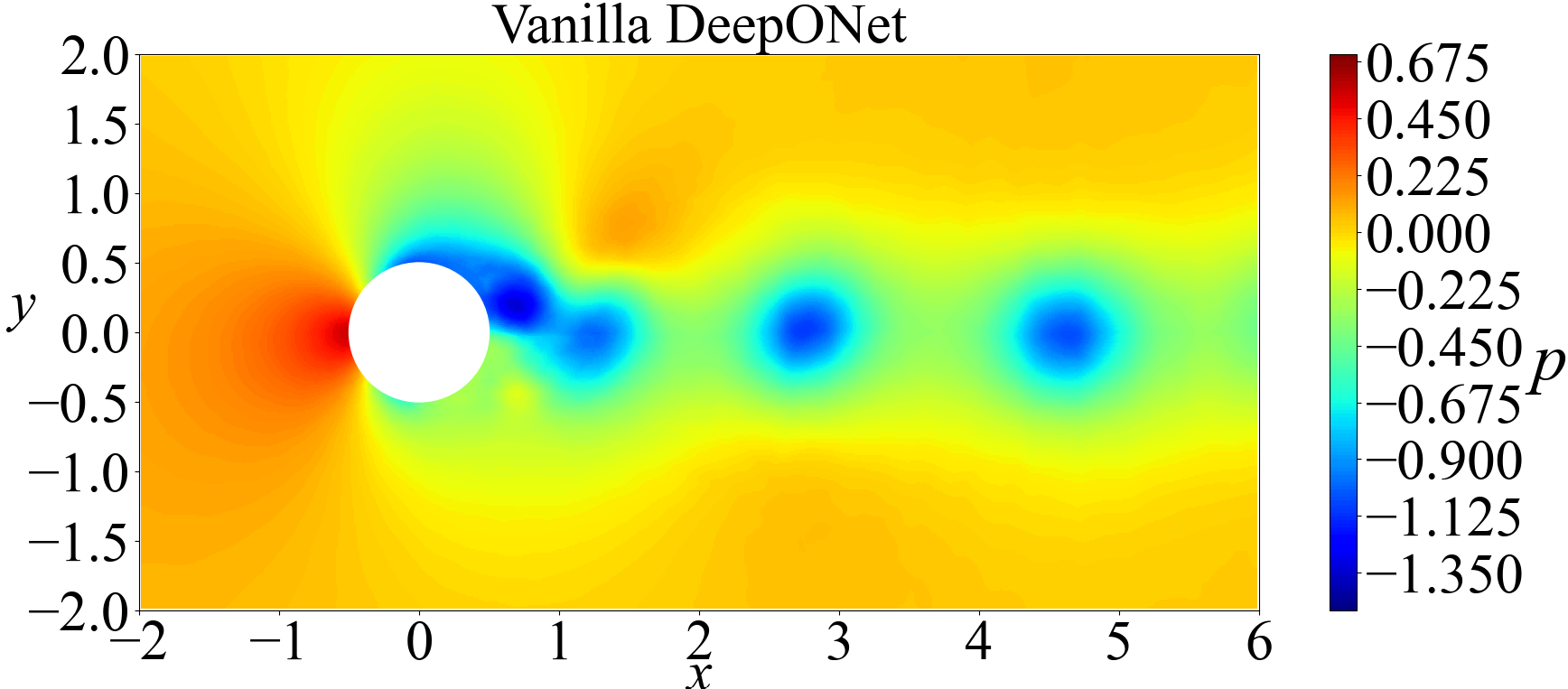}}
  \subfigure[]{
  \label{Pdeltaorigin}
  \includegraphics[scale=0.178]{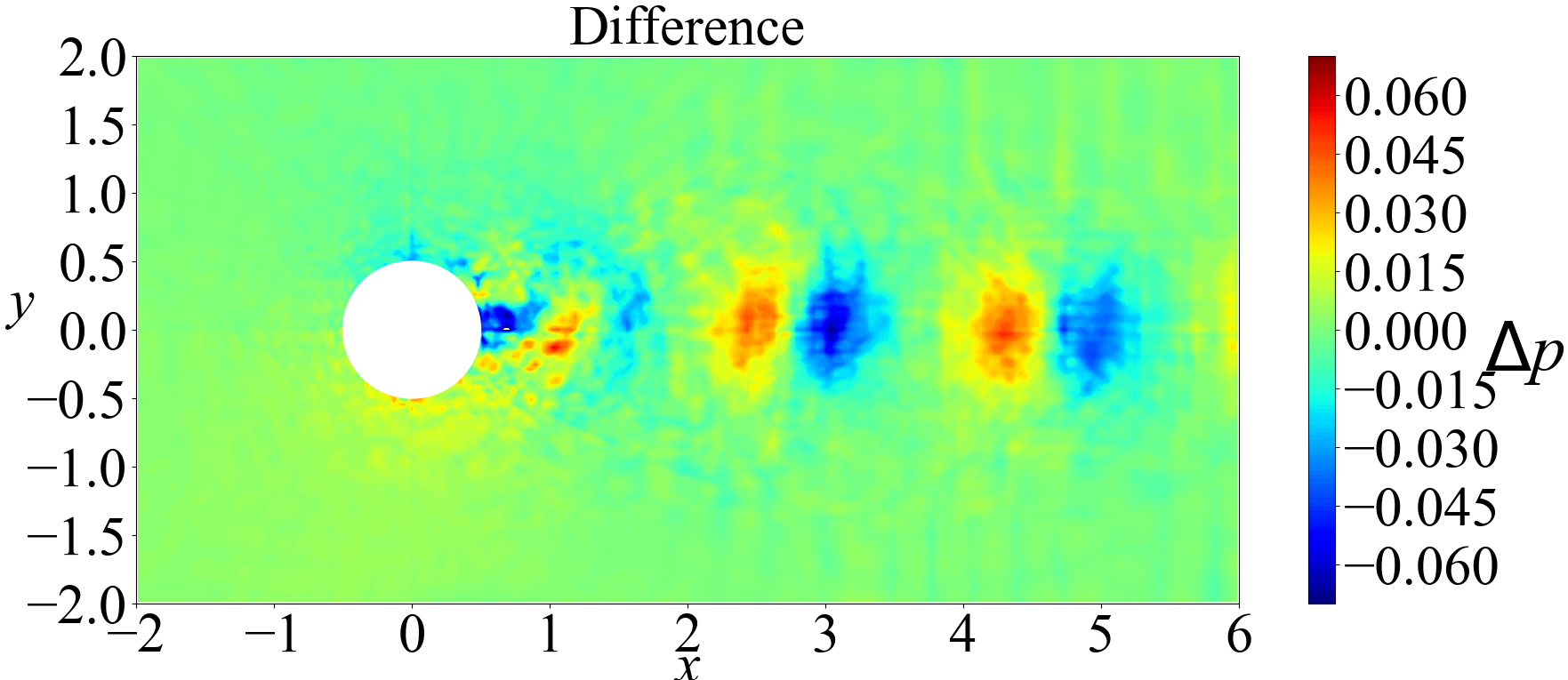}}
 \caption{Comparison of $u$, $v$ and $p$ at $1000th$ extrapolation snapshot between the DNS data and the prediction of the vanilla DeepONet. (a) $u_{DNS}$. (b) $u_{pre}$. (c) $u_{DNS}-u_{pre}$. (d) $v_{DNS}$. (e) $v_{pre}$. (f) $v_{DNS}-v_{pre}$. (g) $p_{DNS}$. (h) $p_{pre}$. (i) $p_{DNS}-p_{pre}$.}\label{origincontour compare}
\end{figure*}
The MSEs for the prediction over $4000$ extrapolation snapshots based on the vanilla DeepONet were already shown in Fig.~\ref{MSE of origin DeepONet}. 
Here the flow field for the $1000th$~($1000\Delta t$) extrapolation snapshot are further displayed and compared with the DNS data in Fig.~\ref{origincontour compare}.
The vanilla DeepONet successfully captures the evolution of the flow fields. 
Even with extrapolation for $1000$ snapshots, visually discerning between the DNS data~(on the first column) and the predictions of DeepONet~(on the second column) remains challenging. 
When the discrepancy between the two sets of results are calculated
and presented in Fig.~\ref{Udeltaorigin}, Fig.~\ref{Vdeltaorigin} and Fig.~\ref{Pdeltaorigin}, it indicates a non-negligible error of the latter,
especially in the wake flow at the downstream of the cylinder.
The error increases significantly as the number of extrapolation snapshots increases as depicted in Fig.~\ref{MSE of origin DeepONet}. 
Initially, the error is on the order of $10^{-5}$, while after $4000th$ extrapolation snapshots the error reaches the order of $10^{-3}$. 
However, it should be noted that when extrapolating with fewer than $500$ snapshots,
the error remains within the order of $10^{-5}$.

\subsection{DeepONet with multi-head non-local block} 
\label{CNNDeepONetMultihead}

The vanilla DeepONet, as demonstrated in Section \ref{originDNO}, exhibits a notable accumulation of error when extrapolating the flow fields.
One potential approach to improve the performance of DeepONet is to enhance the feature extraction capability of the CNN-based branch network
by a multi-head non-local block.
The basic principles have been elaborated in detail in Section \ref{MHnl}. In this section, it is integrated into the DeepONet as an auxiliary module. 

\begin{figure*}[!htb] 
 \centering  
  \subfigure[]{
  \label{UMSEbarnonlocal}
  \includegraphics[scale=0.27]{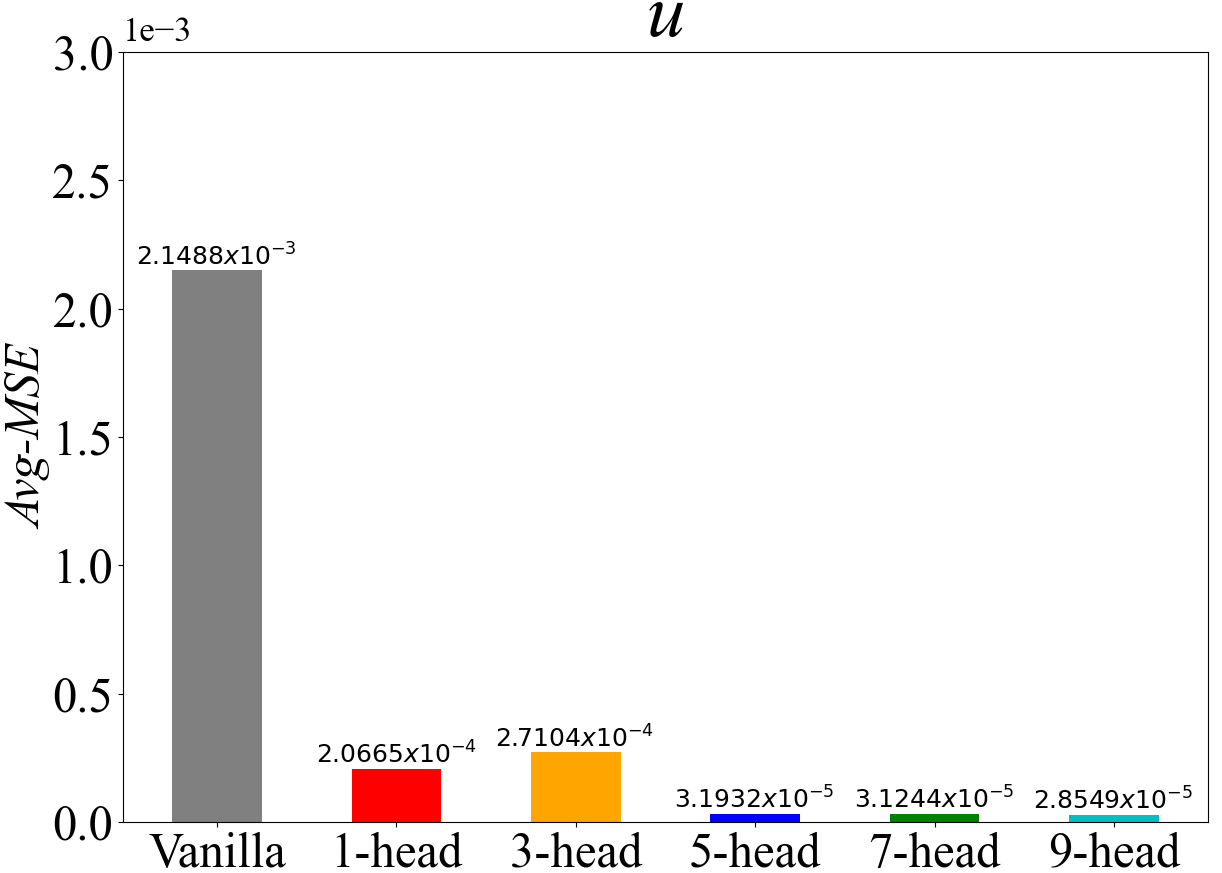}}
  \subfigure[]{
  \label{VMSEbarnonlocal}
  \includegraphics[scale=0.27]{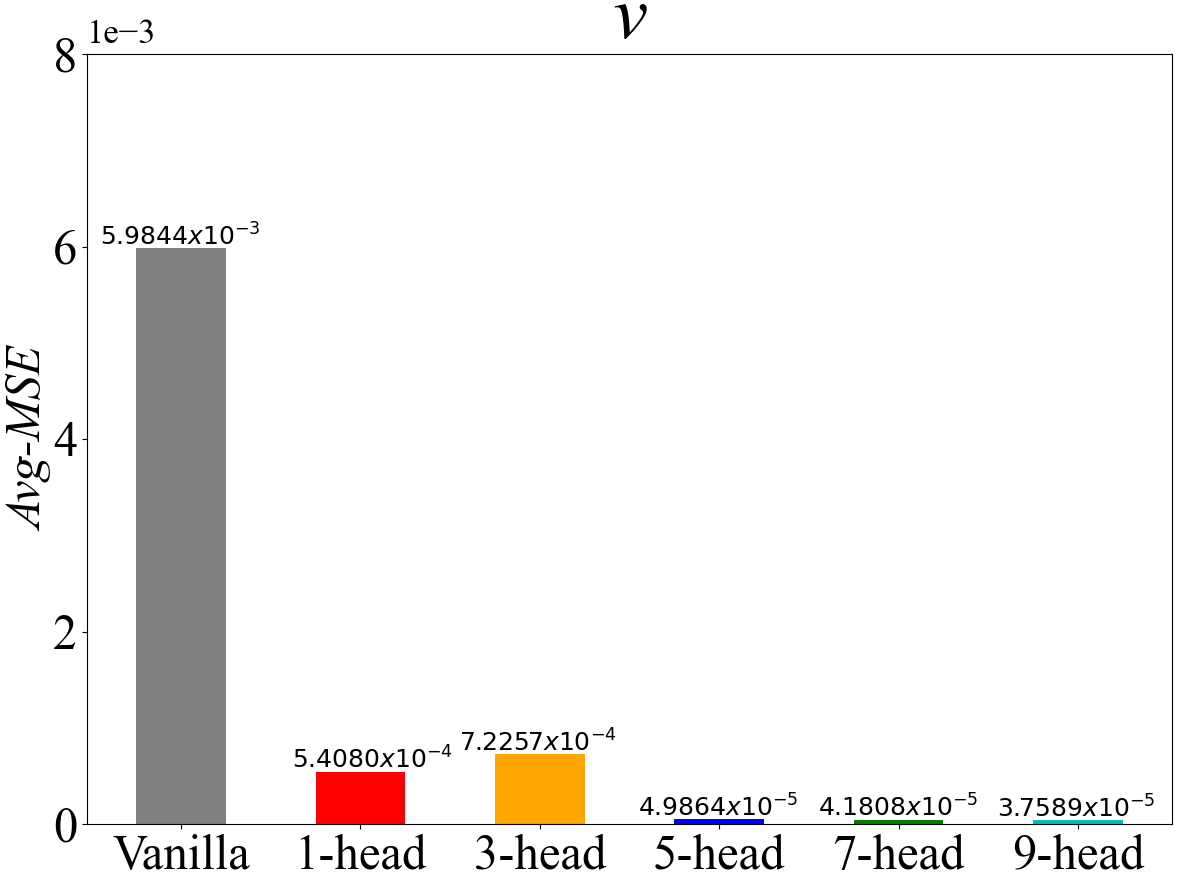}}
  \subfigure[]{
  \label{PMSEbarnonlocal}
  \includegraphics[scale=0.27]{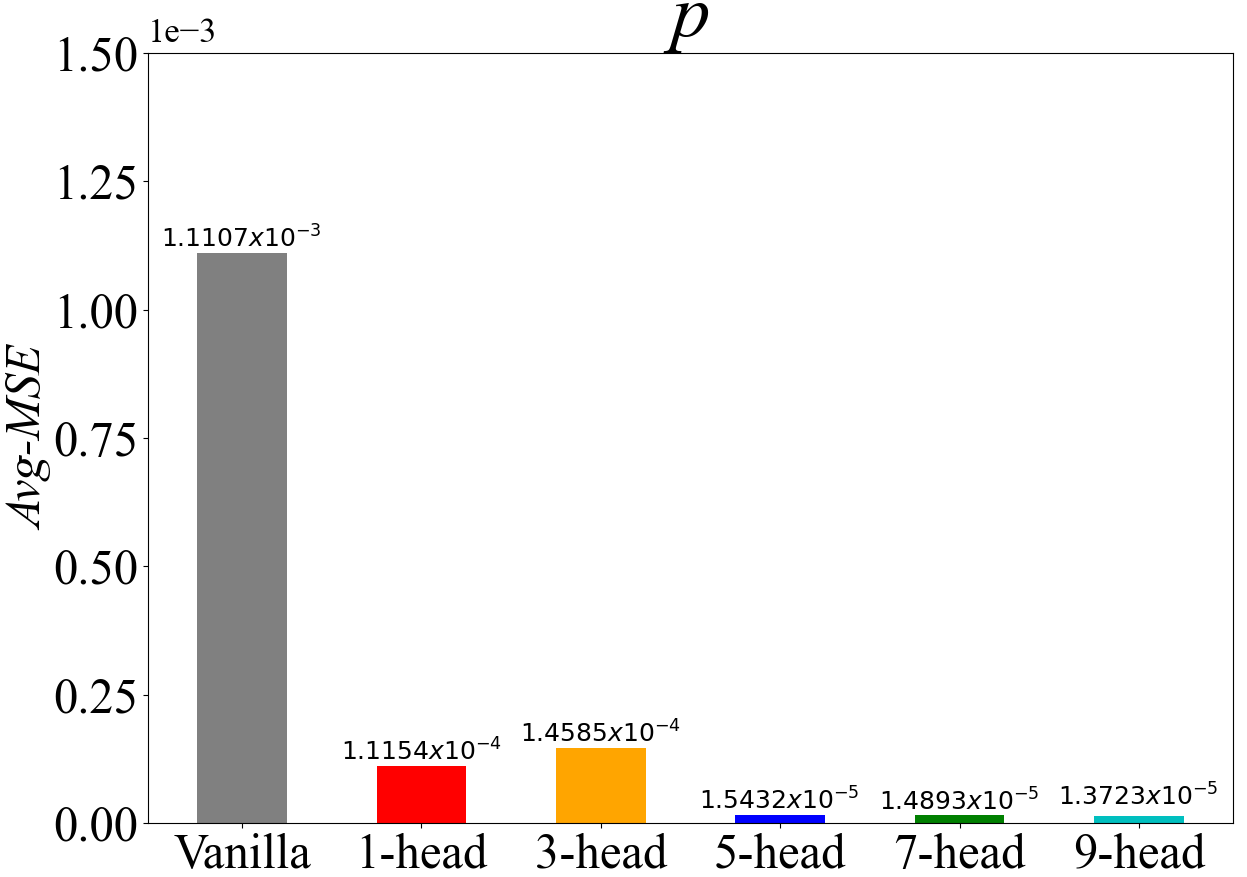}}
 \caption{The MSEs of $u$, $v$ and $p$ between the DNS data and the predictions of vanilla DeepONet, DeepONet + non-local block and DeepONet + multi-head non-local block averaged from $3500th$ to $4000th$ extrapolation snapshots. (a) $u_{MSE}$. (b) $v_{MSE}$. (c) $p_{MSE}$.}
 \label{MSE of origin and non-local}
\end{figure*}

The multi-head non-local block with $1$, $3$, $5$, $7$ and $9$-head are utilized respectively, where the 1-head corresponds to the non-local block. The MSEs of $u$, $v$ and $p$ between the DNS data and the extrapolation results of vanilla DeepONet, DeepONet + non-local block and DeepONet + multi-head non-local block are averaged from $3500th$ to $4000th$ snapshots and are illustrated in Fig.~\ref{UMSEbarnonlocal}-\ref{PMSEbarnonlocal}.

The accuracy of extrapolating the flow field is higher for the DeepONet + 1-head case compared to the vanilla DeepONet case, with an improvement from an order of $10^{-3}$ to $10^{-4}$. This suggests that integrating the non-local block enhances the capability of capturing the long-range dependencies, resulting in a more accurate extraction of flow field information. 

\begin{figure*}[htbp] 
 \centering  
  \subfigure[]{
  \label{UDNSm9}
  \includegraphics[scale=0.178]{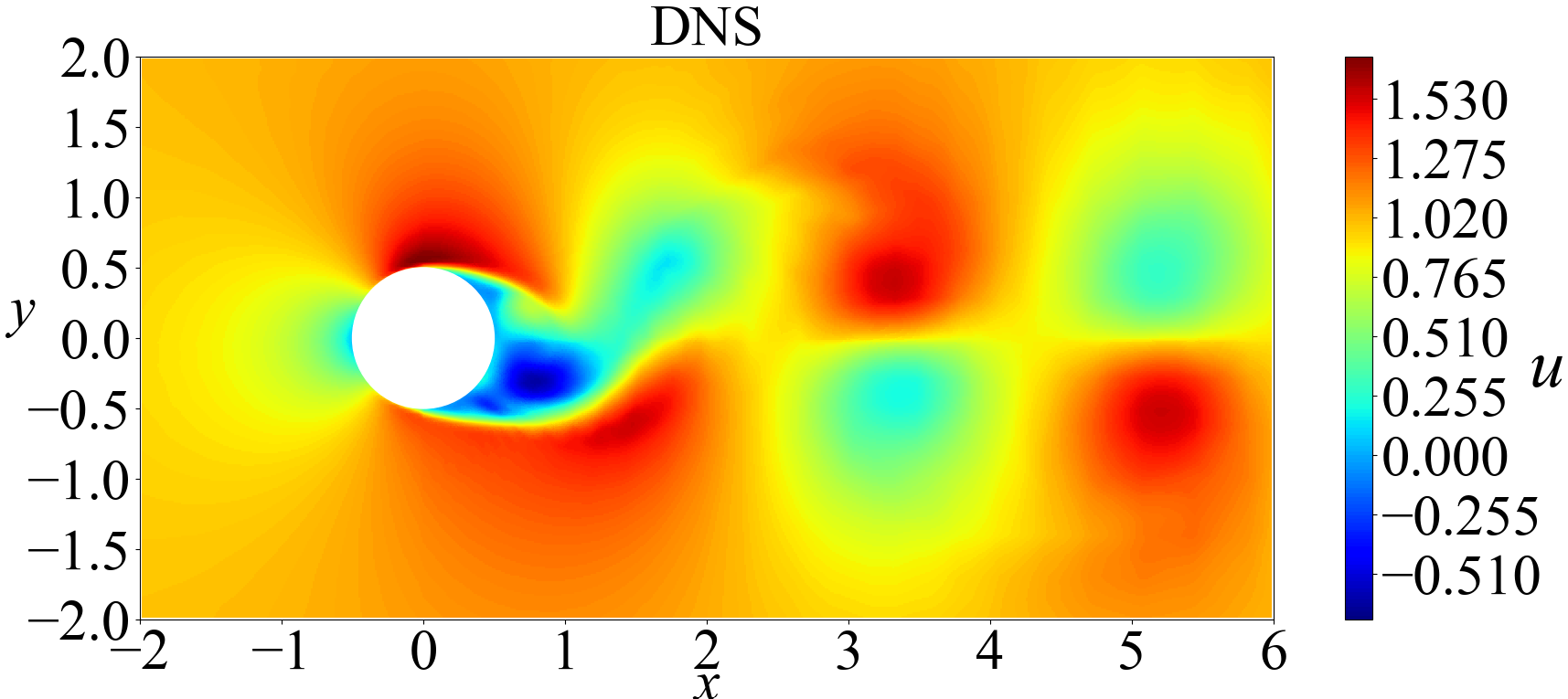}}
  \subfigure[]{
  \label{Uprem9}
  \includegraphics[scale=0.178]{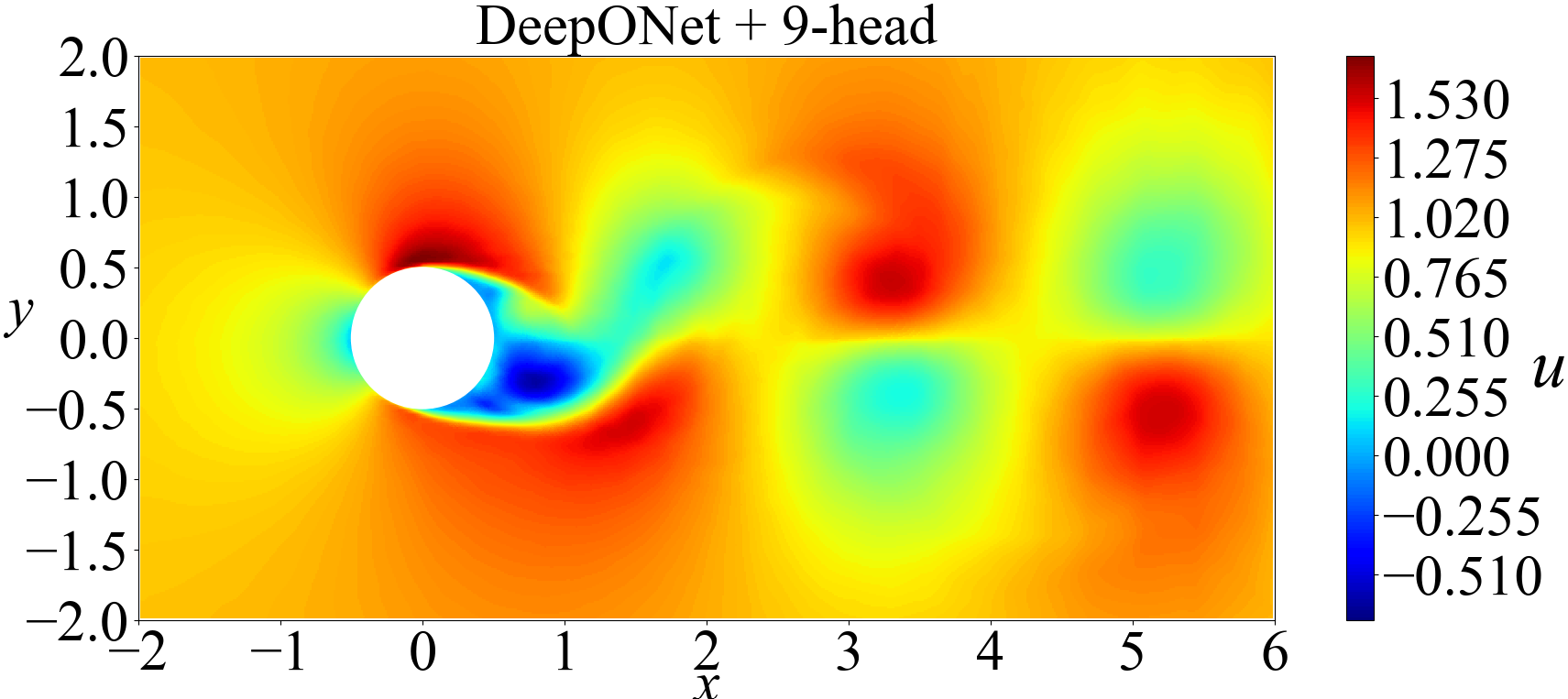}}
  \subfigure[]{
  \label{Udeltam9}
  \includegraphics[scale=0.178]{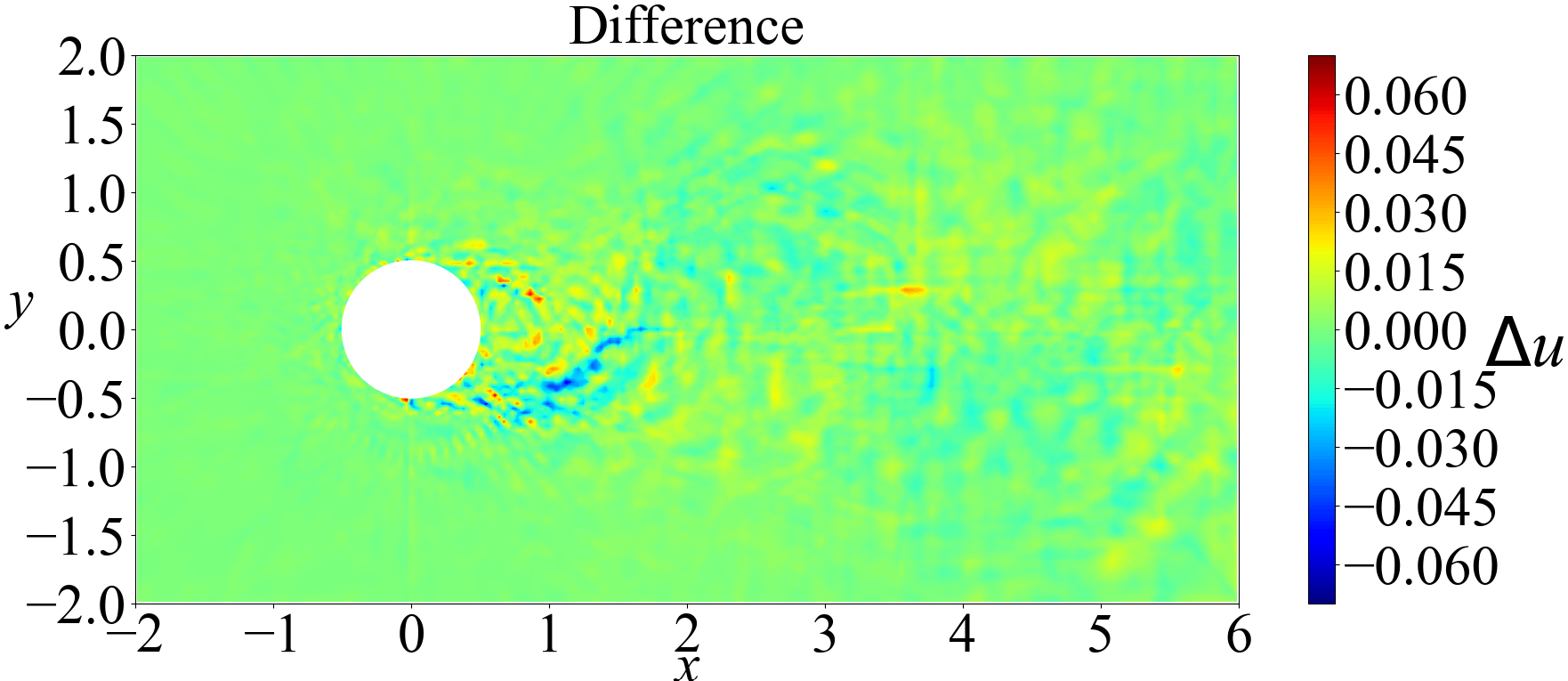}}
  \subfigure[]{
  \label{VDNSm9}
  \includegraphics[scale=0.178]{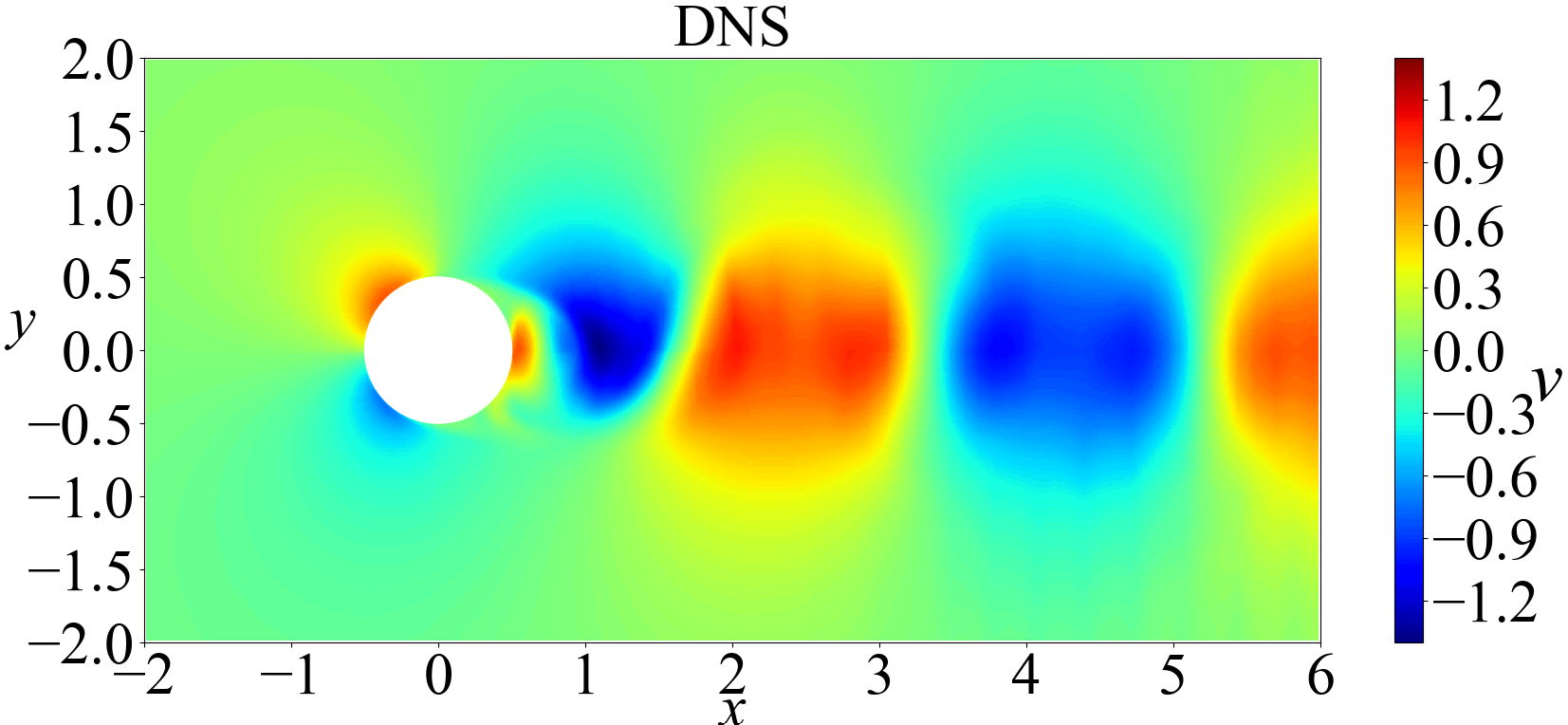}}
  \subfigure[]{
  \label{Vprem9}
  \includegraphics[scale=0.178]{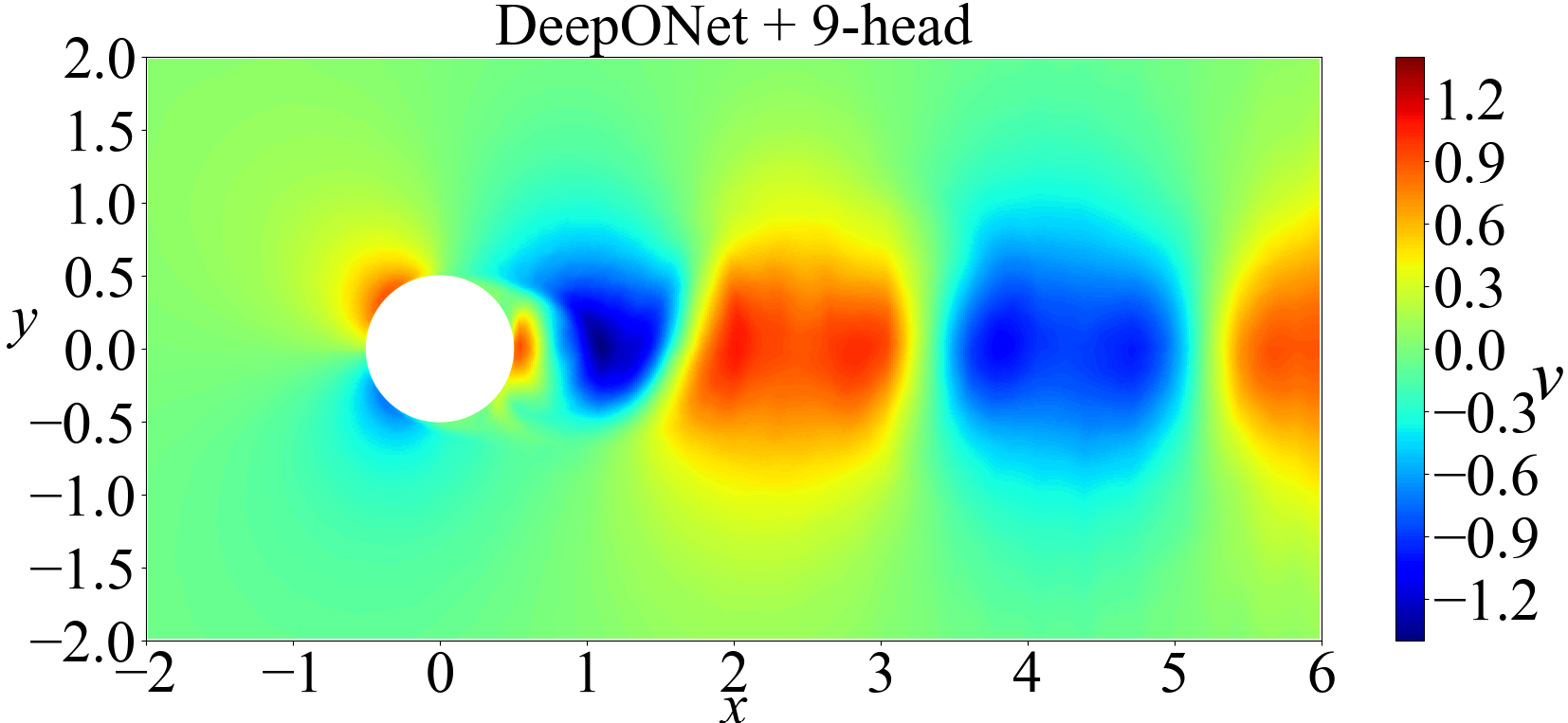}}
  \subfigure[]{
  \label{Vdeltam9}
  \includegraphics[scale=0.178]{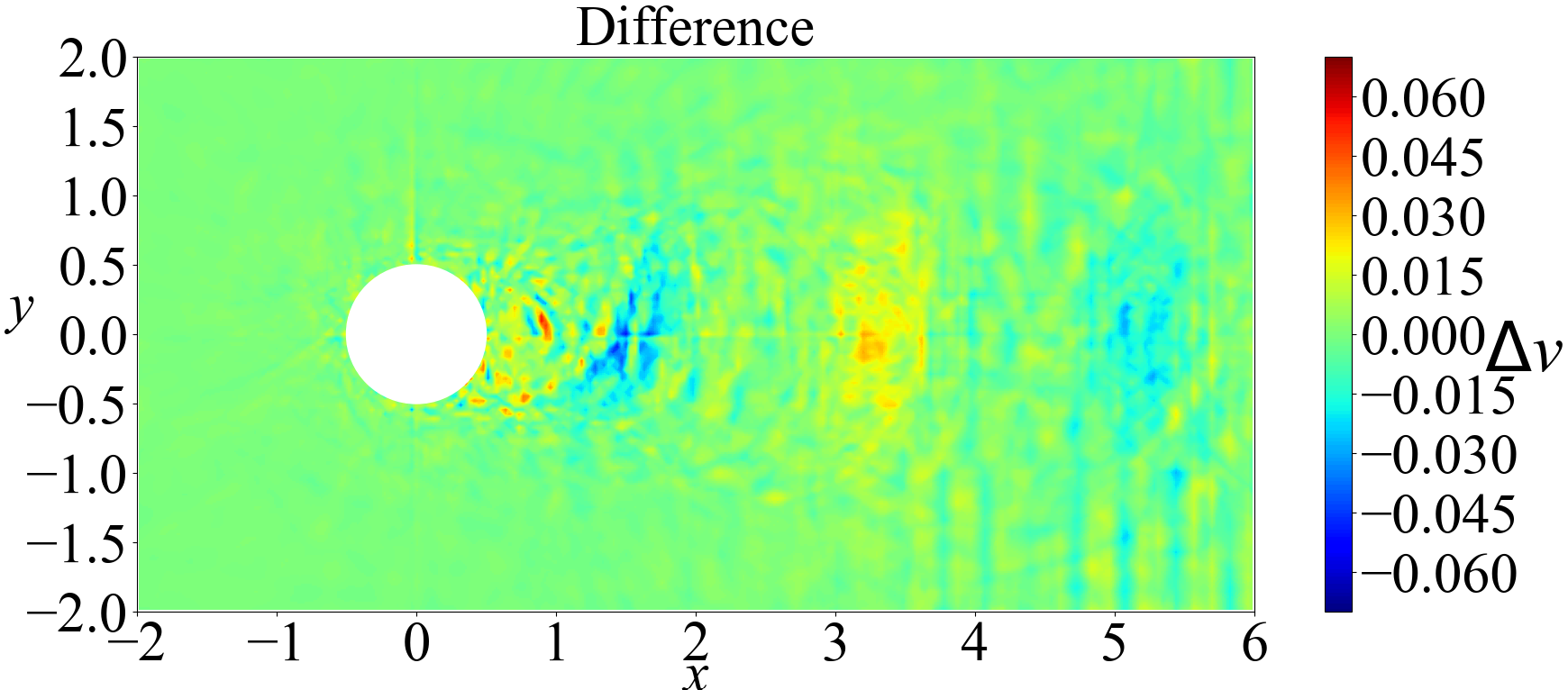}}
  \subfigure[]{
  \label{PDNSm9}
  \includegraphics[scale=0.178]{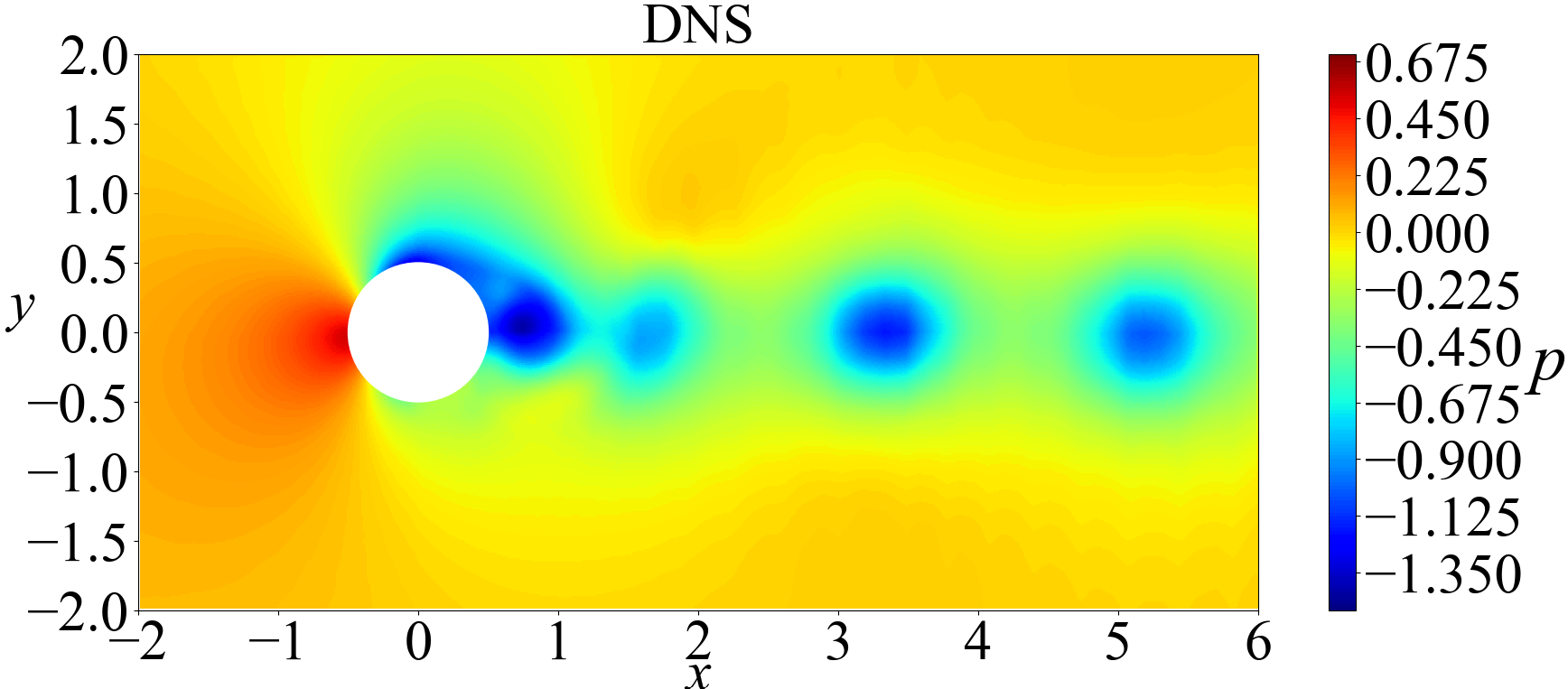}}
  \subfigure[]{
  \label{Pprem9}
  \includegraphics[scale=0.178]{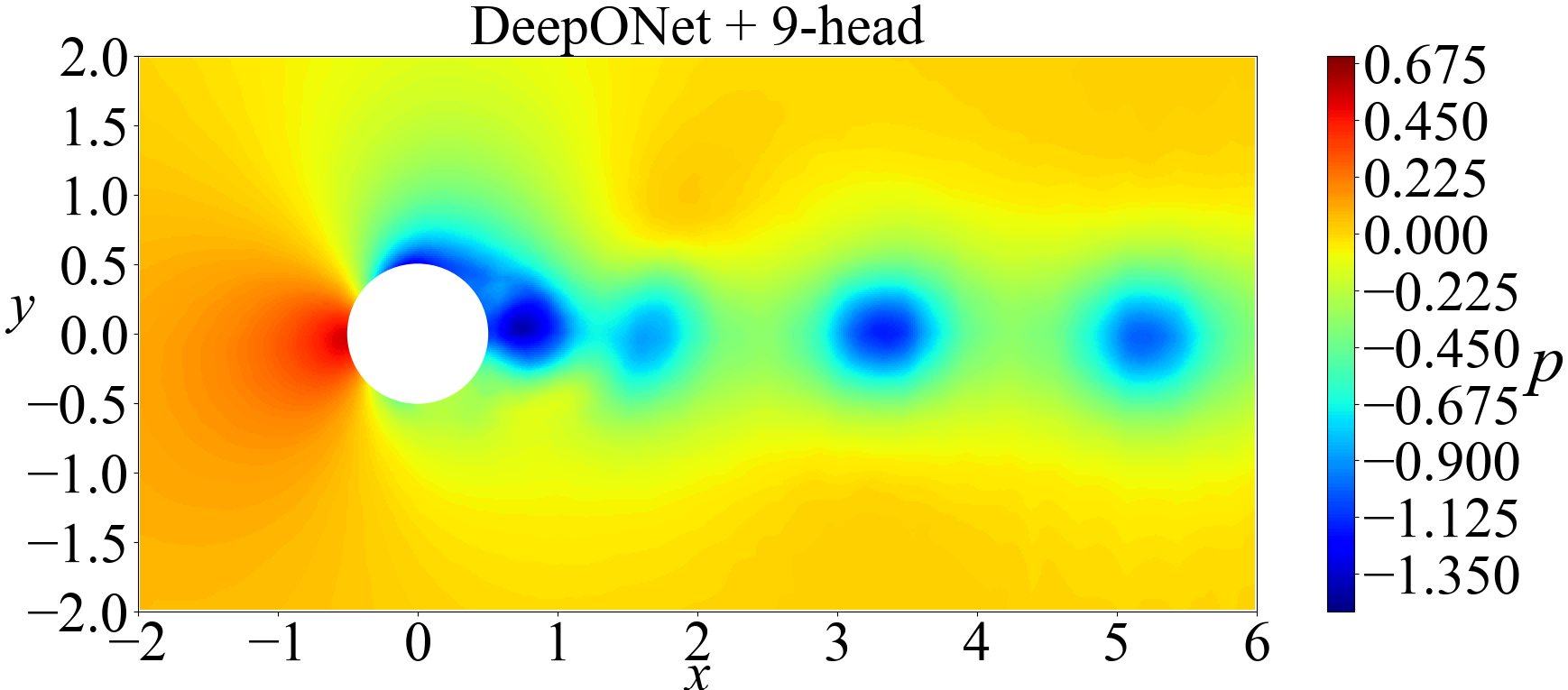}}
  \subfigure[]{
  \label{Pdelta9}
  \includegraphics[scale=0.178]{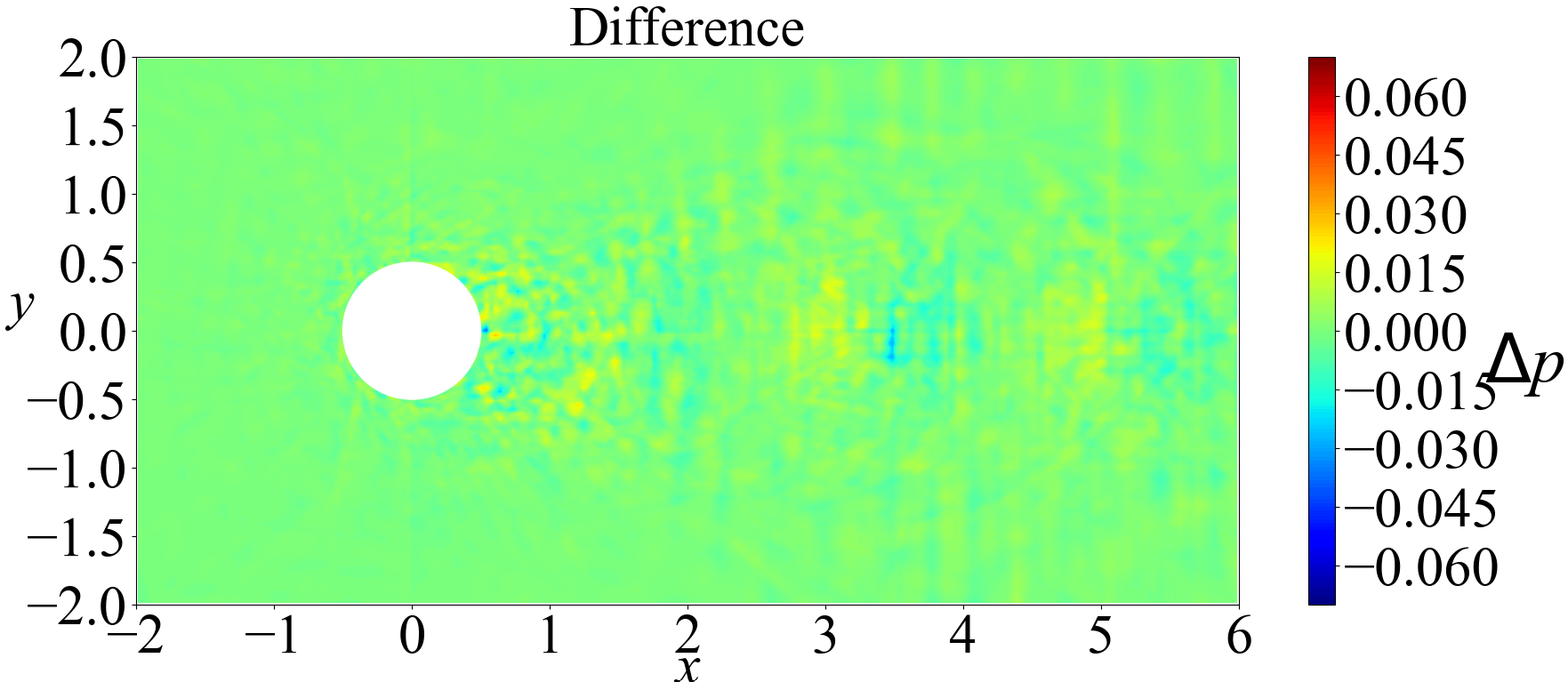}}
  \caption{Comparison of $u$, $v$ and $p$ at $4000th$ extrapolation snapshot step between the DNS data and the predictions of DeepONet + 9-head. (a) $u_{DNS}$. (b) $u_{pre}$. (c) $u_{DNS}-u_{pre}$. (d) $v_{DNS}$. (e) $v_{pre}$. (f) $v_{DNS}-v_{pre}$. (g) $p_{DNS}$. (h) $p_{pre}$. (i) $p_{DNS}-p_{pre}$.}\label{9headscontour compare}
\end{figure*}

The DeepONet also experiences a substantial decrease in the extrapolation error when the multi-head non-local block is integrated, as compared to the vanilla DeepONet. The discrepancy in the extrapolation error between DeepONet + 3-head and DeepONet + 1-head cases is negligible. However, increasing the number of heads to $5$, $7$ and $9$ reveals a noteworthy decrease in the extrapolation error compared to both the vanilla DeepONet and DeepONet + 1-head cases, and the magnitude of the error is reduced to $10^{-5}$. We select only the flow field at the $4000th$ extrapolation snapshot from the DeepONet + 9-head as representative and compare it with the DNS data in Fig.~\ref{9headscontour compare}.

\begin{figure*}[htbp] 
 \centering  
  \subfigure[]{
  \label{UMSEm9}
  \includegraphics[scale=0.255]{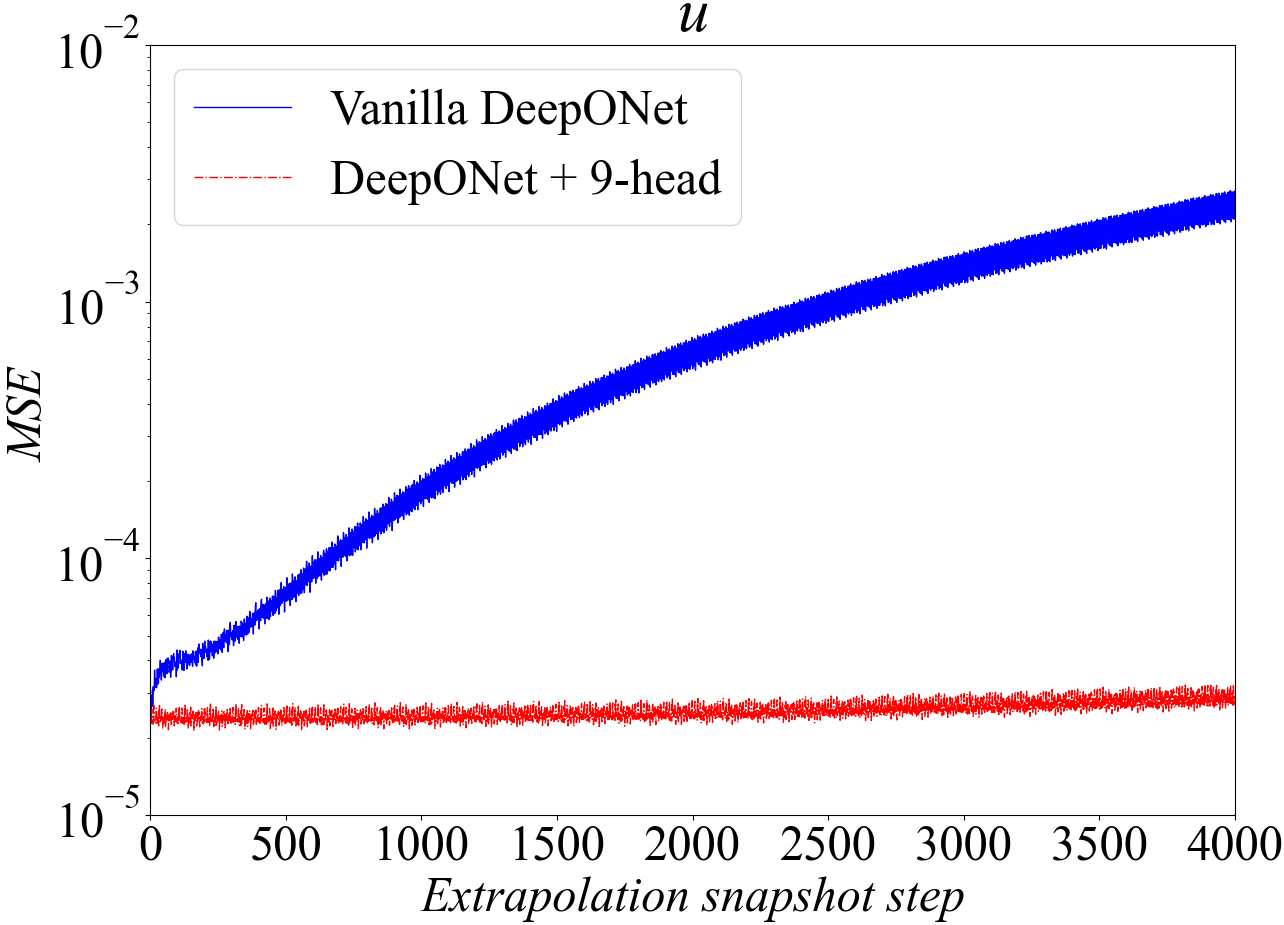}}
  \subfigure[]{
  \label{VMSEm9}
  \includegraphics[scale=0.255]{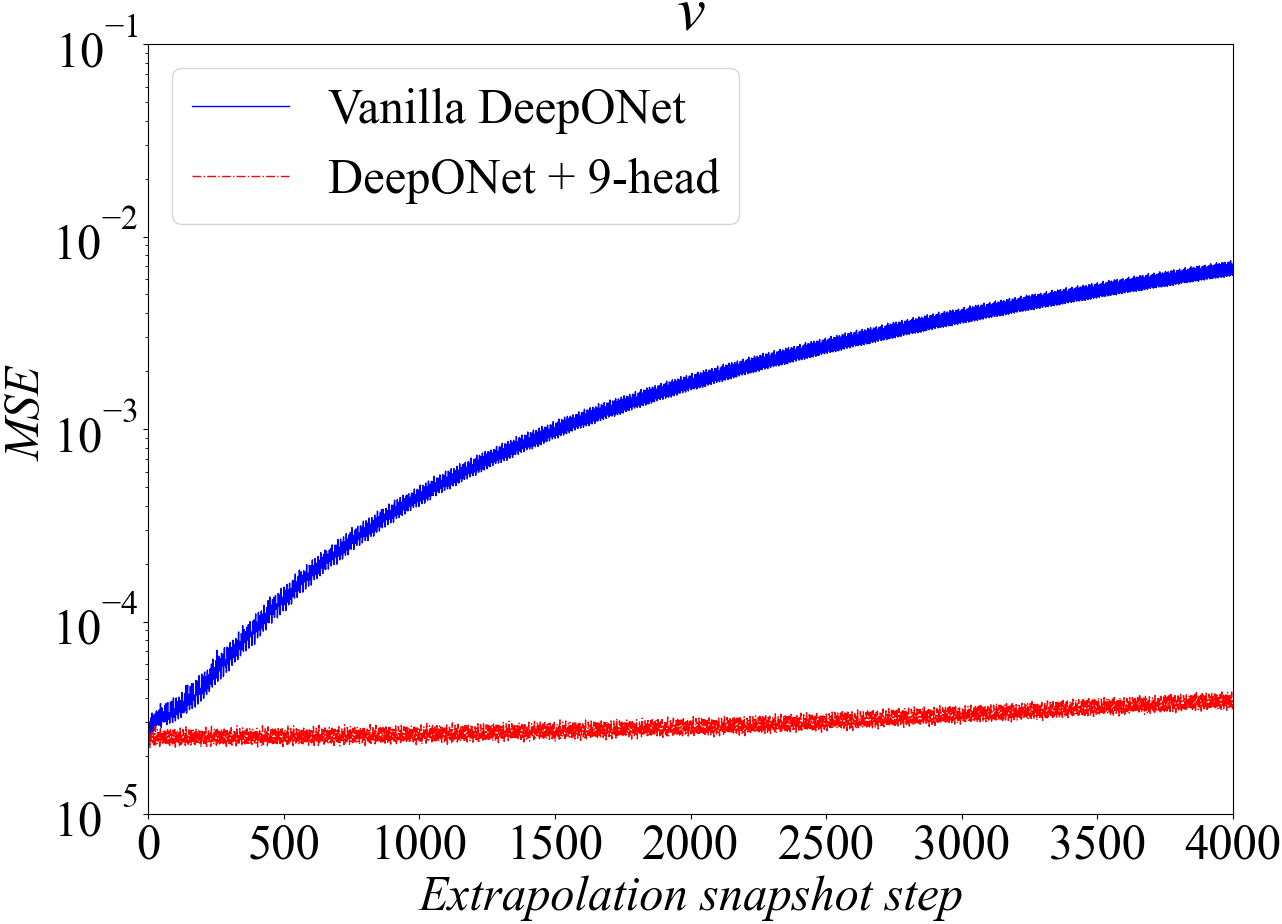}}
  \subfigure[]{
  \label{PMSEm9}
  \includegraphics[scale=0.255]{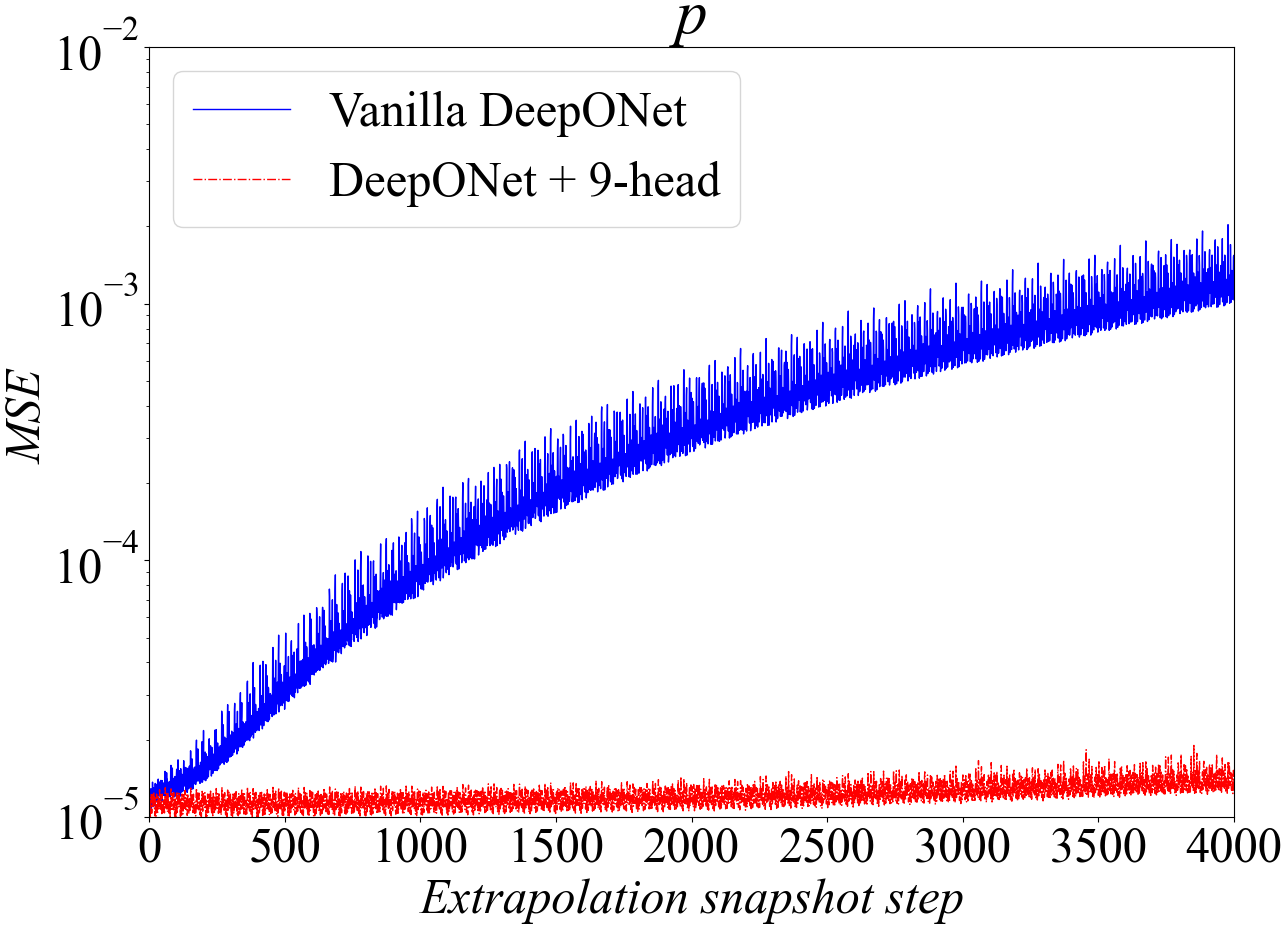}}
 \caption{The evolution of MSEs between the DNS data and the predictions of vanilla DeepONet/DeepONet + 9-head. (a) $u_{MSE}$. (b) $v_{MSE}$. (c) $p_{MSE}$.}
 \label{MSE line non-local compare}
\end{figure*}

The results demonstrate the benefits of integrating the multi-head non-local block in reducing extrapolation errors across the entire flow region compared to the vanilla DeepONet. The evolution of the MSEs between the DNS data and the predictions of DeepONet + 9-head/vanilla DeepONet over $4000$ extrapolation snapshots are summarized in Fig.~\ref{MSE line non-local compare},
where the enhanced version of DeepOnet has a clear superior performance.

\begin{figure*}[htbp] 
 \centering  
  \subfigure[]{
  \label{uutruenonlocal}
  \includegraphics[scale=0.045]{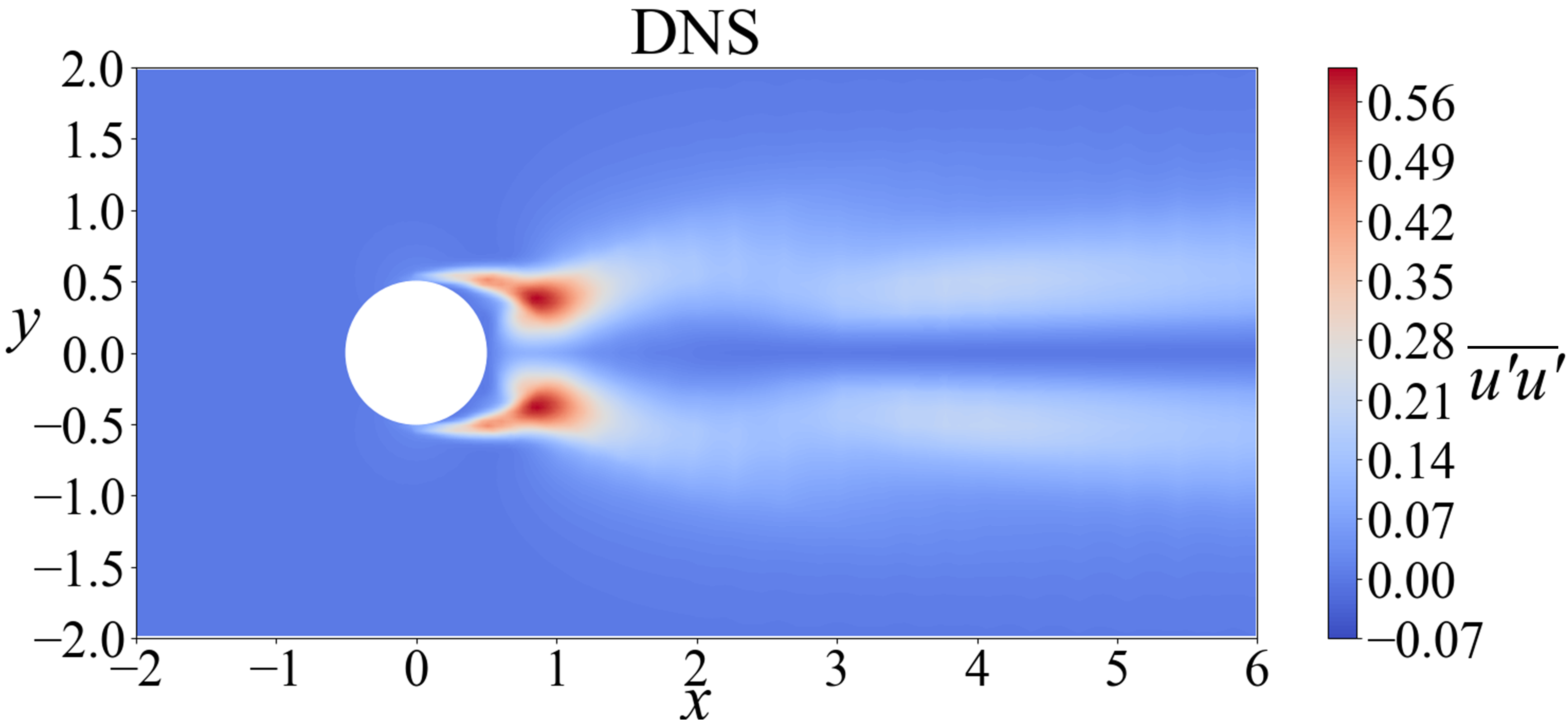}}
  \subfigure[]{
  \label{UUpre9head}
  \includegraphics[scale=0.0453]{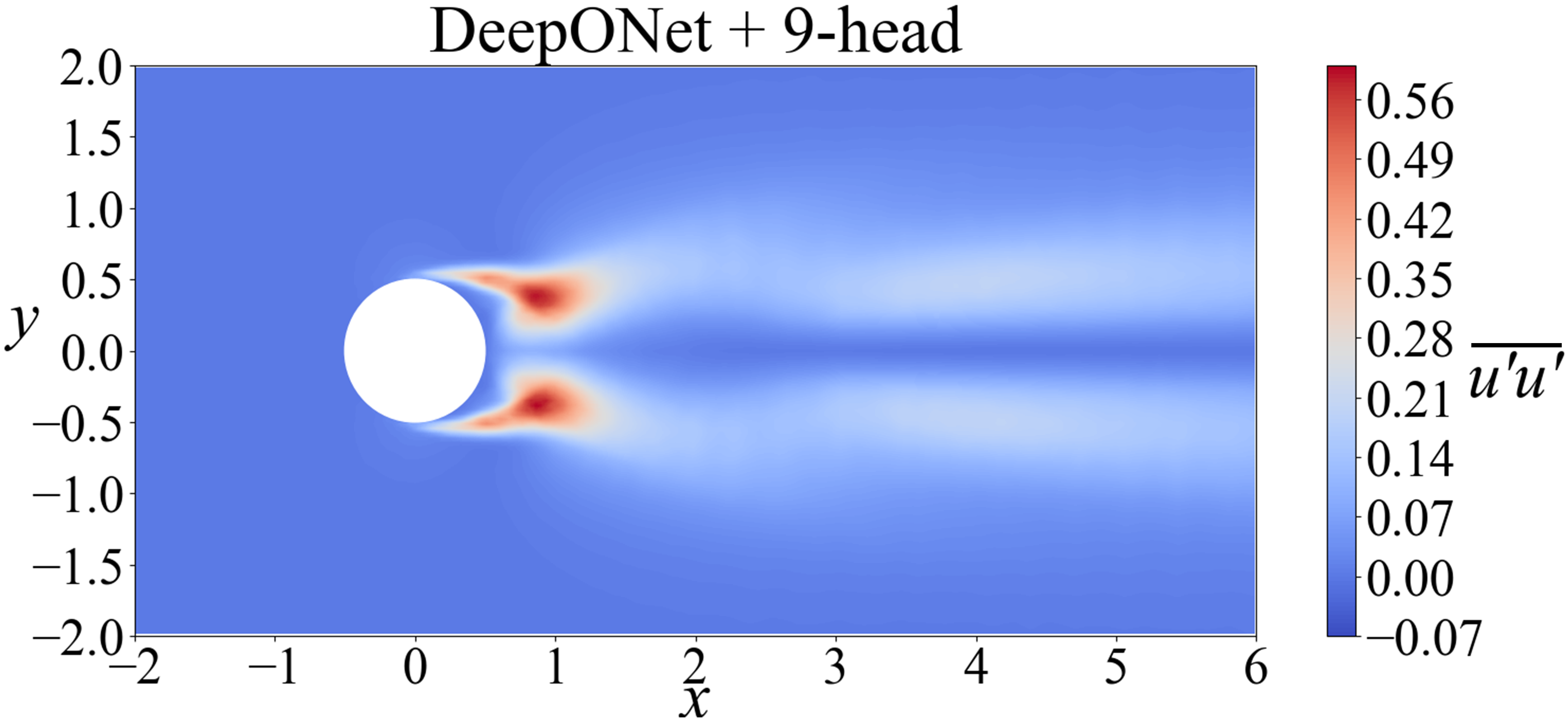}}
  \subfigure[]{
  \label{UUdelta9head}
  \includegraphics[scale=0.045]{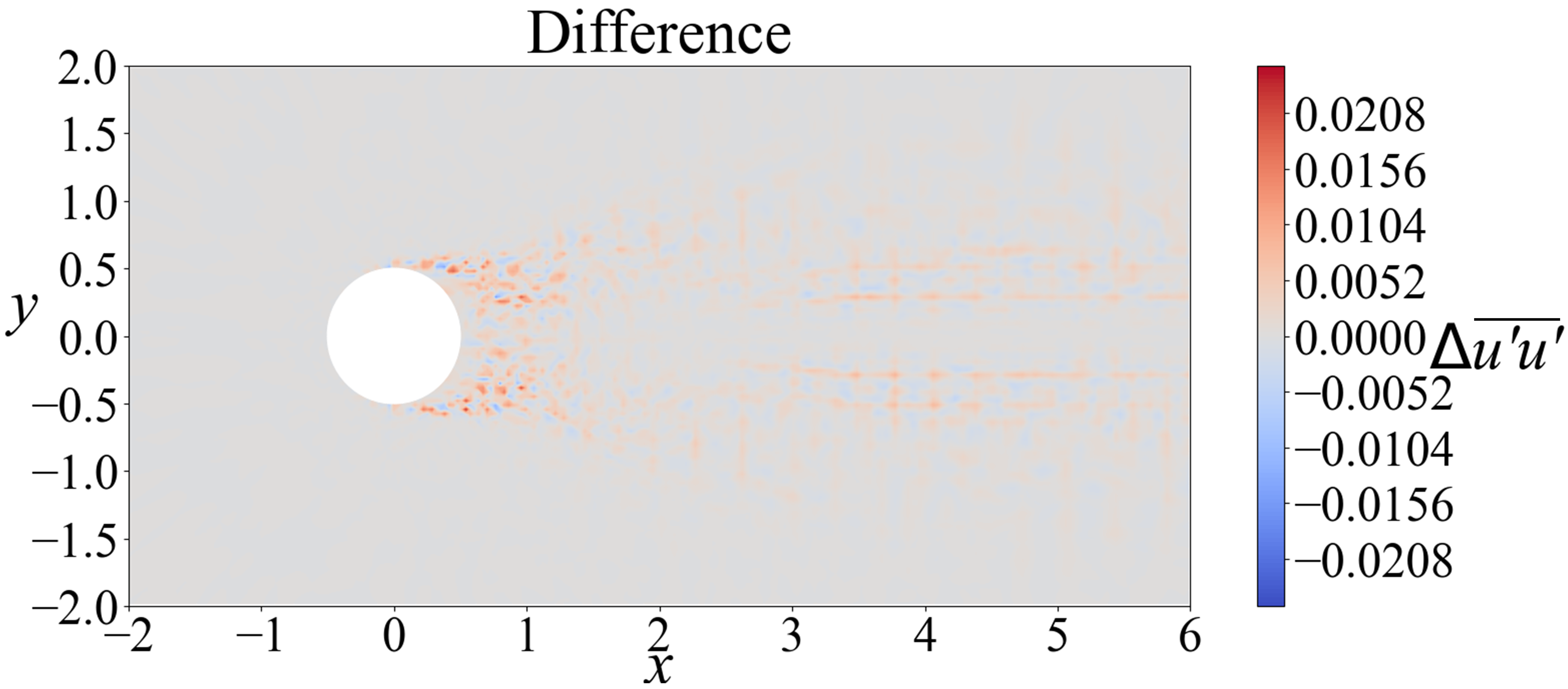}}
  \subfigure[]{
  \label{vvtruenonlocal}
  \includegraphics[scale=0.045]{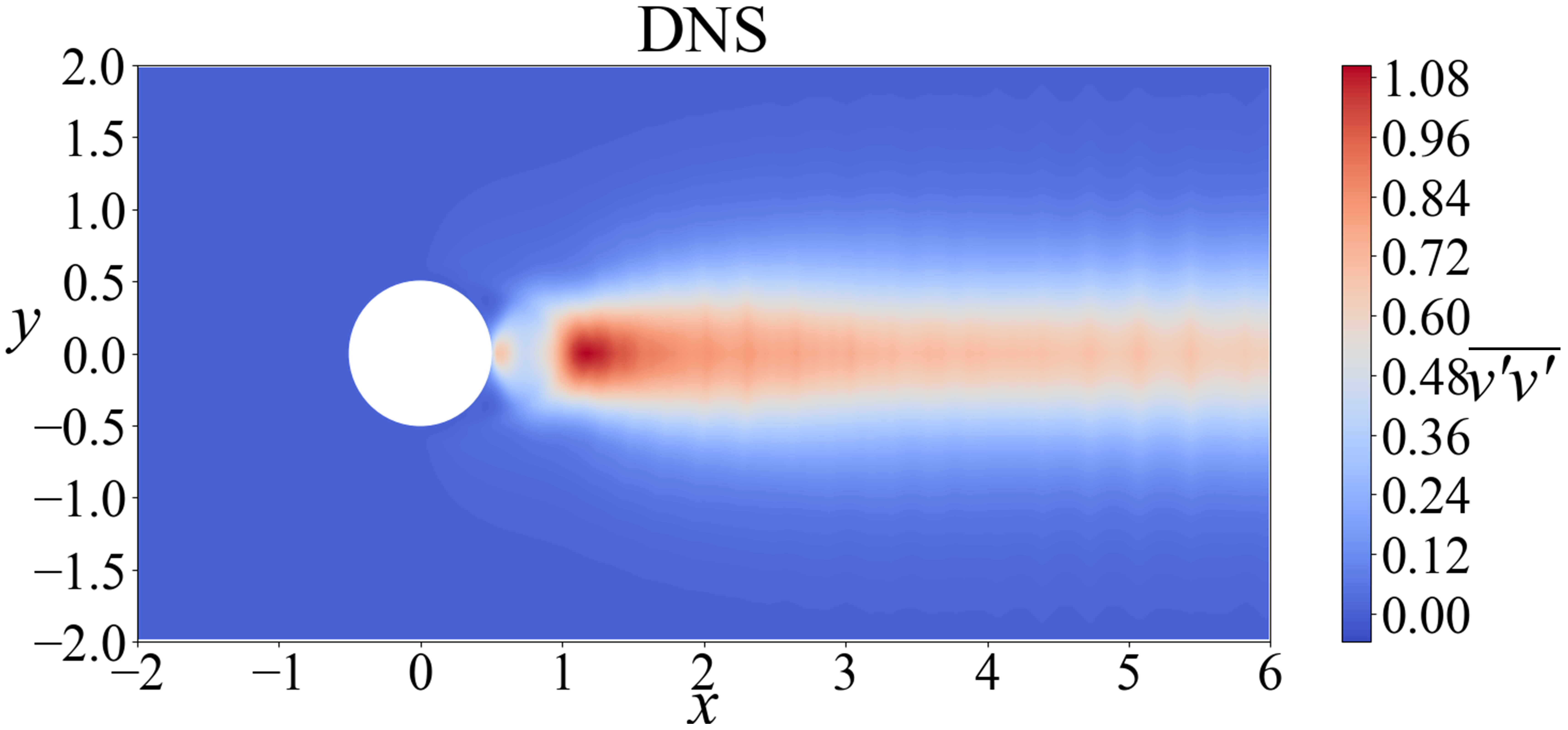}}
  \subfigure[]{
  \label{VVpre9head}
  \includegraphics[scale=0.0453]{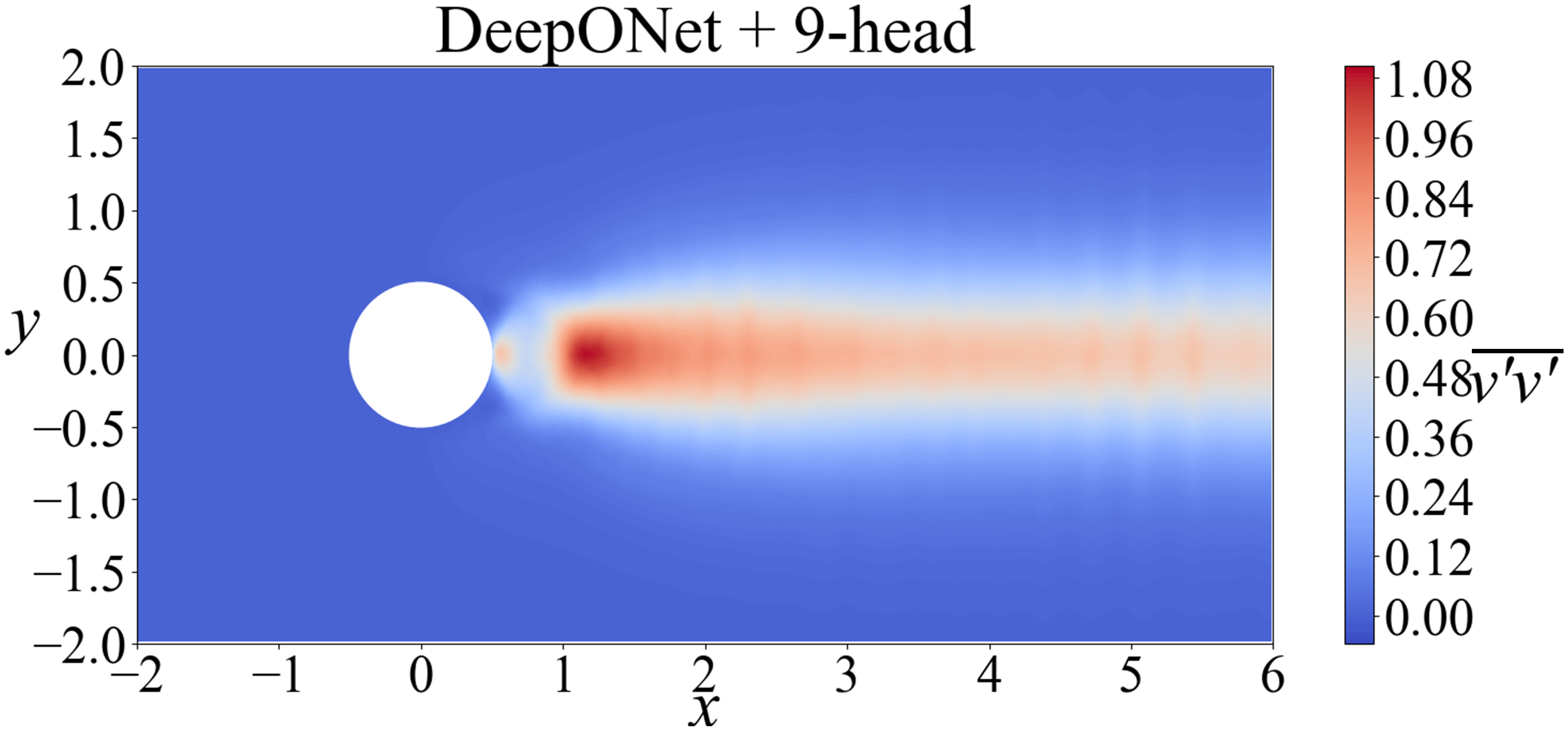}}
  \subfigure[]{
  \label{VVdelta9head}
  \includegraphics[scale=0.045]{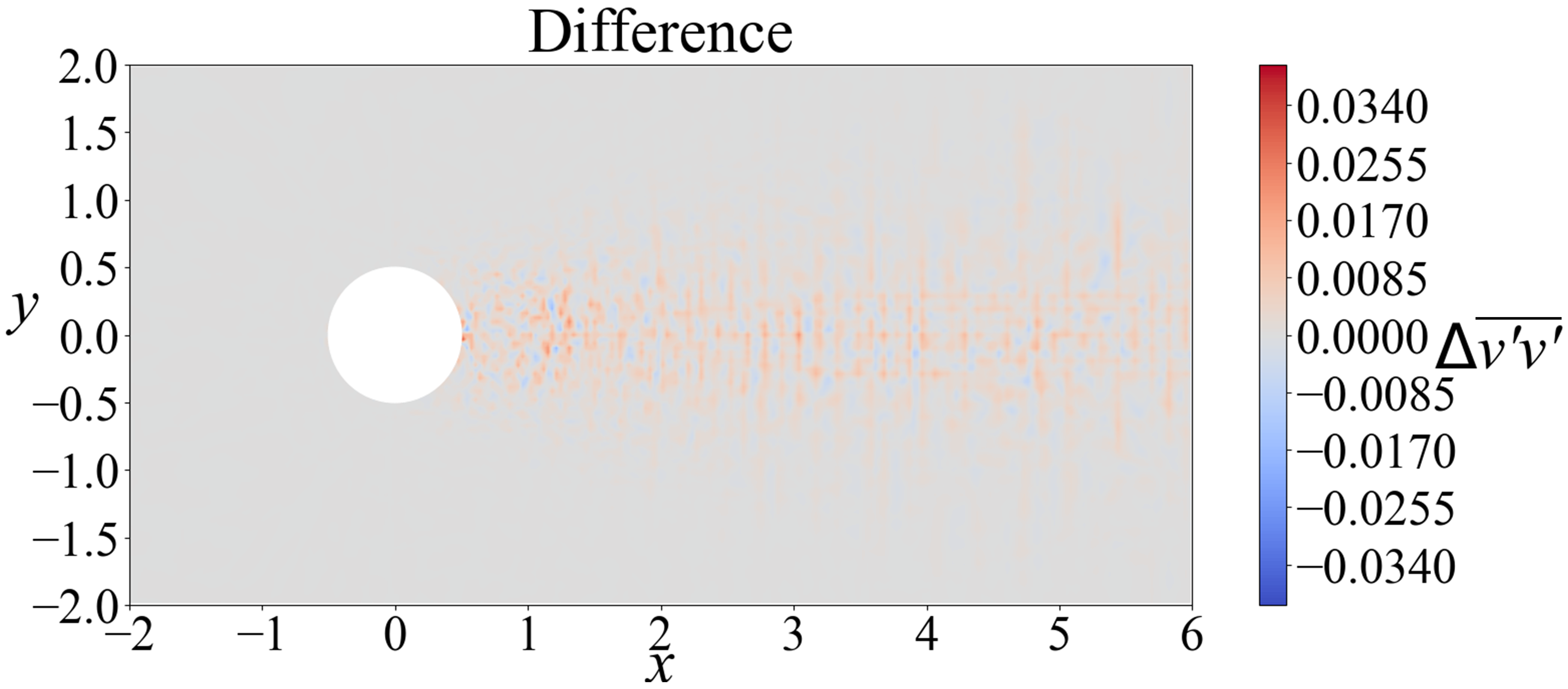}}
  \subfigure[]{
  \label{uvtruenonlocal}
  \includegraphics[scale=0.045]{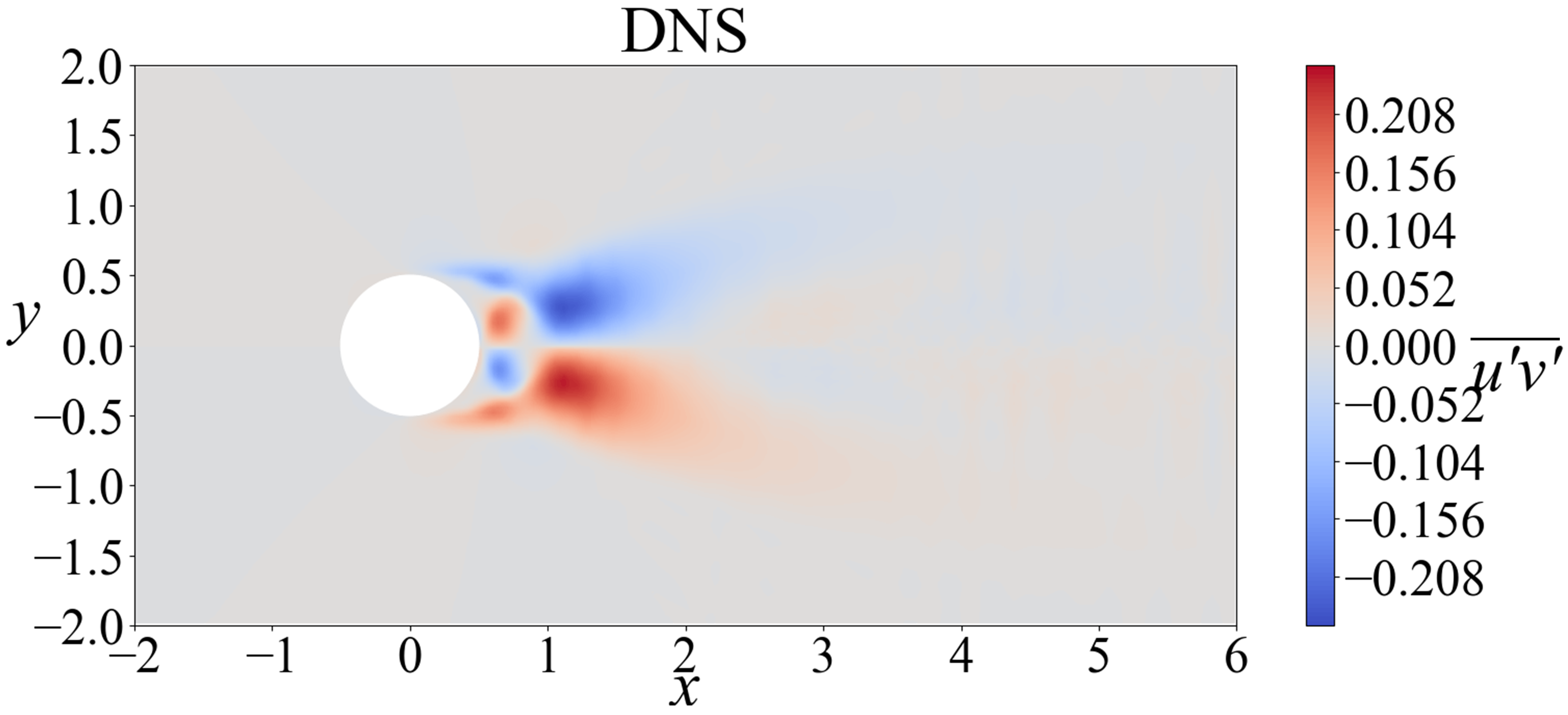}}
  \subfigure[]{
  \label{UVpre9head}
  \includegraphics[scale=0.0453]{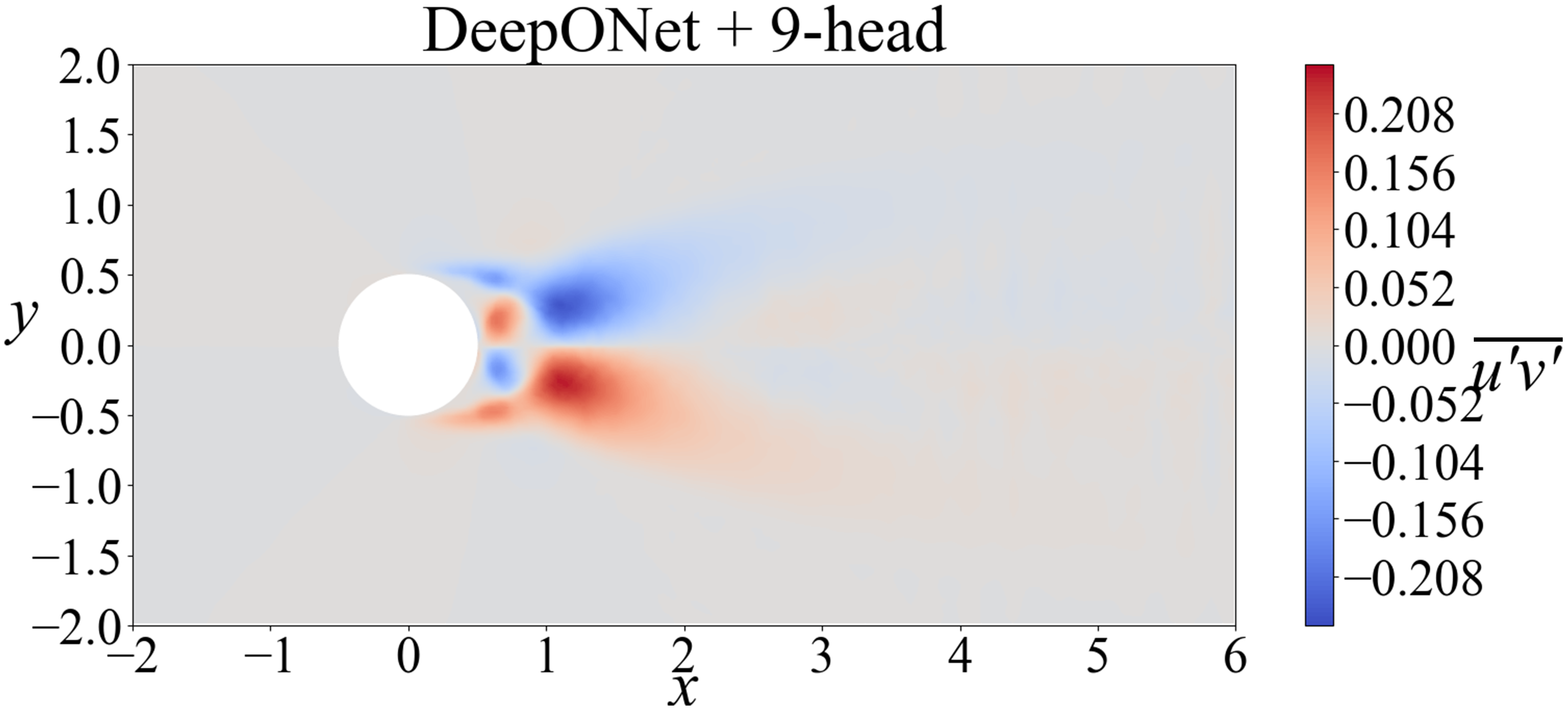}}
  \subfigure[]{
  \label{UVdelta9head}
  \includegraphics[scale=0.045]{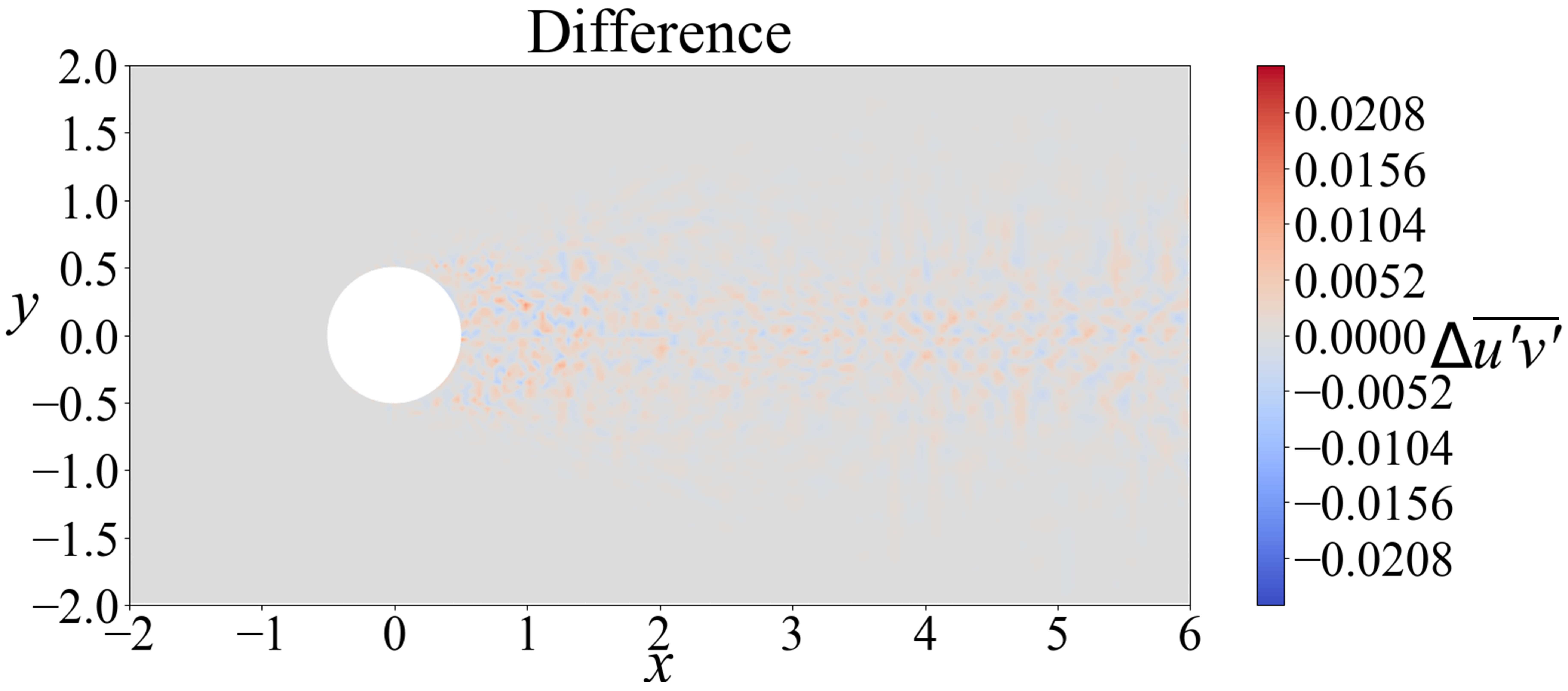}}
  \caption{Comparison of the components of the time-averaged Reynolds stress between the DNS data and the predictions of DeepONet + 9-head. (a) $\overline{u^{'}u^{'}}_{DNS}$. (b) $\overline{u^{'}u^{'}}_{9-head}$. (c) $\overline{u^{'}u^{'}}_{DNS}-\overline{u^{'}u^{'}}_{9-head}$.
  (d) $\overline{v^{'}v^{'}}_{DNS}$. (e) $\overline{v^{'}v^{'}}_{9-head}$. (f) $\overline{v^{'}v^{'}}_{DNS}-\overline{v^{'}v^{'}}_{9-head}$.
  (g) $\overline{u^{'}v^{'}}_{DNS}$. (h) $\overline{u^{'}v^{'}}_{9-head}$. (i) $\overline{u^{'}v^{'}}_{DNS}-\overline{u^{'}v^{'}}_{9-head}$.}\label{non local Re stress compare}
\end{figure*}

The Reynolds stress reflects the pulsating momentum transport in the flow field
and therefore, its three components are computed based on the $4000$ consecutive snapshots of the DNS data and the corresponding predicted snapshots, as presented in Fig.~\ref{non local Re stress compare}. 
The comparison demonstrated a high accuracy of the prediction,
even for the wake flow at the downstream of the cylinder.

\begin{figure*}[htbp] 
 \centering  
  \subfigure[]{
  \label{9headsUPSDxD1}
  \includegraphics[scale=0.30]{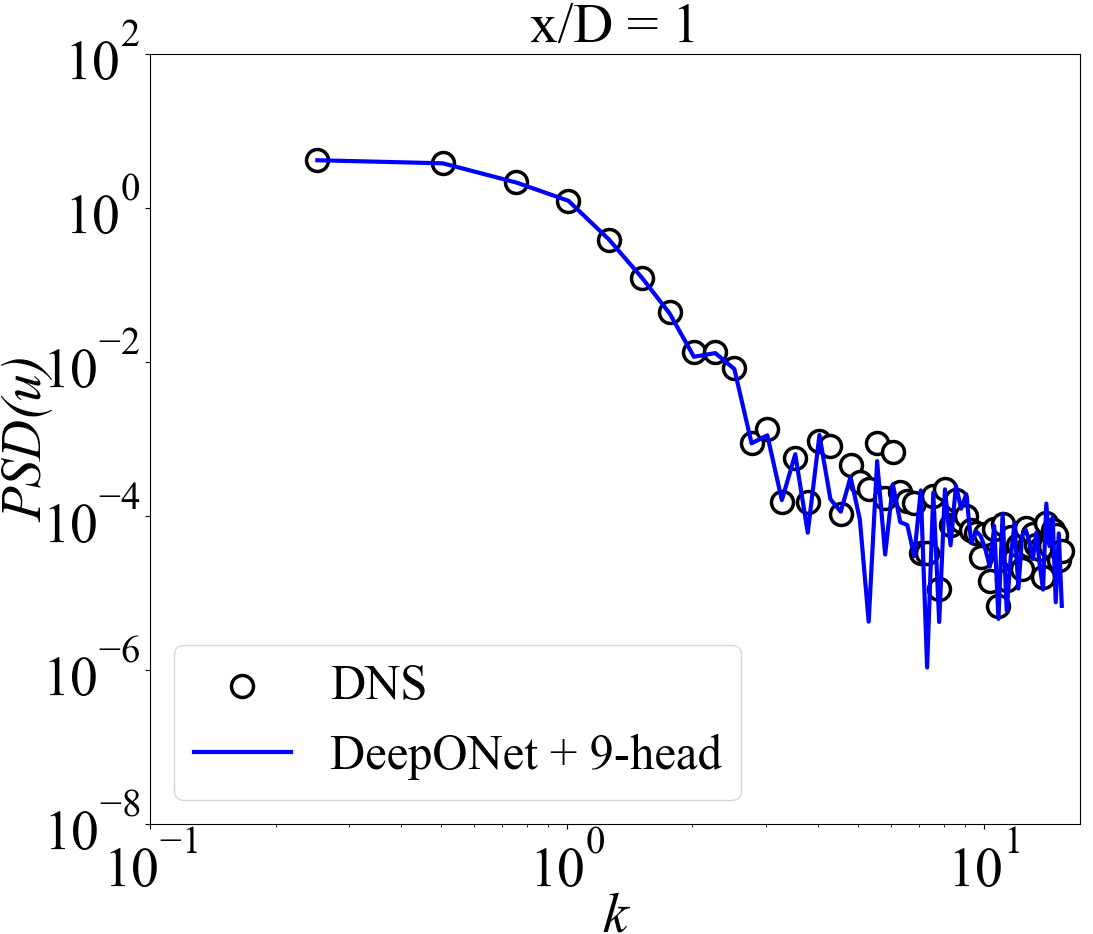}}
  \subfigure[]{
  \label{9headsUPSDxD1.5}
  \includegraphics[scale=0.30]{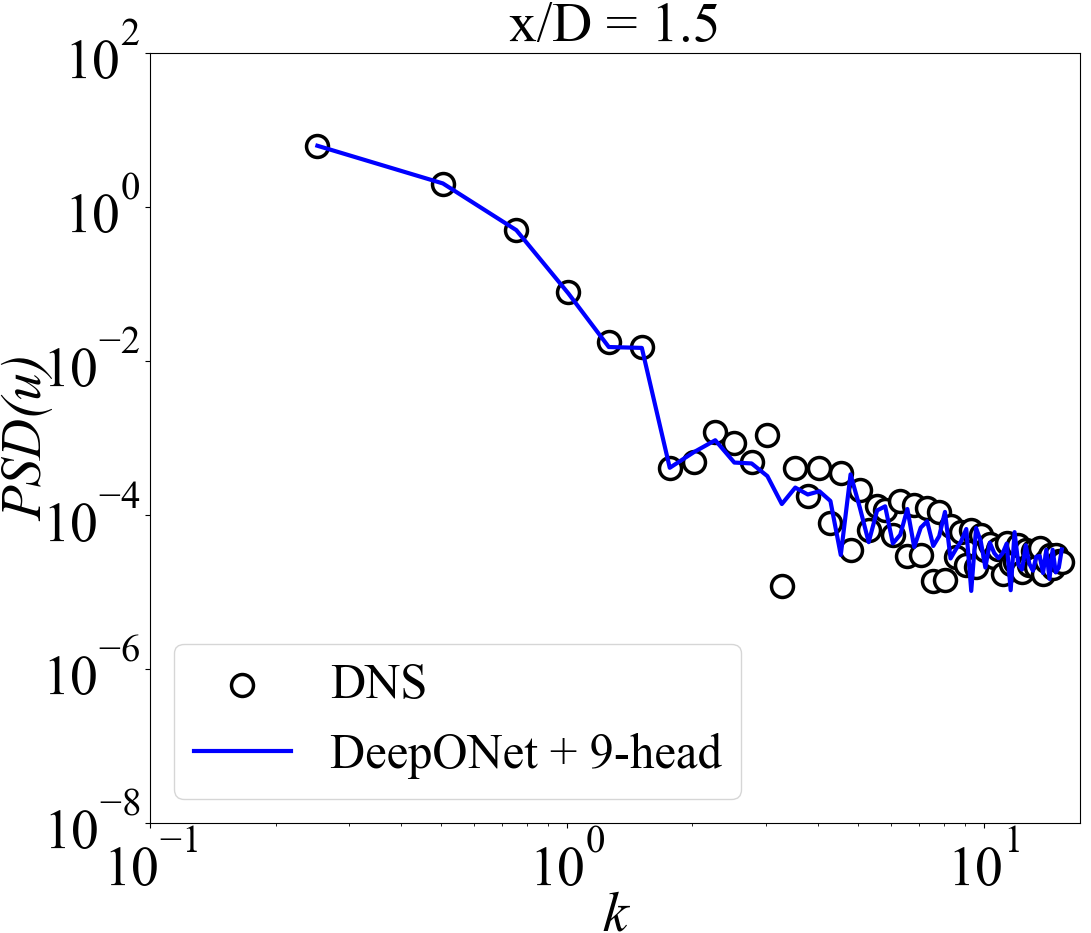}}
  \subfigure[]{
  \label{9headsUPSDxD2}
  \includegraphics[scale=0.30]{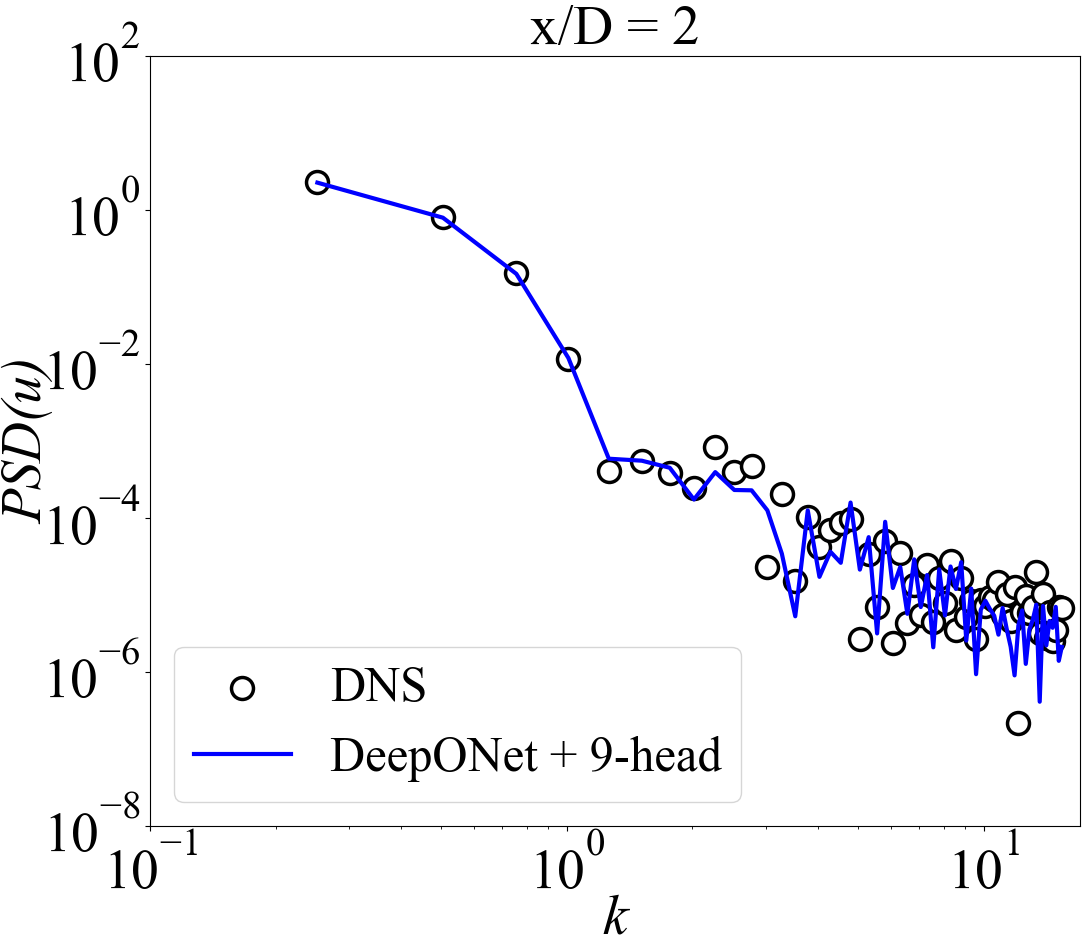}}
  \subfigure[]{
  \label{9headsUPSDxD2.5}
  \includegraphics[scale=0.30]{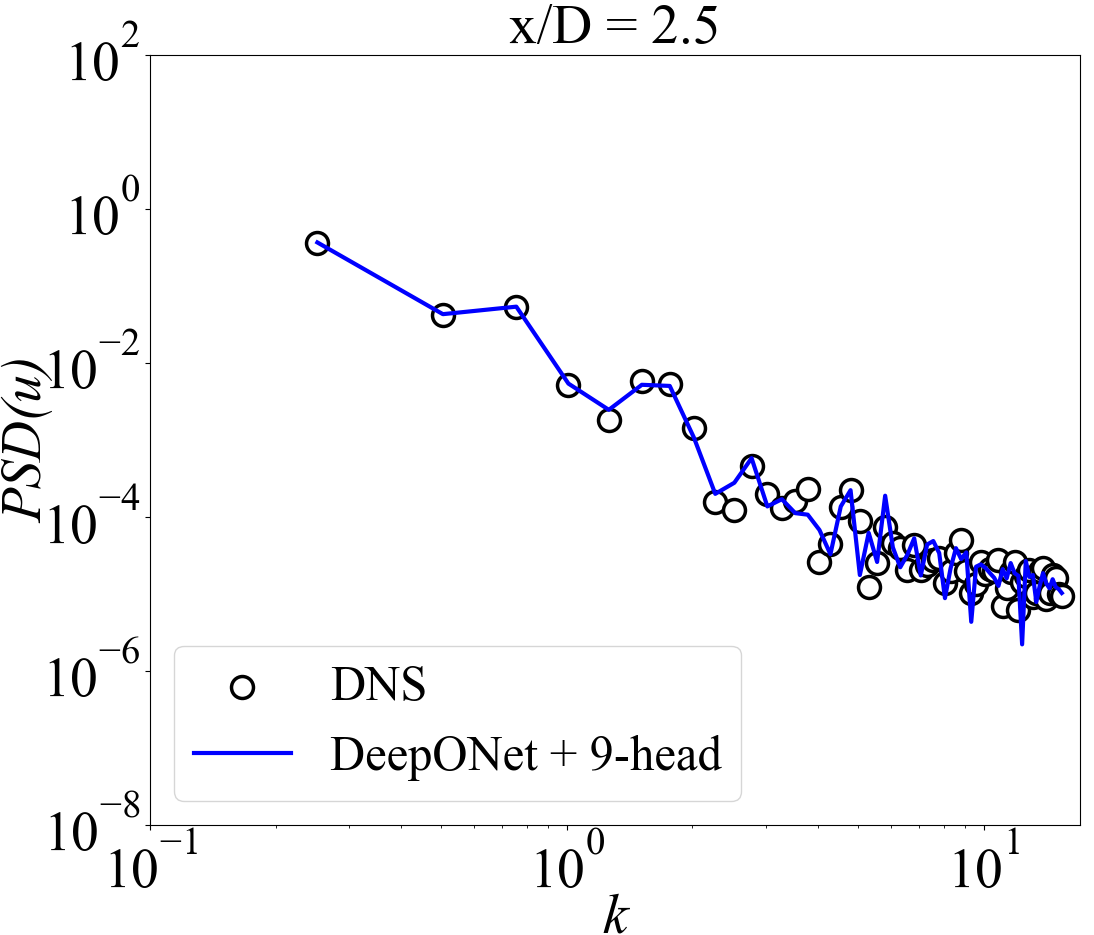}}
  \subfigure[]{
  \label{9headUPSDxD3}
  \includegraphics[scale=0.30]{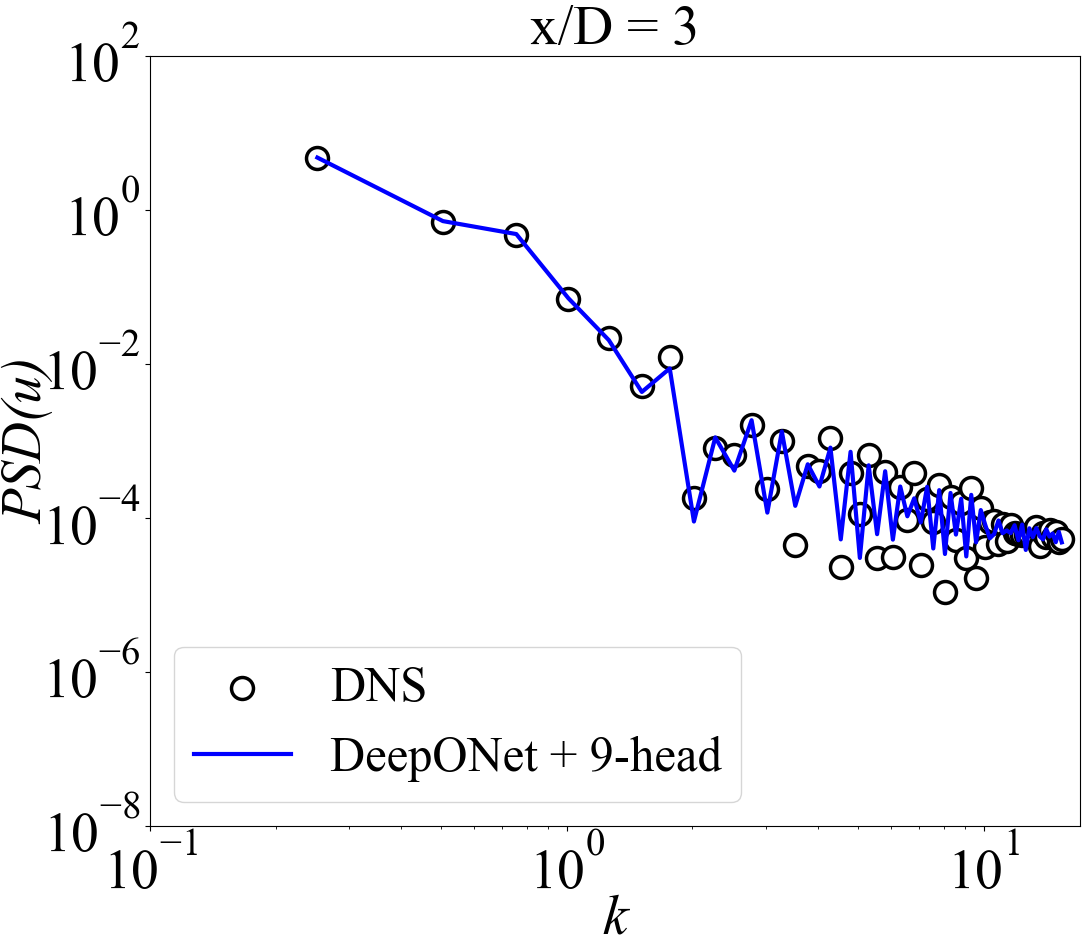}}
  \subfigure[]{
  \label{9headUPSDxD3.5}
  \includegraphics[scale=0.30]{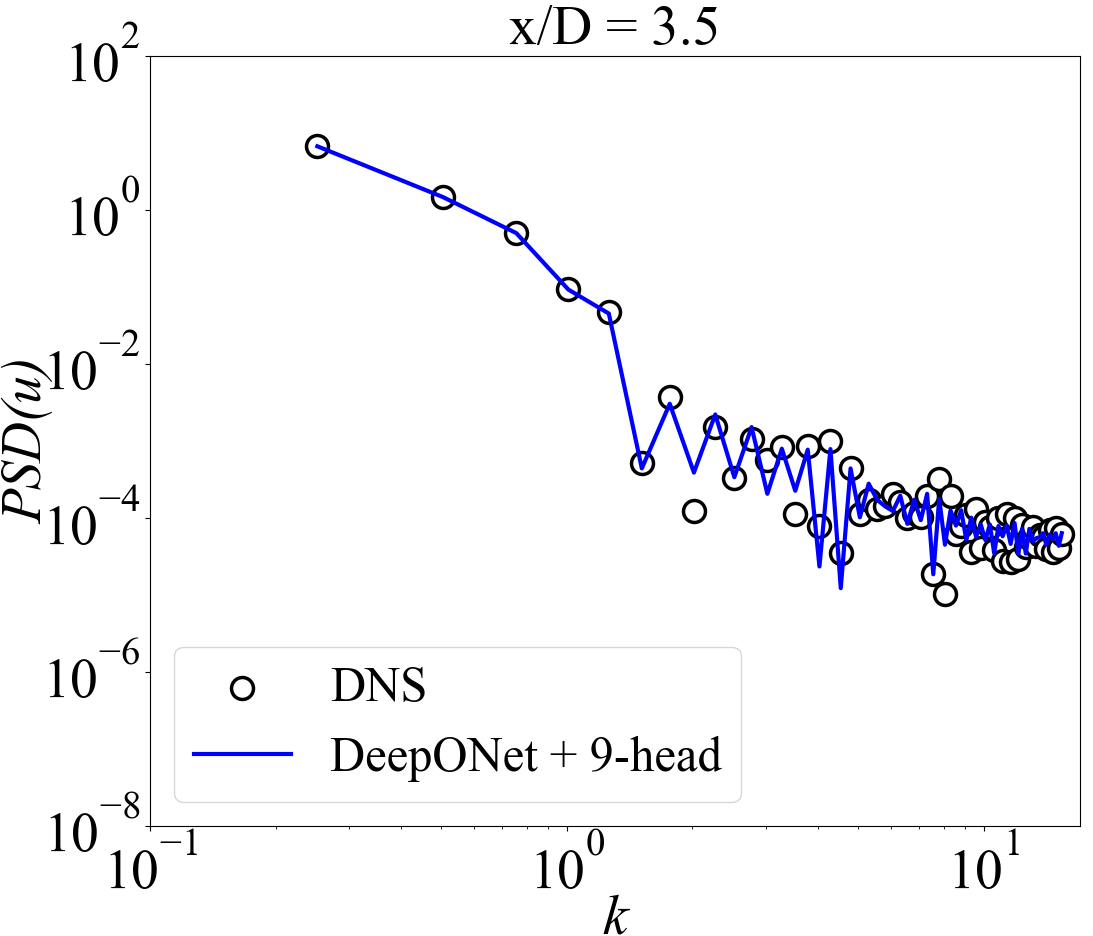}}
  \subfigure[]{
  \label{9headVPSDxD1}
  \includegraphics[scale=0.30]{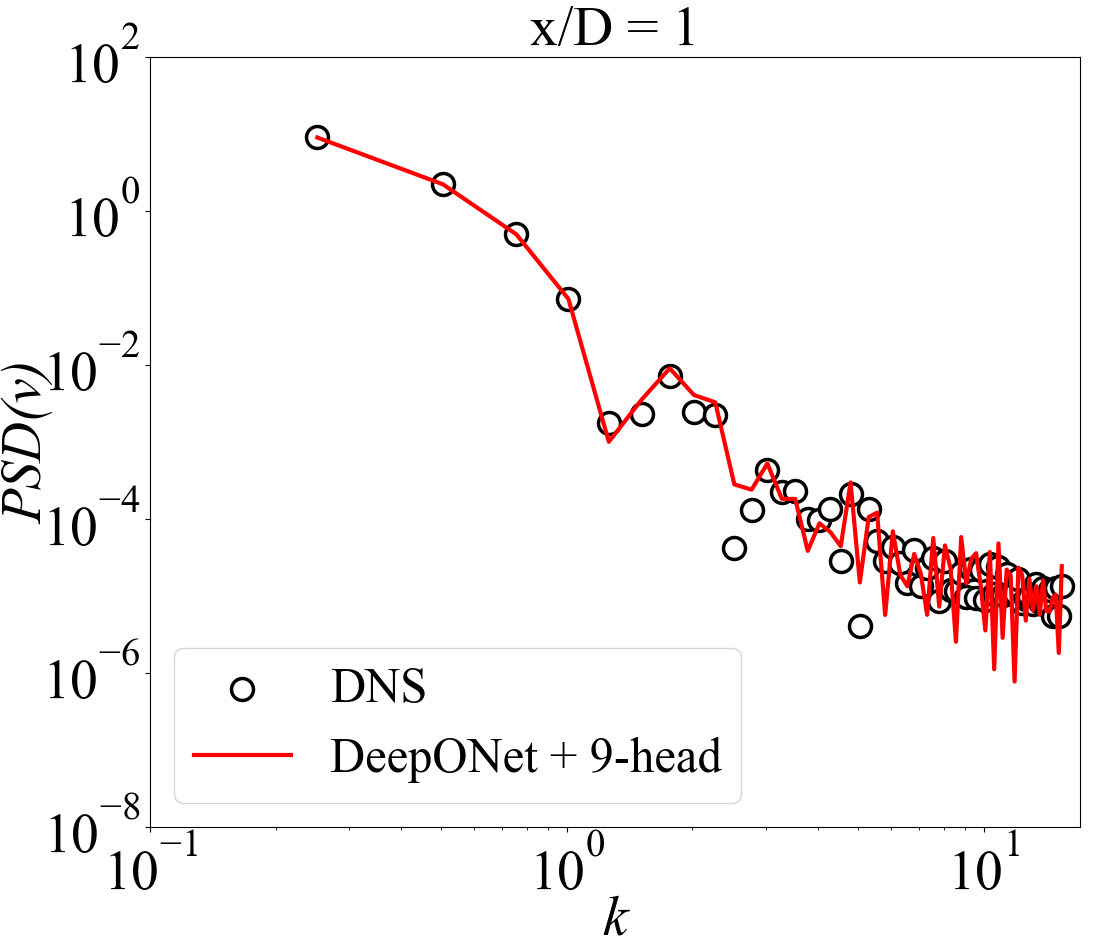}}
  \subfigure[]{
  \label{9headVPSDxD1.5}
  \includegraphics[scale=0.30]{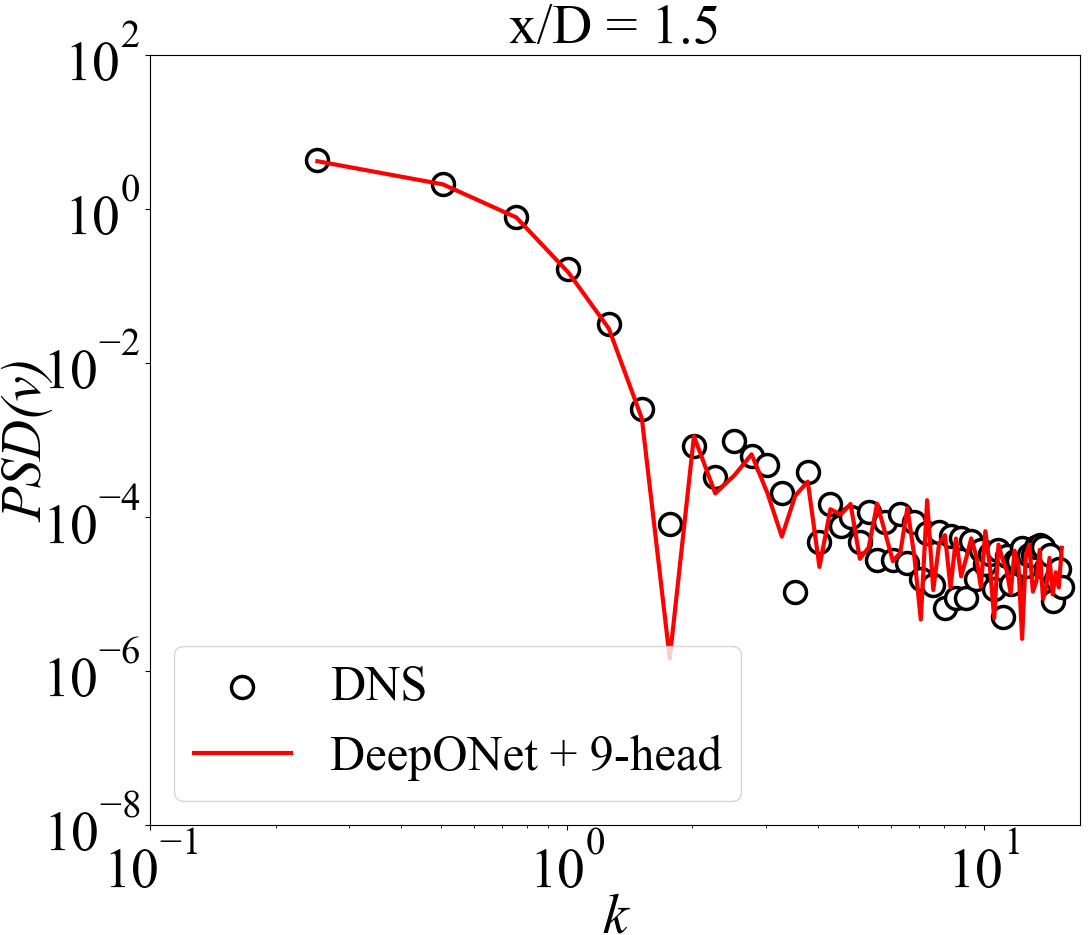}}
  \subfigure[]{
  \label{9headVPSDxD2}
  \includegraphics[scale=0.30]{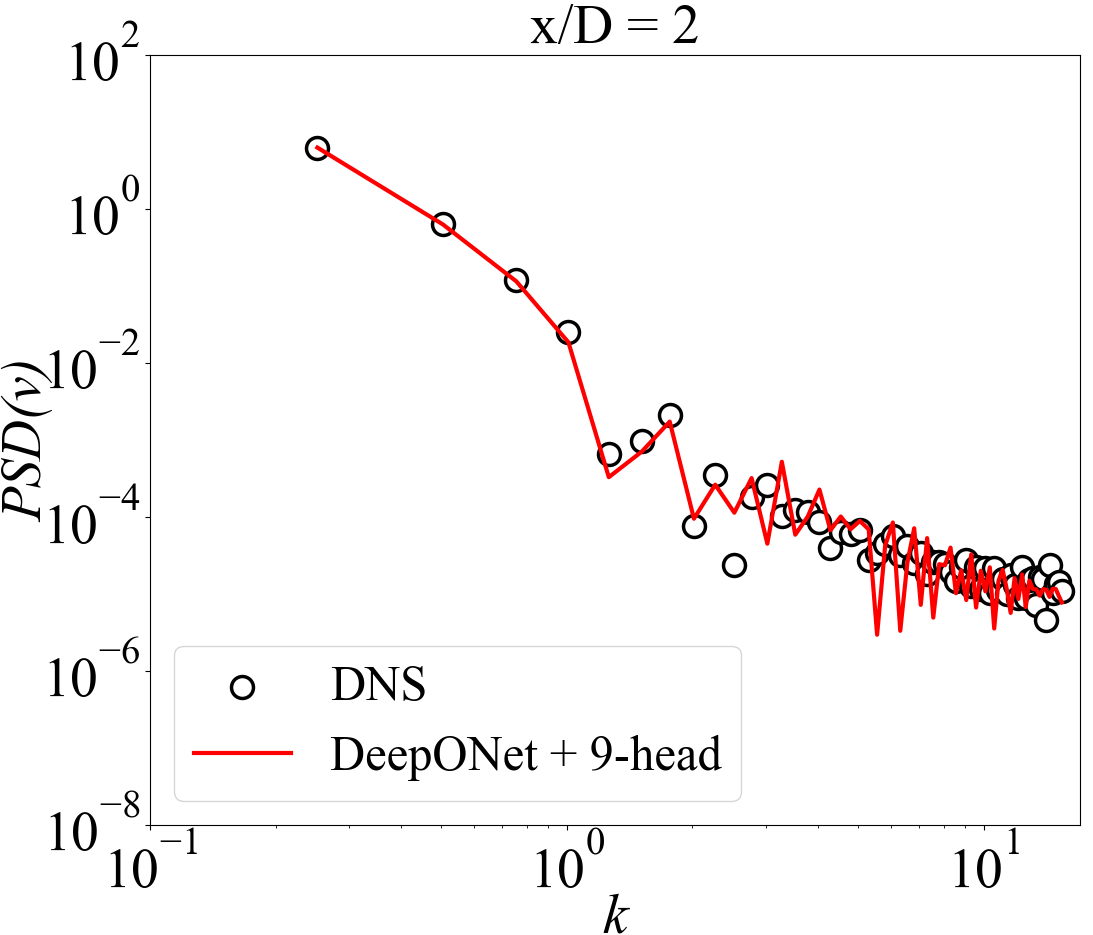}}
  \subfigure[]{
  \label{9headVPSDxD2.5}
  \includegraphics[scale=0.30]{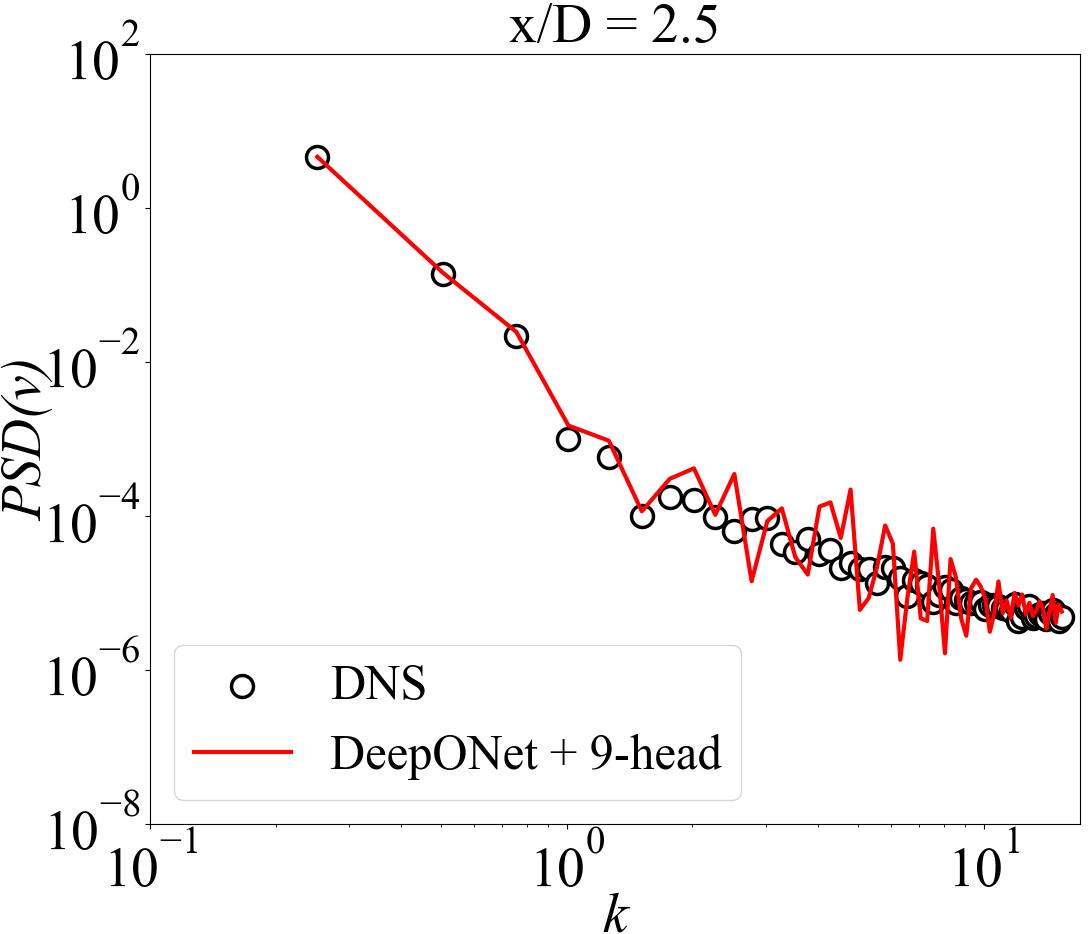}}
  \subfigure[]{
  \label{9headVPSDxD3}
  \includegraphics[scale=0.30]{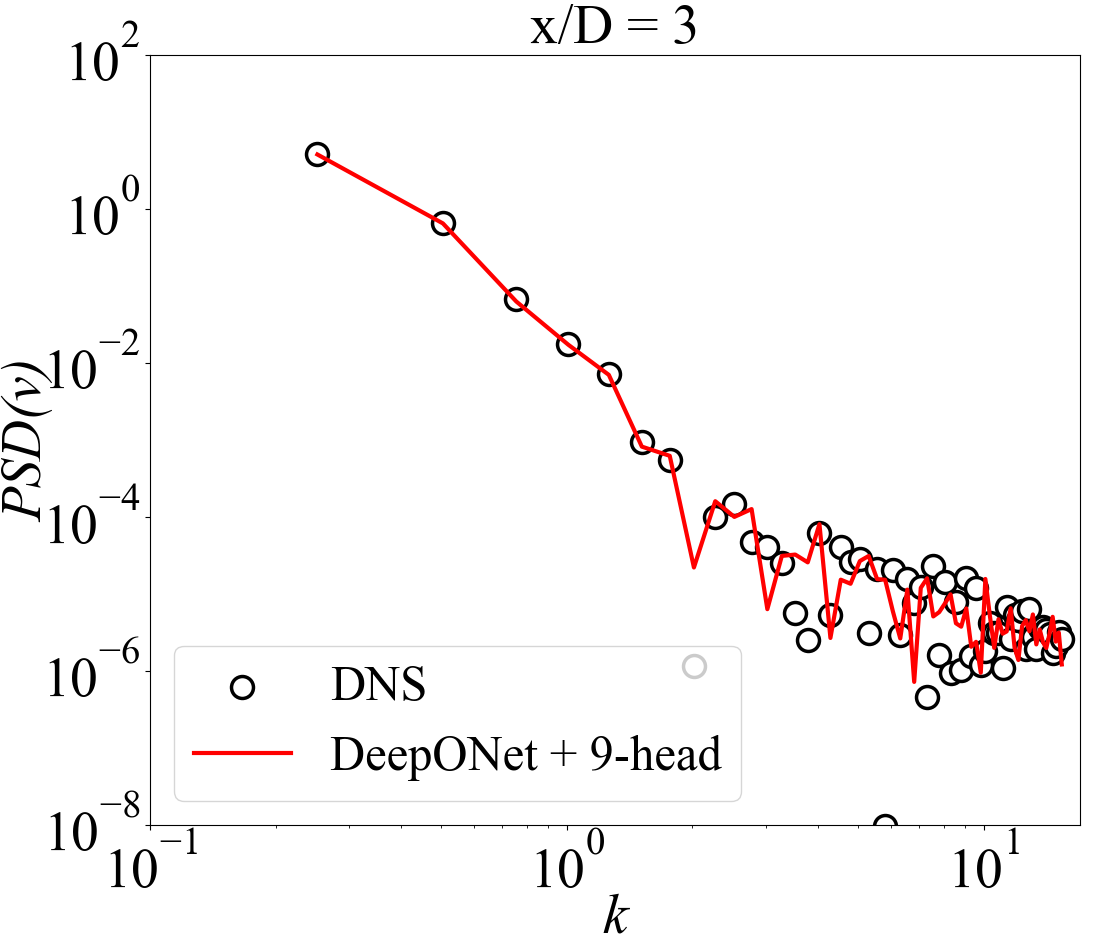}}
  \subfigure[]{
  \label{9headVPSDxD3.5}
  \includegraphics[scale=0.30]{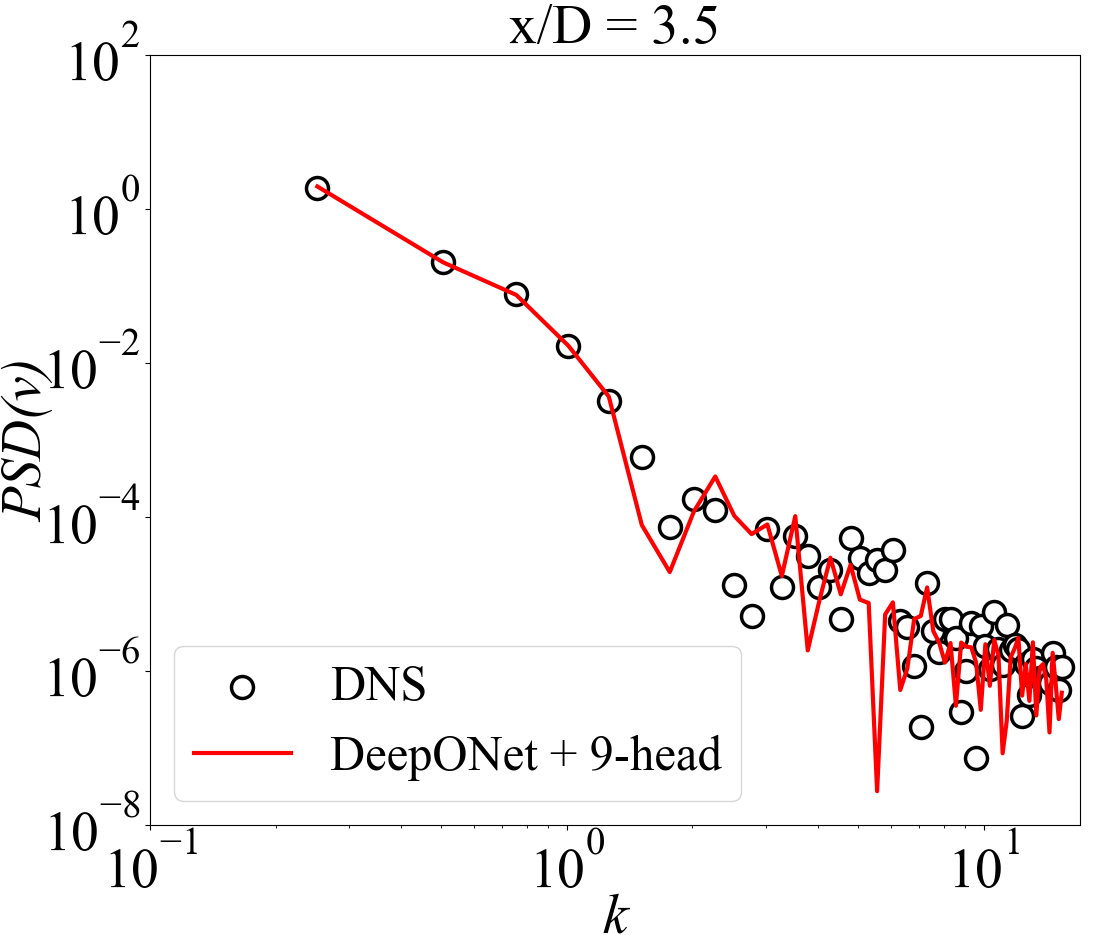}}
 \caption{The power spectral density~(PSD) of $u$ and $v$ at different downstream locations based on the DNS data and the predictions of DeepONet + 9-head at $4000th$ extrapolation snapshot. (a)-(f): $PSD(u)$ for $x/D=1$, $1.5$, $2$, $2.5$, $3$ and $3.5$. (g)-(l): $PSD(v)$ for $x/D=1$, $1.5$, $2$, $2.5$, $3$ and $3.5$. }
 \label{UVPSD non-local}
\end{figure*}

Furthermore, the accuracy of the DeepONet + multi-head non-local block is examined by computing the power spectral density (PSD) of $u$ and $v$ along the vertical axis ($y$ direction) at different downstream distances from the cylinder at the $4000th$ extrapolation snapshot using the fast Fourier transform~(FFT). The results are illustrated in Fig.~\ref{UVPSD non-local}.
The DeepONet + multi-head non-local block demonstrates a high level of accuracy in predicting the low wavenumber components of the flow field. However, "spectral bias"~\cite{xu2019spectralbiasr52} highlights a limitation in the performance of the neural network. The multi-head non-local block primarily emphasizes global information extraction, thereby failing to improve the extraction capacity of high wavenumber components.

The integration of the multi-head non-local block into the DeepONet leads to an expansion of network parameters, thereby inevitably increases the computational overhead for flow field extrapolations. The computation time of the vanilla DeepONet, DeepONet + non-local block and DeepONet + multi-head non-local block during the online forecasting phase over $4000$ extrapolation snapshots are compared in Table~\ref{computate time online forecasting}.
The enhancement improves the accuracy of flow field extrapolations,
which comes at the expense of a significantly longer computational time.
Considering the results in Fig~\ref{MSE of origin and non-local} and Table~\ref{computate time online forecasting} together:
if the accuracy is acceptable on the order of $10^{-4}$,
DeepONet + 1 head non-local block is the most cost effective setup.

\begin{table*}[htbp]
\caption{\label{tab:table1}Computation time during the online forecasting phase.}
\begin{ruledtabular}
\begin{tabular}{ccccccc}
Case  & Vanilla DeepONet & 1-head & 3-head & 5-head & 7-head & 9-head\\
\hline
Computation time  & $26s$ & $45s$ & $91s$ & $131s$ & $173s$ & $215s$\\
\end{tabular}
\label{computate time online forecasting}
\end{ruledtabular}
\end{table*}

\subsection{DeepONet with local smooth optimization technique}

Refining neural network optimization method presents another approach to enhance the performance of DeepONet.
In Section \ref{SAMGSAMsection}, the fundamental principles of the SAM and GSAM algorithms are elaborated in detail. These algorithms aim to improve the local smoothness of the loss landscape during neural network training. The objective in this section is to investigate whether integrating the SAM/GSAM algorithm enhances the performance of the DeepONet.

\subsubsection{DeepONet with SAM}

The SAM algorithm involves a crucial hyper-parameter, denoted as $\rho$, which determines the maximum perturbation~($\epsilon$) of neural network parameters utilized for calculating the sharpness, represented as $\Vert \epsilon \Vert_{2} \leq \rho$. The selection of this parameter significantly influences the effectiveness of the SAM algorithm. Therefore, this section initially compares the impact of different $\rho$ values on the extrapolation performance of DeepONet.

\begin{table}[htbp]
\caption{\label{tab:table2}Values of hyper-parameter $\rho$ in SAM algorithm.}
\begin{ruledtabular}
\begin{tabular}{cccccc}
Case  & $I$ & $II$ & $III$ & $IV$ & $V$ \\
\hline
$\rho$  & 0.00005 & 0.0001 & 0.00015 & 0.0002 & 0.0003\\
\end{tabular}
\label{rho value}
\end{ruledtabular}
\end{table}

\begin{figure*}[htbp] 
 \centering  
  \subfigure[]{
  \label{Udiffrho}
  \includegraphics[scale=0.228]{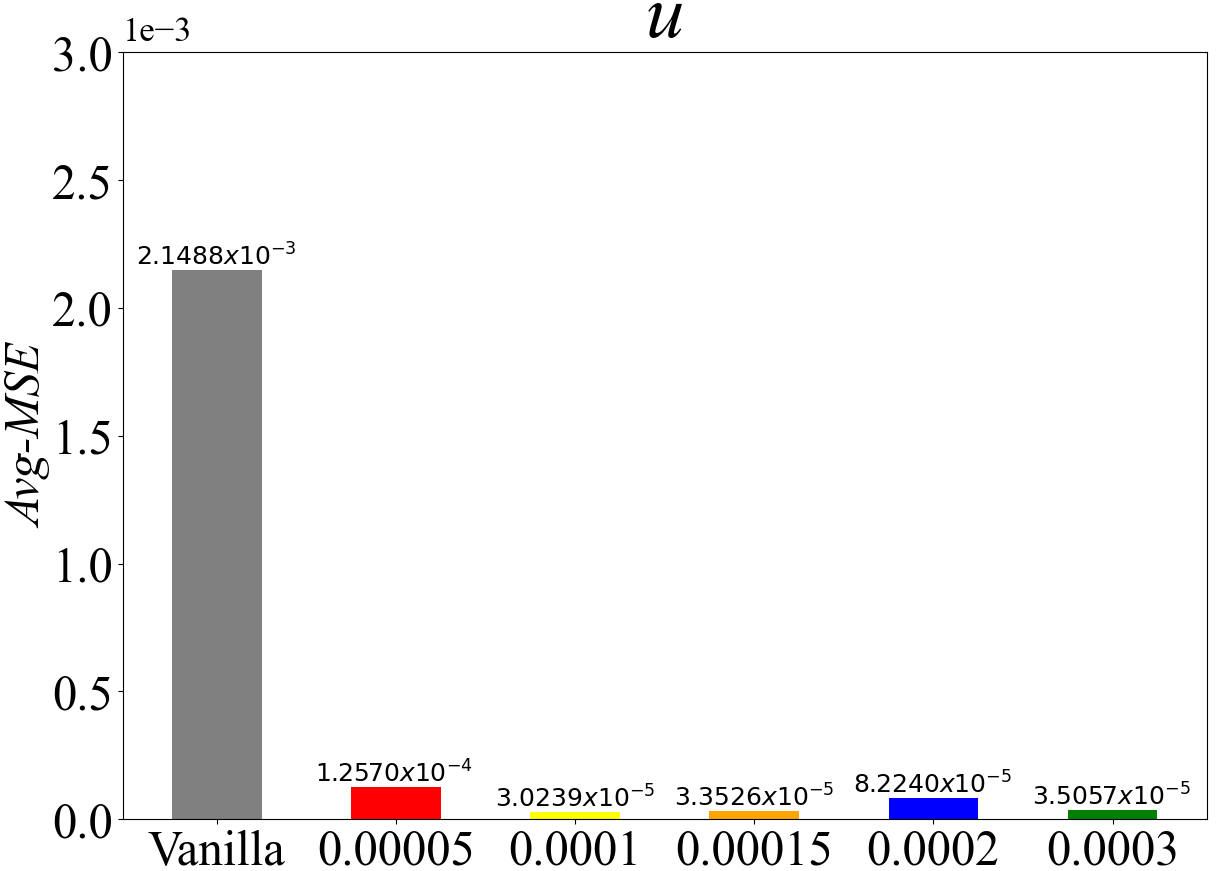}}
  \subfigure[]{
  \label{Vdiffrho}
  \includegraphics[scale=0.228]{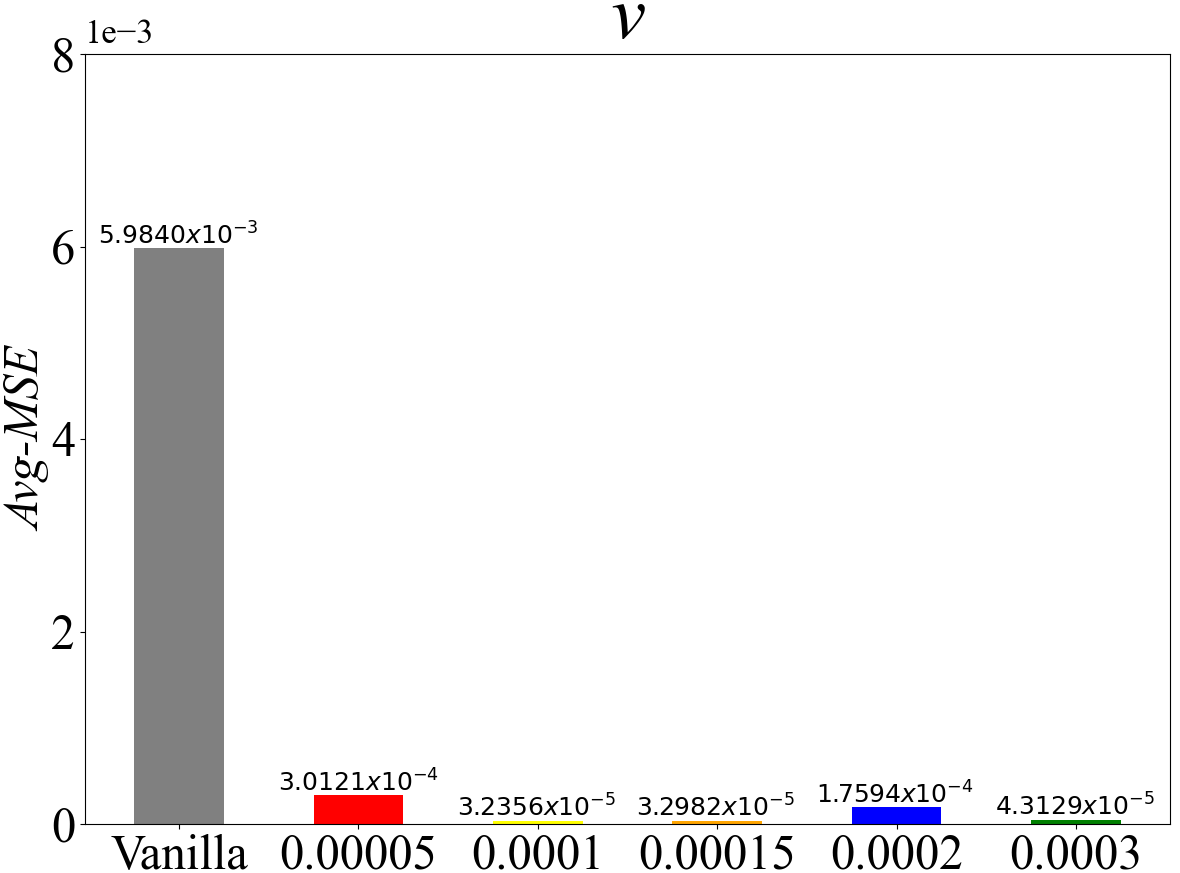}}
  \subfigure[]{
  \label{Pdiffrho}
  \includegraphics[scale=0.228]{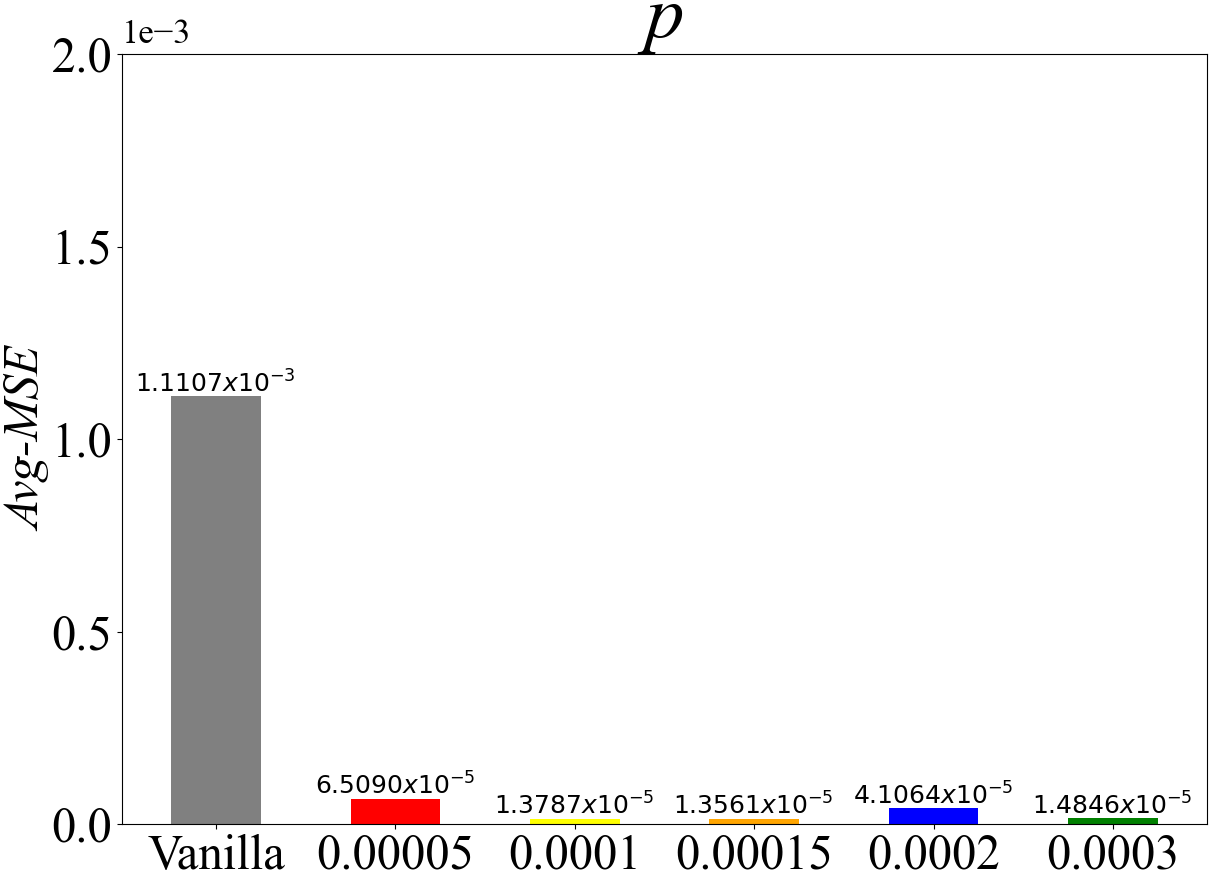}}
 \caption{The average MSEs of $u$,$v$ and $p$ between the DNS data and the predictions of DeepONet + SAM for various $\rho$ calculated from $3500th$ to $4000th$ extrapolation snapshots. (a) $u_{MSE}$. (b) $v_{MSE}$. (c) $p_{MSE}$.}
 \label{MSE diff rho}
\end{figure*}

The selected $\rho$ values are summarized in Table \ref{rho value}.
The MSEs of $u$,$v$ and $p$ between the DNS data and the predictions of DeepONet + SAM averaged from $3500th$ to $4000th$ extrapolation snapshots for various $\rho$ values are illustrated in Fig.~\ref{MSE diff rho}.
The prediction accuracy of the DeepONet + SAM significantly improves when $\rho = 0.00005$, compared to the vanilla DeepONet. This observation suggests that the SAM algorithm enhances the performance of network parameter optimization, leading to a more accurate representation of the spatial-temporal evolution of the flow field. Furthermore, increasing the value of $\rho$ results in further reduction in the prediction error of the DeepONet and superior results are achieved when using both $\rho=0.0001$ and $\rho=0.00015$. The computation results presented in Fig.~\ref{MSE diff rho} support the utilization of $\rho=0.0001$ in the latter part of this work.

The only distinction between the DeepONet + SAM and the vanilla DeepONet lies in their offline training phase, while their online forecasting phase is identical. Thus, the DeepONet + SAM exhibits notably faster prediction speed than DeepONet + multi-head non-local block, as already demonstrated in Table \ref{computate time online forecasting}.

\begin{figure*}[htbp] 
 \centering  
  \subfigure[]{
  \label{UDNSSAM}
  \includegraphics[scale=0.178]{Utrue4000step.png}}
  \subfigure[]{
  \label{UpreSAM}
  \includegraphics[scale=0.178]{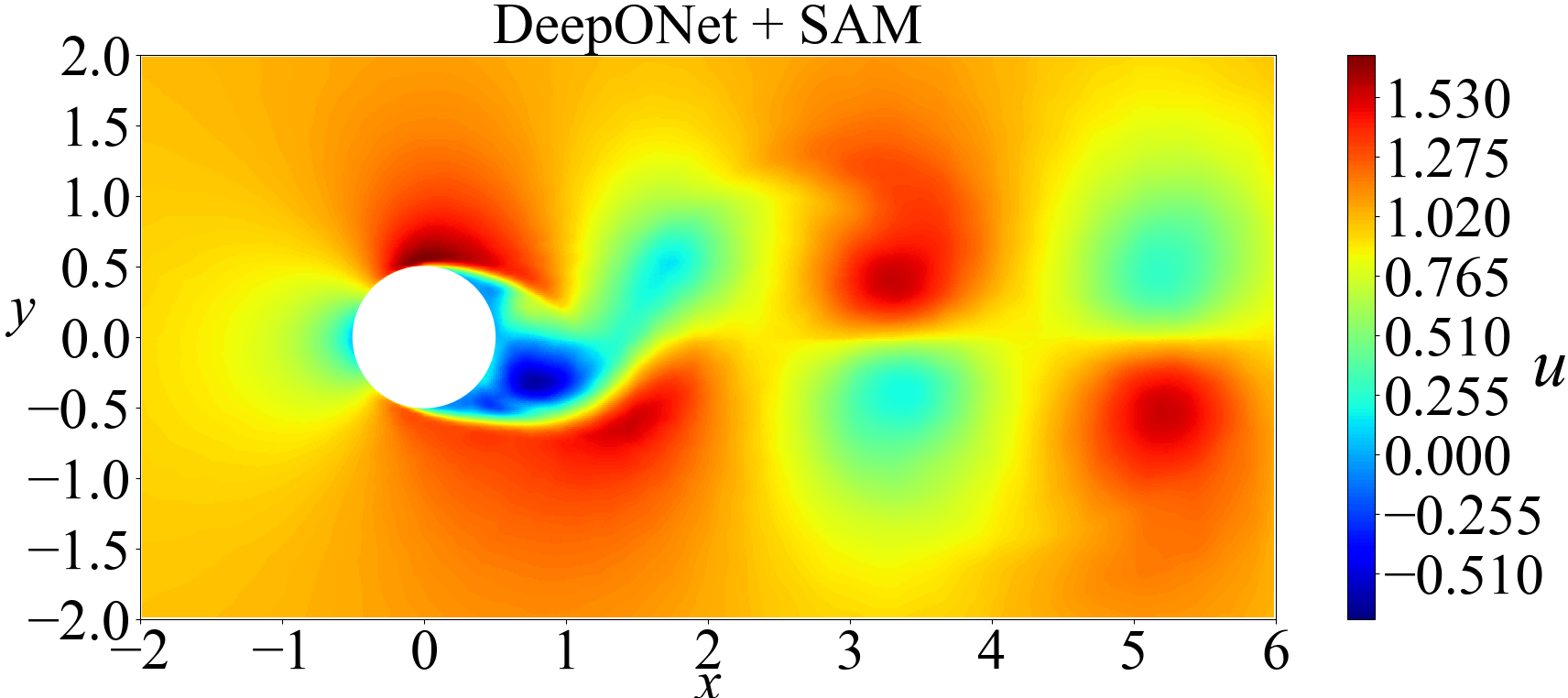}}
  \subfigure[]{
  \label{UdeltaSAM}
  \includegraphics[scale=0.178]{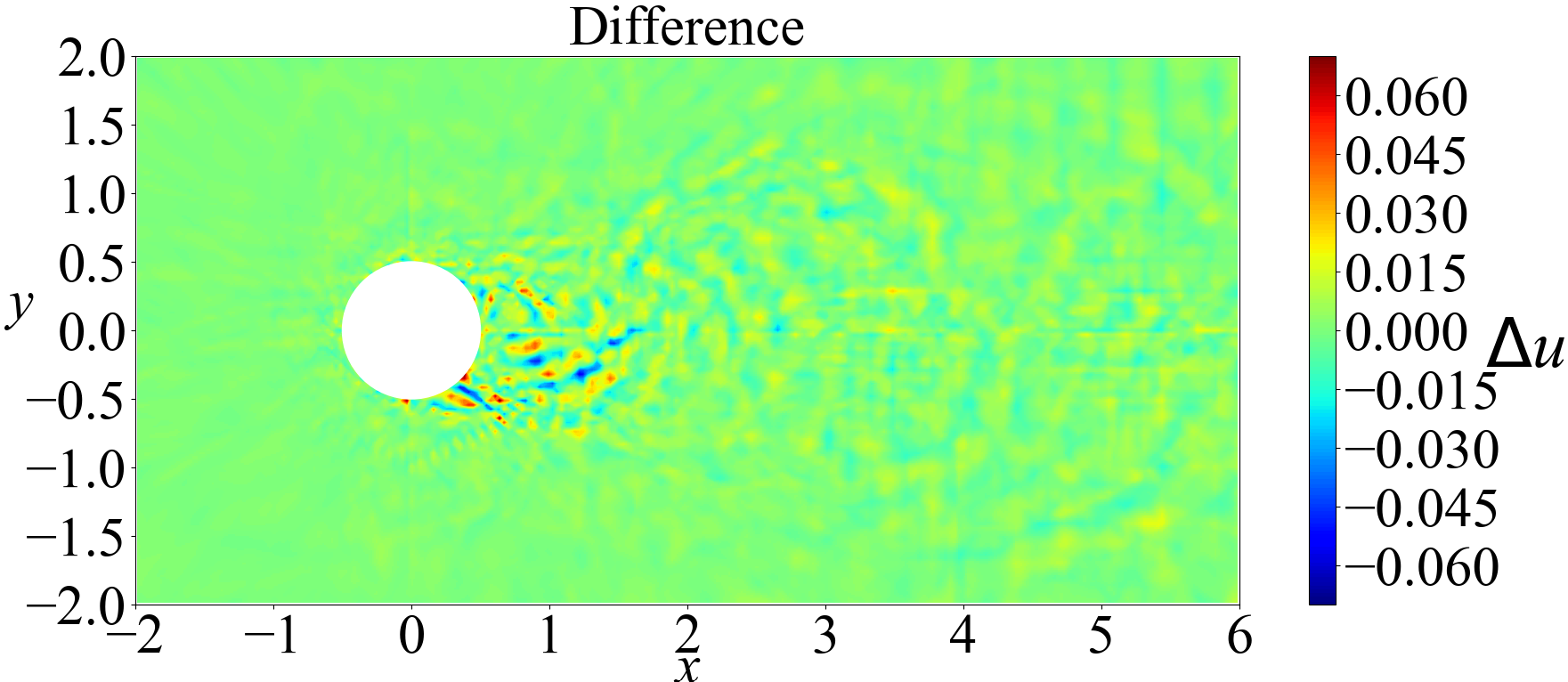}}
  \subfigure[]{
  \label{VDNSSAM}
  \includegraphics[scale=0.178]{Vtrue4000step.png}}
  \subfigure[]{
  \label{VpreSAM}
  \includegraphics[scale=0.178]{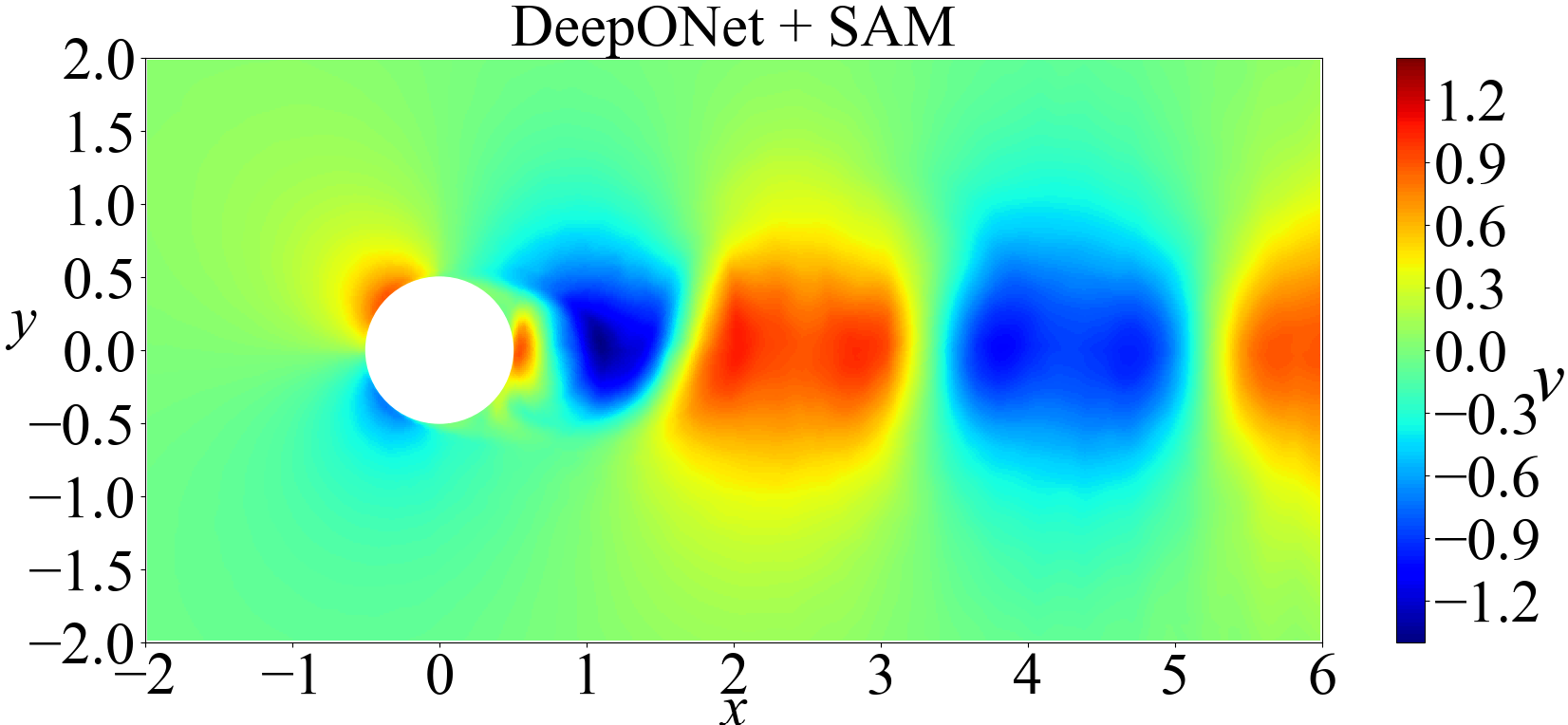}}
  \subfigure[]{
  \label{VdeltaSAM}
  \includegraphics[scale=0.178]{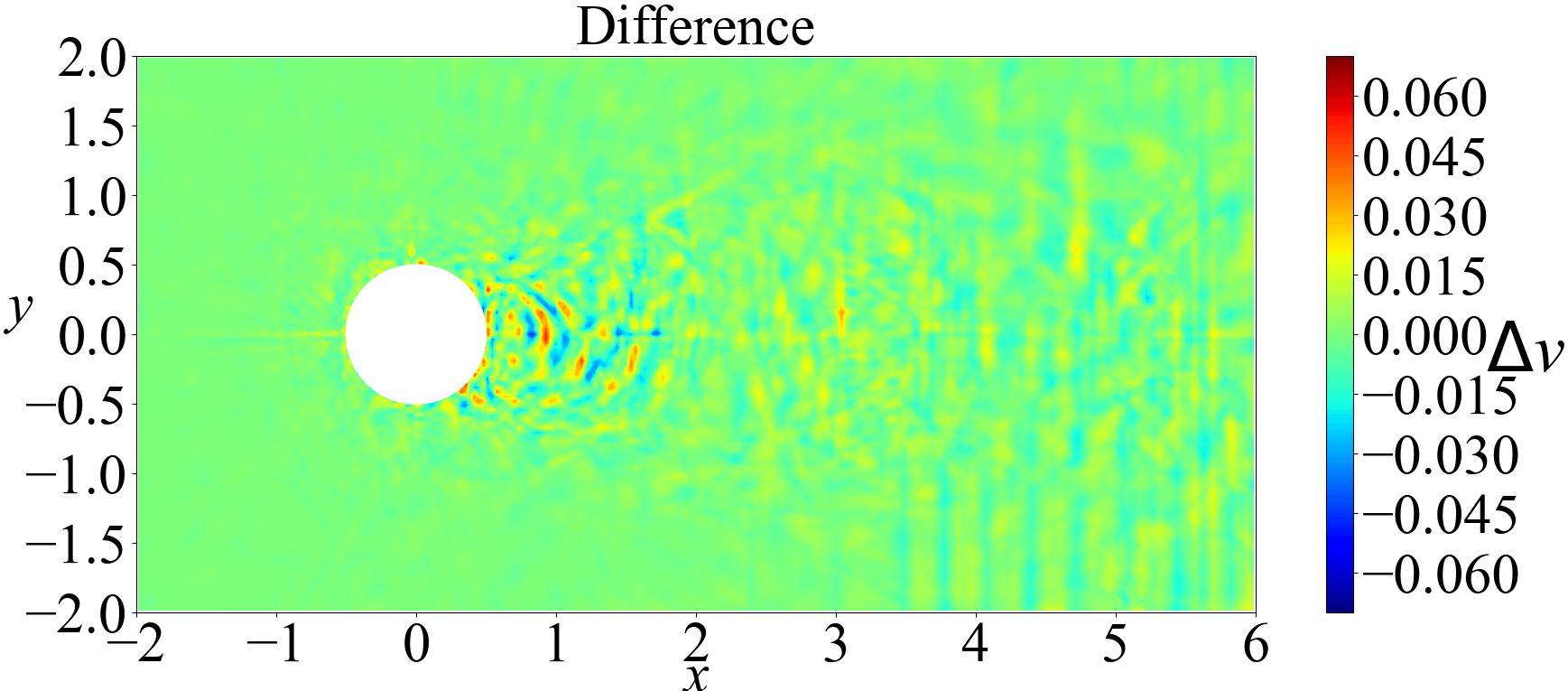}}
  \subfigure[]{
  \label{PDNSSAM}
  \includegraphics[scale=0.178]{Ptrue4000step.png}}
  \subfigure[]{
  \label{PpreSAM}
  \includegraphics[scale=0.178]{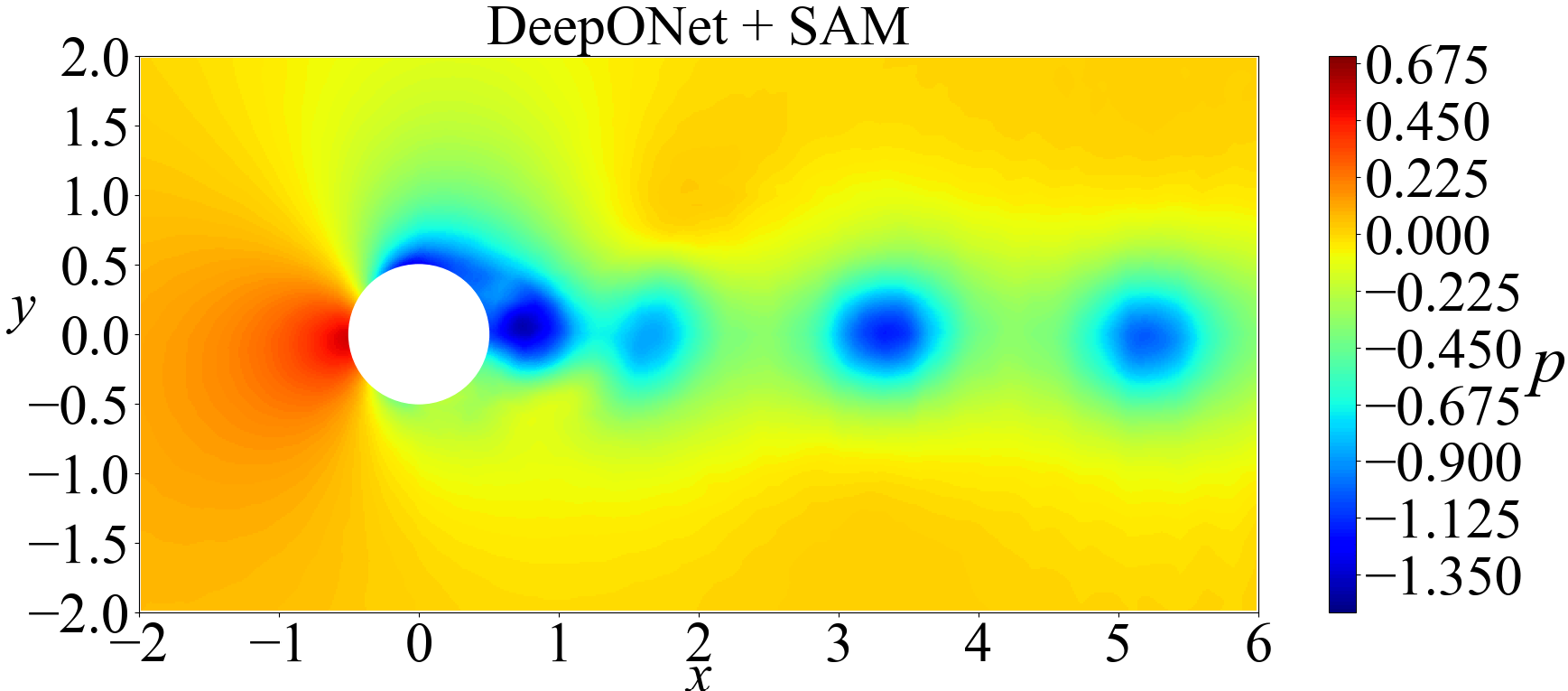}}
  \subfigure[]{
  \label{PdeltaSAM}
  \includegraphics[scale=0.178]{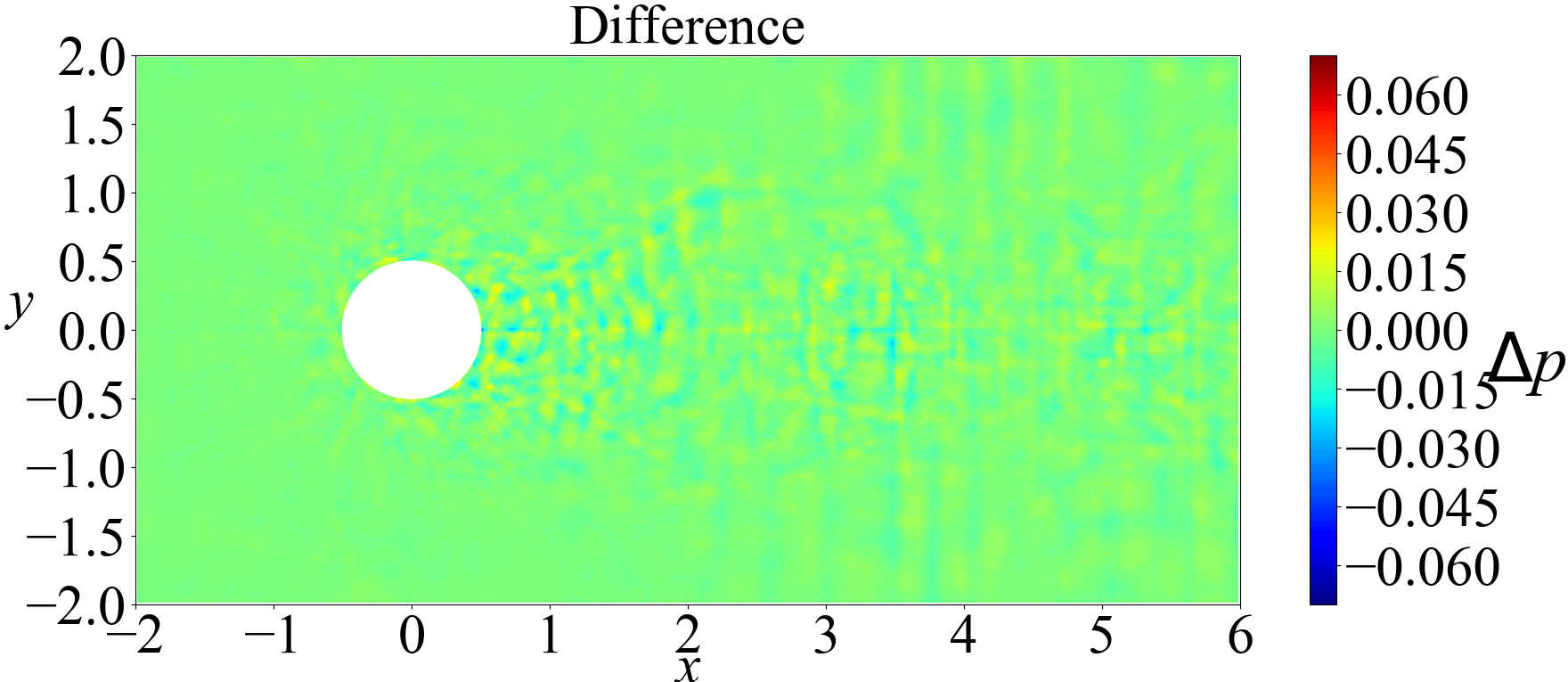}}
  \caption{Comparison of $u$, $v$ and $p$ at $4000th$ snapshot between the DNS data and the prediction of DeepONet + SAM. (a) $u_{DNS}$. (b) $u_{pre}$. (c) $u_{DNS}-u_{pre}$. (d) $v_{DNS}$. (e) $v_{pre}$. (f) $v_{DNS}-v_{pre}$. (g) $p_{DNS}$. (h) $p_{pre}$. (i) $p_{DNS}-p_{pre}$.}\label{SAMcontour compare}
\end{figure*}

The flow field of the $4000th$ extrapolation snapshot based on DeepONet + SAM, are depicted in Fig.~\ref{SAMcontour compare}. These results show similarities to the scenario using DeepONet + 9-head as depicted in Fig.~\ref{9headscontour compare}. The predictions of $u$, $v$ and $p$ exhibit a high level of consistency with the DNS data as illustrated in Fig.~\ref{UdeltaSAM}, \ref{VdeltaSAM} and \ref{PdeltaSAM}.

\begin{figure*}[htbp] 
 \centering  
  \subfigure[]{
  \label{UMSElineSAM}
  \includegraphics[scale=0.255]{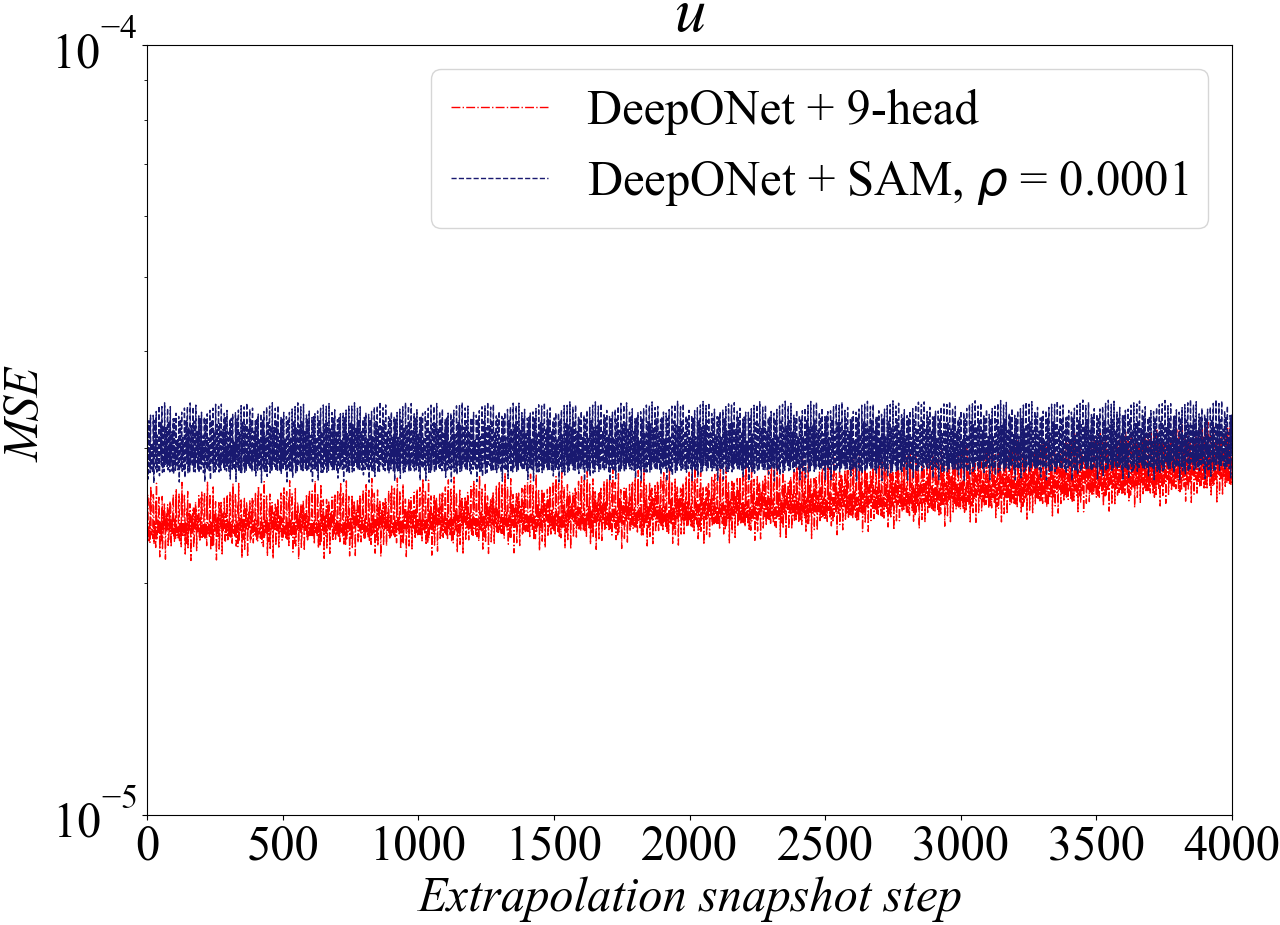}}
  \subfigure[]{
  \label{VMSElineSAM}
  \includegraphics[scale=0.255]{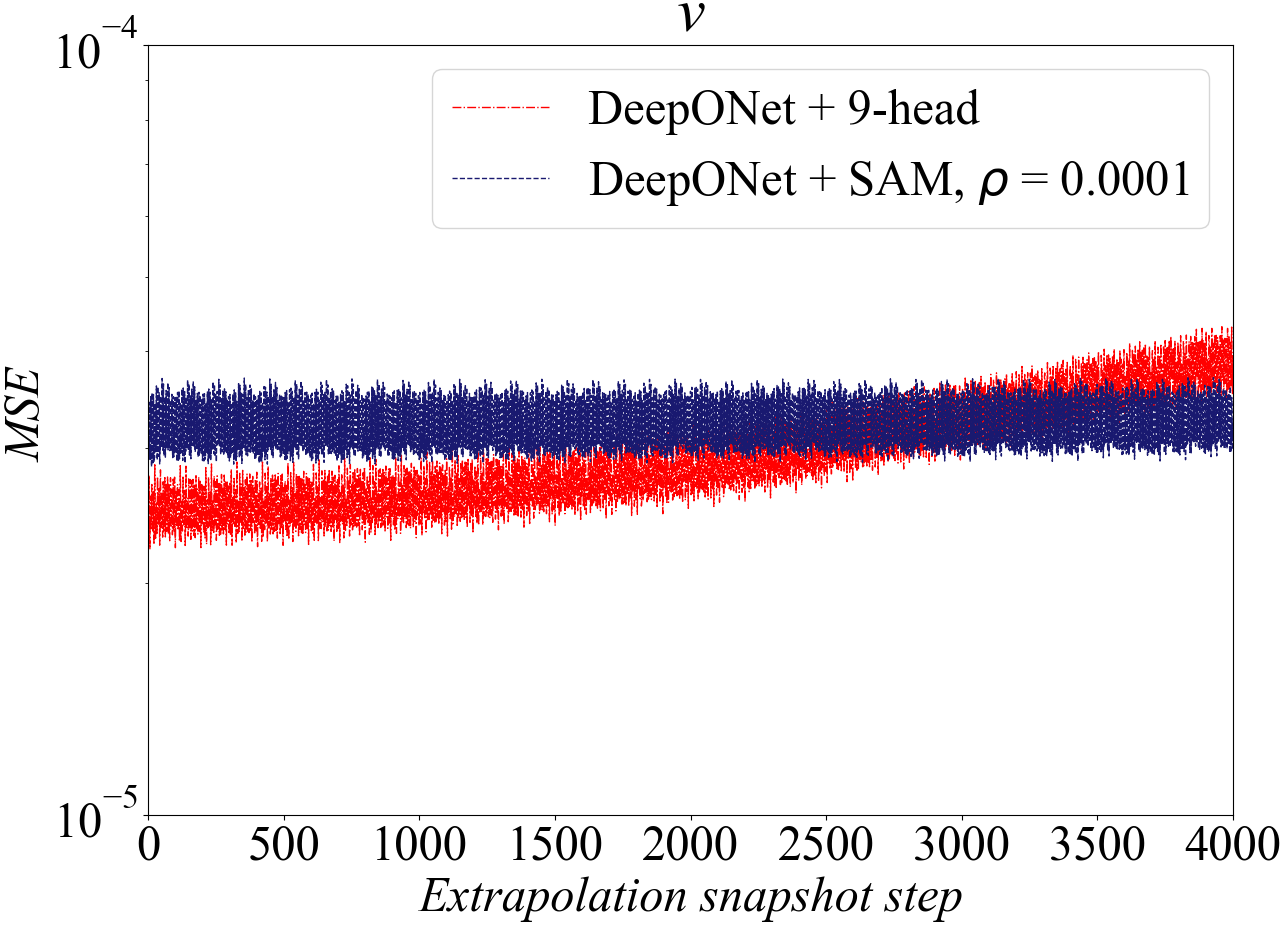}}
  \subfigure[]{
  \label{PMSElineSAM}
  \includegraphics[scale=0.255]{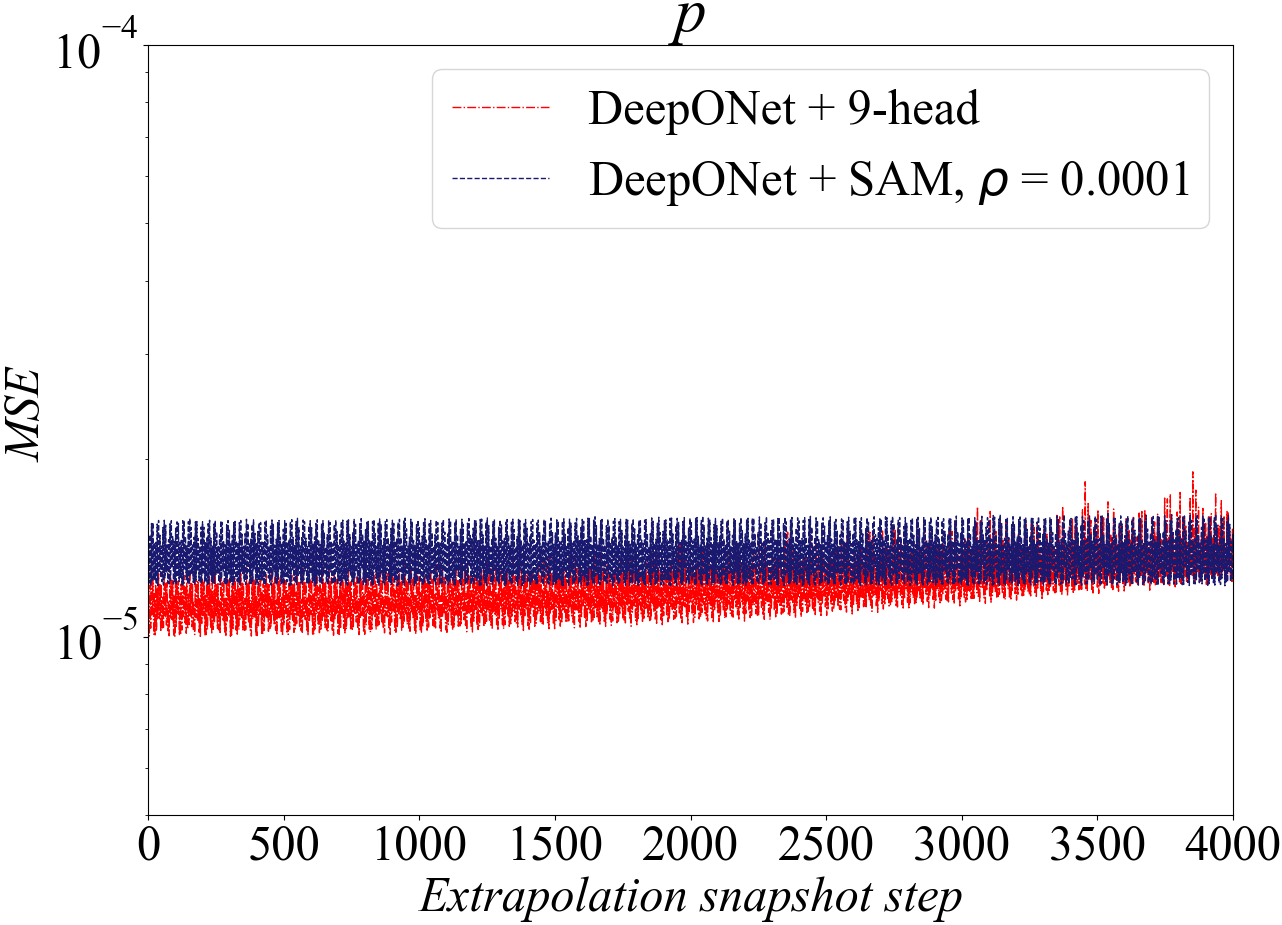}}
 \caption{The evolution of MSEs between the DNS data and the predictions of DeepONet + SAM and DeepONet + 9-head. (a) $u_{MSE}$. (b) $v_{MSE}$. (c) $p_{MSE}$.}
 \label{MSE line of SAM}
\end{figure*}

Furthermore, the evolution of the MSEs between the DNS data and the predictions of DeepONet + SAM over $4000$ extrapolation snapshots are depicted in Fig.~\ref{MSE line of SAM}, along with the corresponding results of DeepONet + 9-head.
The flow fields such as $u$, $v$, and $p$ from DeepONet + SAM do not exhibit apparent accumulation of errors with increasing the number of extrapolation snapshots.
The magnitudes of errors from DeepONet+SAM and DeepONet + 9-head
are comparable, which indicates their equivalent effectiveness in enhancing the accuracy of neural network predictions. 
These results also demonstrate that despite the constraints imposed by the convolution kernel size in CNN, enhancing the optimization approach can still enhance the ability of CNN to extract flow field information. 
It should be noted that although the initial error of the extrapolation predicted by the DeepONet + 9-head is slightly lower, the error accumulation becomes more noticeable as the extrapolation snapshot increases.

\subsubsection{DeepONet with GSAM}
The SAM algorithm does not guarantee a low level of sharpness. To address this issue, the GSAM algorithm is introduced, which is thoroughly explained in section \ref{SAMGSAMsection}. In this section, the GSAM algorithm is combined with Adam optimizer to optimize the neural network parameters of the DeepONet and explore the potential for further enhancements in predicting spatial-temporal evolution of flow fields. $\rho = 0.0001$ is utilized in this section.

\begin{figure*}[htbp] 
 \centering  
  \subfigure[]{
  \label{Udiffalpha}
  \includegraphics[scale=0.27]{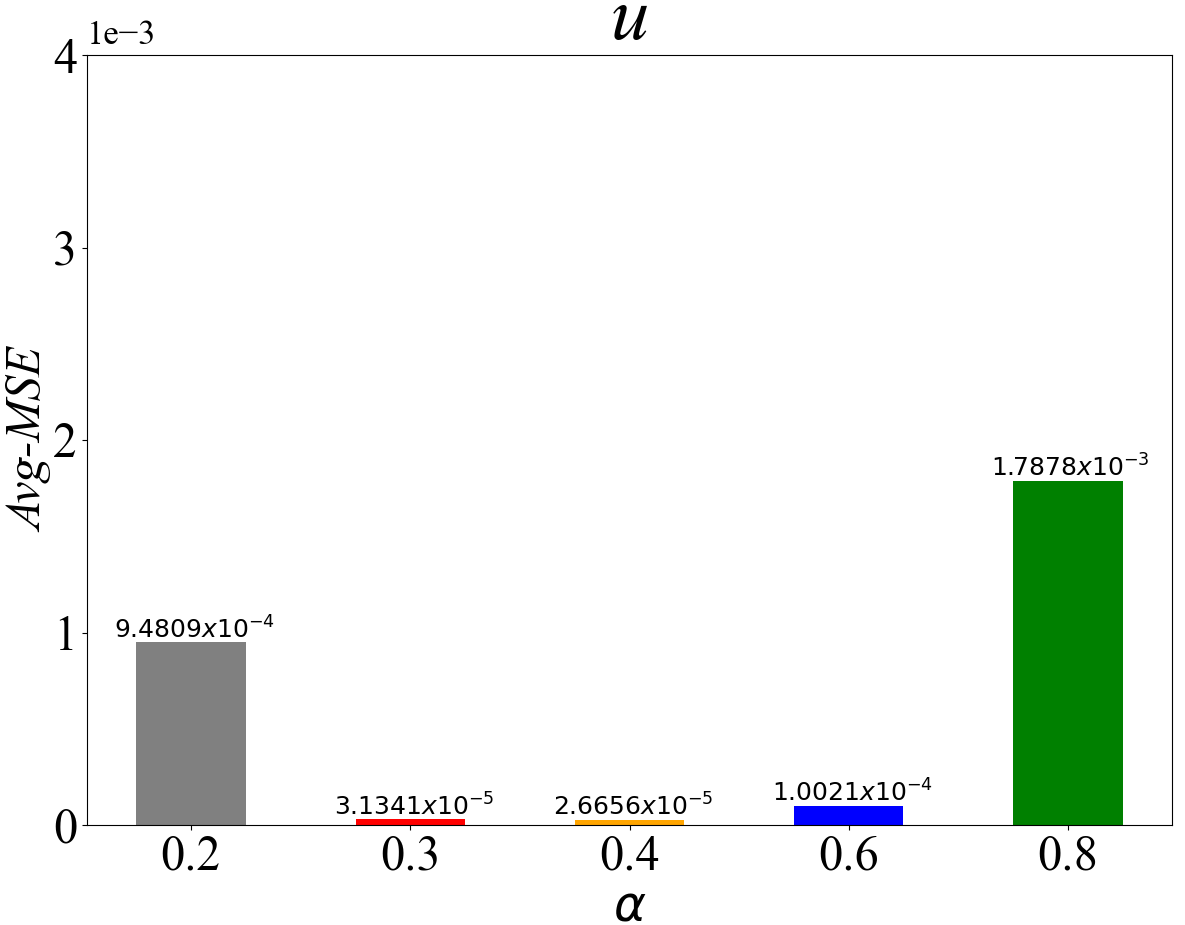}}
  \subfigure[]{
  \label{Vdiffalpha}
  \includegraphics[scale=0.27]{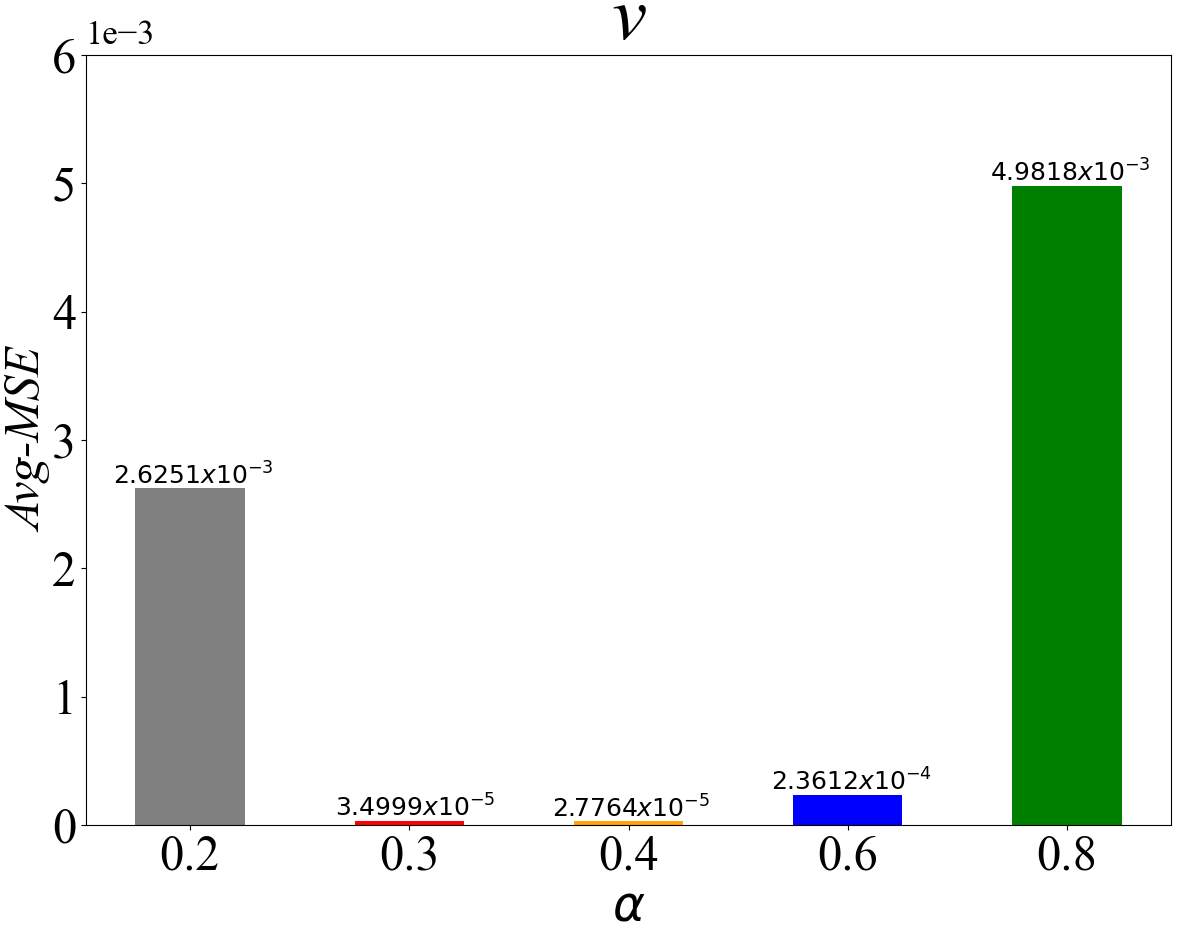}}
  \subfigure[]{
  \label{Pdiffalpha}
  \includegraphics[scale=0.27]{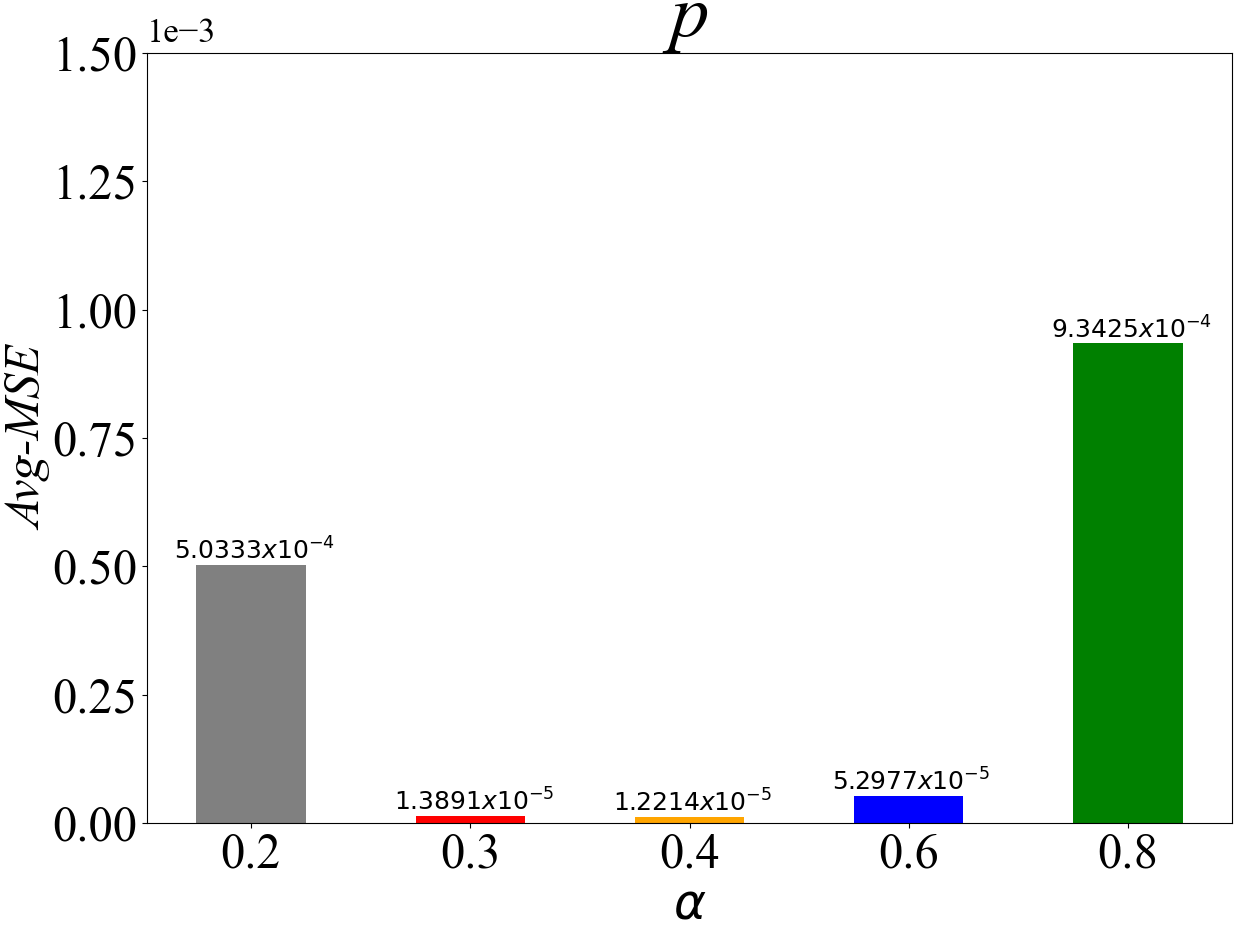}}
 \caption{The MSEs of $u$,$v$ and $p$ between the DNS data and the predictions of DeepONet + GSAM for various $\alpha$ averaged from $3500th$ to $4000th$ extrapolation snapshots. (a) $u_{MSE}$. (b) $v_{MSE}$. (c) $p_{MSE}$.}
 \label{MSE diff alpha}
\end{figure*}

The GSAM algorithm incorporates a regularization term known as the surrogate gap into the loss function of the SAM algorithm, as depicted in Eq.~(\ref{GSAMbasic}). The coefficient $\alpha$ plays a crucial role in determining the effectiveness of the GSAM algorithm. Therefore, an extensive parameter analysis is conducted to determine the optimal value of $\alpha$, and its specific values are summarized in Table \ref{alpha value}.
The MSEs of $u$, $v$ and $p$ between the DNS data and the predictions of DeepONet + GSAM averaged from $3500th$ to $4000th$ extrapolation snapshots for various $\alpha$ values are illustrated in Fig.~\ref{MSE diff alpha}.
Notably, when $\alpha$ is assigned with values of $0.3$ and $0.4$, the prediction errors for $u$, $v$, and $p$ are relatively small with MSEs on the order of $10^{-5}$. Based on these results, $\alpha=0.4$ is consistently utilized throughout this study. 

\begin{table}[htbp]
\caption{\label{tab:table3}Values of hyper-parameter $\alpha$ in GSAM algorithm.}
\begin{ruledtabular}
\begin{tabular}{cccccc}
Case  & $I$ & $II$ & $III$ & $IV$ & $V$ \\
\hline
$\alpha$  & 0.2 & 0.3 & 0.4 & 0.6 & 0.8\\
\end{tabular}
\label{alpha value}
\end{ruledtabular}
\end{table}

\begin{figure*}[htbp] 
 \centering  
  \subfigure[]{
  \label{0.4UMSElineGSAM}
  \includegraphics[scale=0.255]{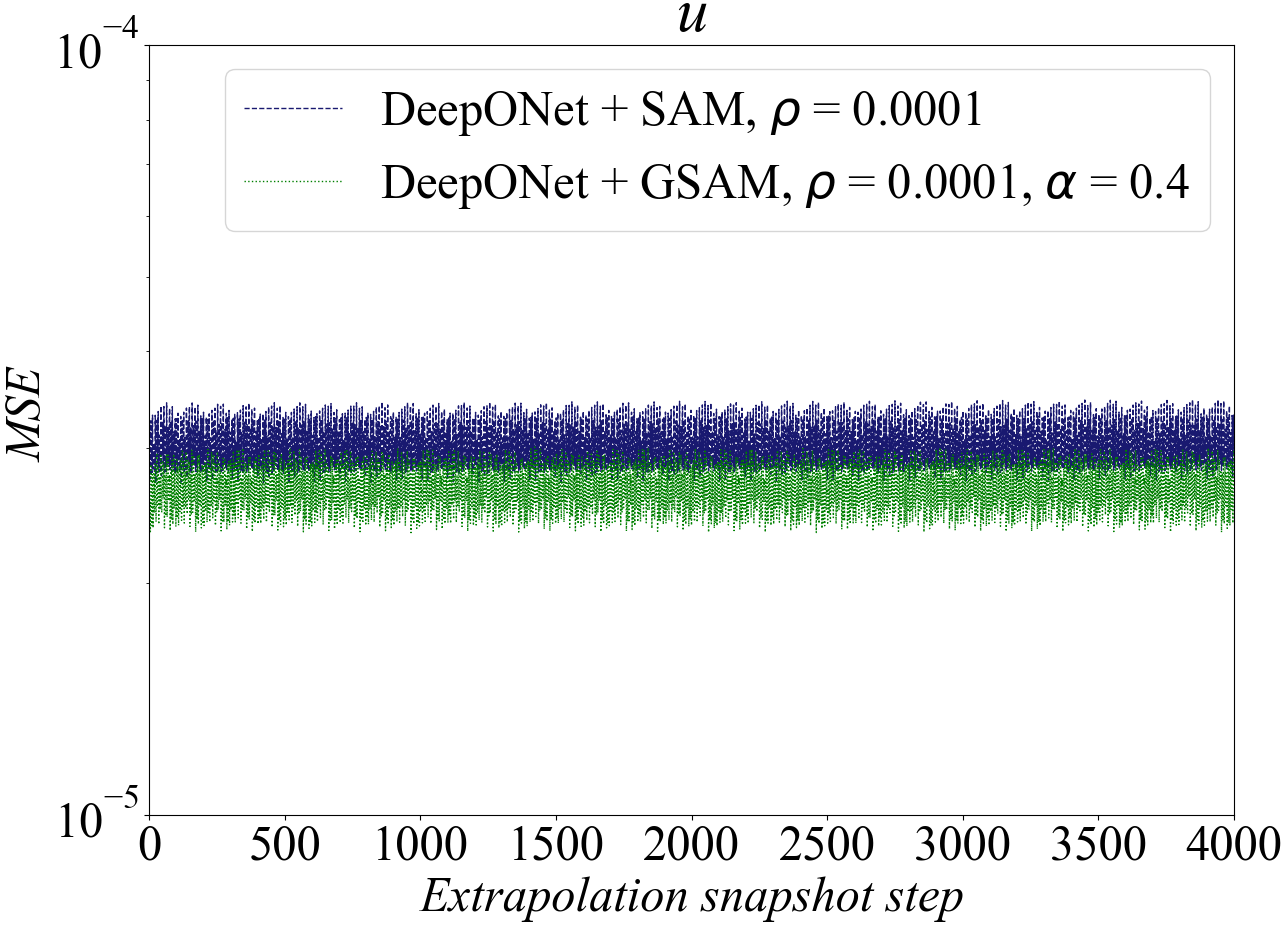}}
  \subfigure[]{
  \label{0.4VMSElineGSAM}
  \includegraphics[scale=0.255]{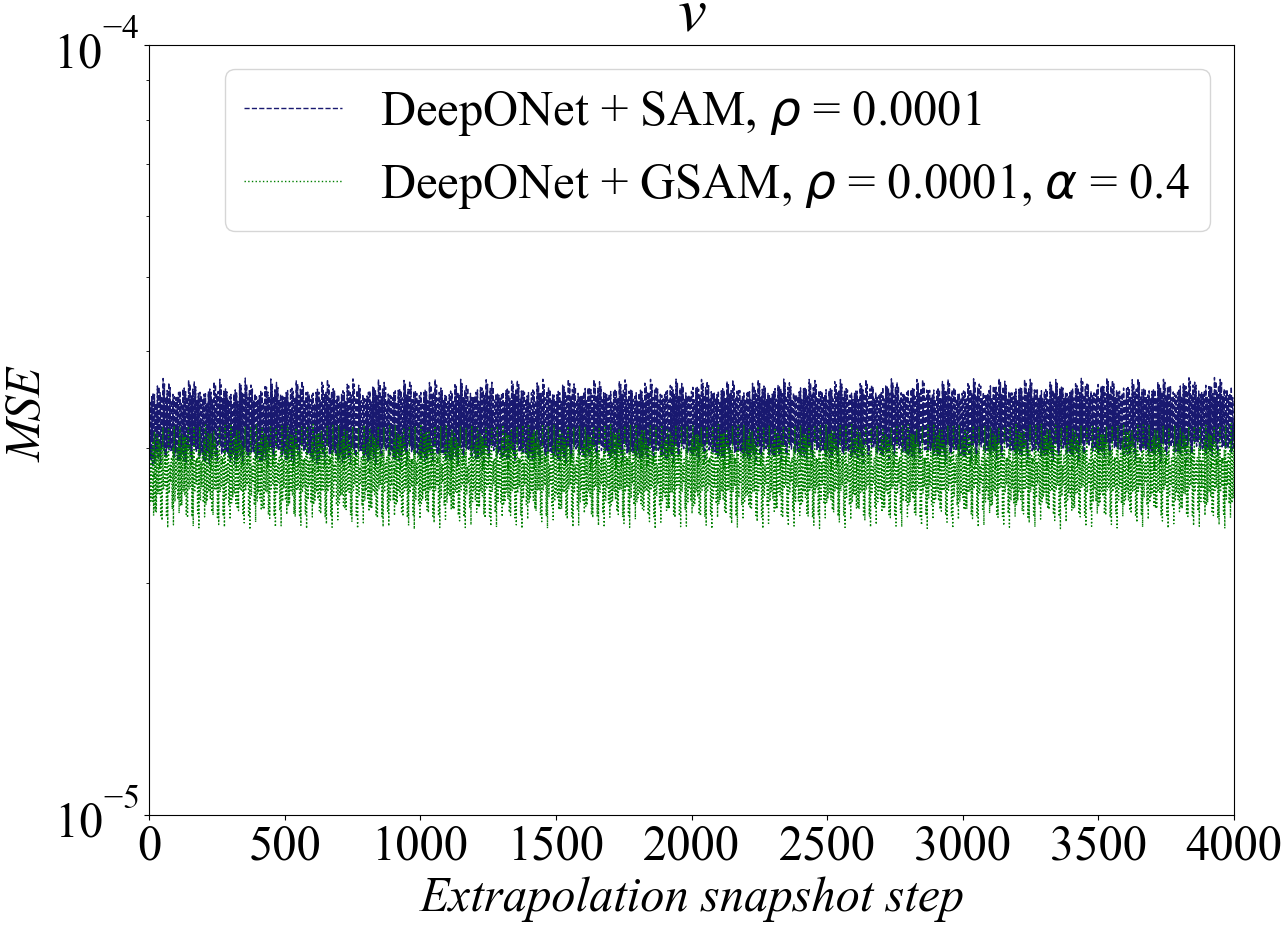}}
  \subfigure[]{
  \label{0.4PMSElineGSAM}
  \includegraphics[scale=0.255]{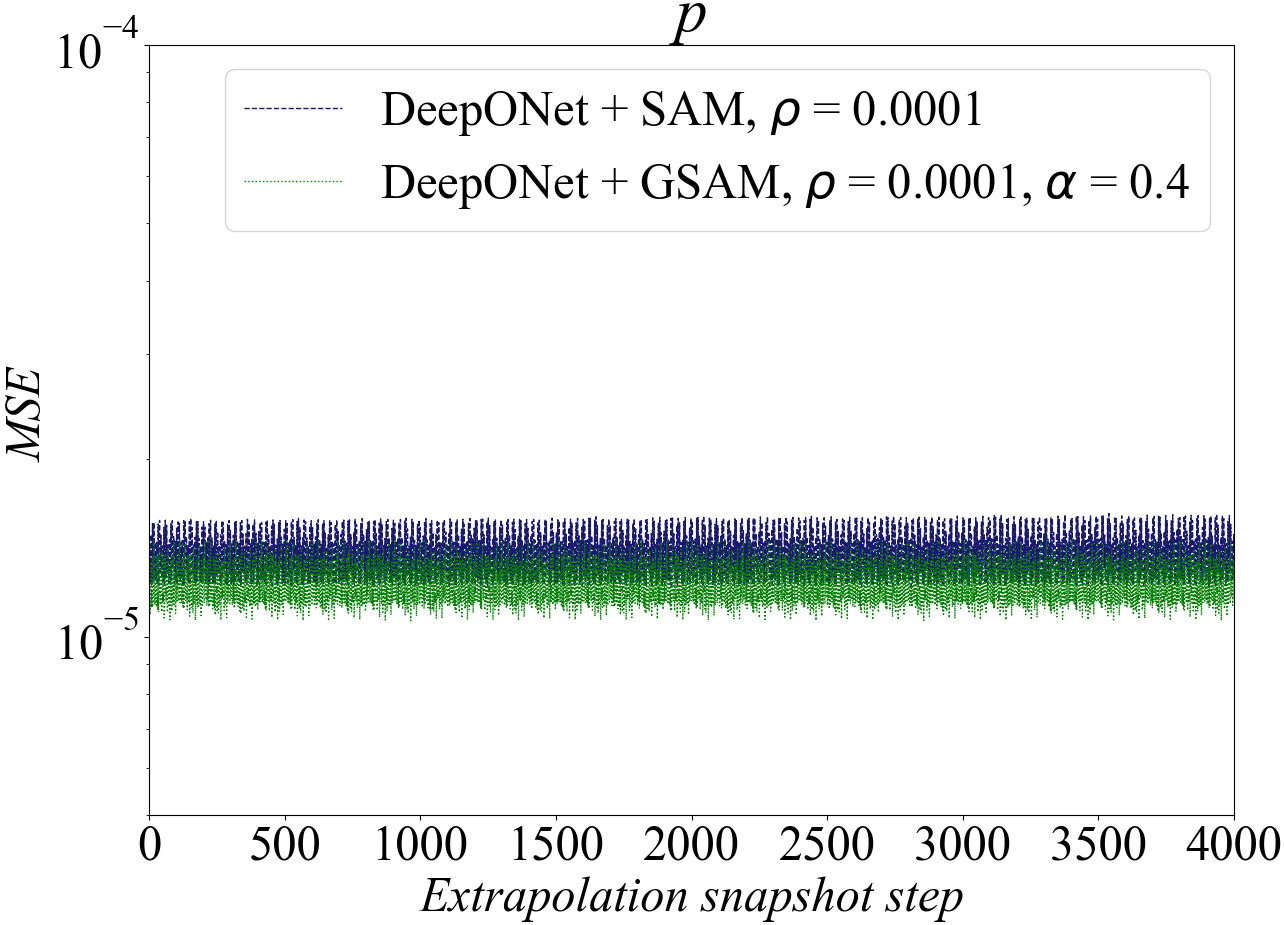}}
 \caption{The evolution of MSEs between the DNS data and the predictions of DeepONet + GSAM and DeepONet + SAM. (a) $u_{MSE}$. (b) $v_{MSE}$. (c) $p_{MSE}$.}
 \label{MSE line of GSAM}
\end{figure*}

The evolution of the MSEs between the predictions of DeepONet + GSAM and the DNS data over $4000$ extrapolation snapshots are depicted in Fig.~\ref{MSE line of GSAM},  along with the corresponding results achieved through DeepONet + SAM.
The prediction errors of $u$, $v$, and $p$ of the DeepONet + GSAM exhibit negligible accumulation during extrapolation, with smaller values compared to those observed in DeepONet + SAM. 
This observation emphasizes the contribution of the GSAM algorithm in enhancing the performance of the vanilla DeepONet in capturing spatial-temporal flow field evolution.

\begin{figure*}[htbp] 
 \centering  
  \subfigure[]{
  \label{UMSE_diffinput}
  \includegraphics[scale=0.227]{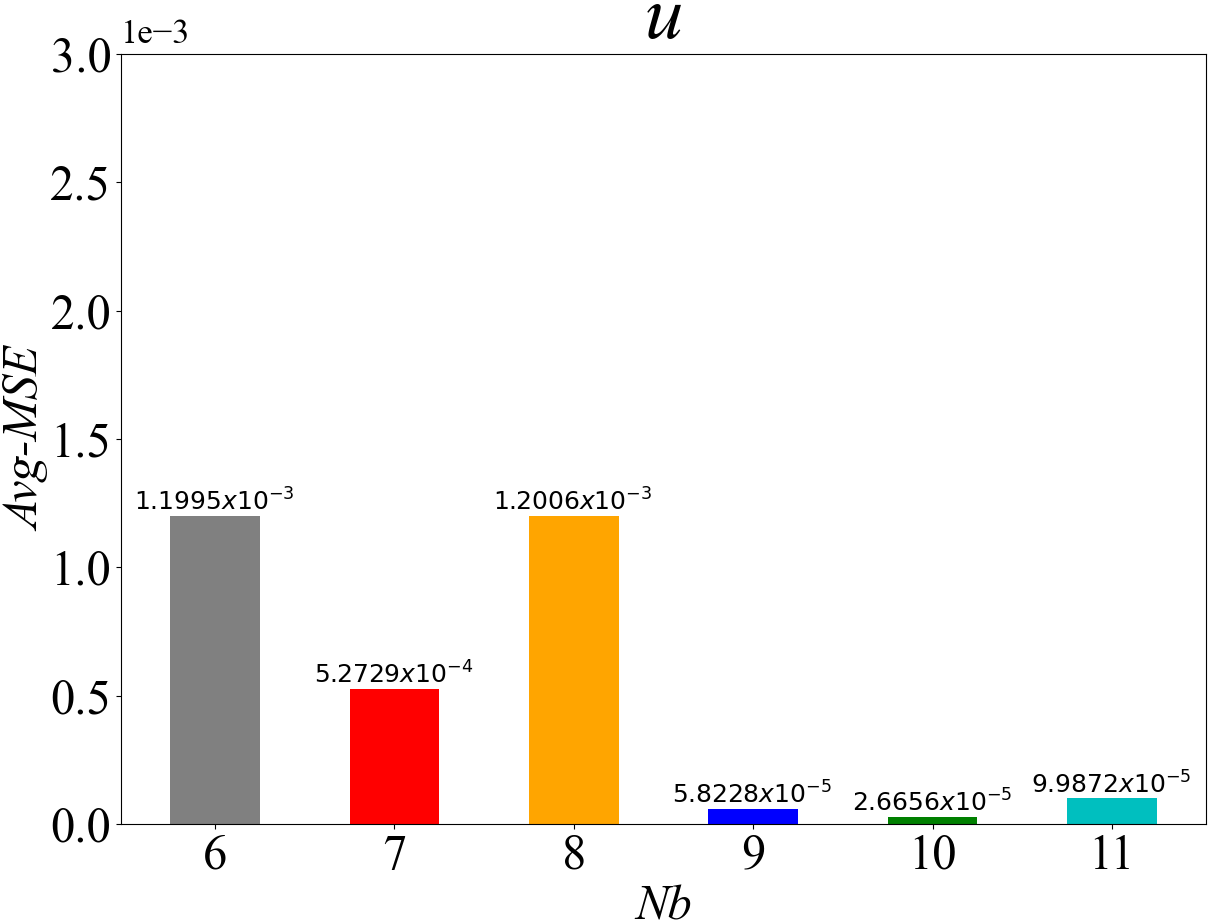}}
  \subfigure[]{
  \label{VMSE_diffinput}
  \includegraphics[scale=0.227]{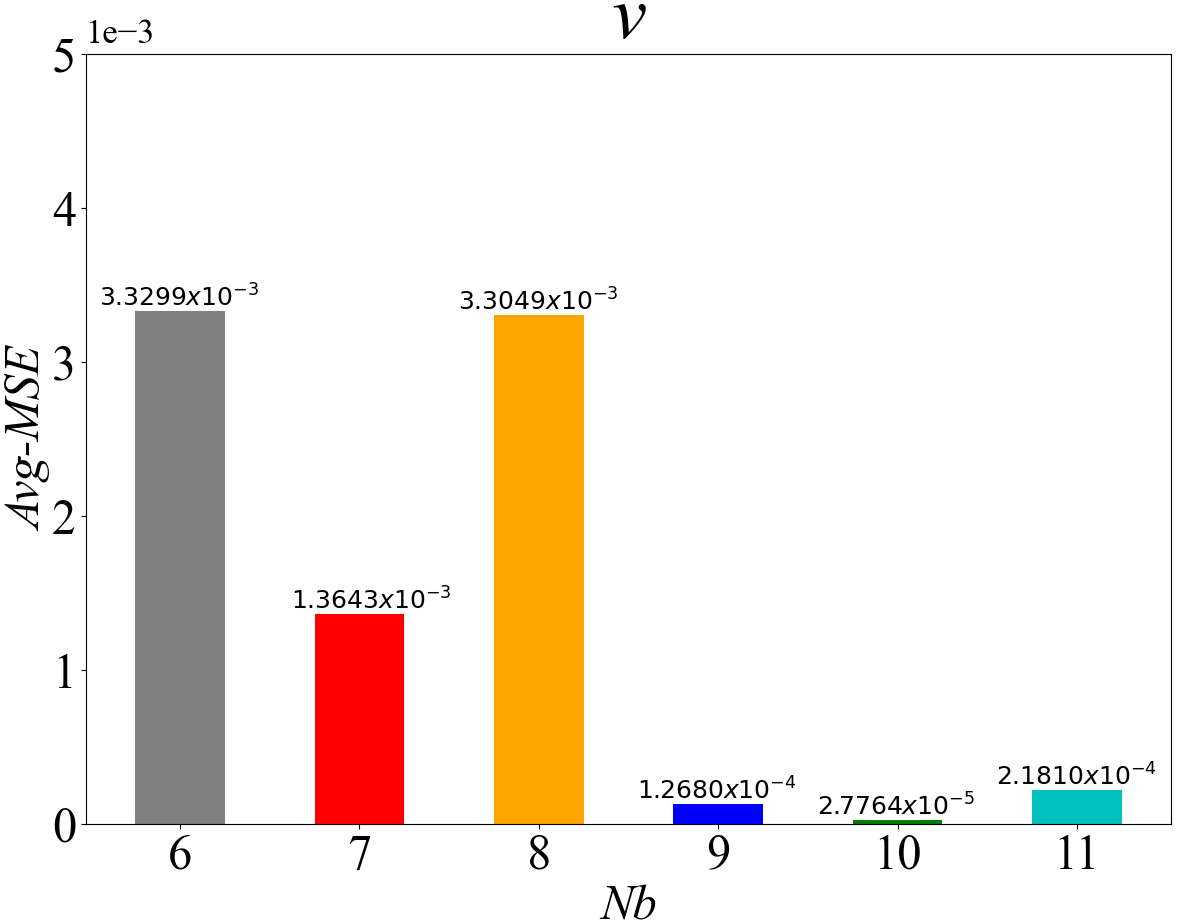}}
  \subfigure[]{
  \label{PMSE_diffinput}
  \includegraphics[scale=0.227]{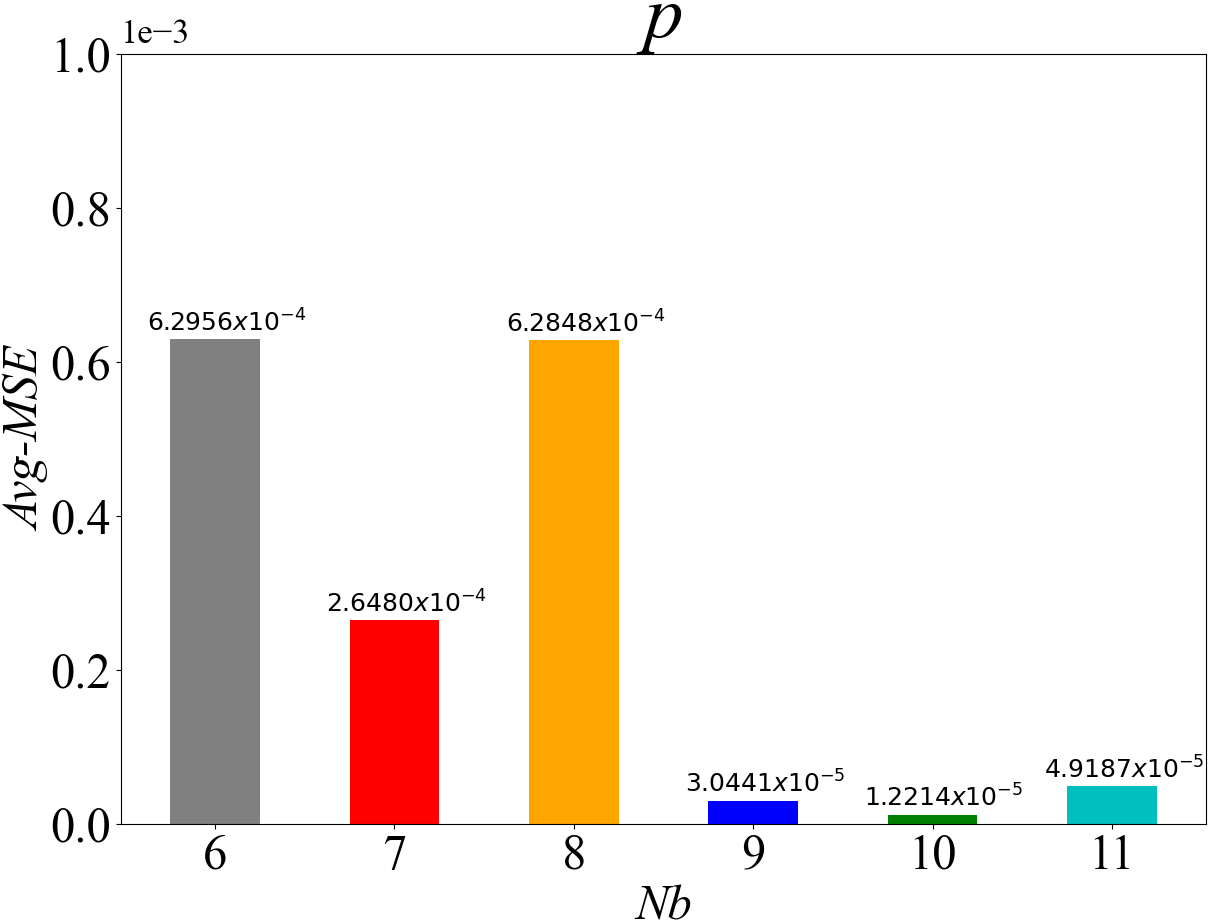}}
 \caption{The MSEs of $u$, $v$ and $p$ between the DNS data and the predictions of DeepONet + GSAM for various $N_{b}$ averaged from $3500th$ to $4000th$ extrapolation snapshots. (a) $u_{MSE}$. (b) $v_{MSE}$. (c) $p_{MSE}$.}
 \label{MSE of diffinput}
\end{figure*}

Based on the DeepONet + GSAM, the impact of different $N_{b}$ on the effectiveness of the DeepONet is investigated. Different values for $N_{b}$ are selected as presented in Table \ref{Nb value}.
The MSEs of $u$,$v$ and $p$ between the DNS data and the predictions of DeepONet + GSAM for various $N_{b}$ averaged from $3500th$ to $4000th$ extrapolation snapshots are illustrated in Fig.~\ref{MSE of diffinput}.

\begin{table}[htbp]
\caption{\label{tab:table4}Values of $N_{b}$.}
\begin{ruledtabular}
\begin{tabular}{ccccccc}
Case  & $I$ & $II$ & $III$ & $IV$ & $V$ & $VI$ \\
\hline
$N_{b}$  & 6 & 7 & 8 & 9 & 10 & 11\\
\end{tabular}
\label{Nb value}
\end{ruledtabular}
\end{table}

The increase in $N_b$ results in a more significant reduction in the long-time extrapolation error of the flow field, particularly for $N_b = 9$, $10$ and $11$. This can be attributed to the fact that the input data for the neural network contains a greater number of features related to the spatial-temporal evolution of the flow field dynamics. However, the long-time extrapolation of flow field based on neural network is a complex problem, with numerous factors influencing prediction accuracy. It is possible that further adjustments to hyper-parameter, improvements in neural network structure can enhance prediction results. However, detailed discussions on these aspects will not be carried out in this work.

The detailed results of DeepONet + GSAM, including the extrapolated flow fields, time-averaged Reynolds stresses and PSD are presented in Appendix~\ref{appendixB}. It is worth noting that the DeepONet + GSAM still fails to accurately predict high-wavenumber flow field components.

\subsubsection{Compared with Dynamic Mode Decomposition}

Dynamic mode decomposition~(DMD)~\cite{r6DMD} is a technique that can capture the spatial-temporal evolution of flow field using high-fidelity data. 
We apply DMD in the same dataset and generate $4000$ extrapolation snapshots.

\begin{figure*}[htbp] 
 \centering  
  \subfigure[]{
  \label{UMSE diff DMDmodes}
  \includegraphics[scale=0.47]{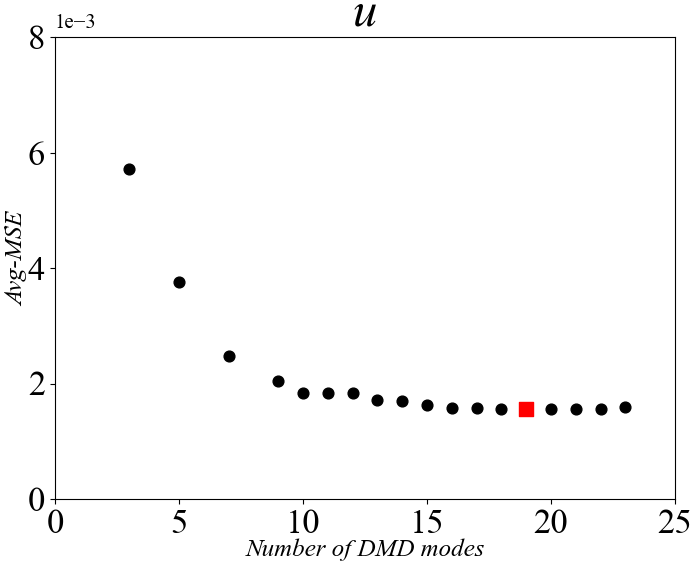}}
  \subfigure[]{
  \label{VMSE diff DMDmodes}
  \includegraphics[scale=0.47]{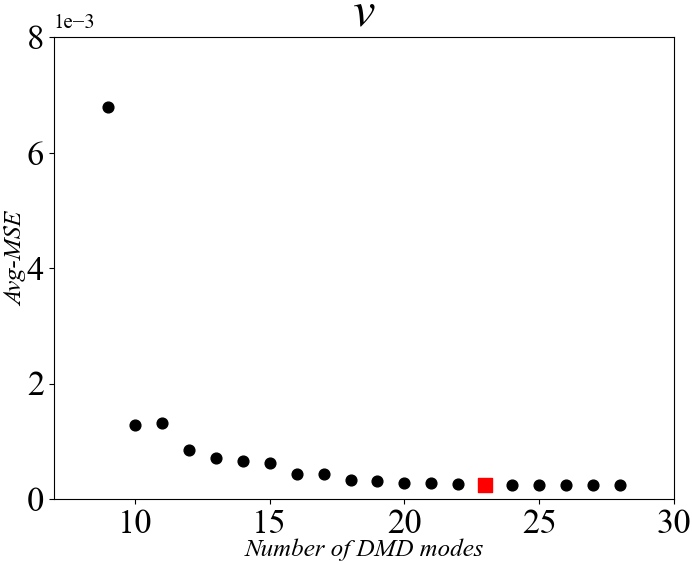}}
  \subfigure[]{
  \label{PMSE diff DMDmodes}
  \includegraphics[scale=0.47]{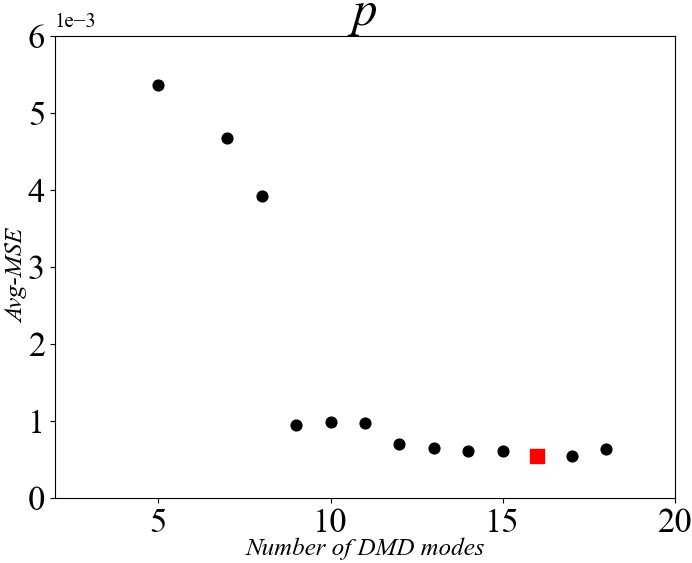}}
 \caption{The MSEs of $u$, $v$ and $p$ between the DNS data and the predictions of DMD with different numbers of modes averaged from $3500th$ to $4000th$ extrapolation snapshots. (a) $u_{MSE}$. (b) $v_{MSE}$. (c) $p_{MSE}$.}
 \label{MSE diff DMDmodes}
\end{figure*}

First, the impact of different numbers of DMD modes on the accuracy is studied. 
The MSEs of $u$, $v$ and $p$ between the DNS data and the predictions of DMD for different DMD modes averaged from $3500th$ to $4000th$ extrapolation snapshots are depicted in Fig.~\ref{MSE diff DMDmodes}. 

\begin{figure*}[htbp] 
 \centering  
  \subfigure[]{
  \label{UDNScompareDMD}
  \includegraphics[scale=0.178]{Utrue4000step.png}}
  \subfigure[]{
  \label{UGSAMcompareDMD}
  \includegraphics[scale=0.178]{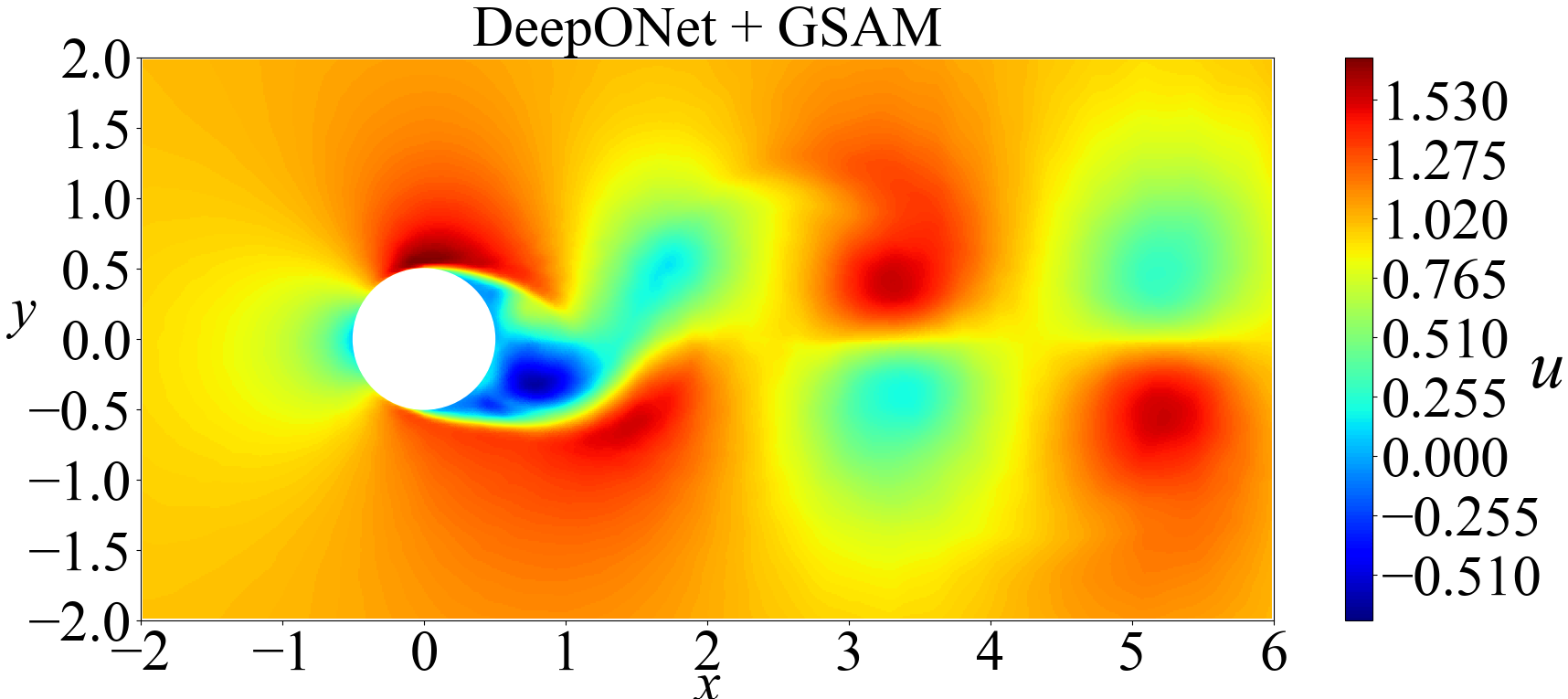}}
  \subfigure[]{
  \label{UDMD}
  \includegraphics[scale=0.178]{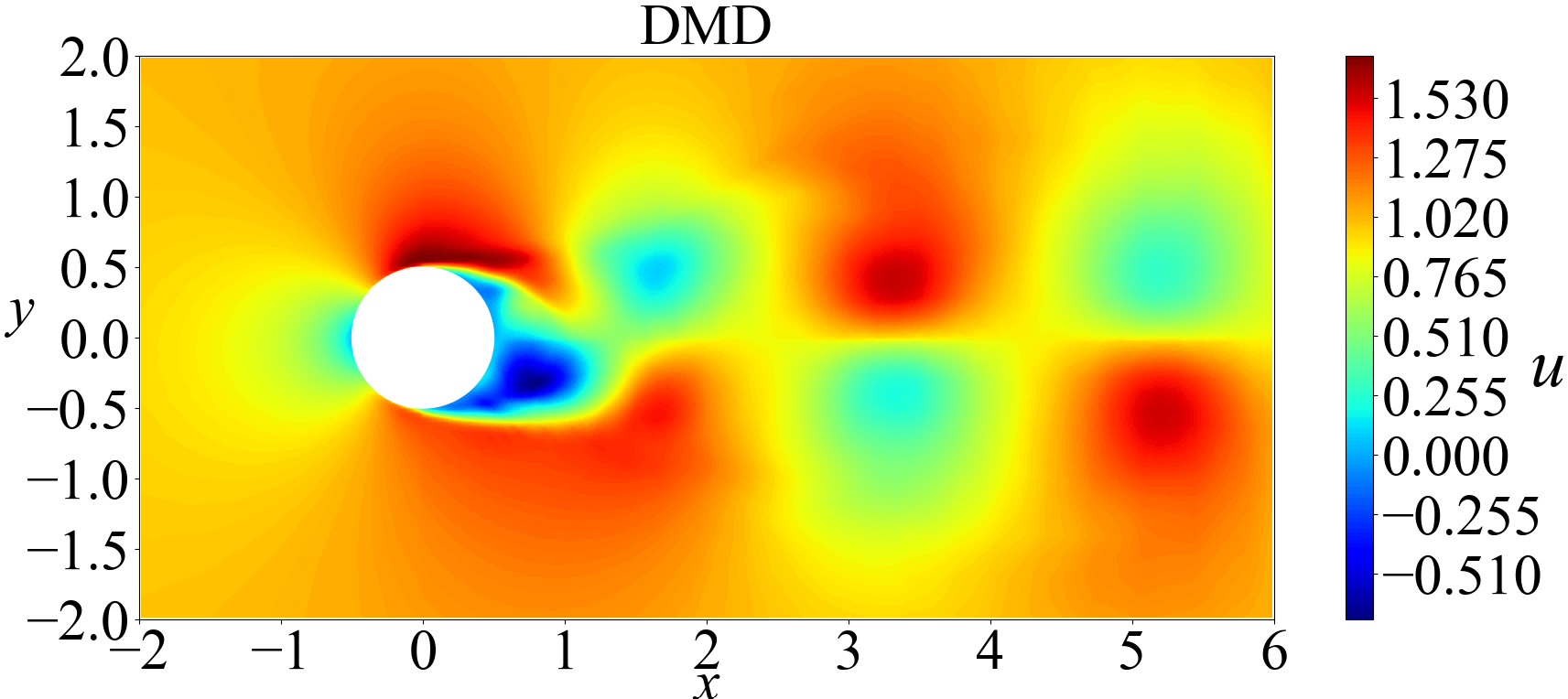}}
  \subfigure[]{
  \label{VDNScompareDMD}
  \includegraphics[scale=0.178]{Vtrue4000step.png}}
  \subfigure[]{
  \label{VGSAMcompareDMD}
  \includegraphics[scale=0.178]{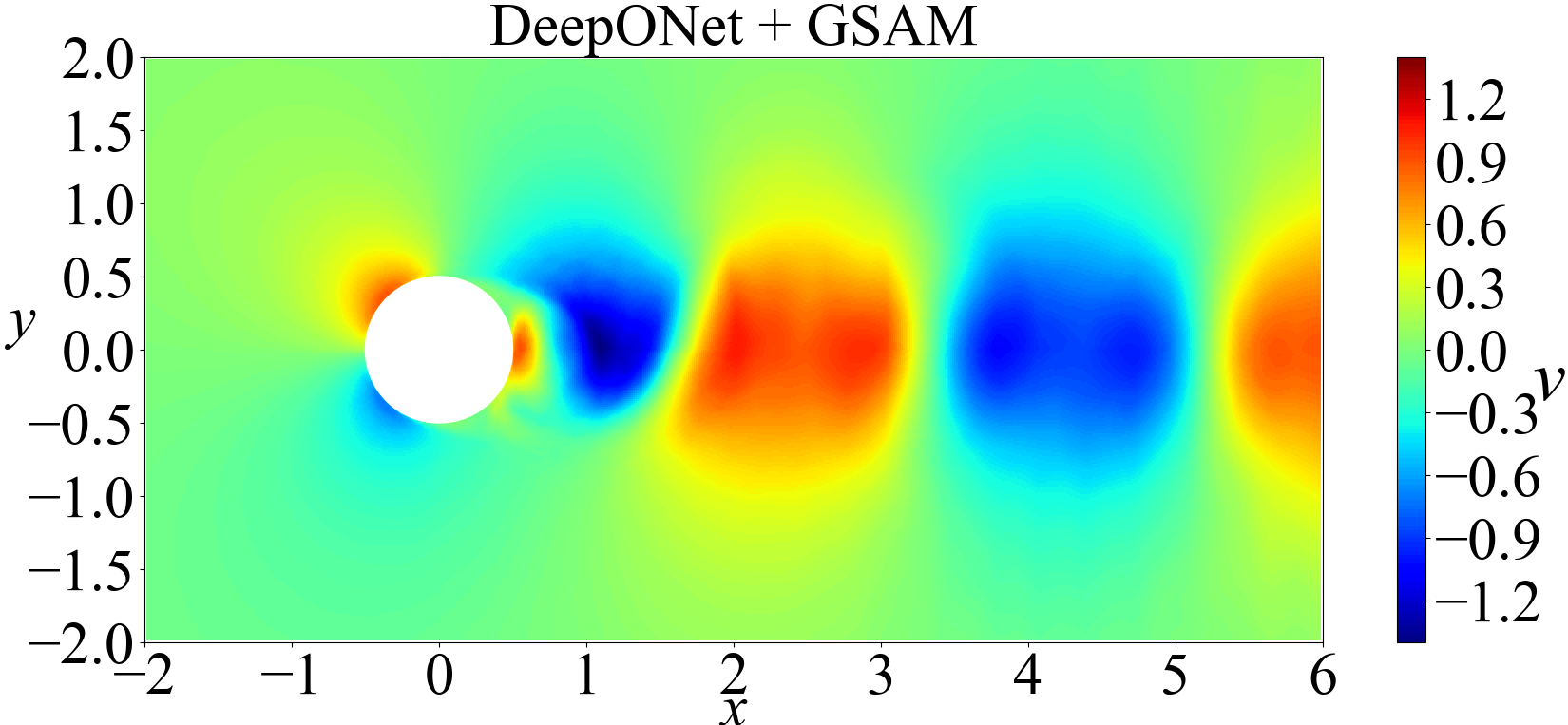}}
  \subfigure[]{
  \label{VDMD}
  \includegraphics[scale=0.178]{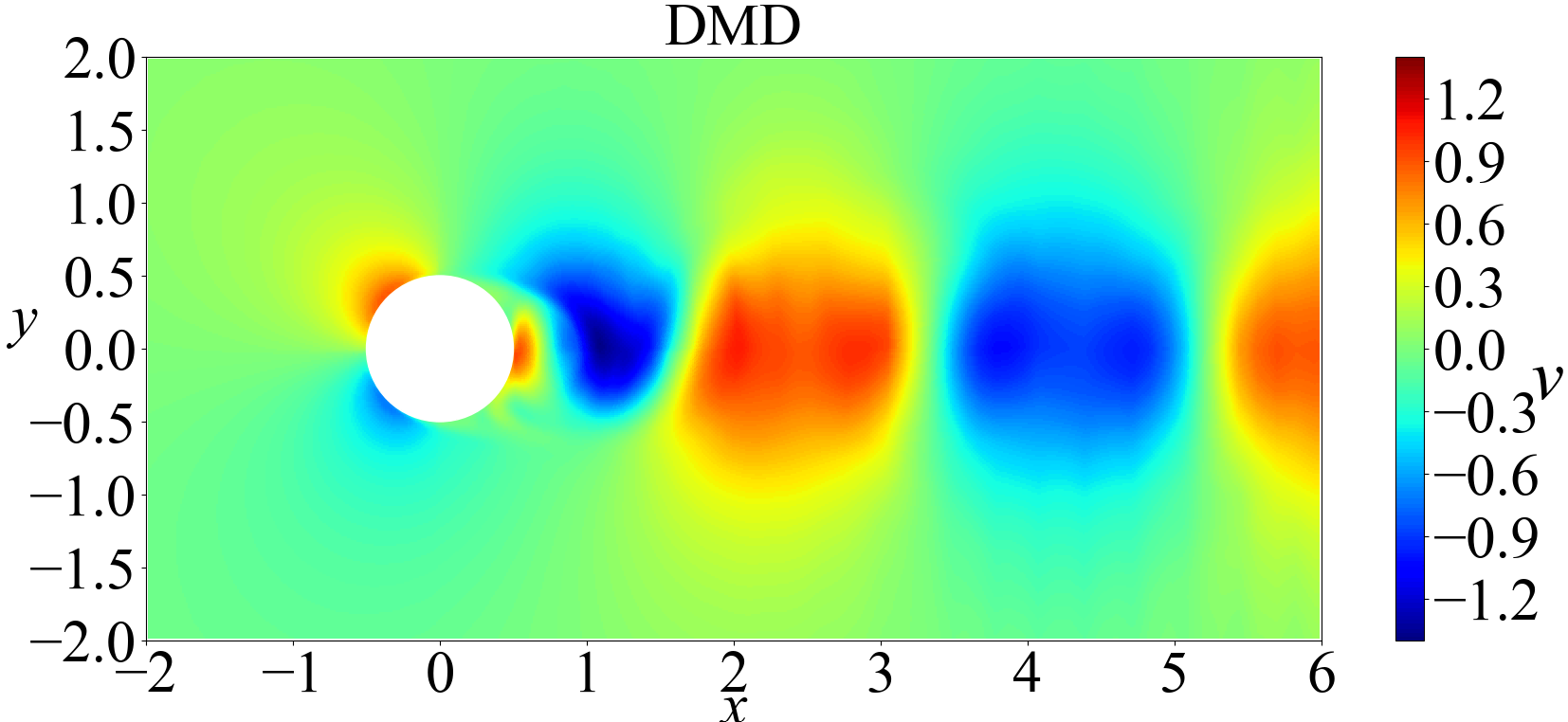}}
  \subfigure[]{
  \label{PDNScoompareDMD}
  \includegraphics[scale=0.178]{Ptrue4000step.png}}
  \subfigure[]{
  \label{PGSAMcompareDMD}
  \includegraphics[scale=0.178]{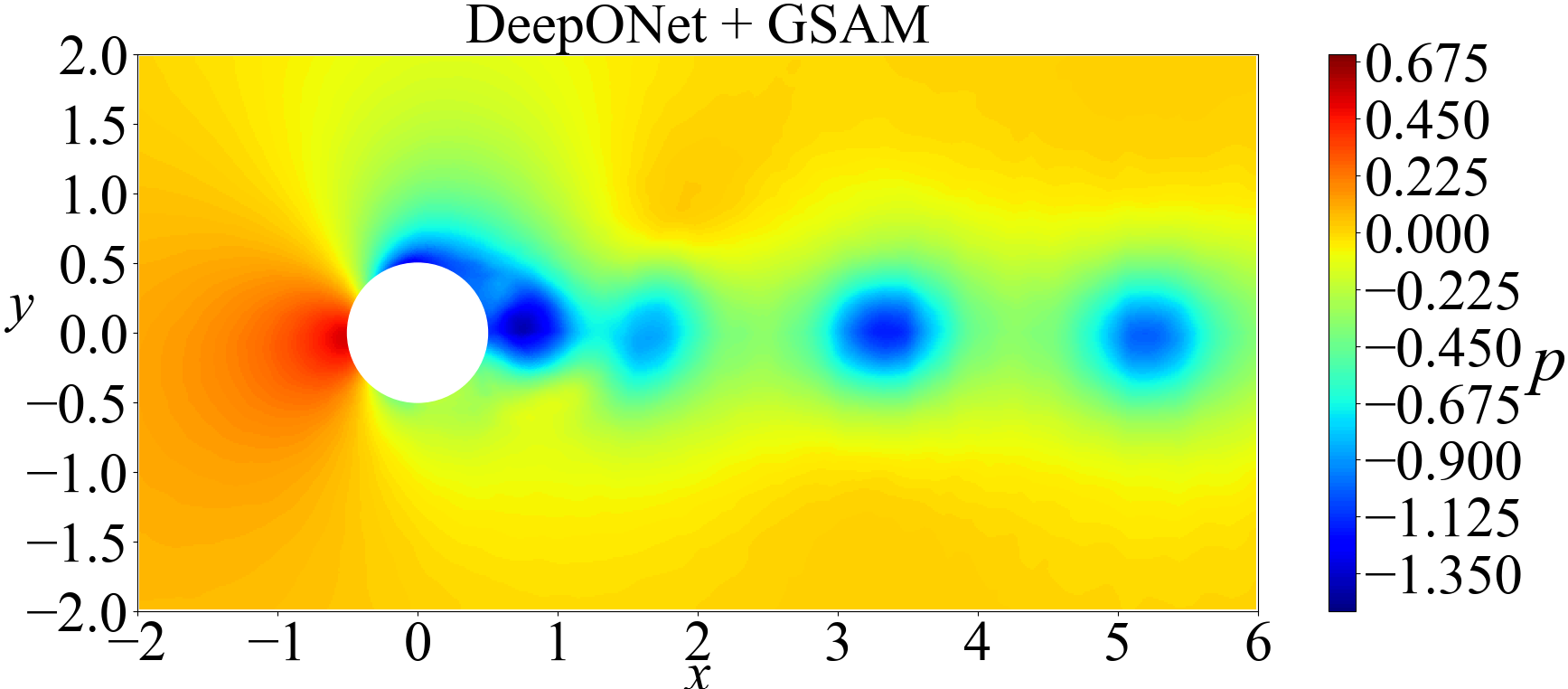}}
  \subfigure[]{
  \label{PDMD}
  \includegraphics[scale=0.178]{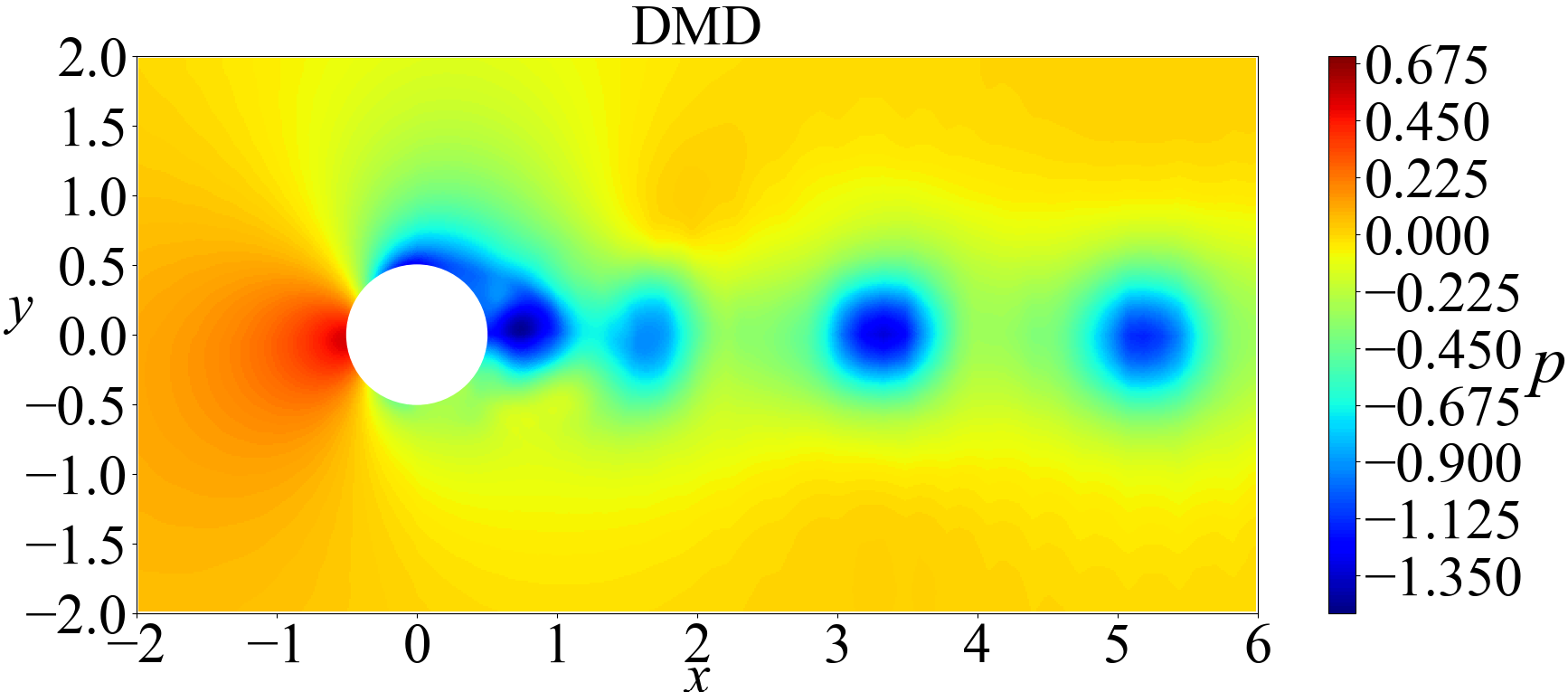}}
  \caption{Comparison of $u$, $v$ and $p$ at $4000th$ extrapolation snapshot between the DNS data, DeepONet + GSAM and DMD. (a) $u_{DNS}$. (b) $u_{preGSAM}$. (c) $u_{preDMD}$. (d) $v_{DNS}$. (e) $v_{preGSAM}$. (f) $v_{preDMD}$. (g) $p_{DNS}$. (h) $p_{preGSAM}$. (i) $p_{preDMD}$.}\label{DMDcontour compare}
\end{figure*}

As the number of DMD modes increases gradually, there is no further reduction in prediction error. The number of DMD modes for $u$, $v$ and $p$ are determined to be 19, 23 and 16 respectively, as indicated by the red square dot in Fig.~\ref{UMSE diff DMDmodes}, \ref{VMSE diff DMDmodes} and \ref{PMSE diff DMDmodes}. The flow field at the $4000th$ extrapolation snapshot calculated using DMD is compared with the results obtained from DNS data and DeepONet + GSAM, as depicted in Fig.~\ref{DMDcontour compare},
where it is difficult to visually note the difference between the results.

\begin{figure*}[htbp] 
 \centering  
  \subfigure[]{
  \label{UDMD compare GSAM}
  \includegraphics[scale=0.27]{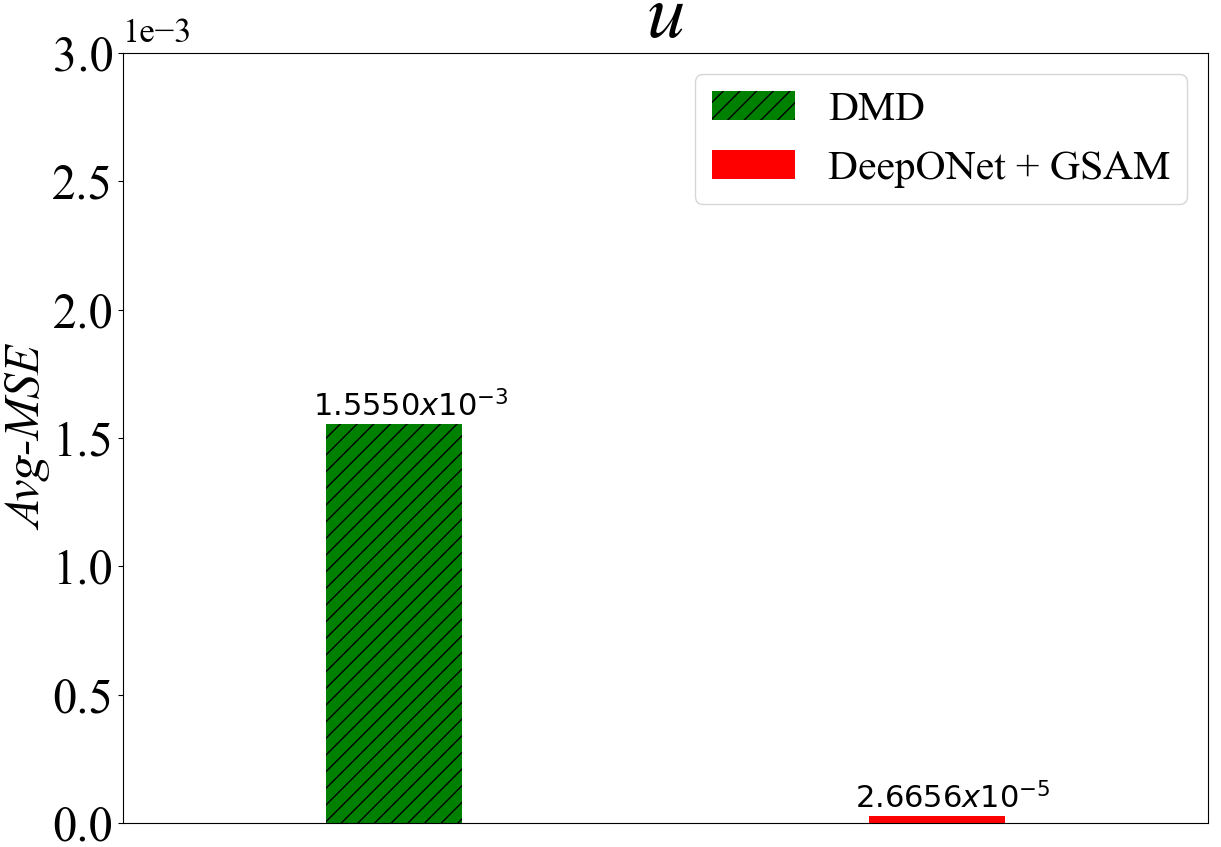}}
  \subfigure[]{
  \label{VDMD compare GSAM}
  \includegraphics[scale=0.27]{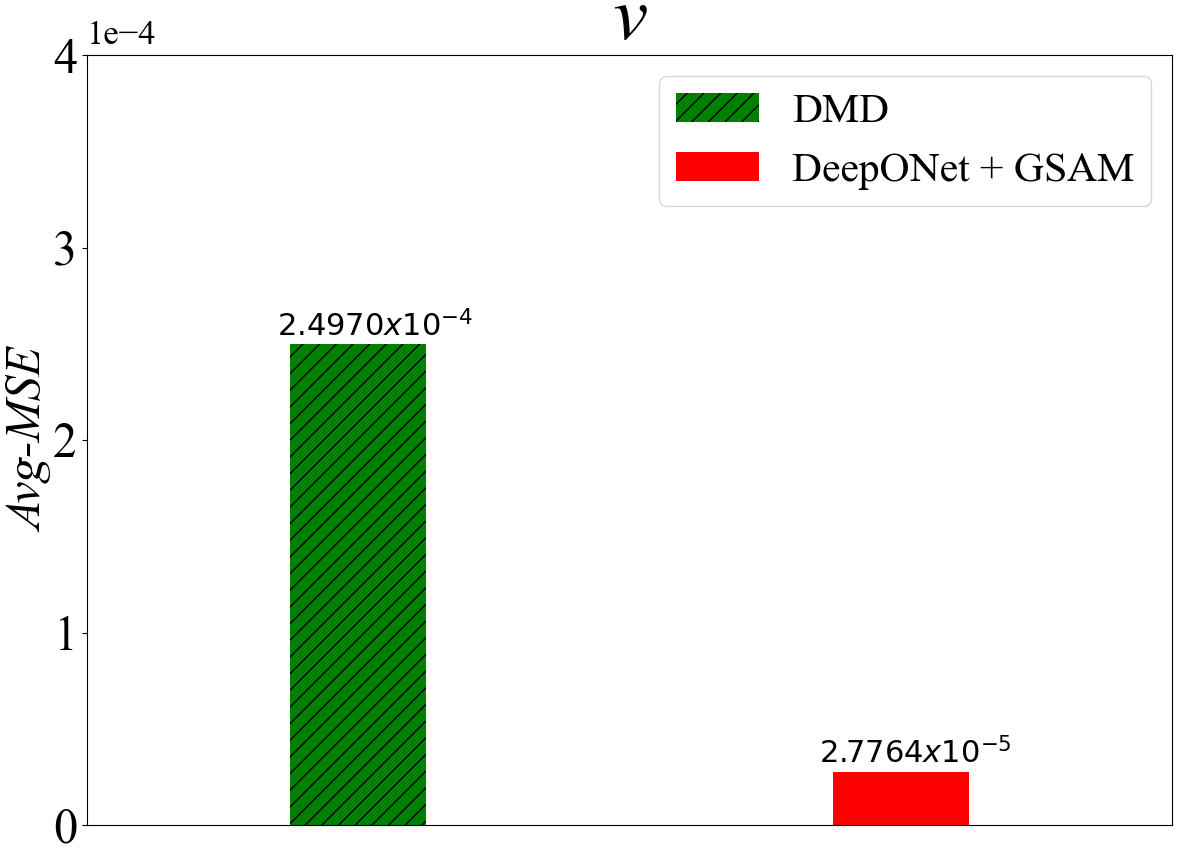}}
  \subfigure[]{
  \label{PDMD compare GSAM}
  \includegraphics[scale=0.27]{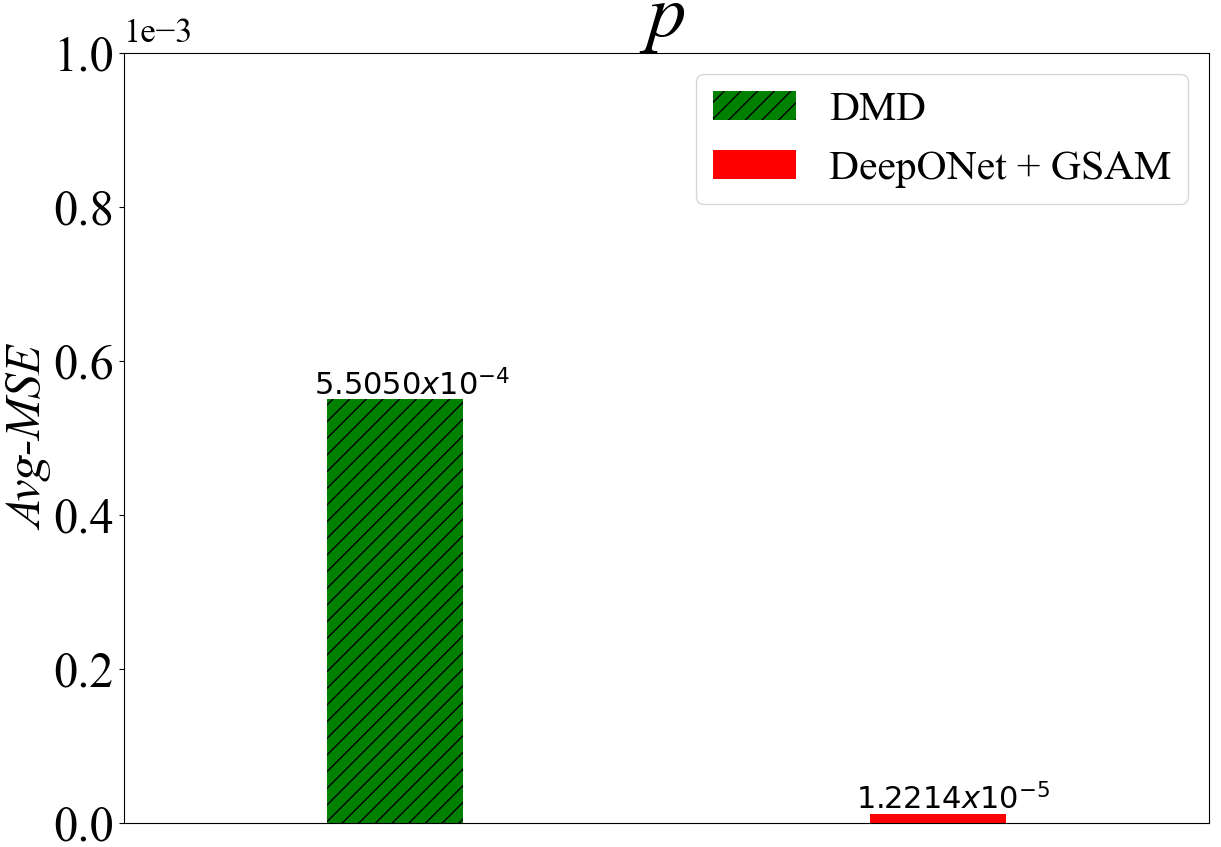}}
 \caption{The MSEs of $u$, $v$ and $p$ between the DNS data and the predictions of DeepONet + GSAM and DMD averaged from $3500th$ to $4000th$ extrapolation snapshots. (a) $u_{MSE}$. (b) $v_{MSE}$. (c) $p_{MSE}$.}
 \label{DMD compare GSAM}
\end{figure*}
Although the dynamic evolution of the flow field can be obtained by DMD, significant discrepancies still exist between the predictions and the DNS data, particularly in terms of $u$. The MSEs of $u$, $v$ and $p$ between the DNS data and the predictions of DeepONet + GSAM and DMD averaged from $3500th$ to $4000th$ extrapolation snapshots are depicted in Fig.~\ref{DMD compare GSAM}.
The prediction error of the DeepONet + GSAM is significantly lower compared to that of DMD. This outcome can be attributed to the fact that DMD assumes linearity in the evolution of flow field dynamics, which is apparently inappropriate for complex flows.

In summary, the DeepONet + GSAM demonstrates exceptional accuracy and rapid prediction speed for the extrapolation of the flow condition utilized in this work.

\section{CONCLUSIONS}
\label{conclusion}

In this study, the DeepONet is employed to model unsteady flow. The branch network employs a CNN structure to extract the spatial-temporal evolution of high-dimensional flow fields, while the trunk network uses an FNN to encode the spatial-temporal coordinates. The proposed DeepONet architecture is trained to forecast the flow field of the 2D flow around a circular cylinder at $Re = 1000$. 

Firstly, the vanilla DeepONet demonstrates a notable increase in the error of the flow field over a few hundreds extrapolation snapshots. This suggests that the vanilla DeepONet is incompetent to effectively capture the spatial-temporal evolution of the flow field over a long time. Therefore, two distinct approaches for improving the performance of the DeepONet are explored.

The first approach aims to enhance the ability of the DeepONet in extracting flow field characteristics from the dataset. The multi-head non-local block structure is proposed and integrated into the branch network as an auxiliary module to emphasize on capturing long-range dependencies in the flow field. The results demonstrate that the DeepONet + multi-head non-local block accurately predicts the unsteady flow fields and maintains a negligible error growth as the extrapolation step increases, especially with a higher number of "heads". This indicates that the enhanced model achieves greater accuracy and more stable network parameters. However, this improvement comes at the cost of increased wall time during the online forecasting phase.

The second approach aims to enhance the network parameters' optimization. Inspired by the notion that the smoothness of the loss landscape impacts the generalization of the neural network, the SAM algorithm in deep learning field is introduced. This algorithm is combined with the Adam optimizer to train the DeepONet. The results demonstrate that this approach can achieve comparable accuracy to the DeepONet + multi-head non-local block, while simultaneously reducing the wall the time during the online forecasting phase. Subsequently, due to the inability of the SAM algorithm to guarantee a low level of “sharpness”, the GSAM algorithm is introduced to explore potential enhancements to the performance of the DeepONet. The results demonstrated a further reduction in prediction error.

By employing the aforementioned methodologies, the PSD of $u$ and $v$ along the vertical axis at various downstream distances from the cylinder at the $4000th$ extrapolation snapshot are computed and compare with the DNS data. The results demonstrate that the extraction of flow field characteristics with low wavenumber are very accurate, while those with high wavenumber are relative poor. 
The results are in line with the property of spectral bias of neural network, which remains difficult to extract features at small scale.

We have not yet investigated the long-term spatial-temporal evolution of more complex and aperiodic flow fields, which is a challenging task for deep learning due to its limitation in extrapolating complex unseen phenomena. However, the idea presented in this work enhances the capability of DNNs to learn physics from high-fidelity data, thus holding promising implications for future research endeavors. We aim to extend this work in future research. For example, by replacing CNN with graph neural network~(GNN) in constructing the branch network to directly handle unstructured data, we may reduce the error introduced by interpolating DNS data onto a uniform grid and expand the application scope of the surrogate model. Additionally, we may utilize meta-learning techniques to enhance the fast adaptability and generalization of the DeepONet when encountering new flow conditions.

\begin{acknowledgments}
X. Bian received the starting grant from 100 talents program of Zhejiang University.
The authors appreciate discussions with Prof. Zhenhua Xia. 
\end{acknowledgments}

\section*{Data Availability Statement}
The data that support the findings of this study are available from the corresponding author upon reasonable request.

\appendix
\section{Verficaiton of numerical simulation}
\label{appendixA}

The 2D flow around a circular cylinder is simulated using the spectral/{\em hp} element method. 
When the flow is fully developed, its average total, pressure and viscous drag coefficients are compared to those of Henderson et al..~\cite{henderson1995physics} 
The comparison is shown in Table~\ref{result compared}, demonstrating a high accuracy of our simulation.

\begin{table}[htbp]
\caption{\label{tab:table5}Comparison of numerical results for the 2D flow around a circular cylinder with $Re = 1000$.}
\begin{ruledtabular}
\begin{tabular}{ccccccc}
    & Total & Pressure & Viscous \\
\hline
Henderson et al.~\cite{henderson1995physics}  & 1.51 & 1.39 & 0.12 \\
Present work  & 1.54 & 1.41 & 0.13 \\
\end{tabular}
\label{result compared}
\end{ruledtabular}
\end{table}

\section{Detailed results of the DeepONet + GSAM}
\label{appendixB}
\begin{figure*}[htbp] 
 \centering  
  \subfigure[]{
  \label{UDNSGSAM}
  \includegraphics[scale=0.178]{Utrue4000step.png}}
  \subfigure[]{
  \label{UpreGSAM}
  \includegraphics[scale=0.178]{GSAMUpre4000step.png}}
  \subfigure[]{
  \label{UdeltaGSAM}
  \includegraphics[scale=0.178]{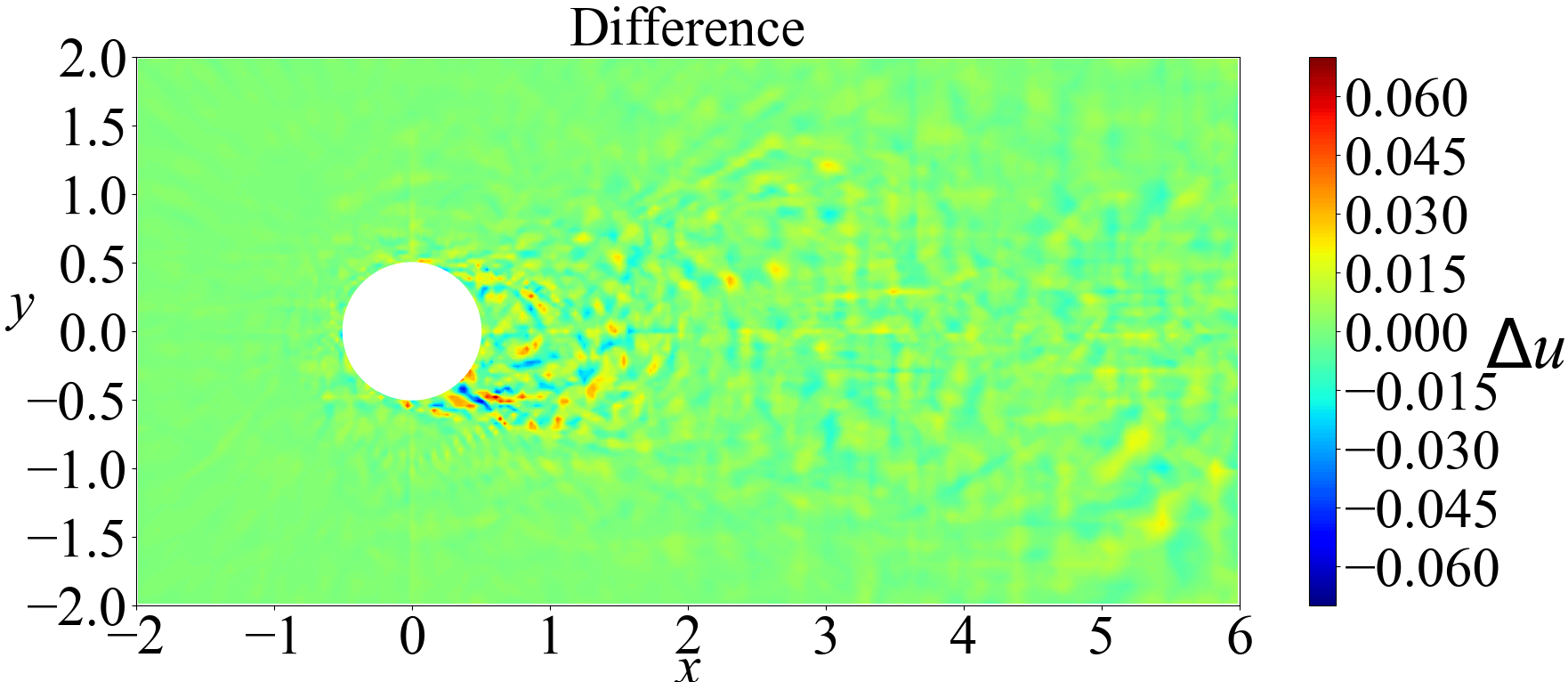}}
  \subfigure[]{
  \label{VDNSGSAM}
  \includegraphics[scale=0.178]{Vtrue4000step.png}}
  \subfigure[]{
  \label{VpreGSAM}
  \includegraphics[scale=0.178]{GSAMVpre4000step.png}}
  \subfigure[]{
  \label{VdeltaGSAM}
  \includegraphics[scale=0.178]{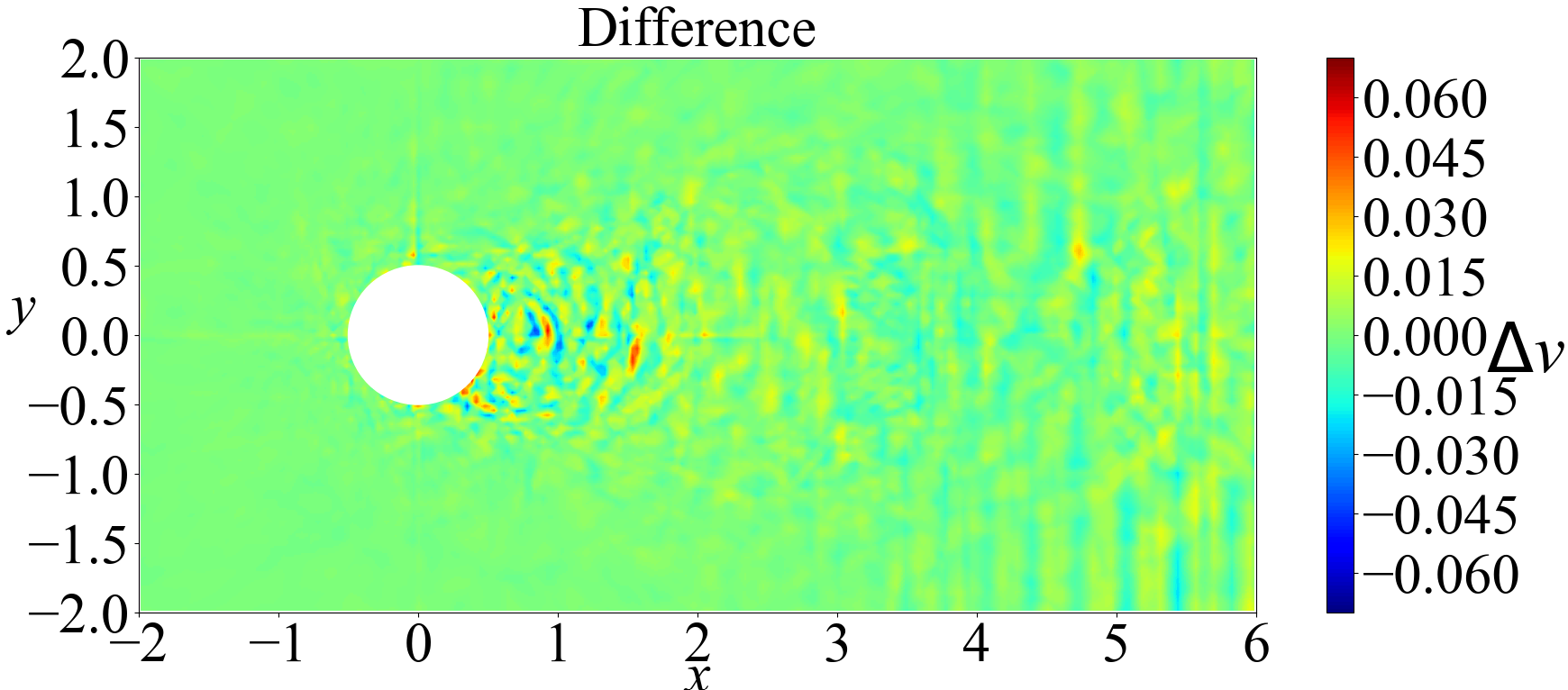}}
  \subfigure[]{
  \label{PDNSGSAM}
  \includegraphics[scale=0.178]{Ptrue4000step.png}}
  \subfigure[]{
  \label{PpreGSAM}
  \includegraphics[scale=0.178]{GSAMPpre4000step.png}}
  \subfigure[]{
  \label{PdeltaGSAM}
  \includegraphics[scale=0.178]{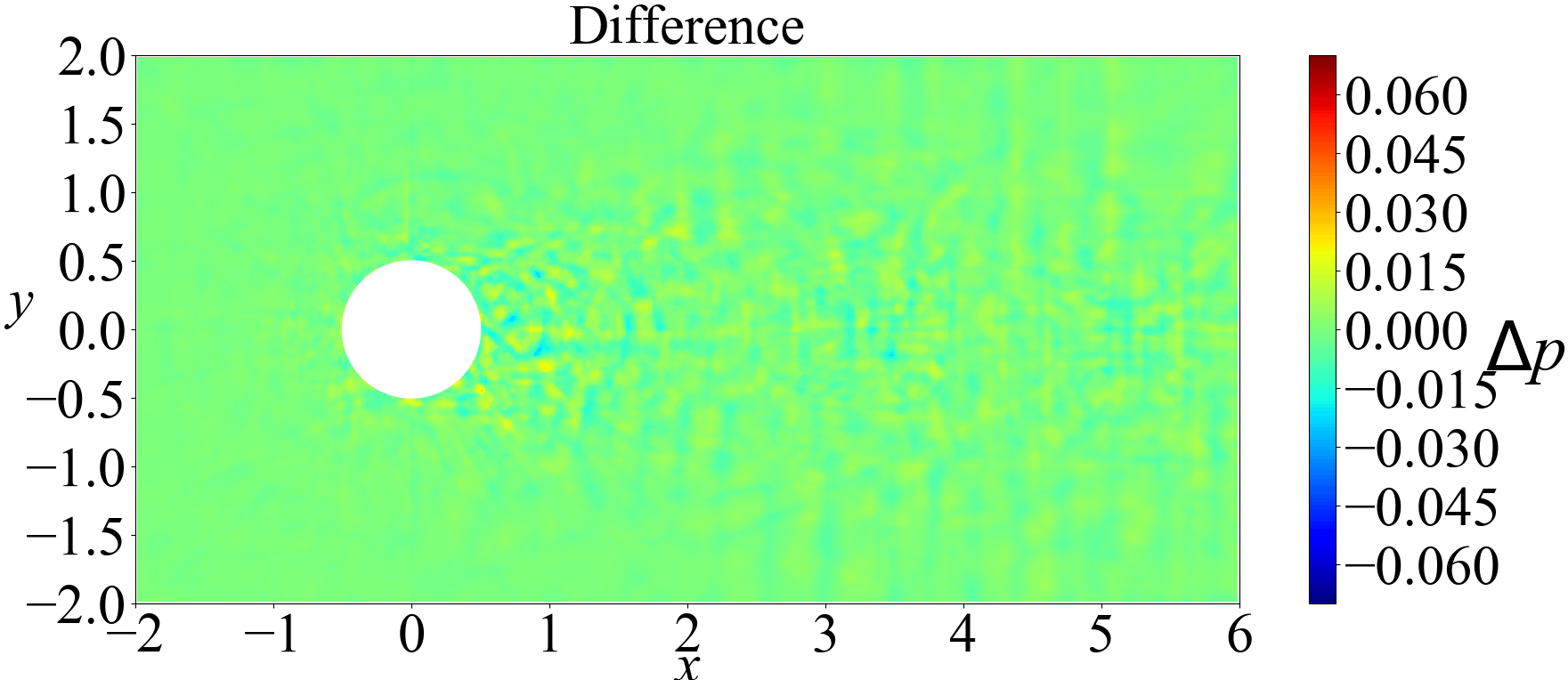}}
  \caption{Comparison of $u$, $v$ and $p$ at $4000th$ extrapolation snapshot between the DNS data and the prediction of DeepONet + GSAM. (a) $u_{DNS}$. (b) $u_{pre}$. (c) $u_{DNS}-u_{pre}$. (d) $v_{DNS}$. (e) $v_{pre}$. (f) $v_{DNS}-v_{pre}$. (g) $p_{DNS}$. (h) $p_{pre}$. (i) $p_{DNS}-p_{pre}$.}\label{GSAMcontour compare}
\end{figure*}

The detailed results of the DeepONet + GSAM are presented in this section.
The flow field of the $4000th$ extrapolation snapshot is depicted in Fig.~\ref{GSAMcontour compare}. The predicted flow field demonstrates a remarkable level of consistency with the DNS data.

\begin{figure*}[htbp] 
 \centering  
  \subfigure[]{
  \label{uutrueGSAM}
  \includegraphics[scale=0.045]{uutrue.png}}
  \subfigure[]{
  \label{UUpreGSAM}
  \includegraphics[scale=0.045]{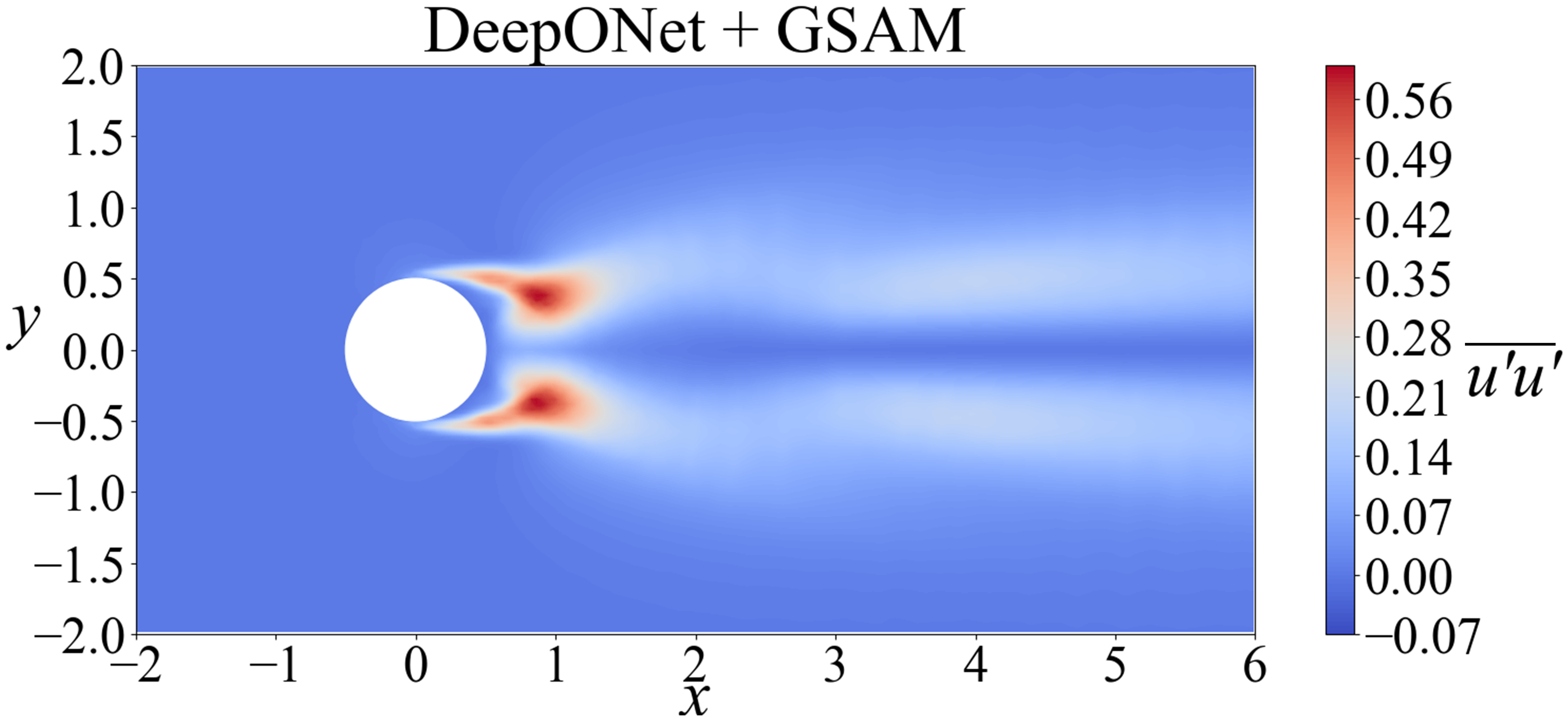}}
  \subfigure[]{
  \label{UUdeltaGSAM}
  \includegraphics[scale=0.045]{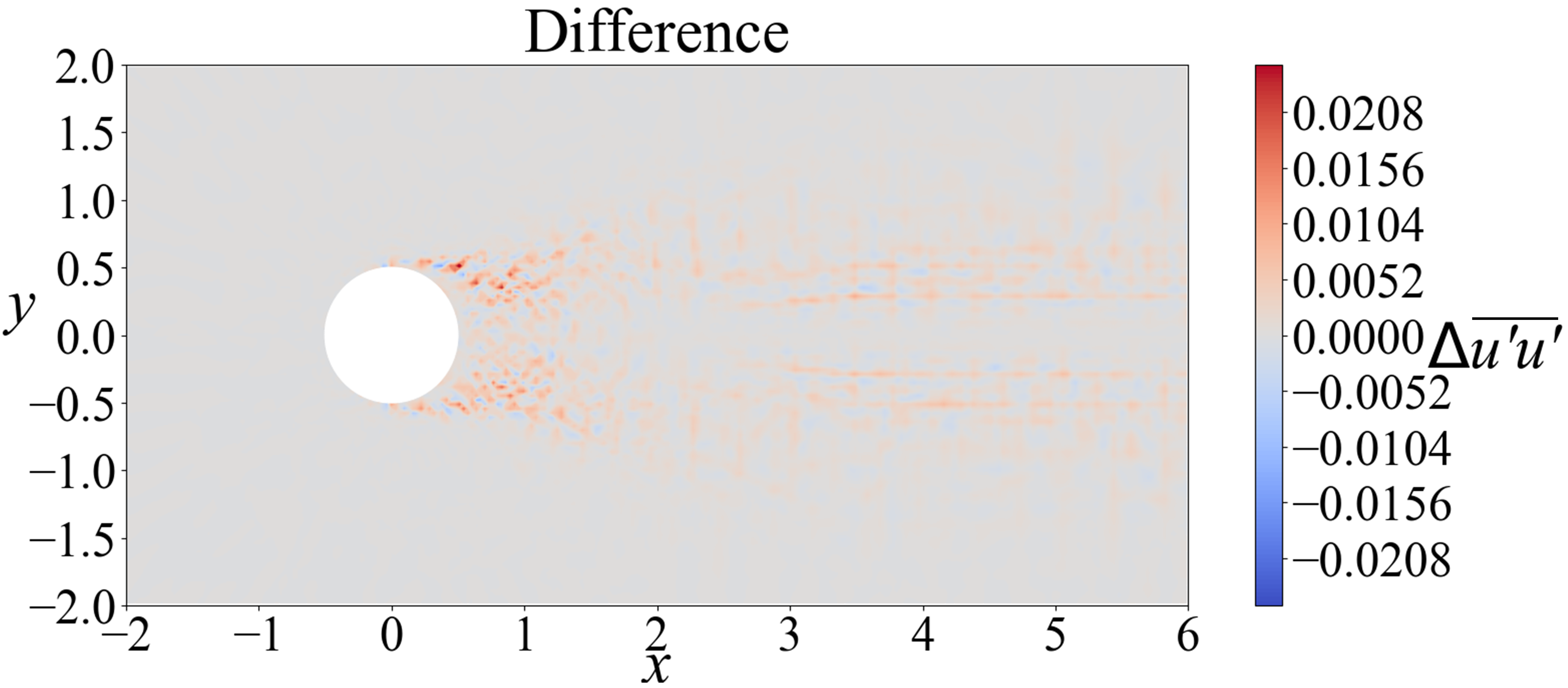}}
  \subfigure[]{
  \label{vvtrueGSAM}
  \includegraphics[scale=0.045]{vvtrue.png}}
  \subfigure[]{
  \label{VVpreGSAM}
  \includegraphics[scale=0.045]{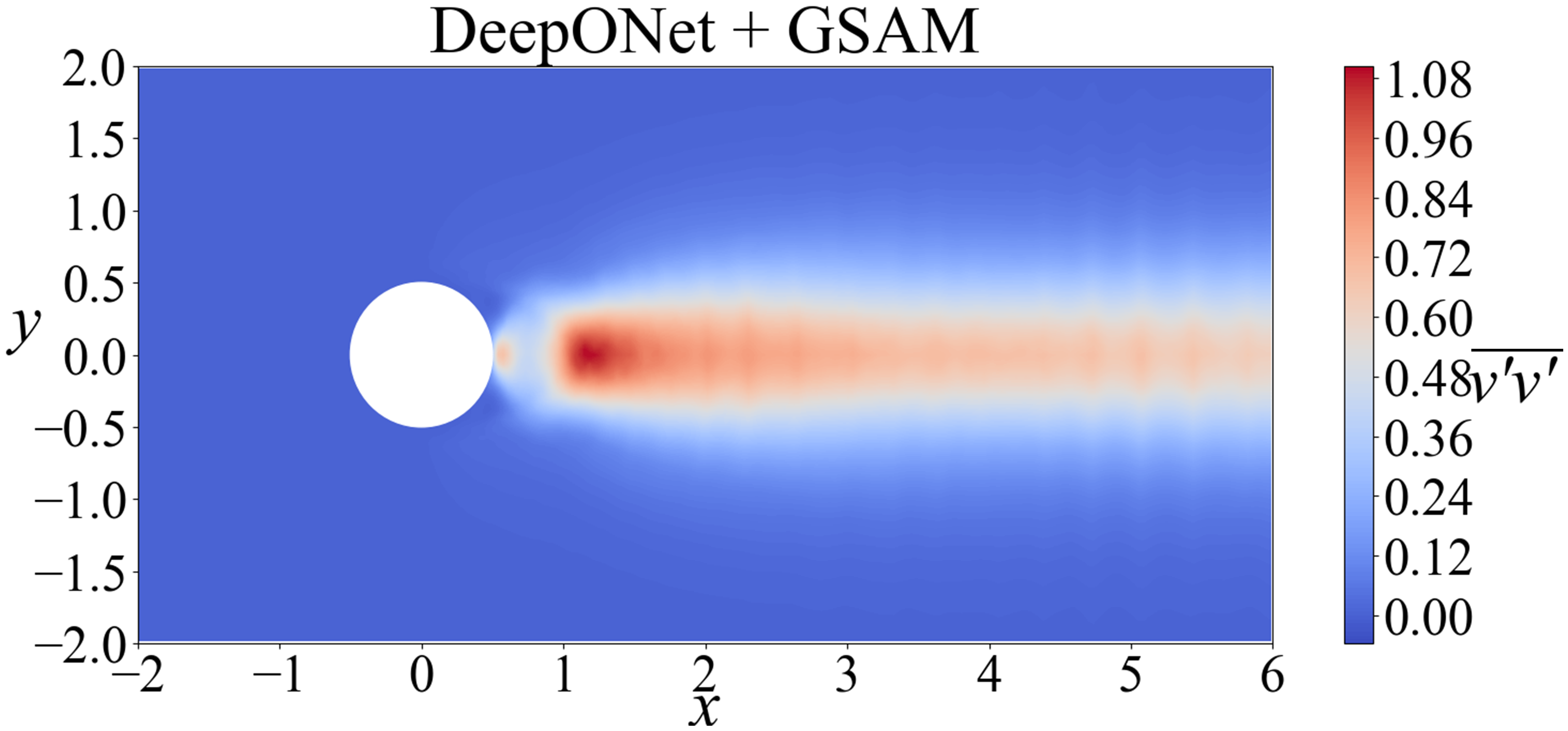}}
  \subfigure[]{
  \label{VVdeltaGSAM}
  \includegraphics[scale=0.045]{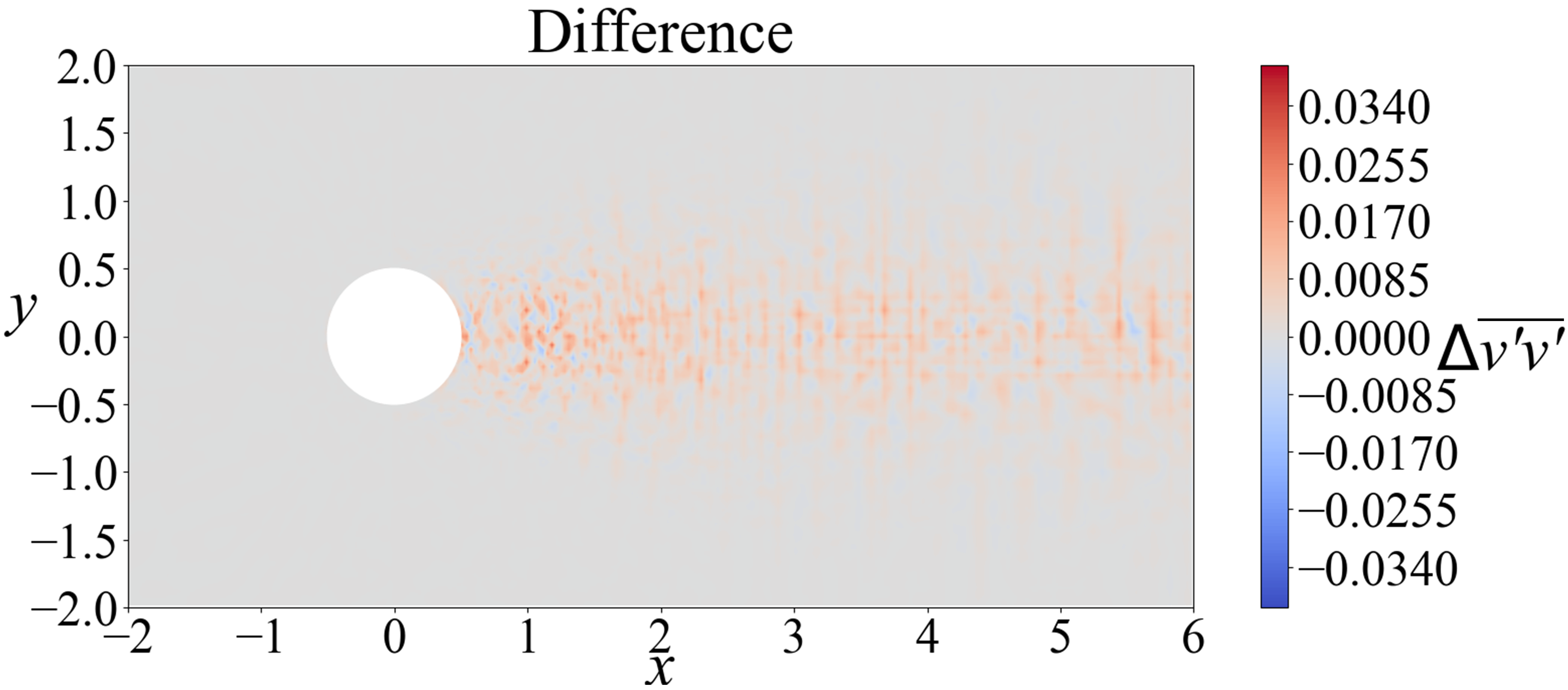}}
  \subfigure[]{
  \label{uvtrueGSAM}
  \includegraphics[scale=0.045]{uvtrue.png}}
  \subfigure[]{
  \label{UVpreGSAM}
  \includegraphics[scale=0.045]{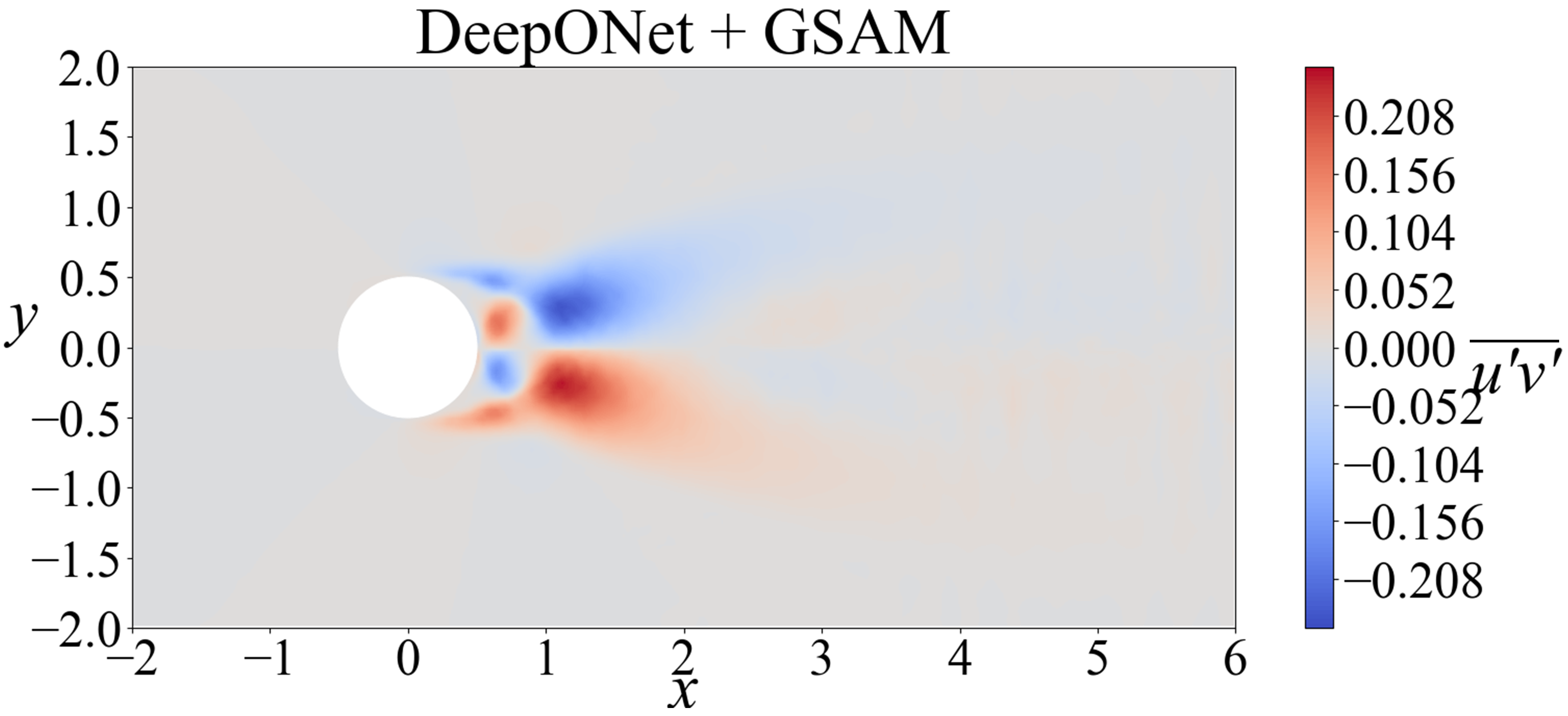}}
  \subfigure[]{
  \label{UVdeltaGSAM}
  \includegraphics[scale=0.045]{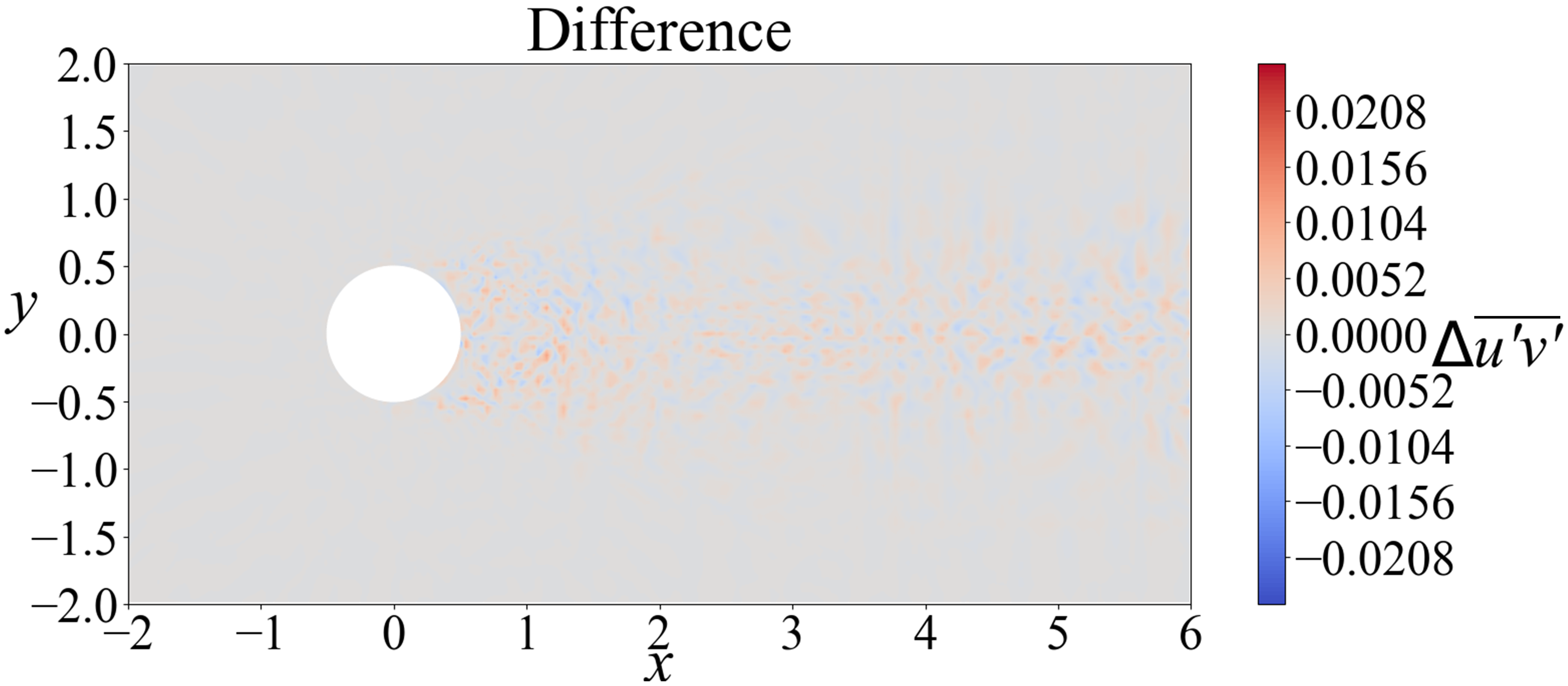}}
  \caption{Comparison of the time-averaged Reynolds stress between the DNS data and the predictions of DeepONet + GSAM. (a) $\overline{u^{'}u^{'}}_{DNS}$. (b) $\overline{u^{'}u^{'}}_{GSAM}$. (c) $\overline{u^{'}u^{'}}_{DNS}-\overline{u^{'}u^{'}}_{GSAM}$.
  (d) $\overline{v^{'}v^{'}}_{DNS}$. (e) $\overline{v^{'}v^{'}}_{GSAM}$. (f) $\overline{v^{'}v^{'}}_{DNS}-\overline{v^{'}v^{'}}_{GSAM}$.
  (g) $\overline{u^{'}v^{'}}_{DNS}$. (h) $\overline{u^{'}v^{'}}_{GSAM}$. (i) $\overline{u^{'}v^{'}}_{DNS}-\overline{u^{'}v^{'}}_{GSAM}$.}\label{Reynold stress GSAM}
\end{figure*}

\begin{figure*}[htbp] 
 \centering  
  \subfigure[]{
  \label{GSAMUPSDxD1}
  \includegraphics[scale=0.3]{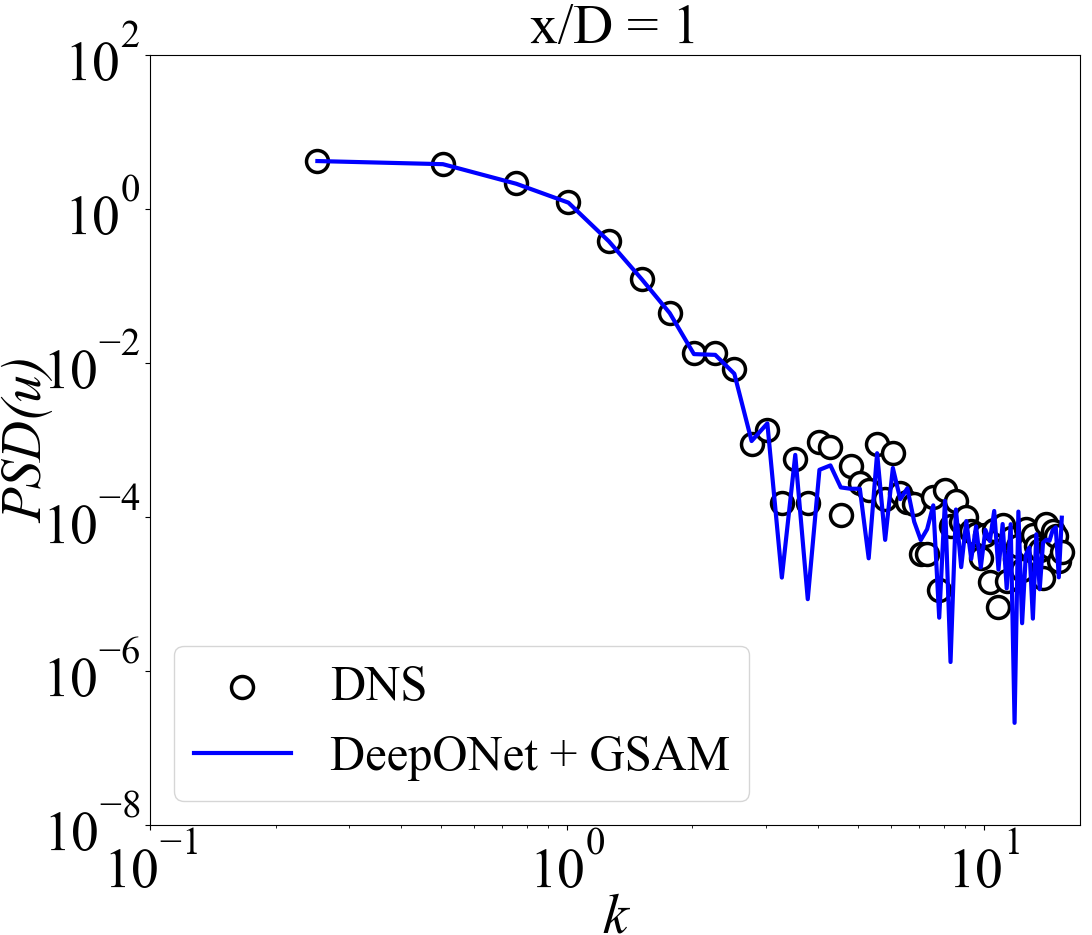}}
  \subfigure[]{
  \label{GSAMUPSDxD1.5}
  \includegraphics[scale=0.3]{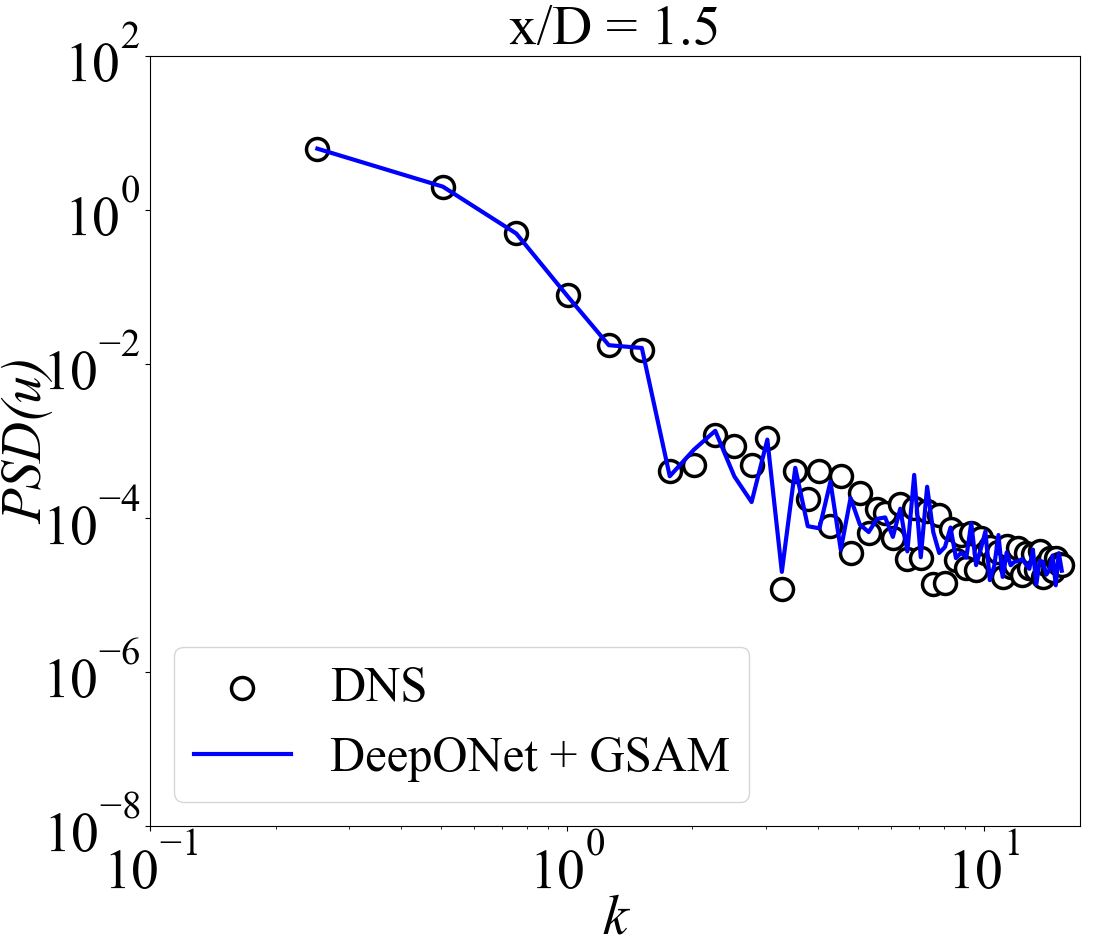}}
  \subfigure[]{
  \label{GSAMUPSDxD2}
  \includegraphics[scale=0.3]{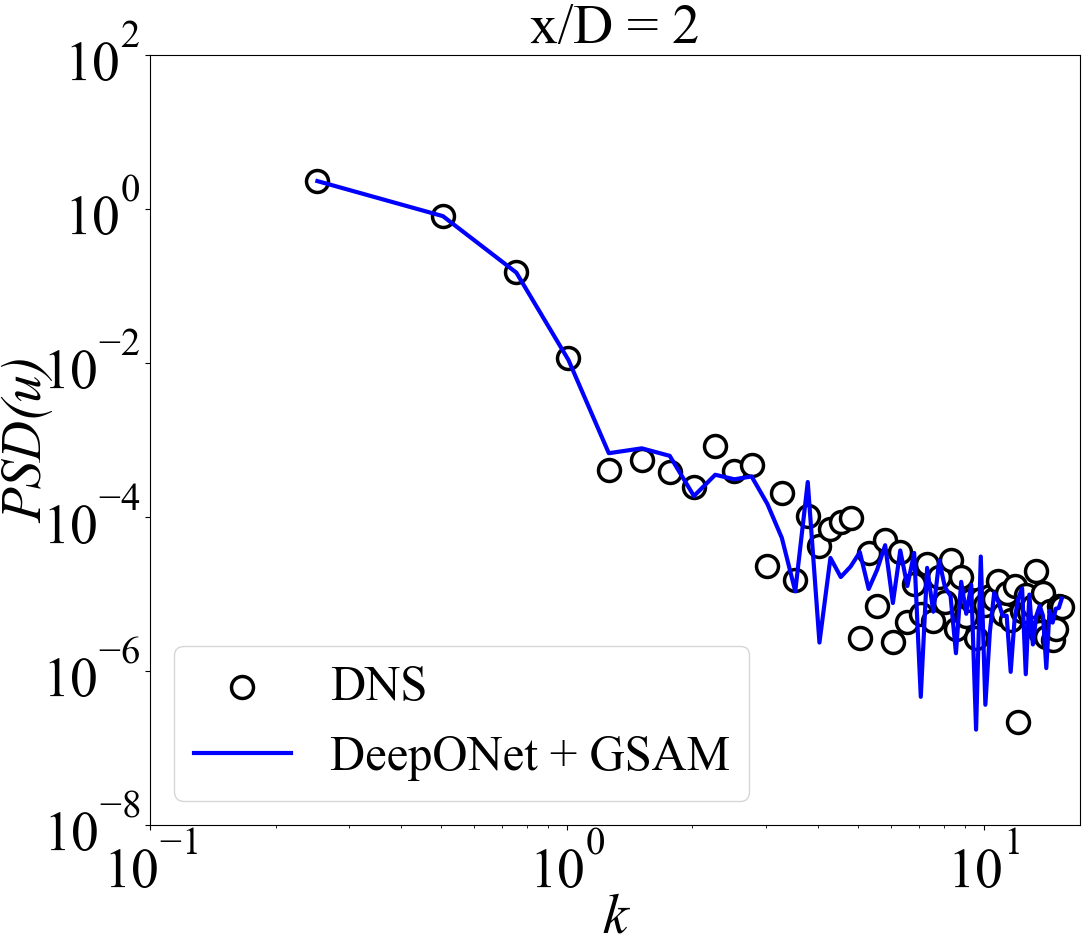}}
  \subfigure[]{
  \label{GSAMUPSDxD2.5}
  \includegraphics[scale=0.3]{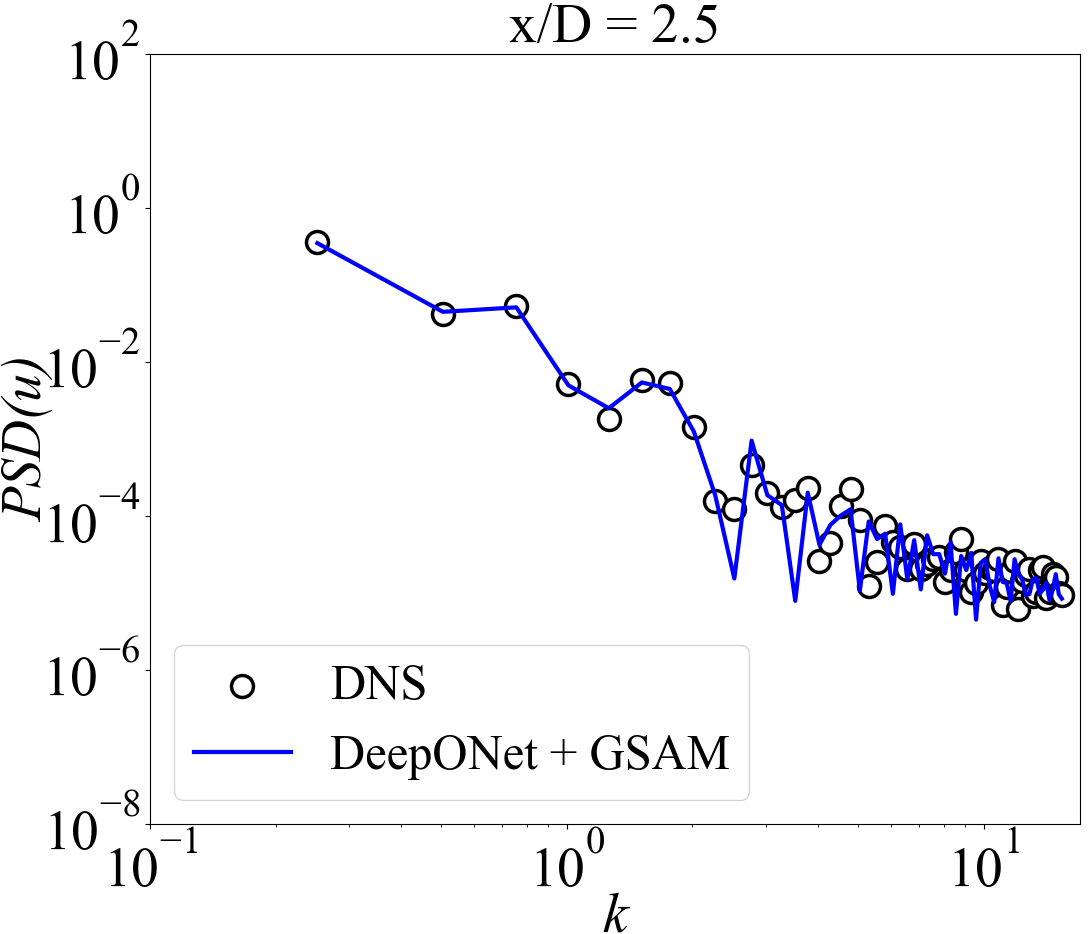}}
  \subfigure[]{
  \label{GSAMUPSDxD3}
  \includegraphics[scale=0.3]{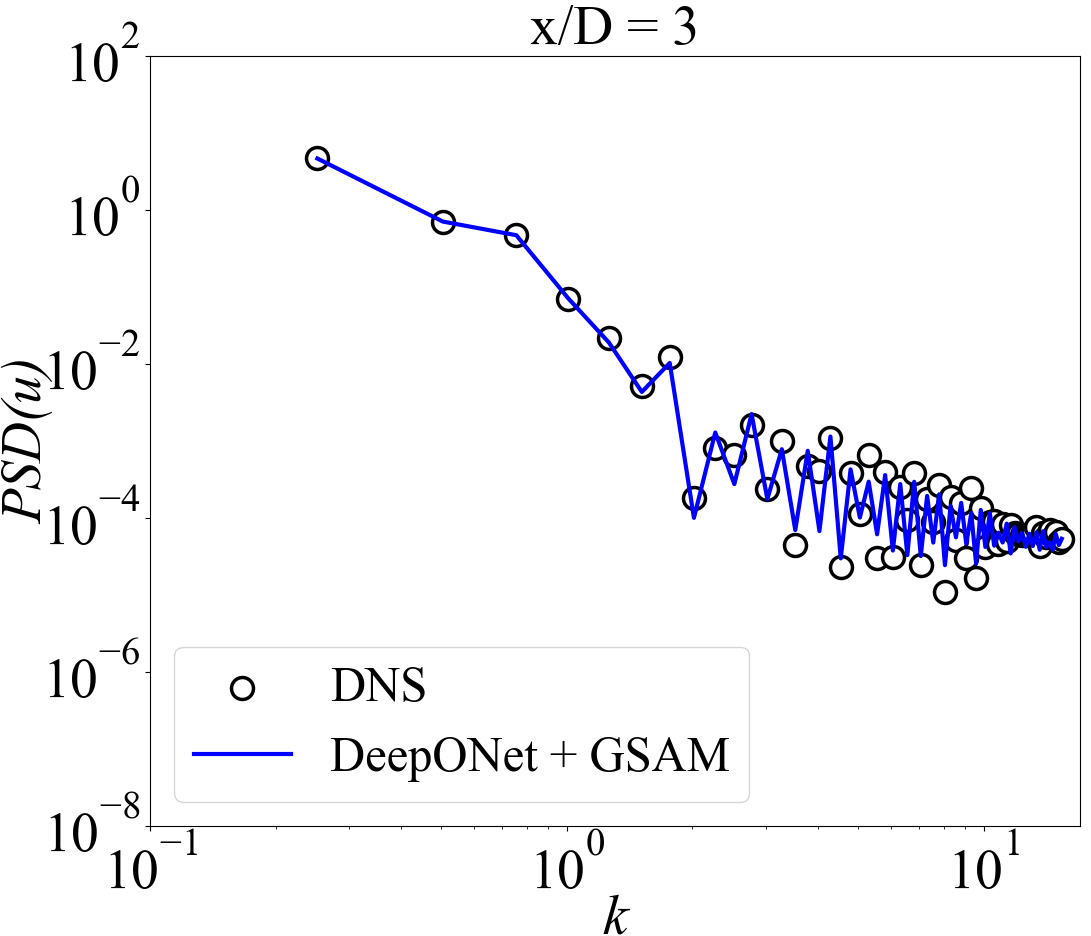}}
  \subfigure[]{
  \label{GSAMUPSDxD3.5}
  \includegraphics[scale=0.3]{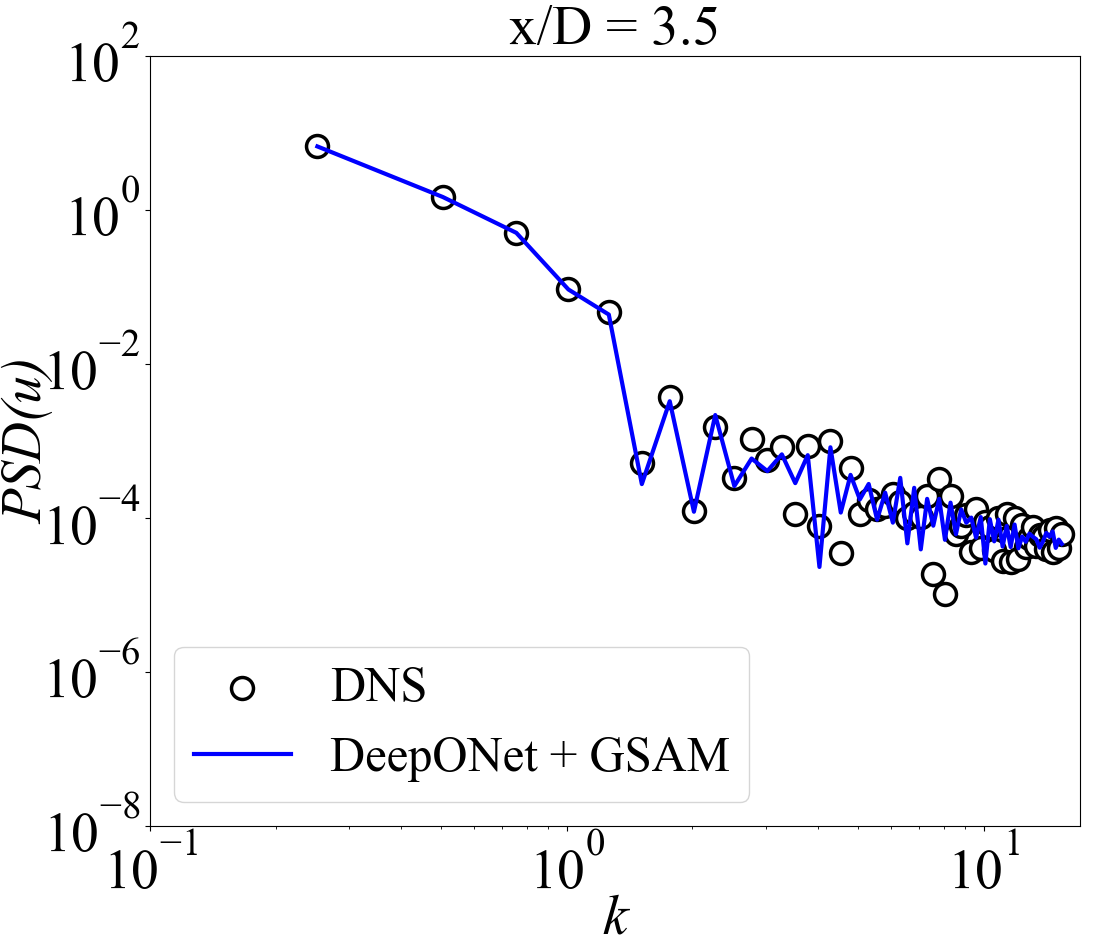}}
  \subfigure[]{
  \label{GSAMVPSDxD1}
  \includegraphics[scale=0.3]{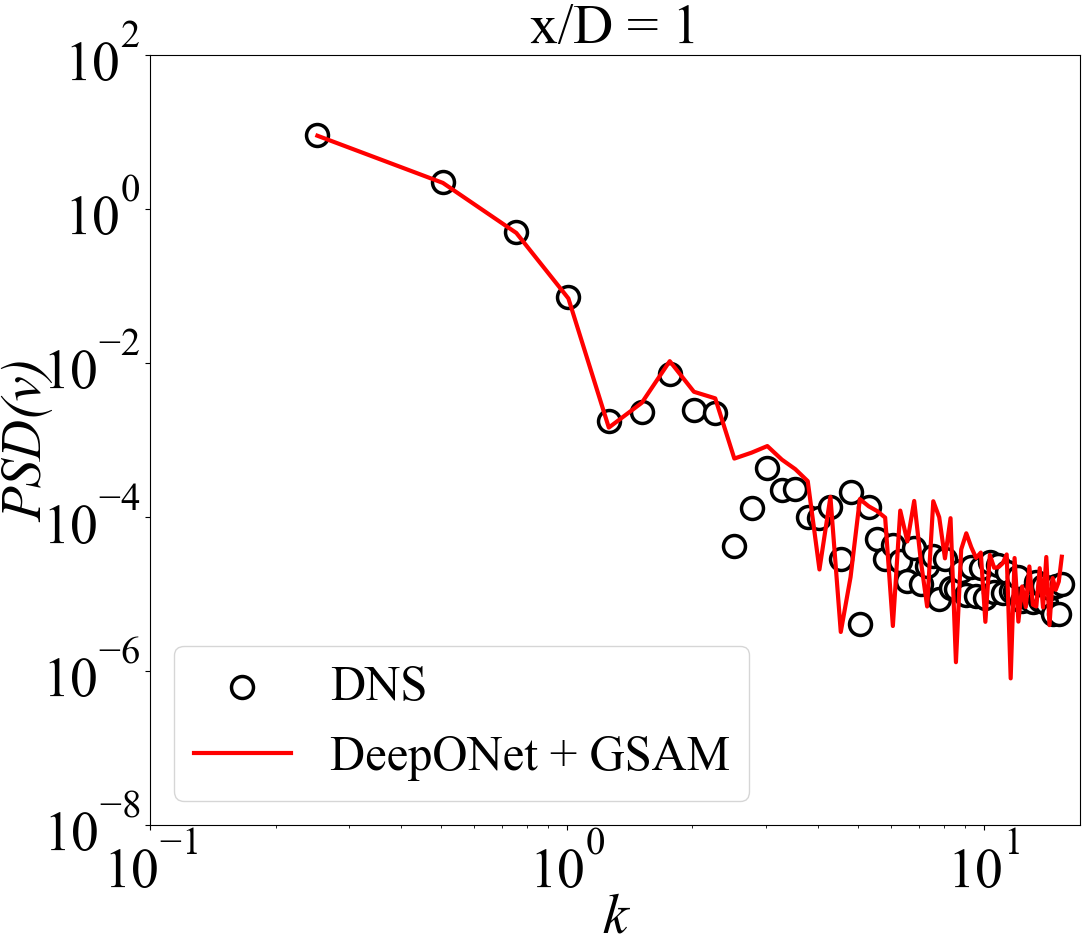}}
  \subfigure[]{
  \label{GSAMVPSDxD1。5}
  \includegraphics[scale=0.3]{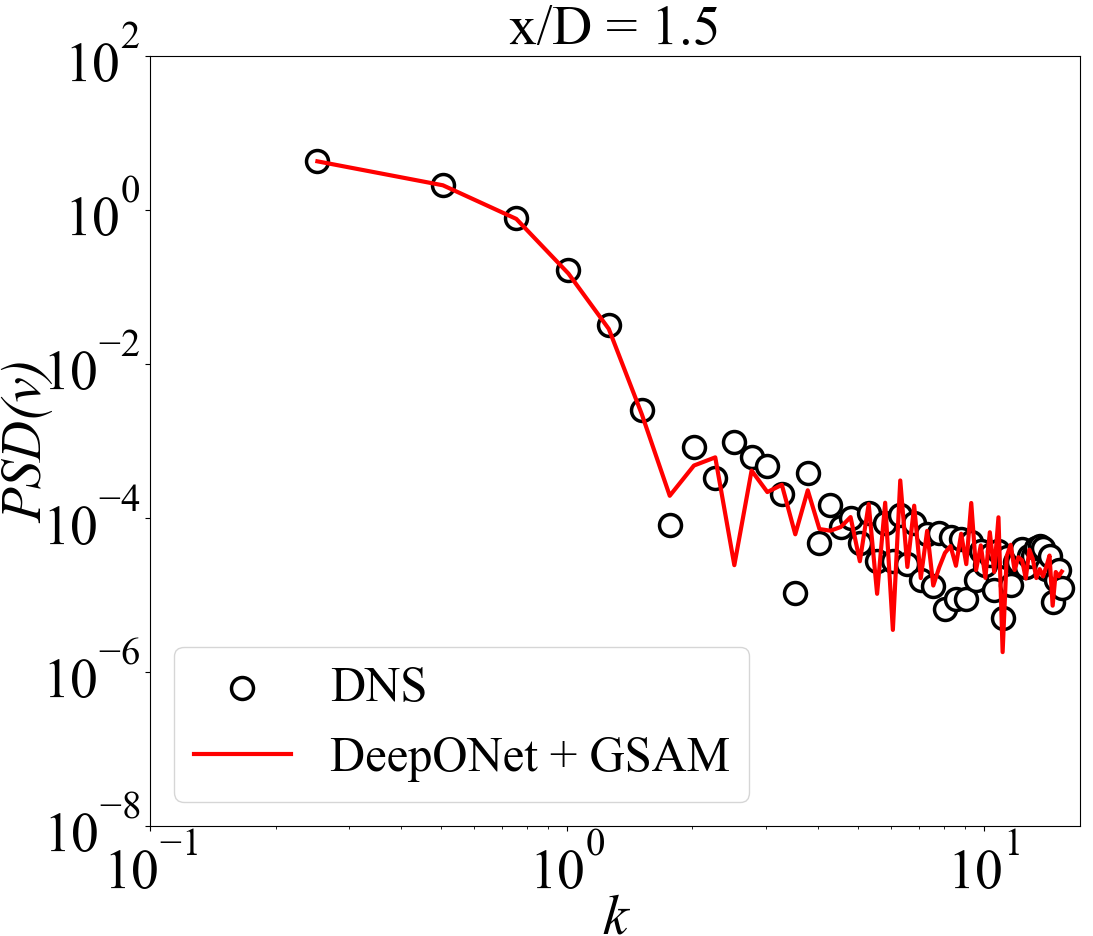}}
  \subfigure[]{
  \label{GSAMVPSDxD2}
  \includegraphics[scale=0.3]{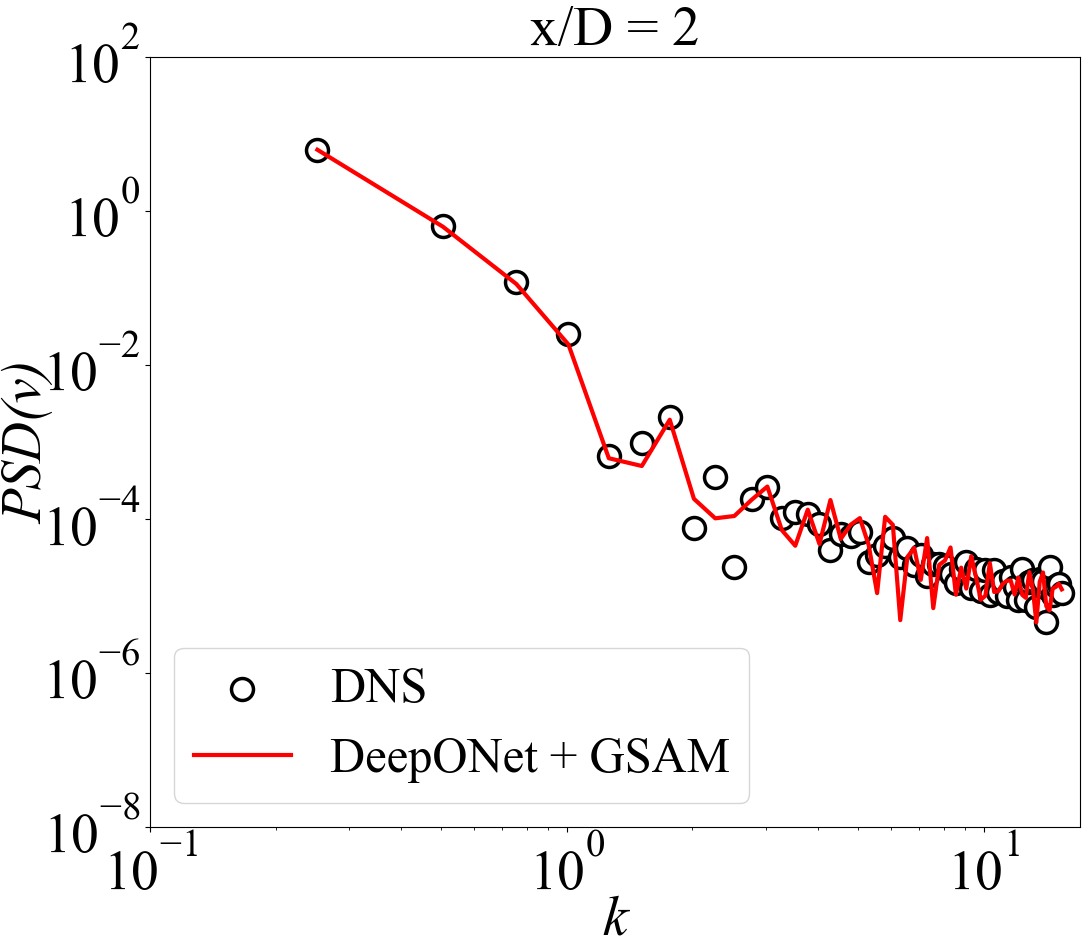}}
  \subfigure[]{
  \label{GSAMVPSDxD2.5}
  \includegraphics[scale=0.3]{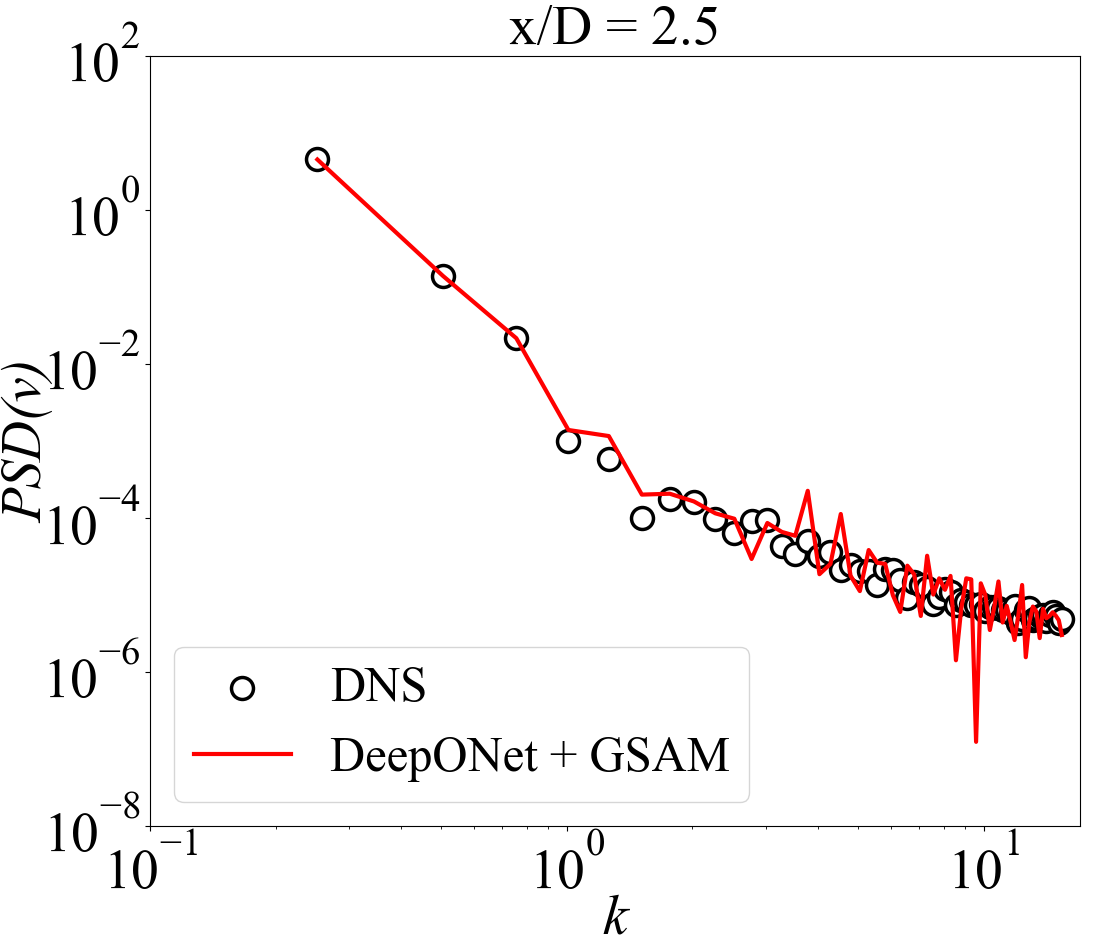}}
  \subfigure[]{
  \label{GSAMVPSDxD3}
  \includegraphics[scale=0.3]{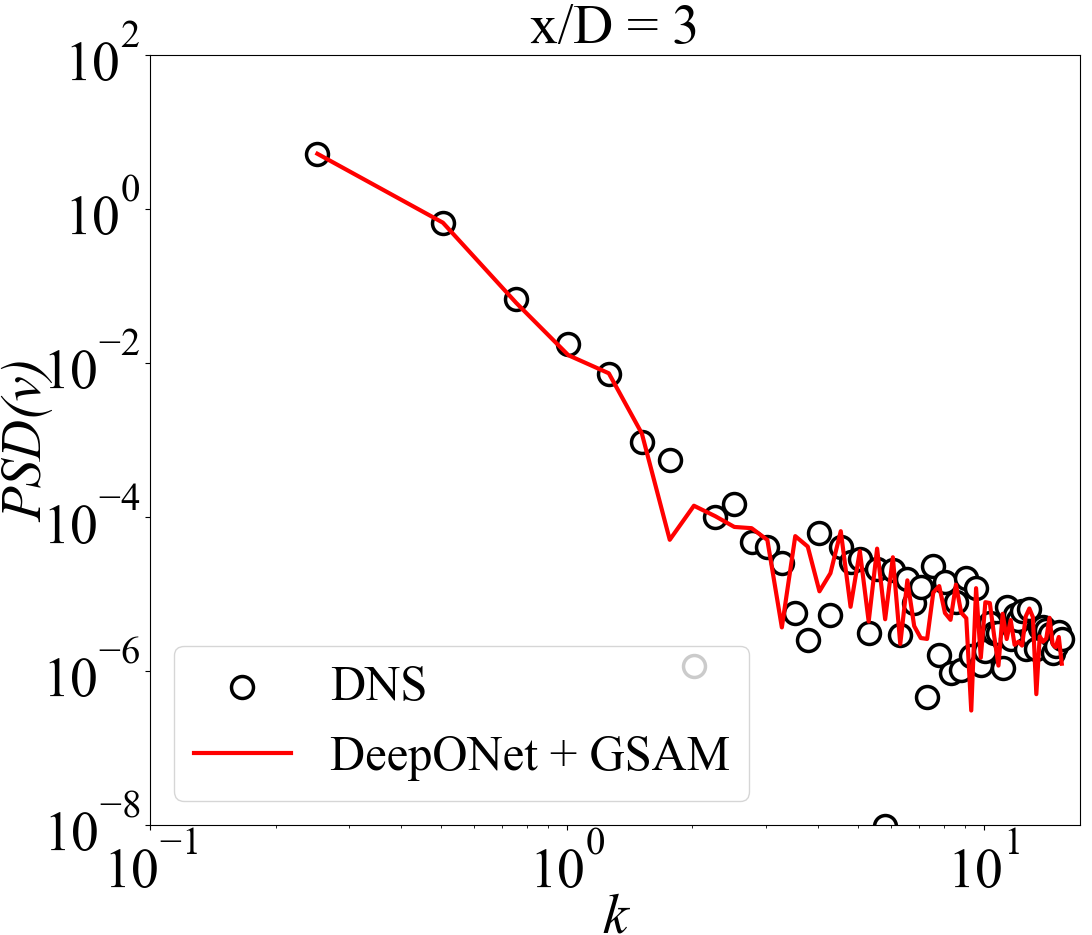}}
  \subfigure[]{
  \label{GSAMVPSDxD3.5}
  \includegraphics[scale=0.3]{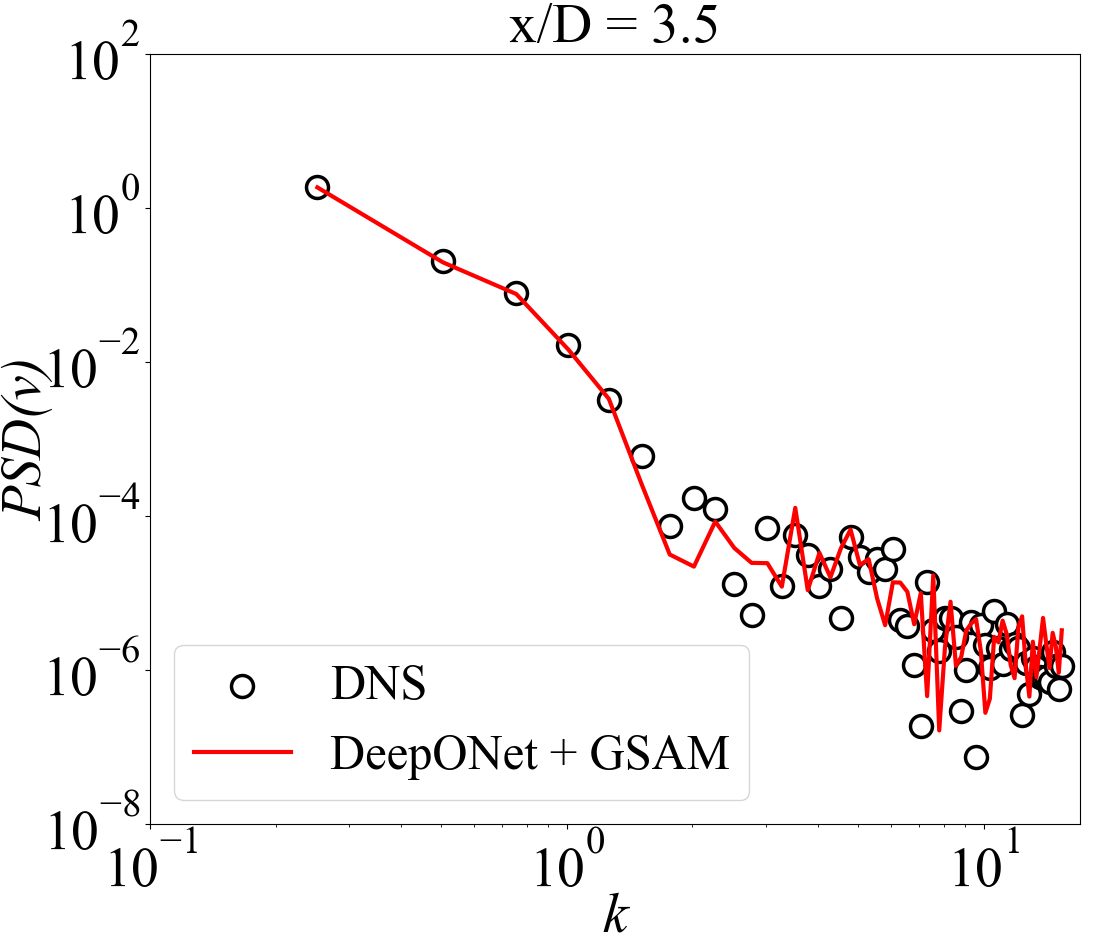}}
 \caption{The PSD of $u$ and $v$ at different downstream locations based on the DNS data and the predictions of DeepONet + GSAM at $4000th$ extrapolation snapshot. 
 (a)-(f): $PSD(u)$ for $x/D=1$, $1.5$, $2$, $2.5$, $3$ and $3.5$. (g)-(l): $PSD(v)$ for $x/D=1$, $1.5$, $2$, $2.5$, $3$ and $3.5$.}
 \label{PSD UV GSAM}
\end{figure*}

The time-averaged Reynolds stresses predicted by DeepONet + GSAM are presented in Fig.~\ref{Reynold stress GSAM}.  Furthermore, the PSD of $u$ and $v$ along the $y$ direction at various downstream distances from the cylinder are shown in Fig.~\ref{PSD UV GSAM}.

Similar to section \ref{CNNDeepONetMultihead}, the DeepONet + GSAM accurately predicts the time-averaged Reynolds stress of the flow field and shows excellent agreement with DNS data. Additionally, the DeepONet + GSAM demonstrates remarkable precision in forecasting the low-wavenumber components of $u$ and $v$. However, it is incompetent to provide accurate predictions for high-wavenumber flow field components.

\section{$N_s = 100$ and $\Delta t = 0.2$}
\label{appendixC}

The DeepONet + GSAM is trained to extrapolate the flow field of 2D flow around a circular cylinder at $Re = 1000$ with $\Delta t = 0.2$ in this section.

\begin{figure*}[htbp]
\centering
  \includegraphics[scale=0.14]{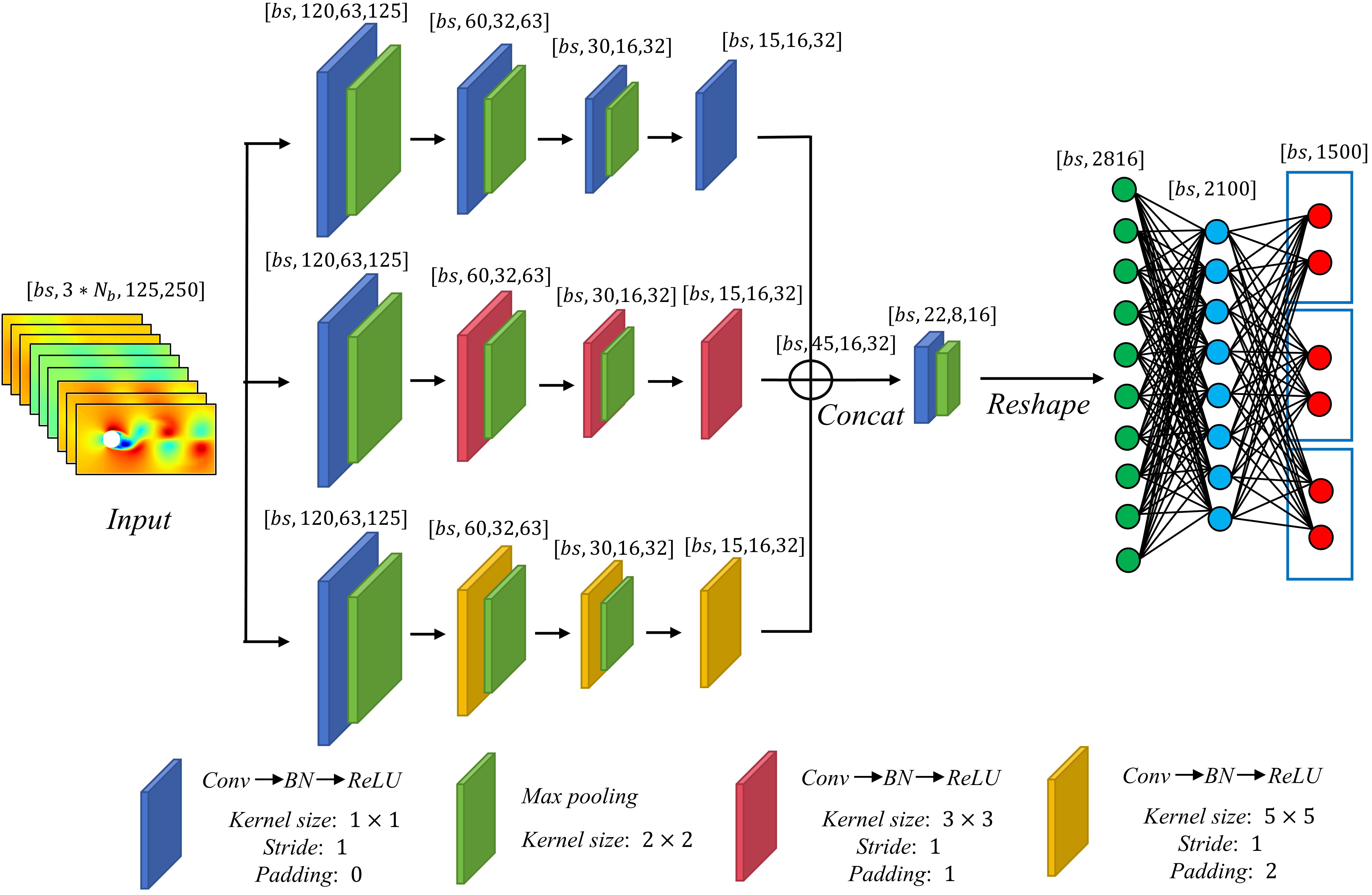}\\
  \caption{The detail architecture and parameters of the branch network for $\Delta t = 0.2$.}\label{Detail 0.2 Branch}
\end{figure*}

$N_b$ is set to 20 in order to encompass adequate flow field evolution information as $\Delta t$ is decreased; $N_t = 2000$; the initial $lr$ is set to 0.0001, and a $cosine$ $lr$ schedule is employed; $epoch = 1000$; $batchsize = 20$; $\rho=0.0001$; $\alpha=0.4$. The parameters of trunk network remain unchanged, while the neural network parameters of branch network are adjusted to enhance the extraction of flow field characteristics, as illustrated in Fig.~\ref{Detail 0.2 Branch}. The specific parameter analysis is no longer presented in exhaustive detail, but rather focuses on the main results.

\begin{figure*}[htbp] 
 \centering  
  \subfigure[]{
  \label{0.2UDNSGSAM}
  \includegraphics[scale=0.178]{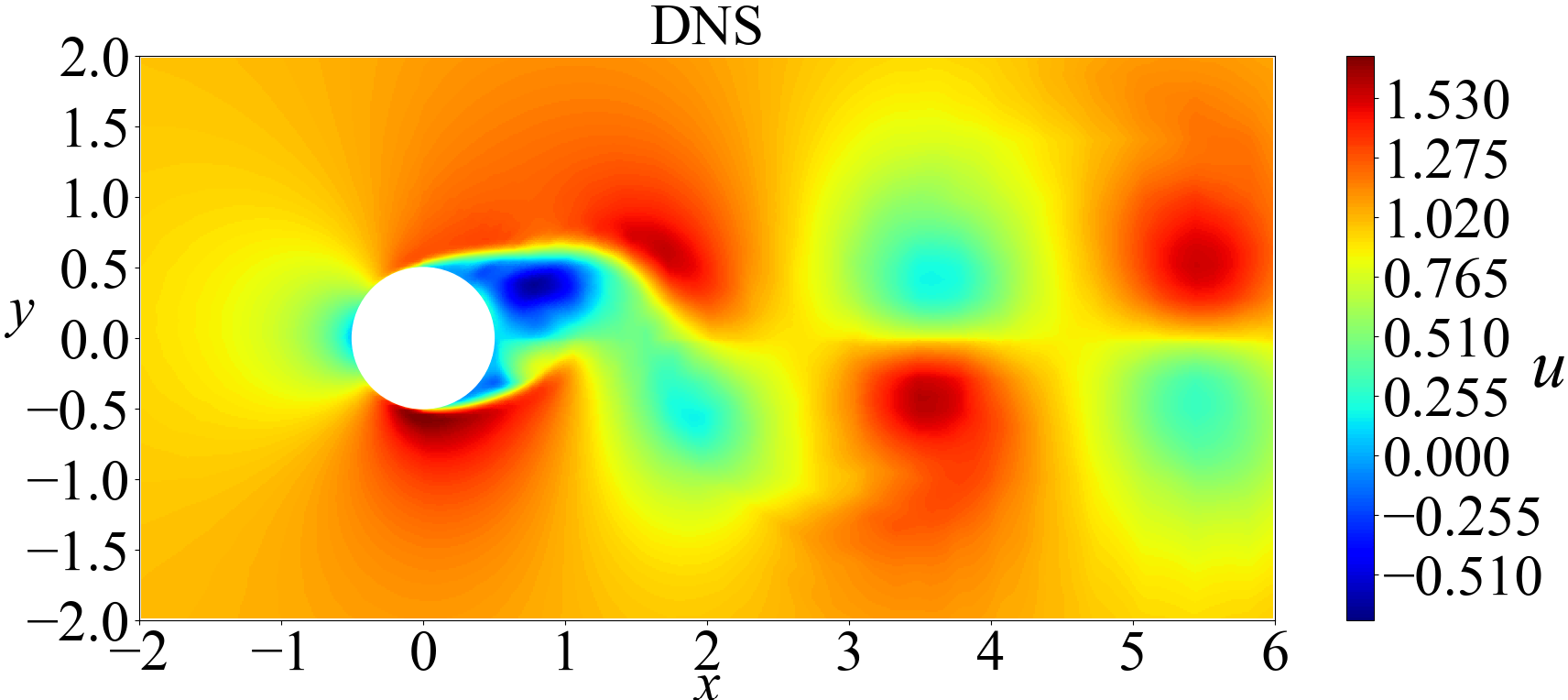}}
  \subfigure[]{
  \label{0.2UpreGSAM}
  \includegraphics[scale=0.178]{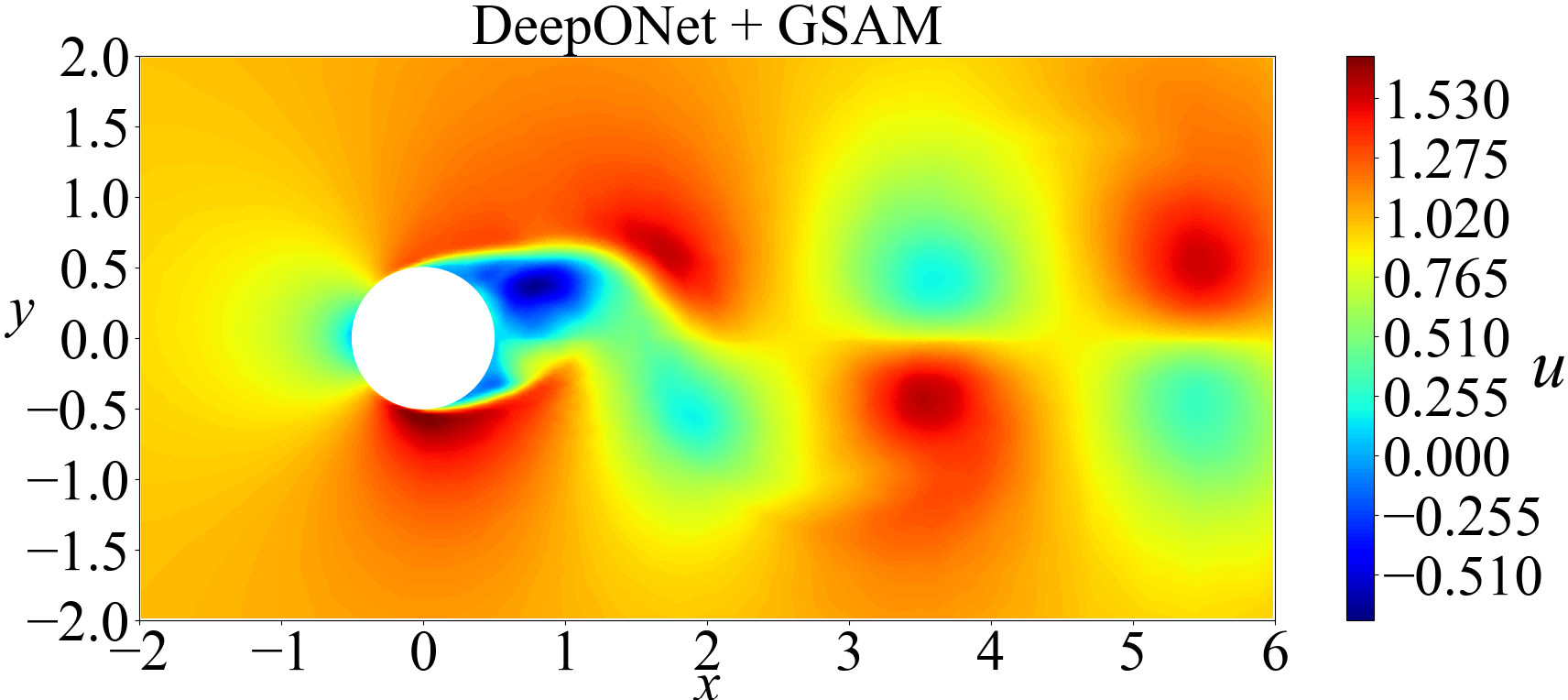}}
  \subfigure[]{
  \label{0.2UdeltaGSAM}
  \includegraphics[scale=0.178]{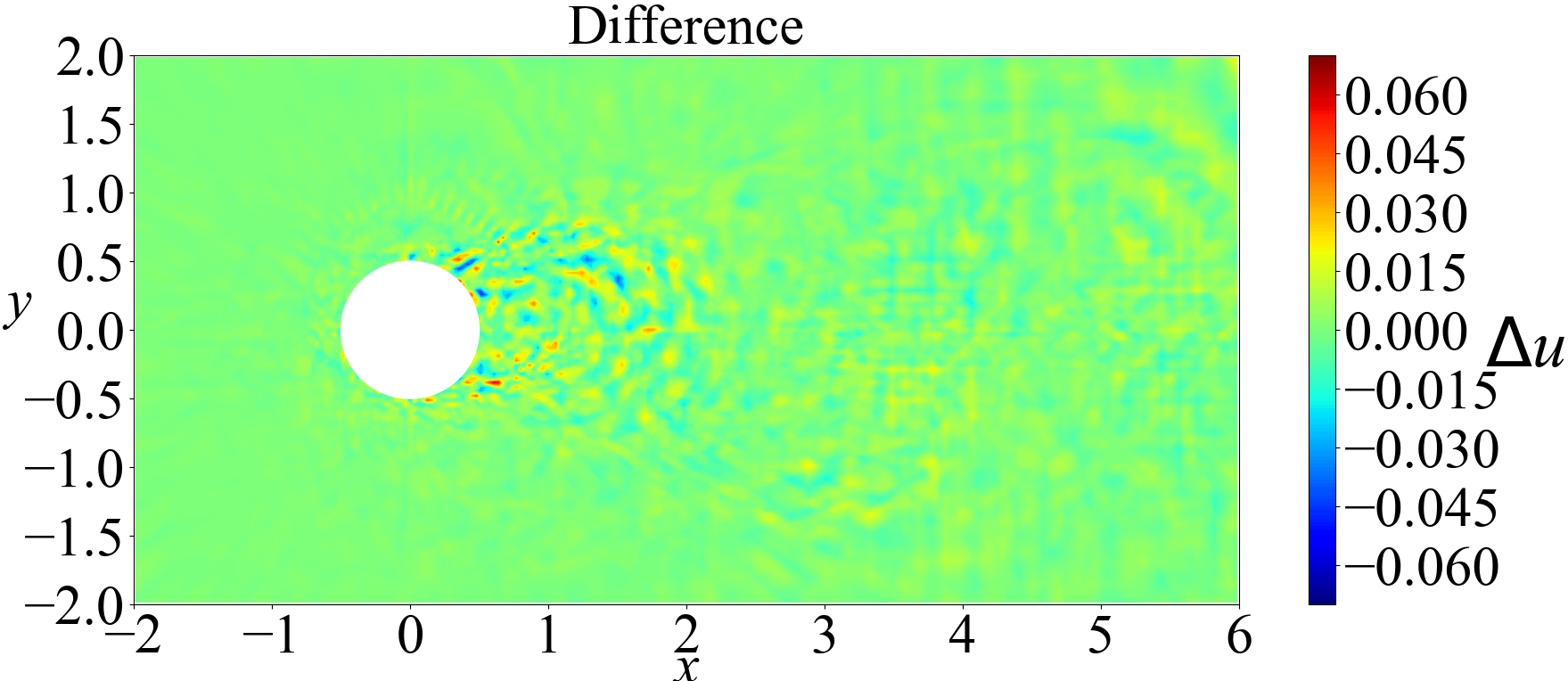}}
  \subfigure[]{
  \label{0.2VDNSGSAM}
  \includegraphics[scale=0.178]{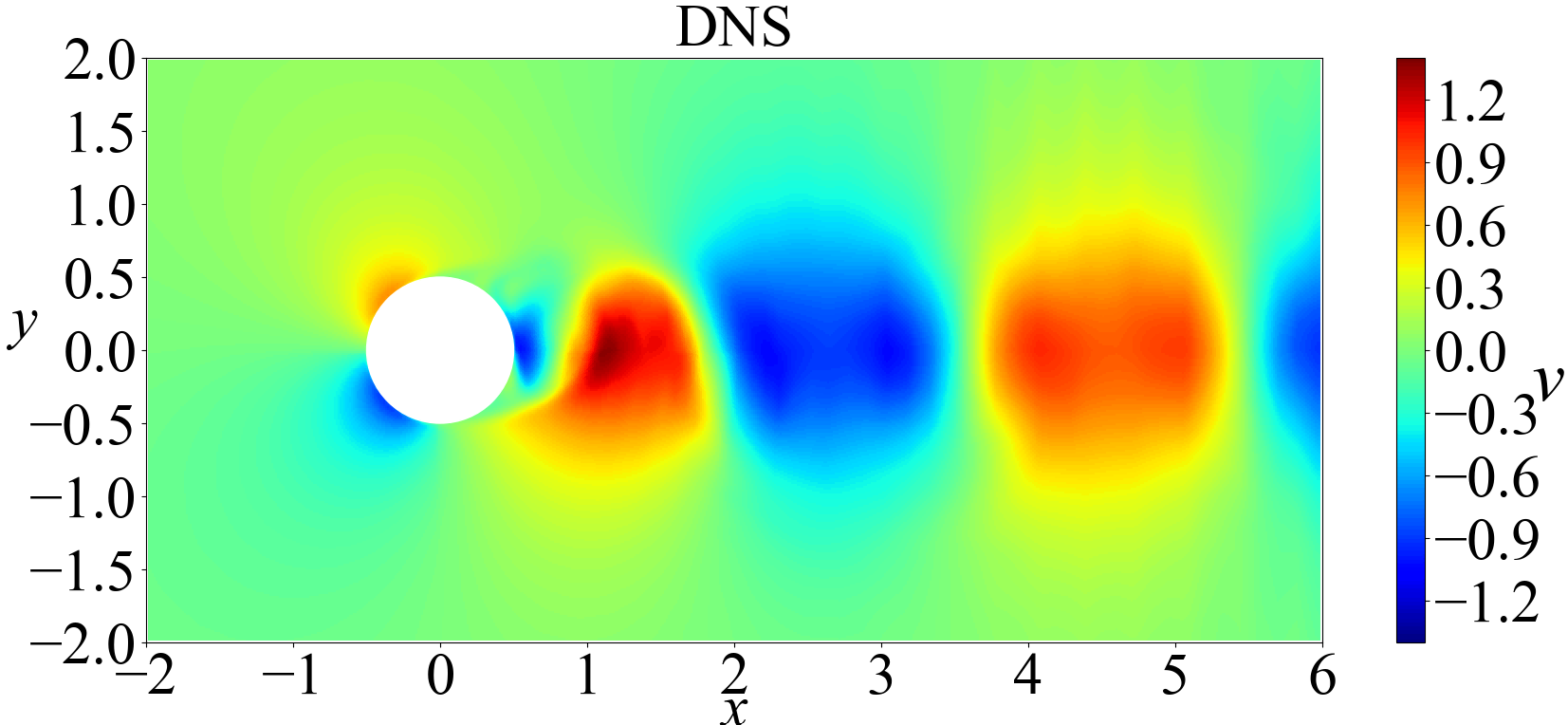}}
  \subfigure[]{
  \label{0.2VpreGSAM}
  \includegraphics[scale=0.178]{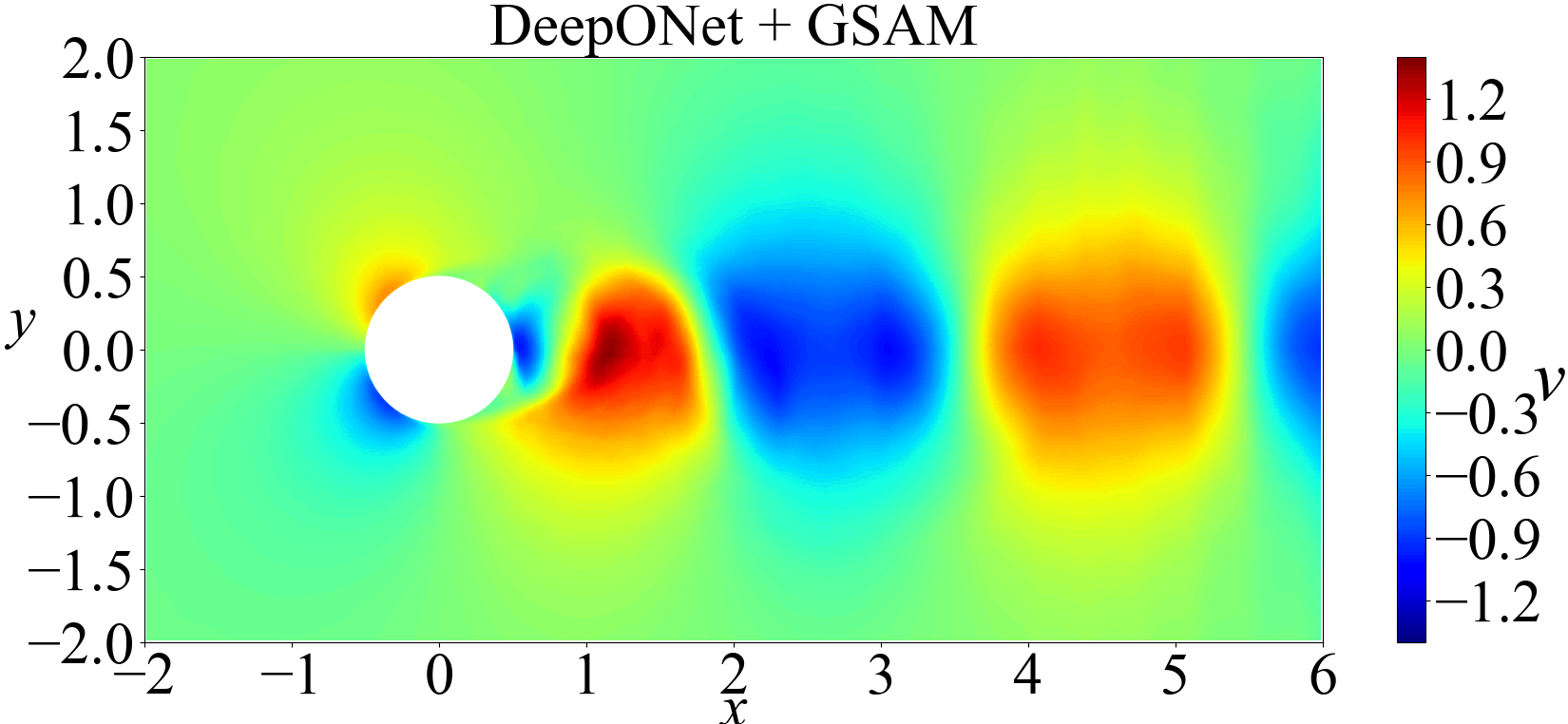}}
  \subfigure[]{
  \label{0.2VdeltaGSAM}
  \includegraphics[scale=0.178]{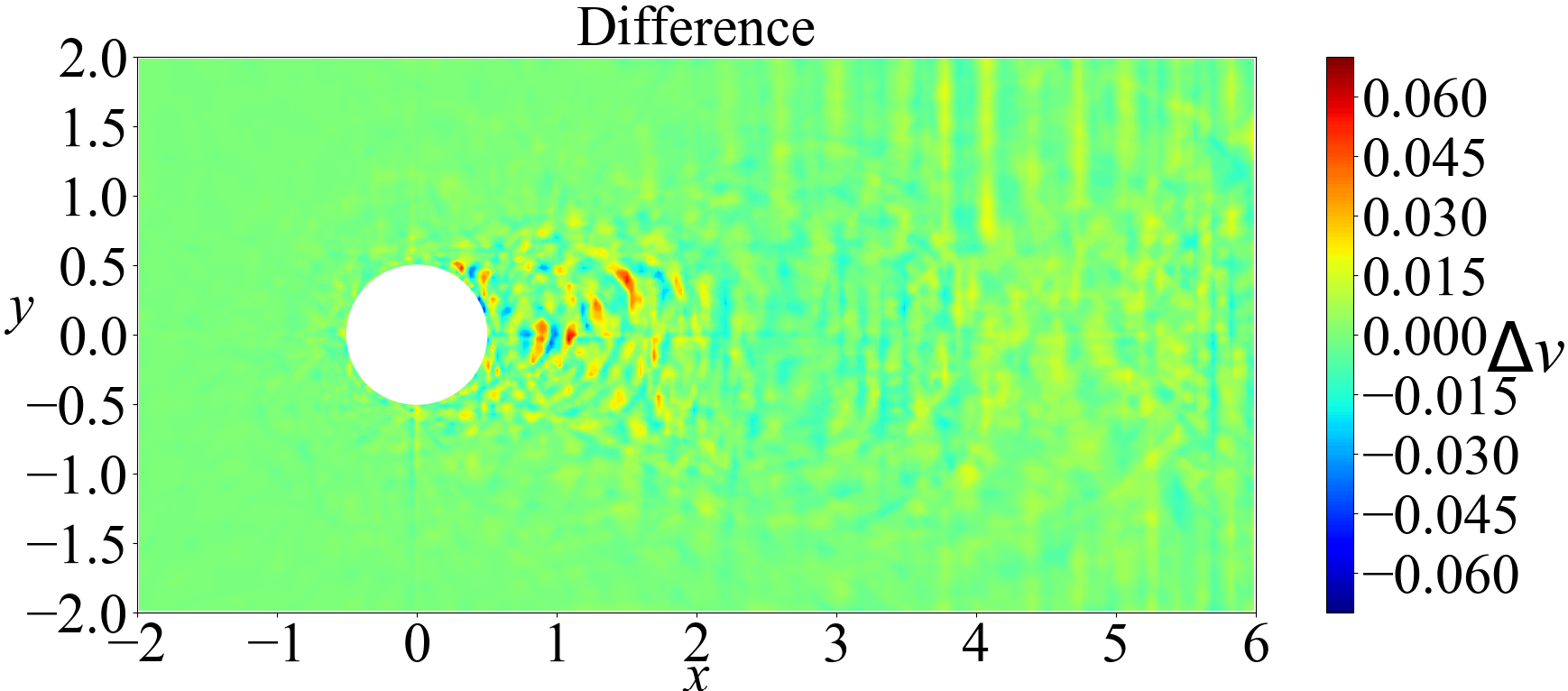}}
  \subfigure[]{
  \label{0.2PDNSGSAM}
  \includegraphics[scale=0.178]{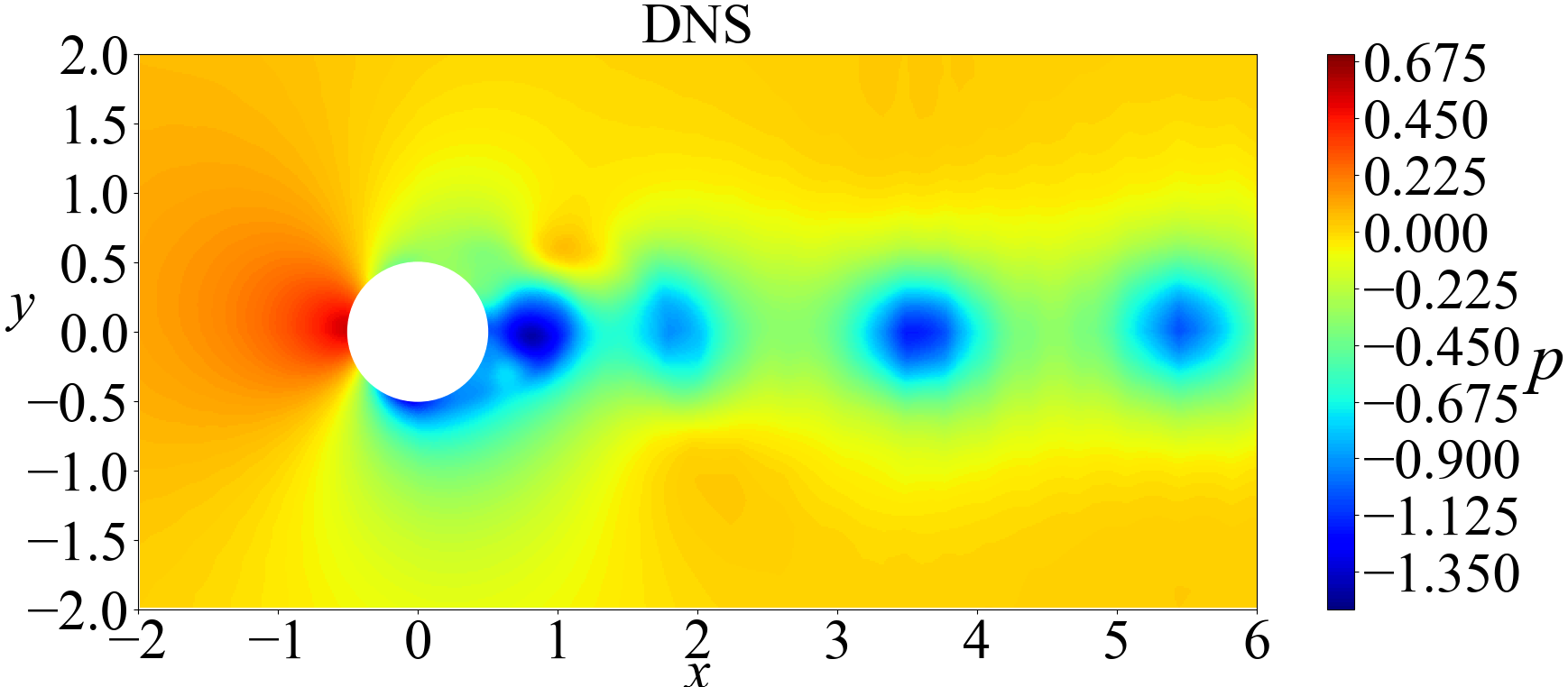}}
  \subfigure[]{
  \label{0.2PpreGSAM}
  \includegraphics[scale=0.178]{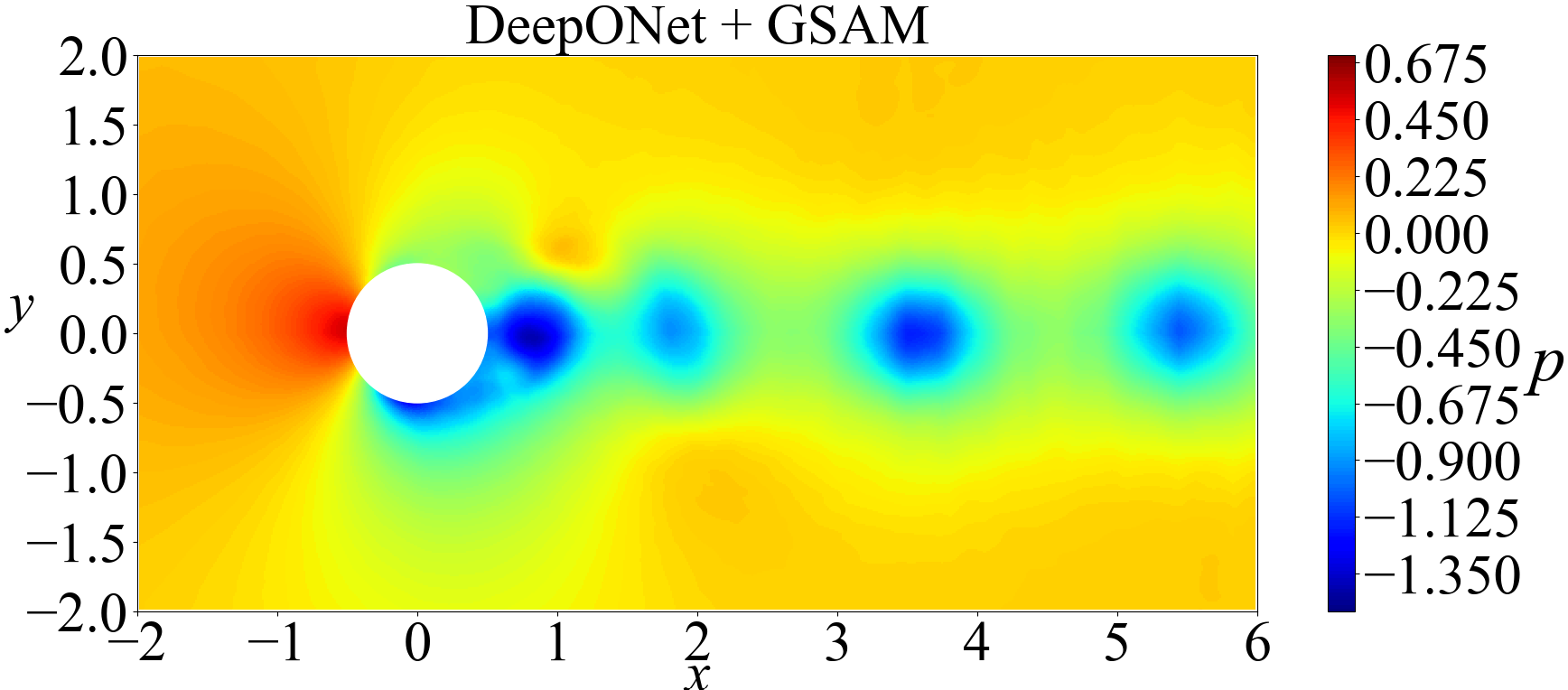}}
  \subfigure[]{
  \label{0.2PdeltaGSAM}
  \includegraphics[scale=0.178]{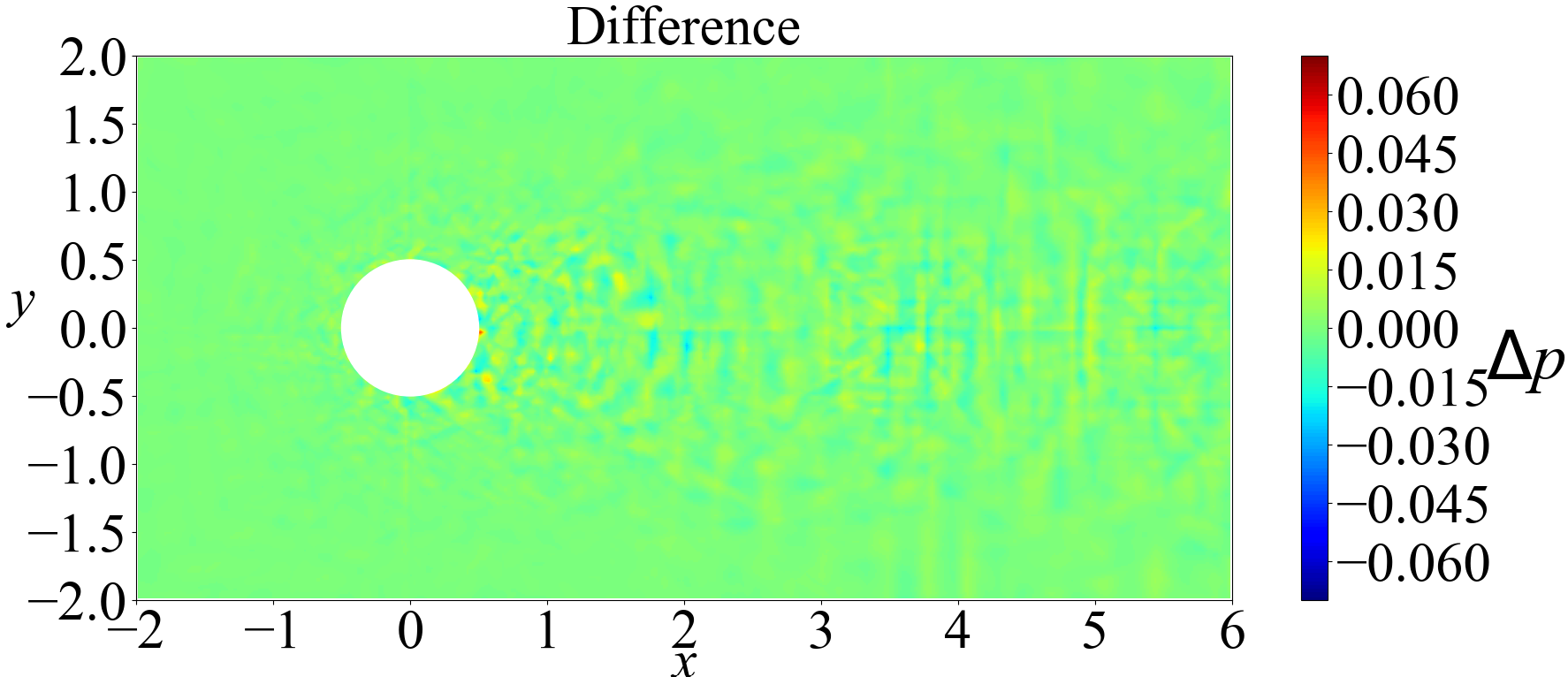}}
  \caption{Comparison of $u$, $v$ and $p$ at $4000th$ extrapolation snapshot between the prediction of DeepONet + GSAM and the DNS data for $\Delta t = 0.2$. (a) $u_{DNS}$. (b) $u_{pre}$. (c) $u_{DNS}-u_{pre}$. (d) $v_{DNS}$. (e) $v_{pre}$. (f) $v_{DNS}-v_{pre}$. (g) $p_{DNS}$. (h) $p_{pre}$. (i) $p_{DNS}-p_{pre}$.}\label{0.2GSAMcontour compare}
\end{figure*}

\begin{figure*}[htbp] 
 \centering  
  \subfigure[]{
  \label{UMSElineGSAM0.2}
  \includegraphics[scale=0.25]{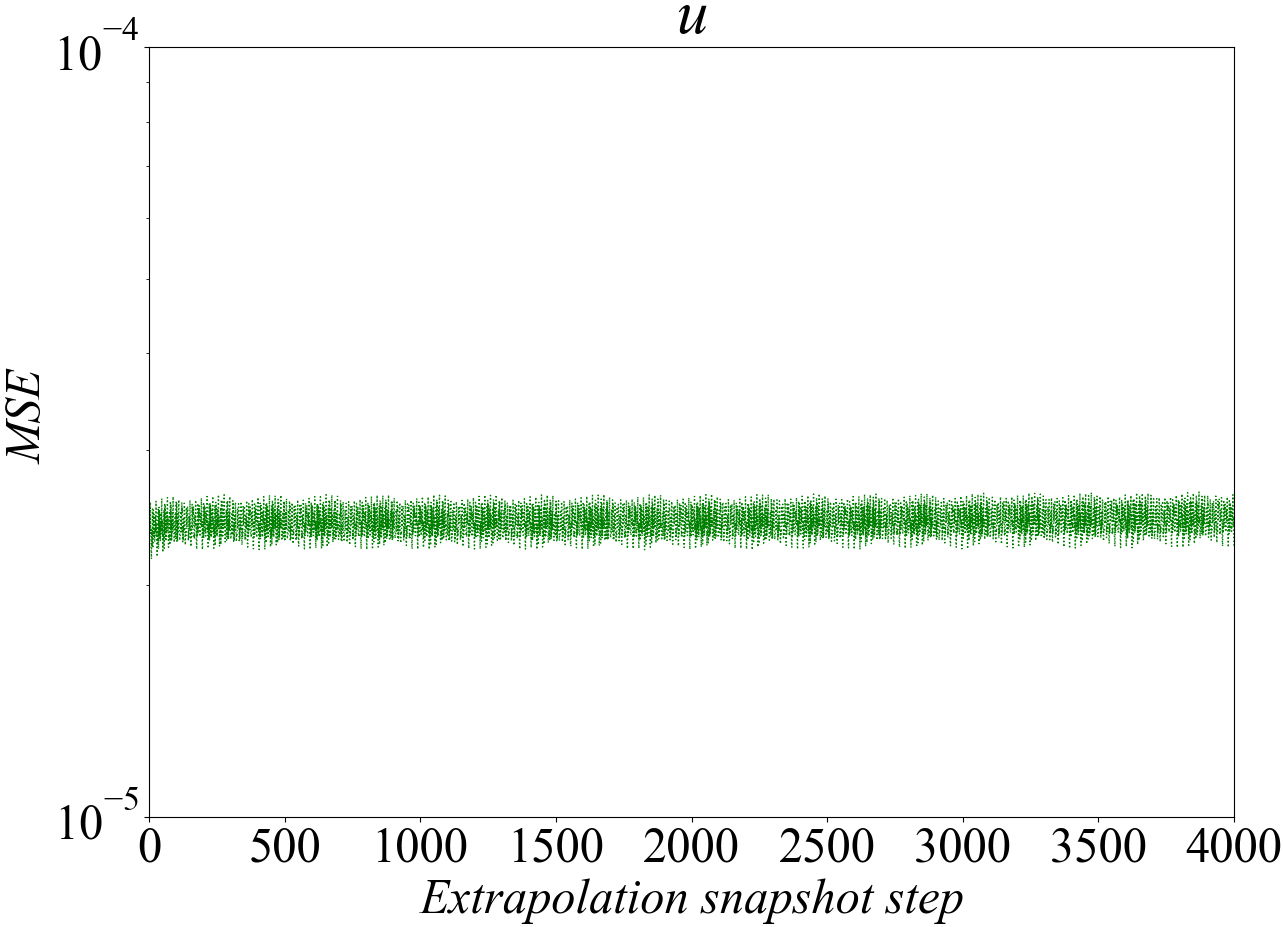}}
  \subfigure[]{
  \label{VMSElineGSAM0.2}
  \includegraphics[scale=0.25]{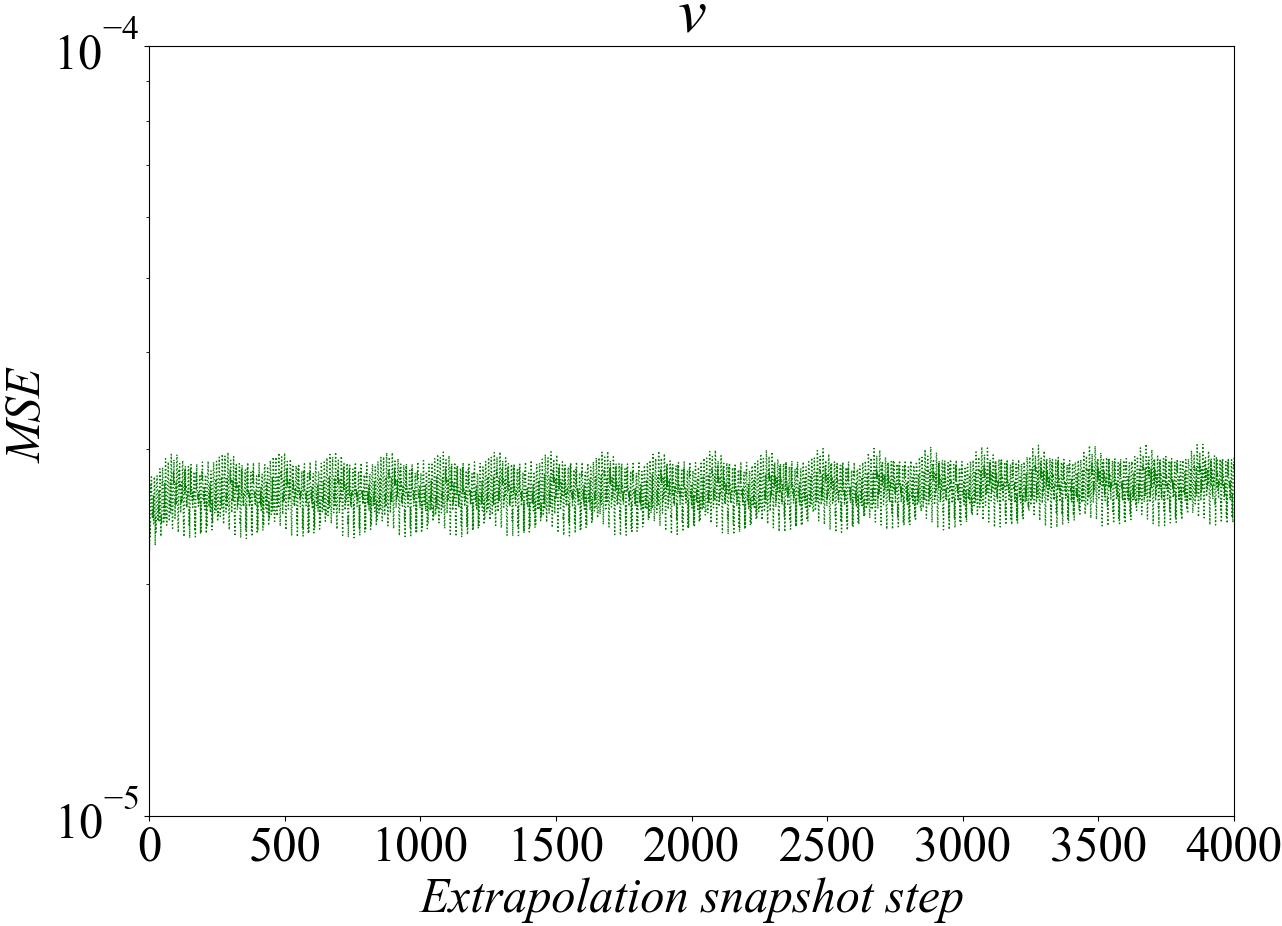}}
  \subfigure[]{
  \label{PMSElineGSAM0.2}
  \includegraphics[scale=0.25]{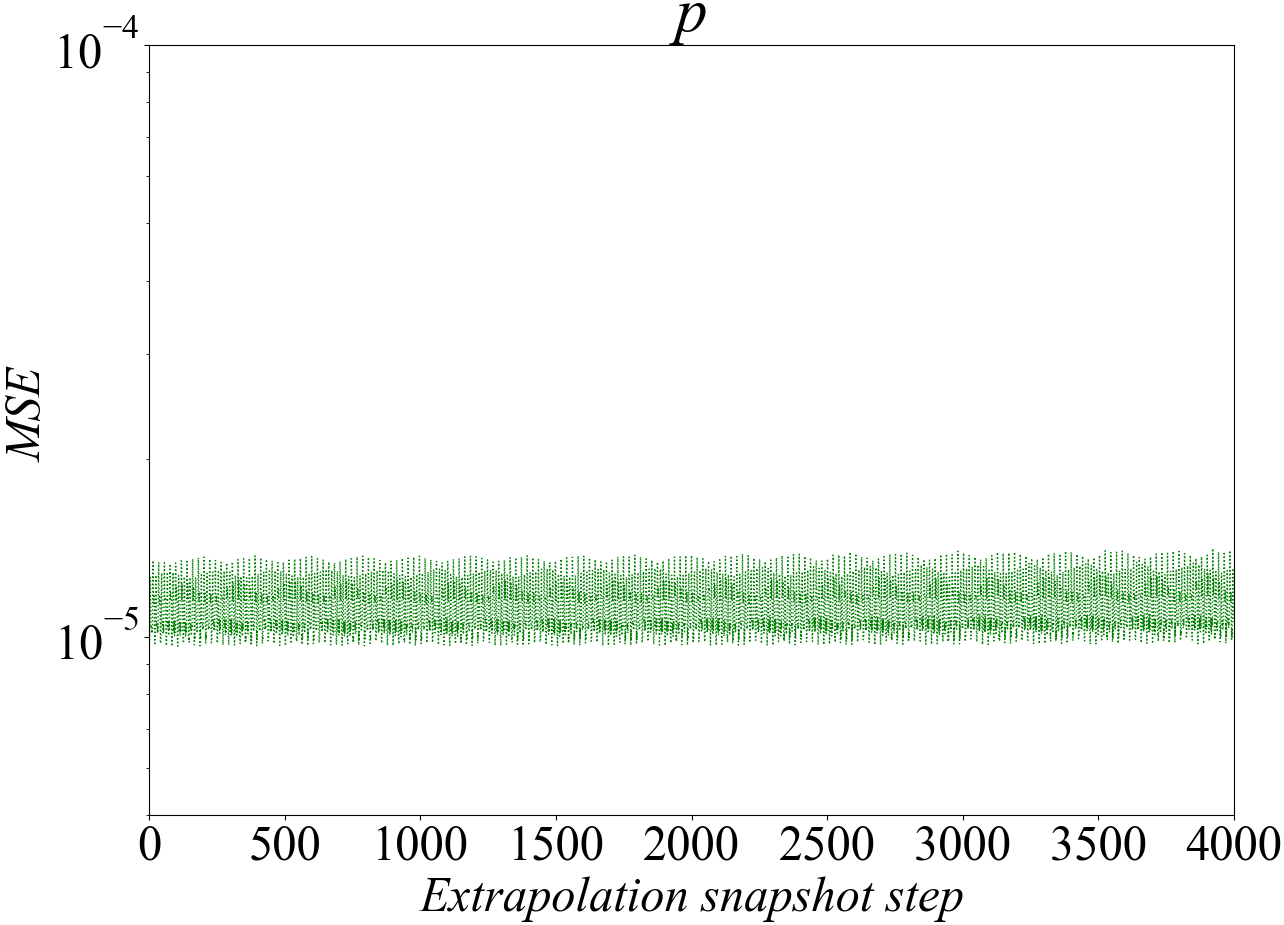}}
 \caption{The evolution of MSEs between the DNS data and the predictions of DeepONet + GSAM for $\Delta t = 0.2$. (a) $u_{MSE}$. (b) $v_{MSE}$. (c) $p_{MSE}$.}
 \label{0.2MSE line of GSAM}
\end{figure*}

The flow field results for the $4000th$ extrapolation snapshot based on the DeepONet + GSAM and the evolution of the MSEs between the DNS data and the predictions of DeepONet + GSAM over 4000 extrapolation snapshots are depicted in Fig.~\ref{0.2GSAMcontour compare} and Fig.~\ref{0.2MSE line of GSAM}, respectively.

The observation reveals that DeepONet + GSAM exhibits remarkable proficiency in capturing the temporal evolution of the flow field when employing smaller data snapshot time step.

\begin{figure*}[htbp] 
 \centering  
  \subfigure[]{
  \label{0.2UMSE diff DMDmodes}
  \includegraphics[scale=0.47]{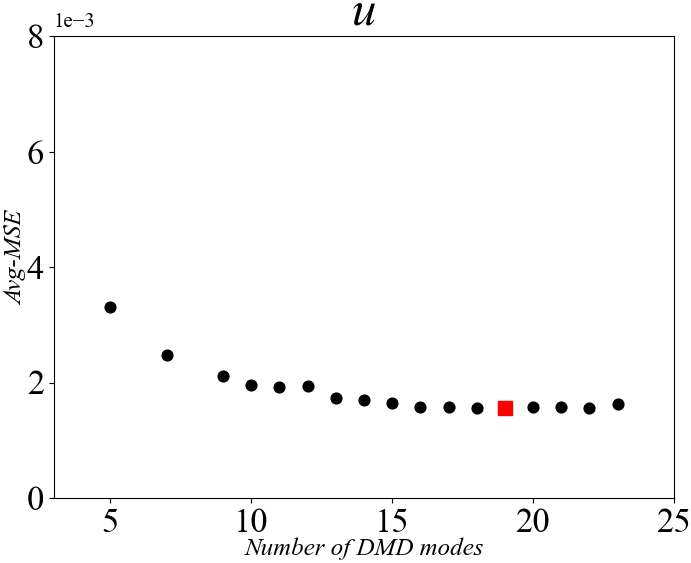}}
  \subfigure[]{
  \label{0.2VMSE diff DMDmodes}
  \includegraphics[scale=0.47]{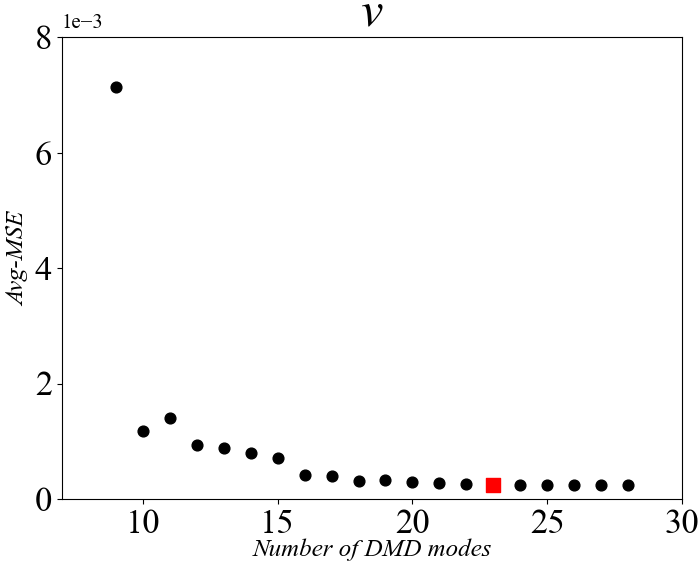}}
  \subfigure[]{
  \label{0.2PMSE diff DMDmodes}
  \includegraphics[scale=0.47]{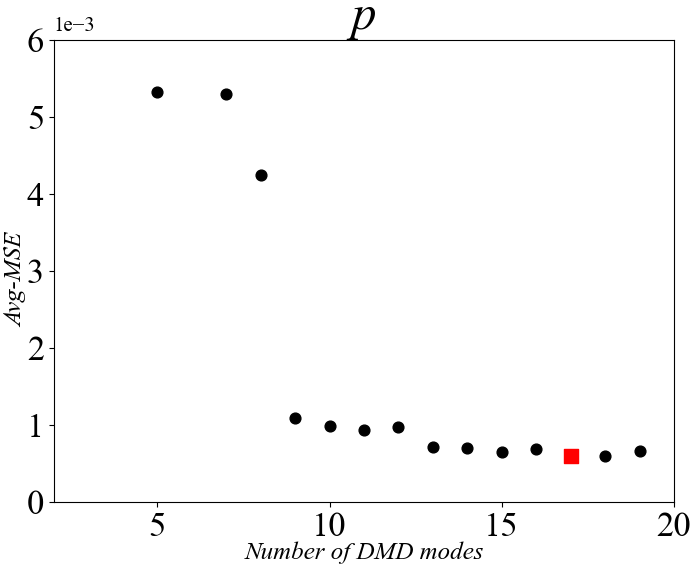}}
 \caption{The MSEs of $u$, $v$ and $p$ between the DNS data and the predictions of DMD for various DMD modes averaged from $3500th$ to $4000th$ extrapolation snapshots for $\Delta t = 0.2$. (a) $u_{MSE}$. (b) $v_{MSE}$. (c) $p_{MSE}$.}
 \label{0.2MSE diff DMDmodes}
\end{figure*}

The prediction results of DeepONet + GSAM are then compared with those of DMD. The number of DMD modes for $u$, $v$ and $p$ are determined to be 19, 23 and 17 respectively, as indicated by the red square dot in Fig.~\ref{0.2UMSE diff DMDmodes}, \ref{0.2VMSE diff DMDmodes} and \ref{0.2PMSE diff DMDmodes}. 

\begin{figure*}[htbp] 
 \centering  
  \subfigure[]{
  \label{0.2UDMD compare GSAM}
  \includegraphics[scale=0.27]{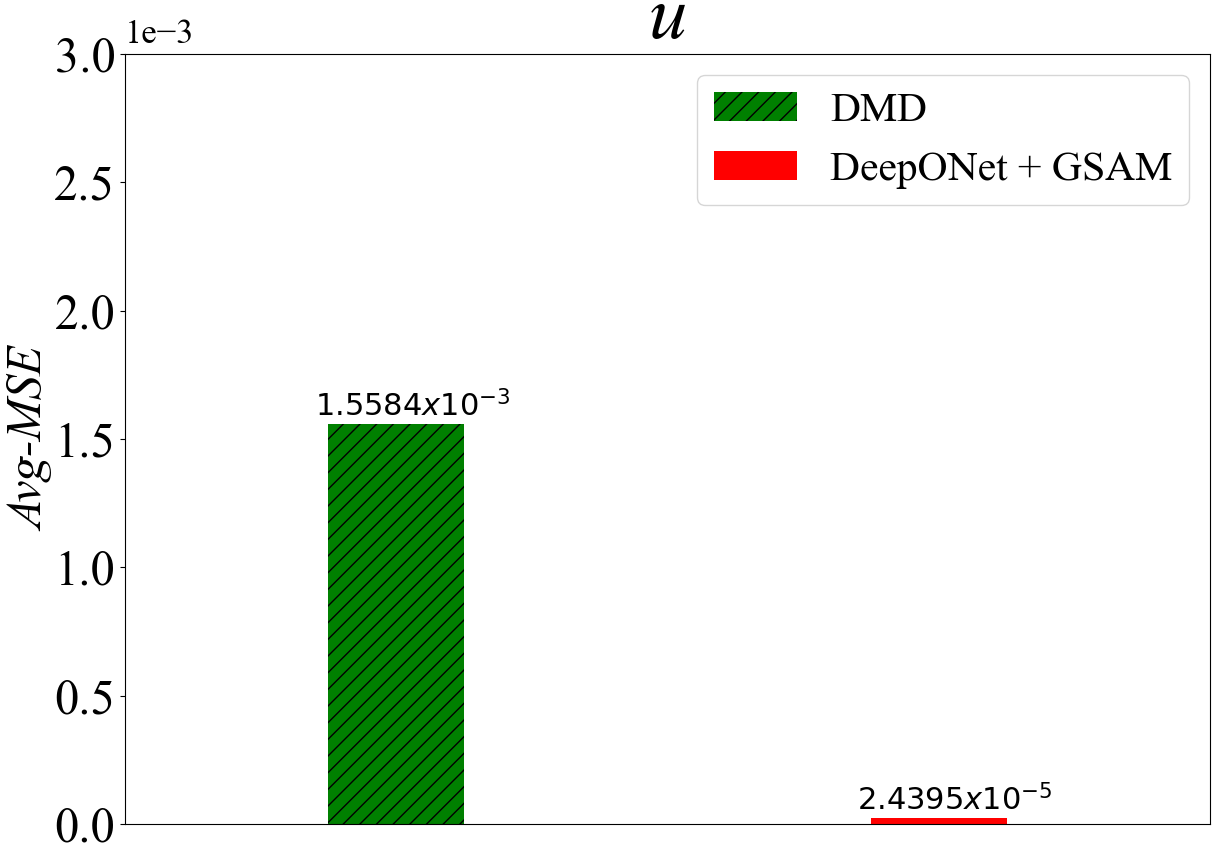}}
  \subfigure[]{
  \label{0.2VDMD compare GSAM}
  \includegraphics[scale=0.27]{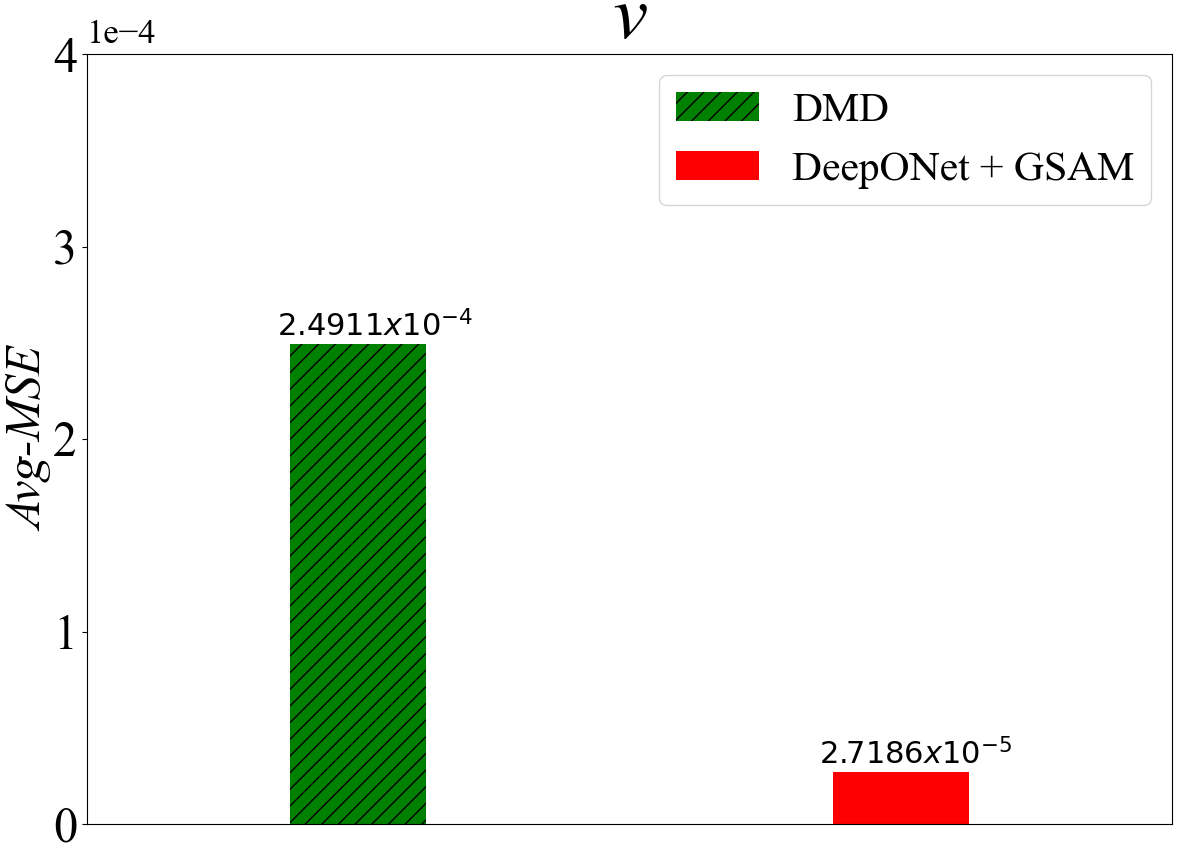}}
  \subfigure[]{
  \label{0.2PDMD compare GSAM}
  \includegraphics[scale=0.27]{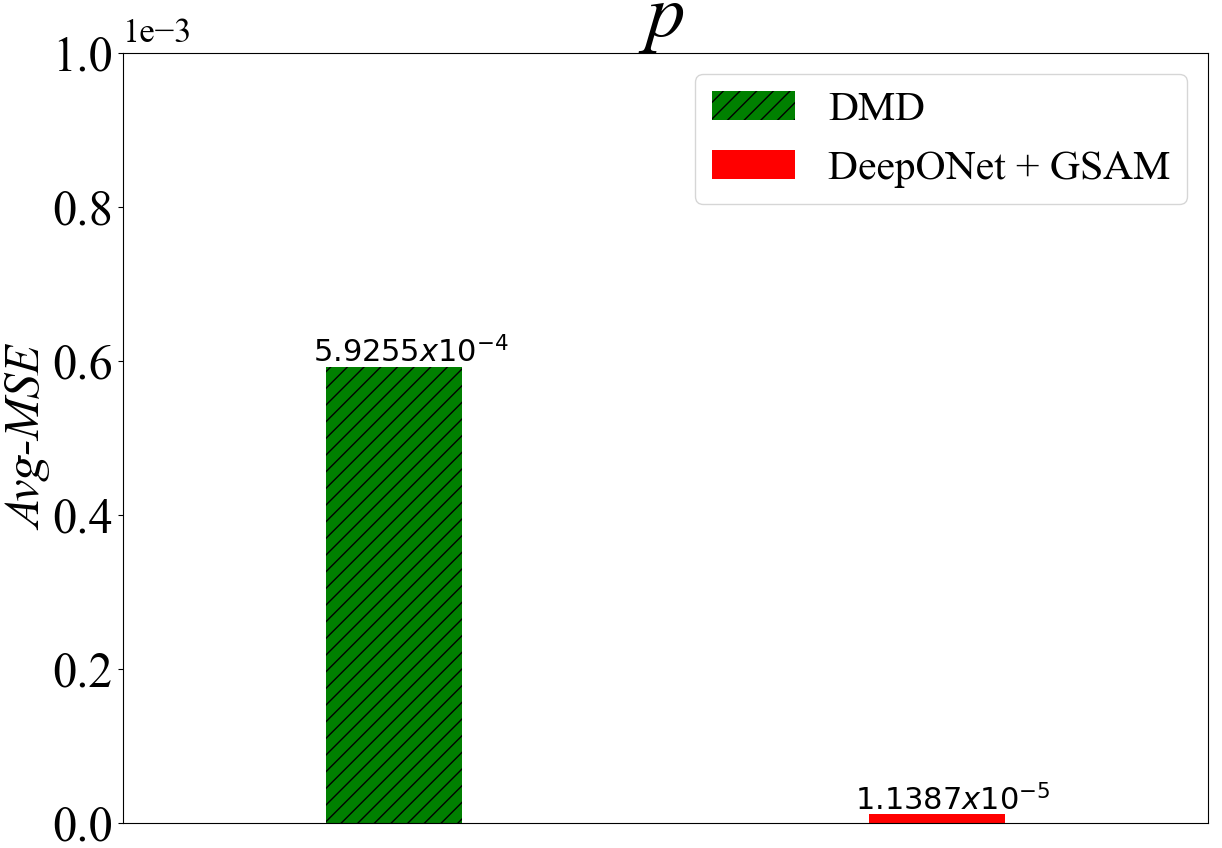}}
 \caption{The MSE of $u$, $v$ and $p$ between the DNS data and the predictions of DeepONet + GSAM and DMD averaged from $3500th$ to $4000th$ extrapolation snapshots for $\Delta t = 0.2$. (a) $u_{MSE}$. (b) $v_{MSE}$. (c) $p_{MSE}$.}
 \label{0.2DMD compare GSAM}
\end{figure*}

The results obtained are presented and Fig.~\ref{0.2DMD compare GSAM}, revealing a clear superiority of the DeepONet + GSAM compared with those generated by DMD.


\section{$N_s = 300$ and $\Delta t = 0.6$}
\label{appendixD}

The DeepONet + GSAM is trained to extrapolate the flow field of 2D flow around a circular cylinder at $Re = 1000$ with $\Delta t = 0.6$ in this section.

$N_{b} = 6$; $N_{t} = 2000$; the initial $lr$ is set to $0.0001$, and a $cosine$ $lr$ schedule is employed; $epoch = 1000$; $batchsize = 20$; $\rho = 0.0001$; $\alpha = 0.4$. The neural network parameters of both the trunk network and branch network remain consistent with those of the $\Delta t = 0.4$ case. The specific parameter analysis is no longer presented in exhaustive detail, but rather focuses on the main results.

\begin{figure*}[htbp] 
 \centering  
  \subfigure[]{
  \label{0.6UDNSGSAM}
  \includegraphics[scale=0.178]{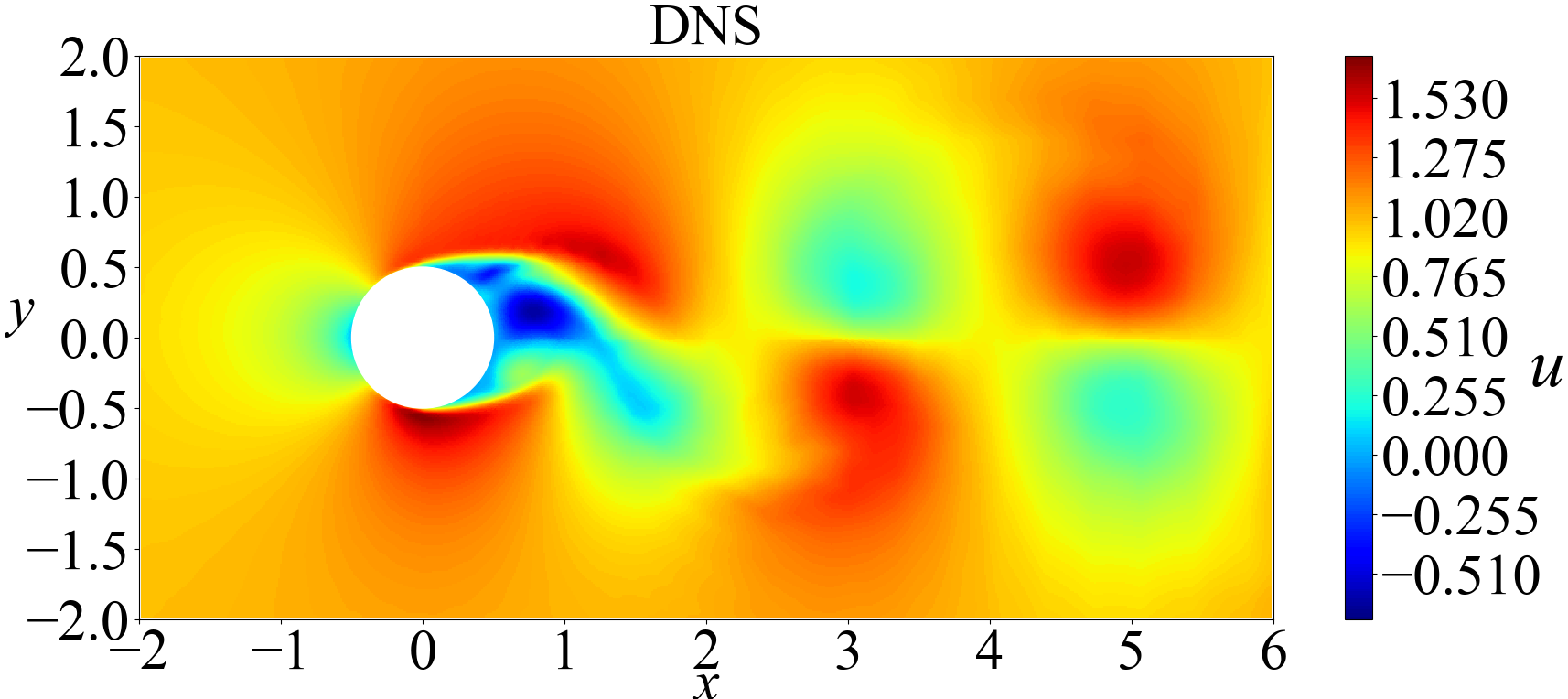}}
  \subfigure[]{
  \label{0.6UpreGSAM}
  \includegraphics[scale=0.178]{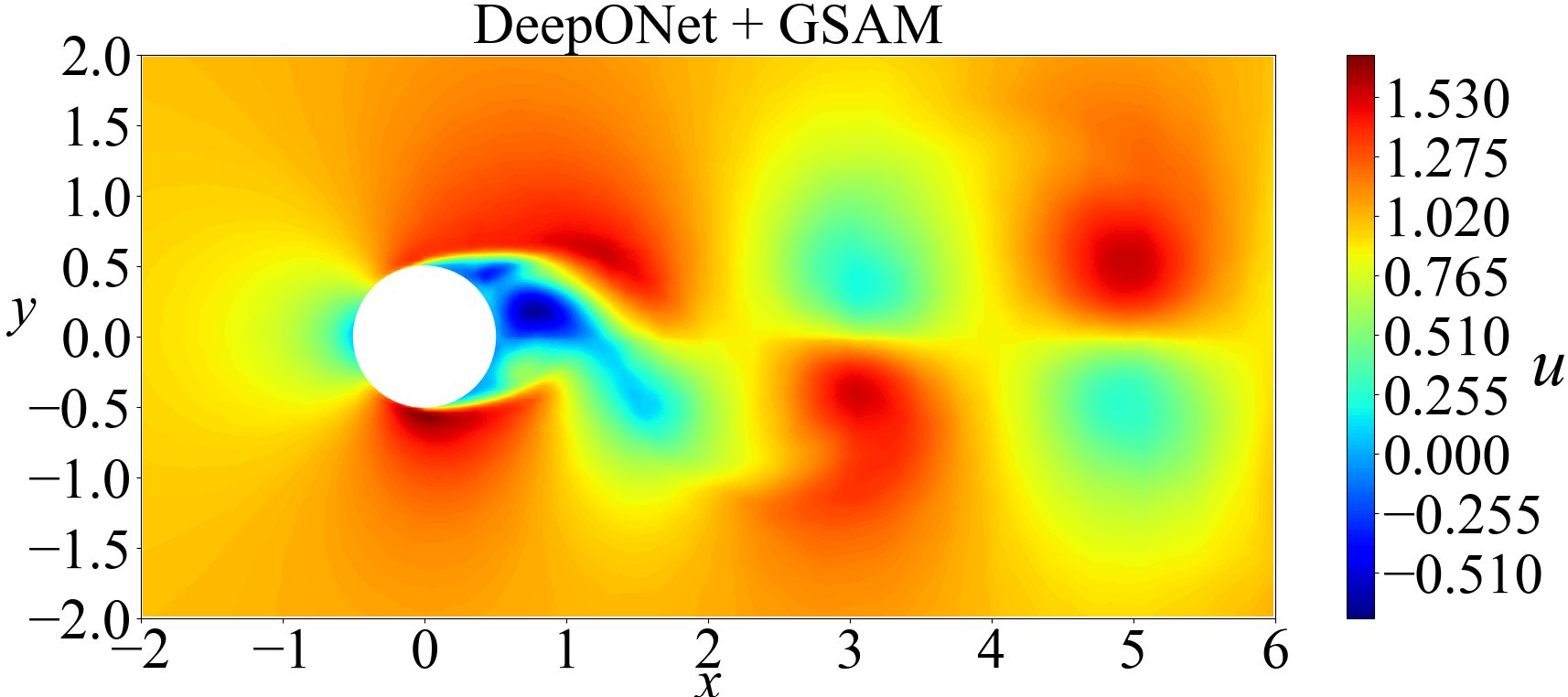}}
  \subfigure[]{
  \label{0.6UdeltaGSAM}
  \includegraphics[scale=0.178]{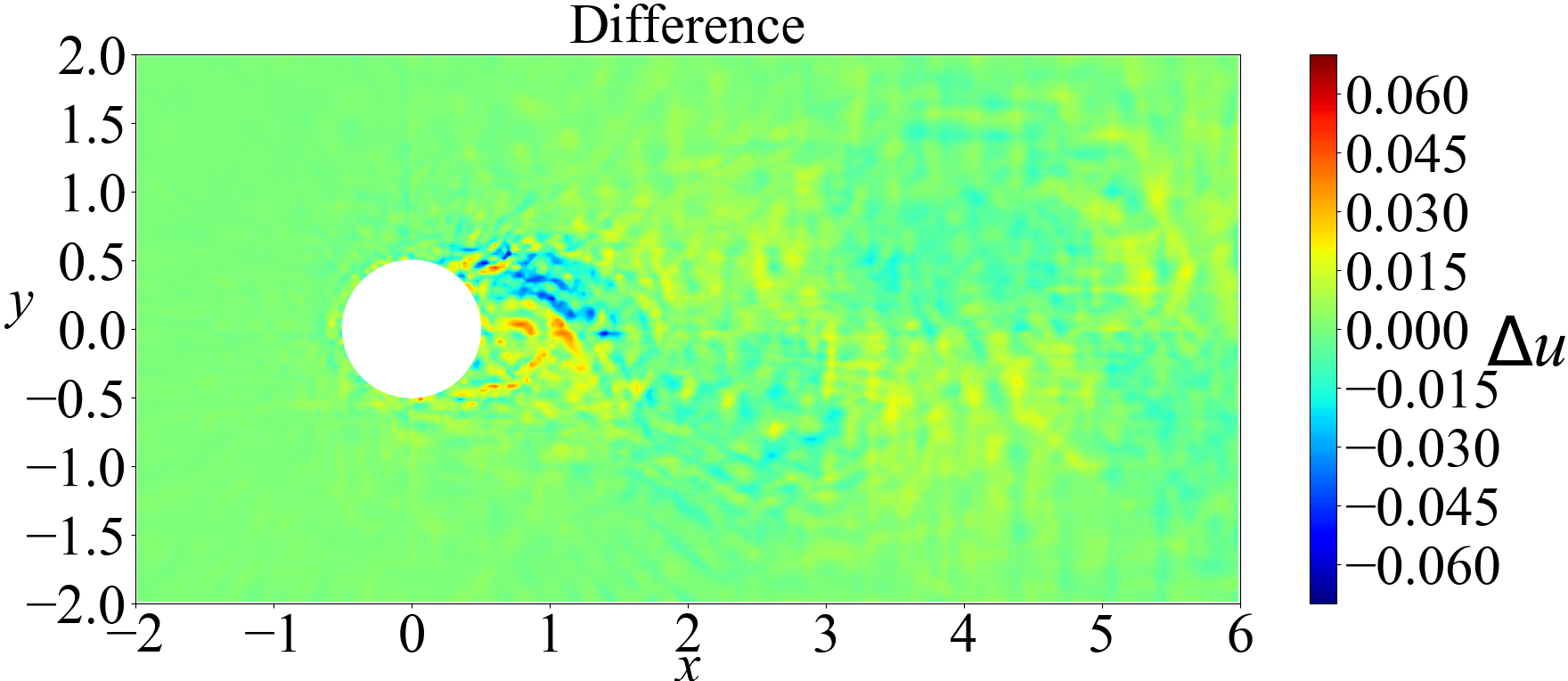}}
  \subfigure[]{
  \label{0.6VDNSGSAM}
  \includegraphics[scale=0.178]{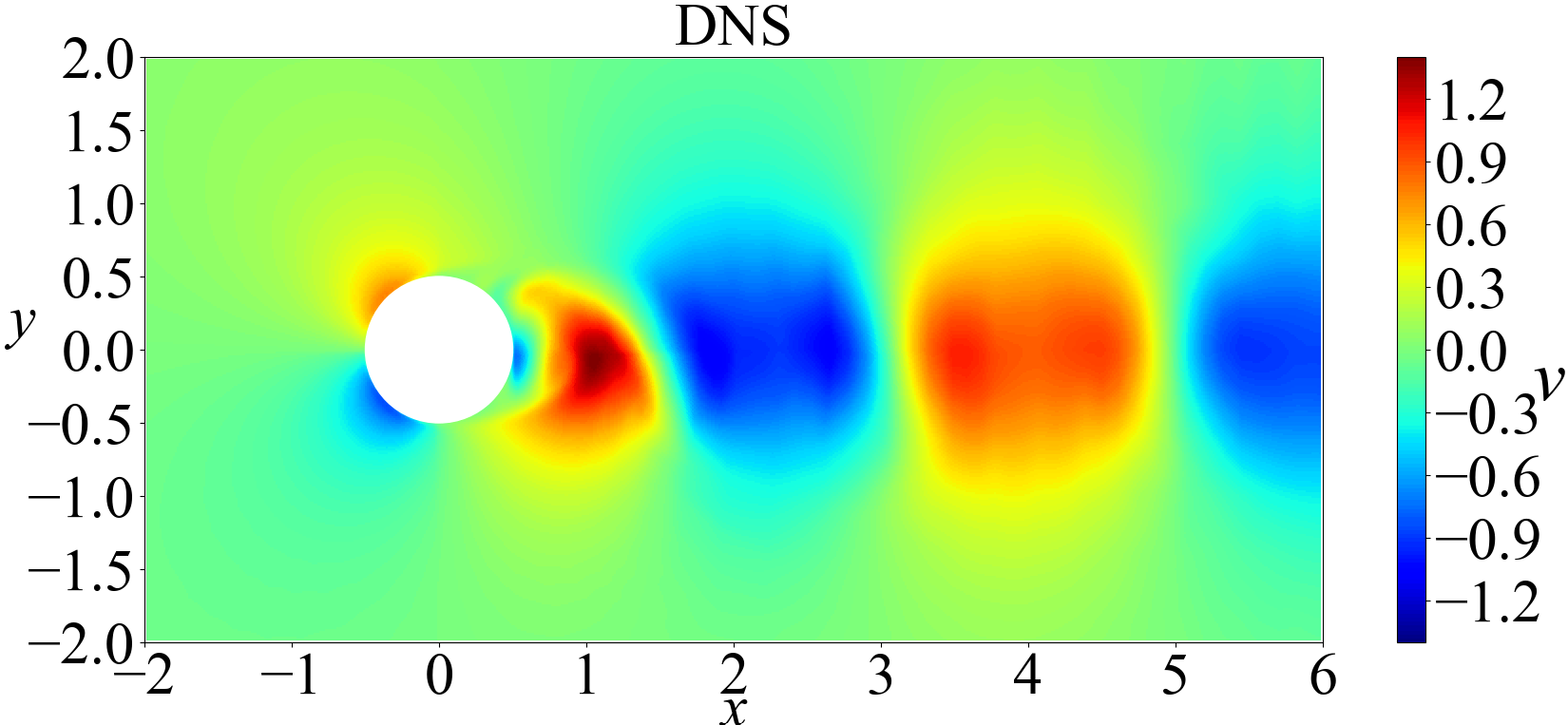}}
  \subfigure[]{
  \label{0.6VpreGSAM}
  \includegraphics[scale=0.178]{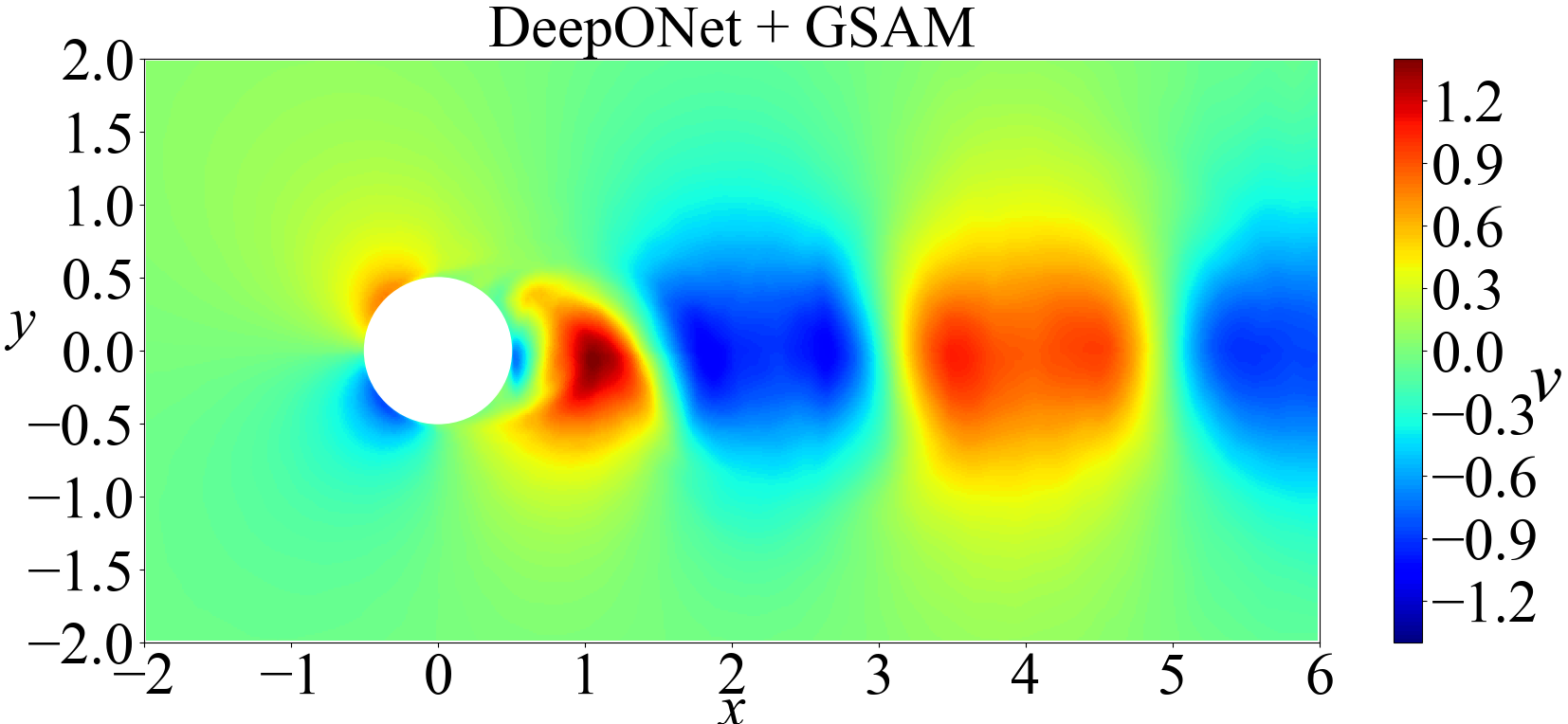}}
  \subfigure[]{
  \label{0.6VdeltaGSAM}
  \includegraphics[scale=0.178]{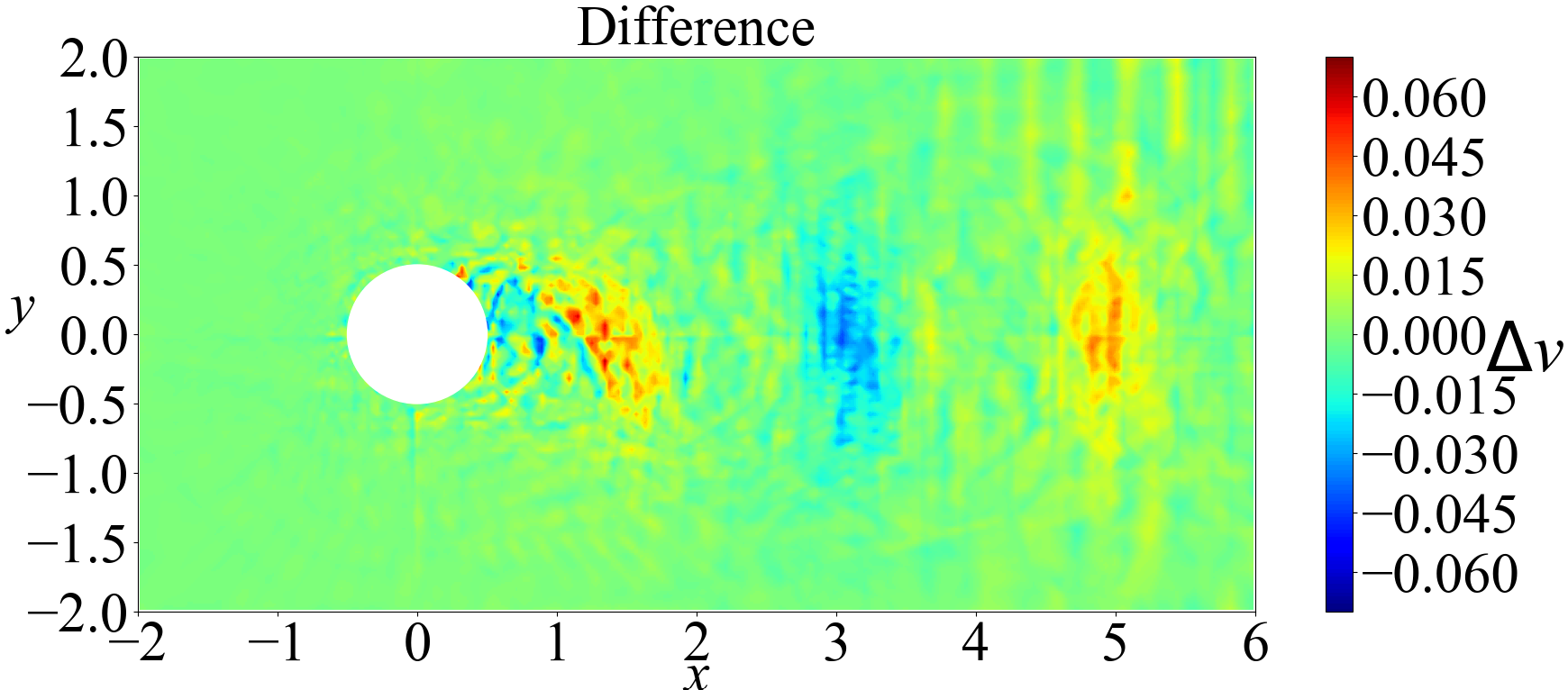}}
  \subfigure[]{
  \label{0.6PDNSGSAM}
  \includegraphics[scale=0.178]{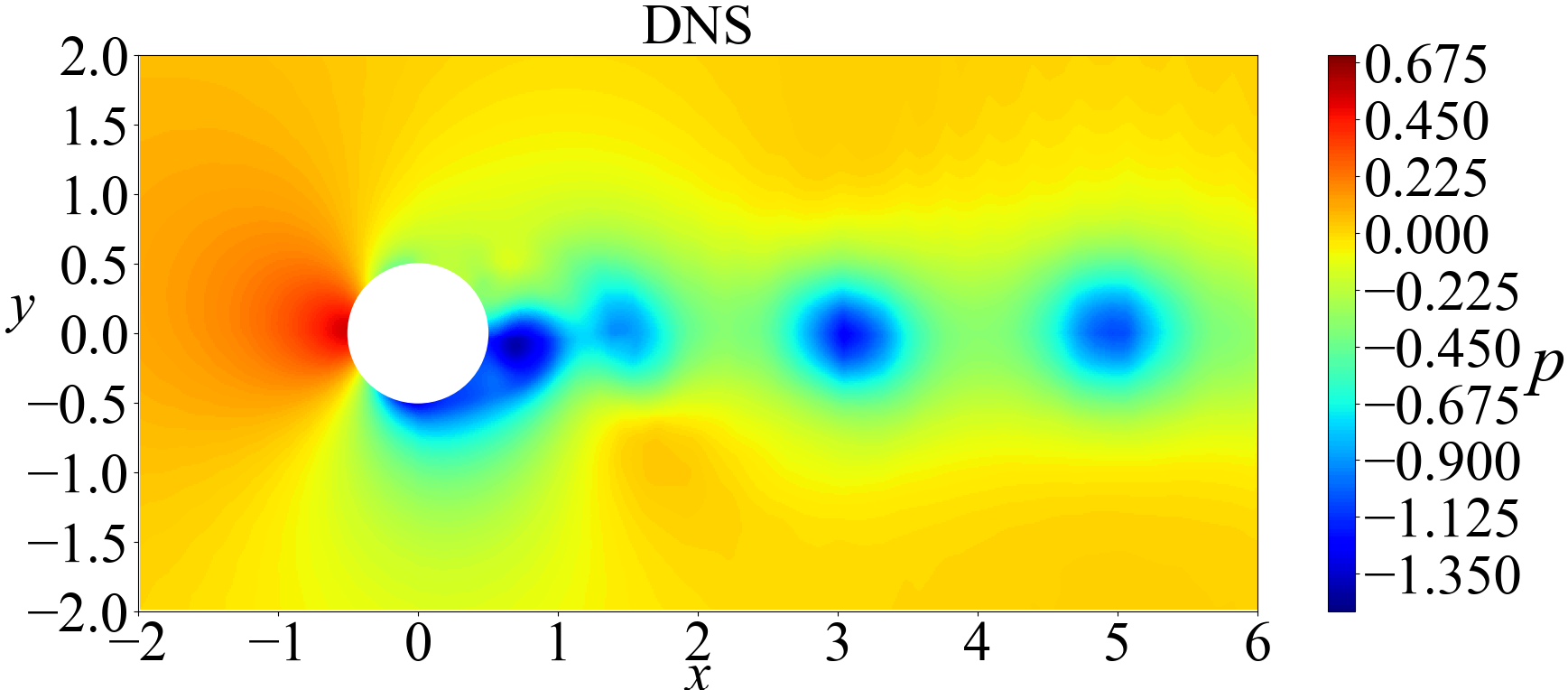}}
  \subfigure[]{
  \label{0.6PpreGSAM}
  \includegraphics[scale=0.178]{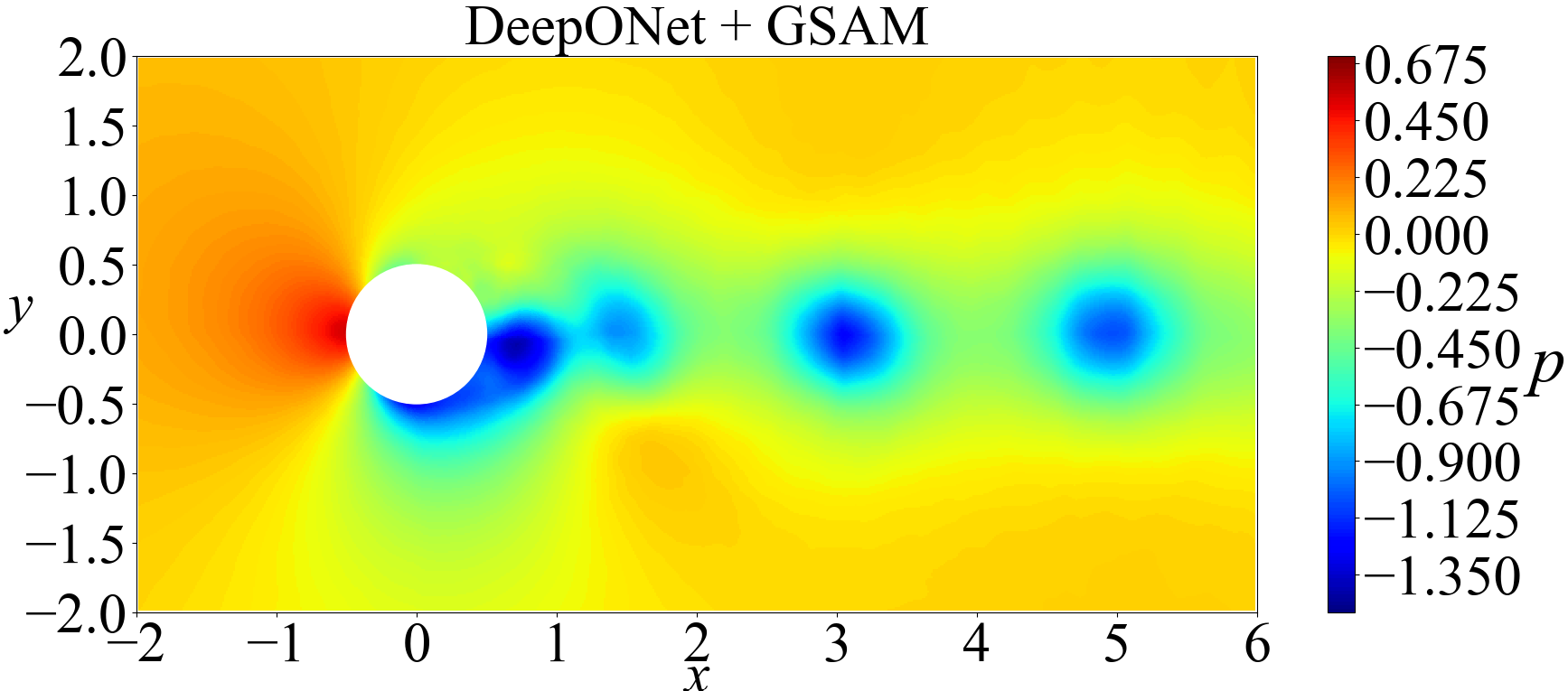}}
  \subfigure[]{
  \label{0.6PdeltaGSAM}
  \includegraphics[scale=0.178]{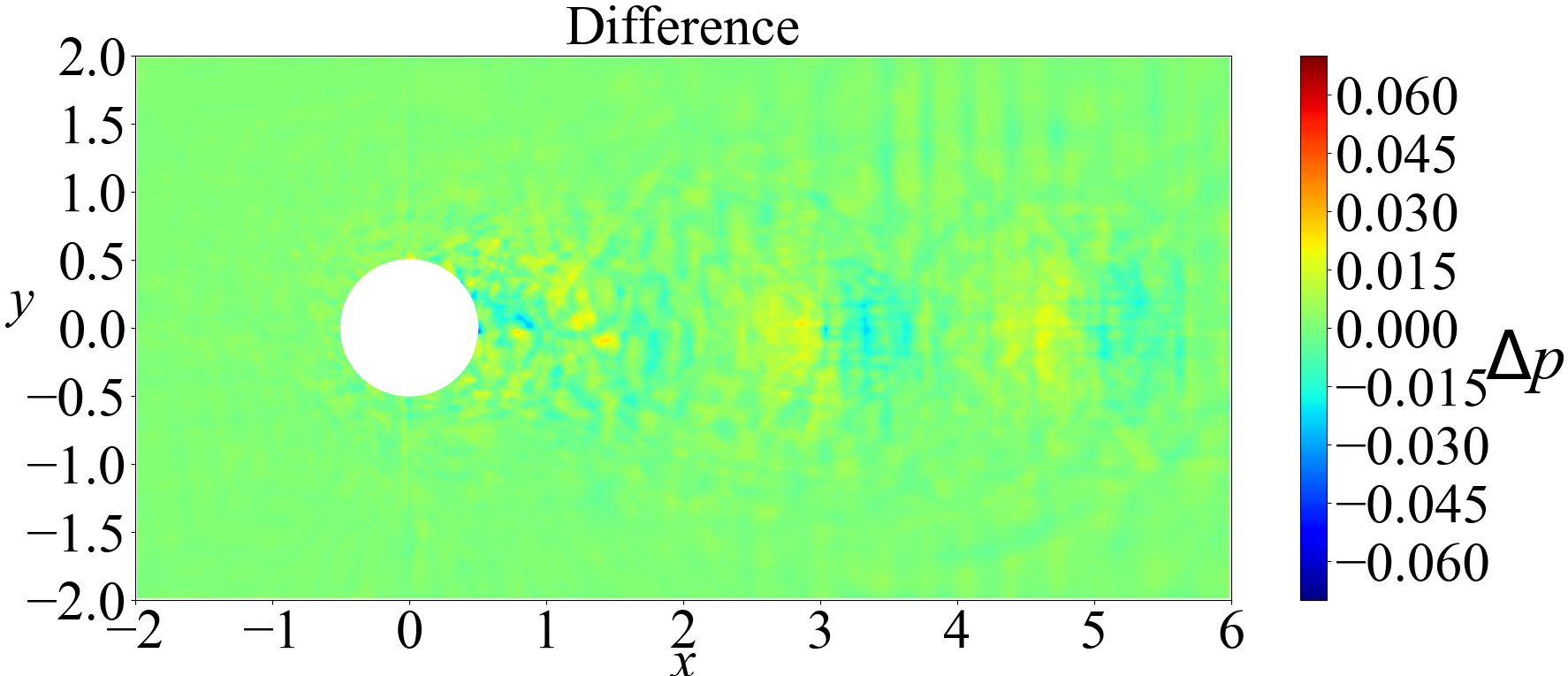}}
  \caption{Comparison of $u$, $v$ and $p$ at $4000th$ extrapolation snapshot between the prediction of DeepONet + GSAM and the DNS data for $\Delta t = 0.6$. (a) $u_{DNS}$. (b) $u_{pre}$. (c) $u_{DNS}-u_{pre}$. (d) $v_{DNS}$. (e) $v_{pre}$. (f) $v_{DNS}-v_{pre}$. (g) $p_{DNS}$. (h) $p_{pre}$. (i) $p_{DNS}-p_{pre}$.}\label{0.6GSAMcontour compare}
\end{figure*}

\begin{figure*}[htbp] 
 \centering  
  \subfigure[]{
  \label{UMSElineGSAM0.6}
  \includegraphics[scale=0.25]{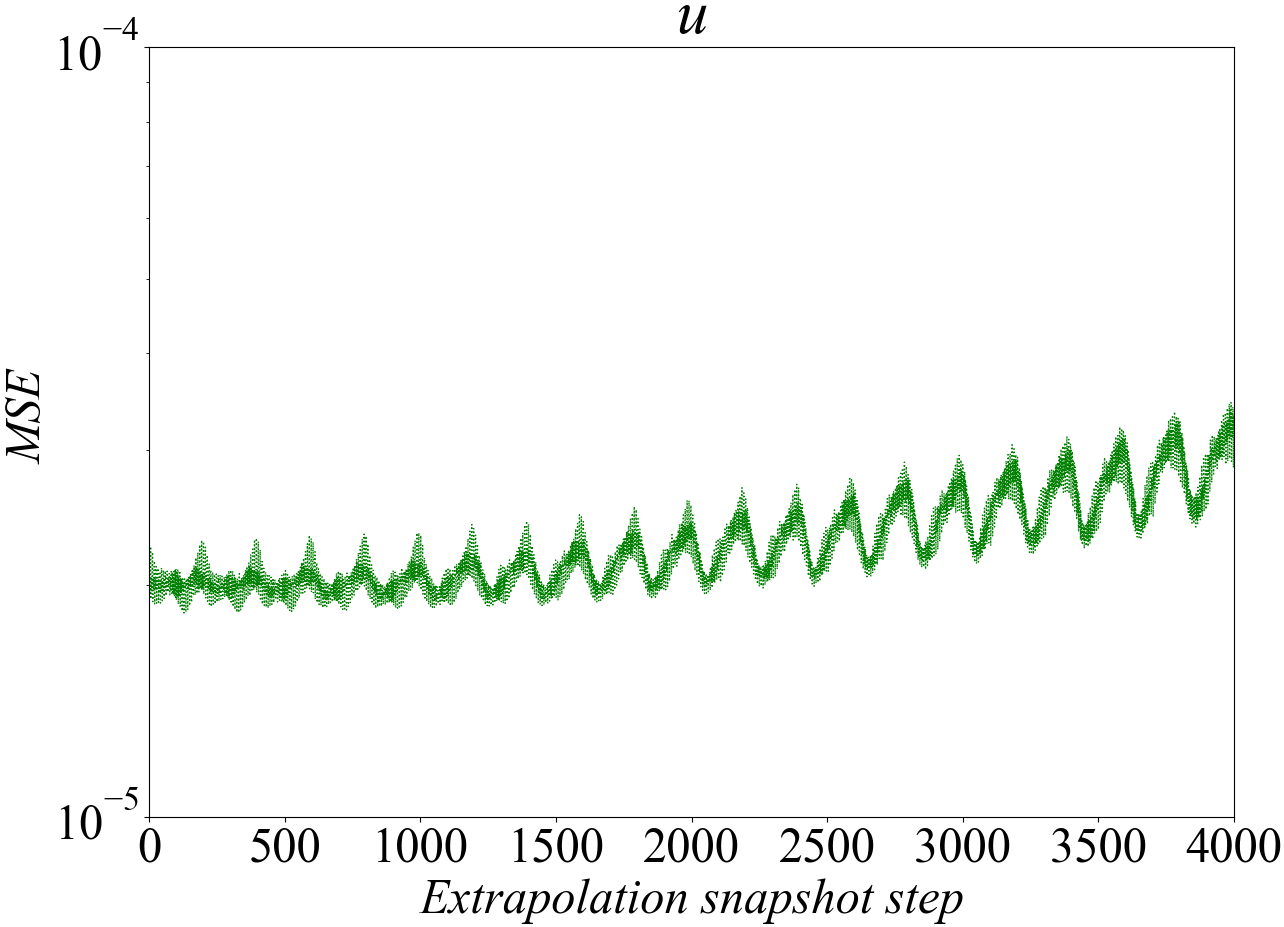}}
  \subfigure[]{
  \label{VMSElineGSAM0.6}
  \includegraphics[scale=0.25]{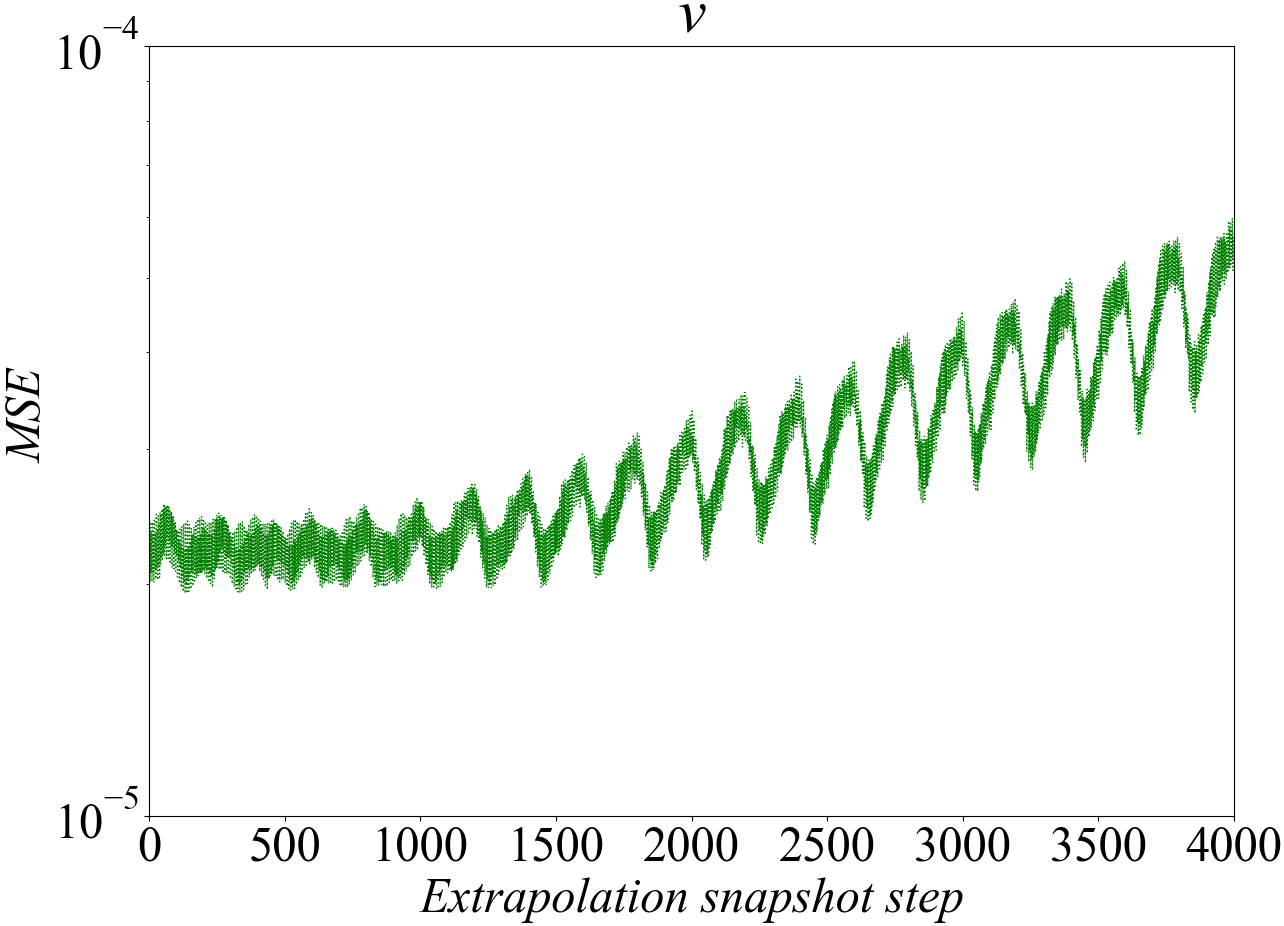}}
  \subfigure[]{
  \label{PMSElineGSAM0.6}
  \includegraphics[scale=0.25]{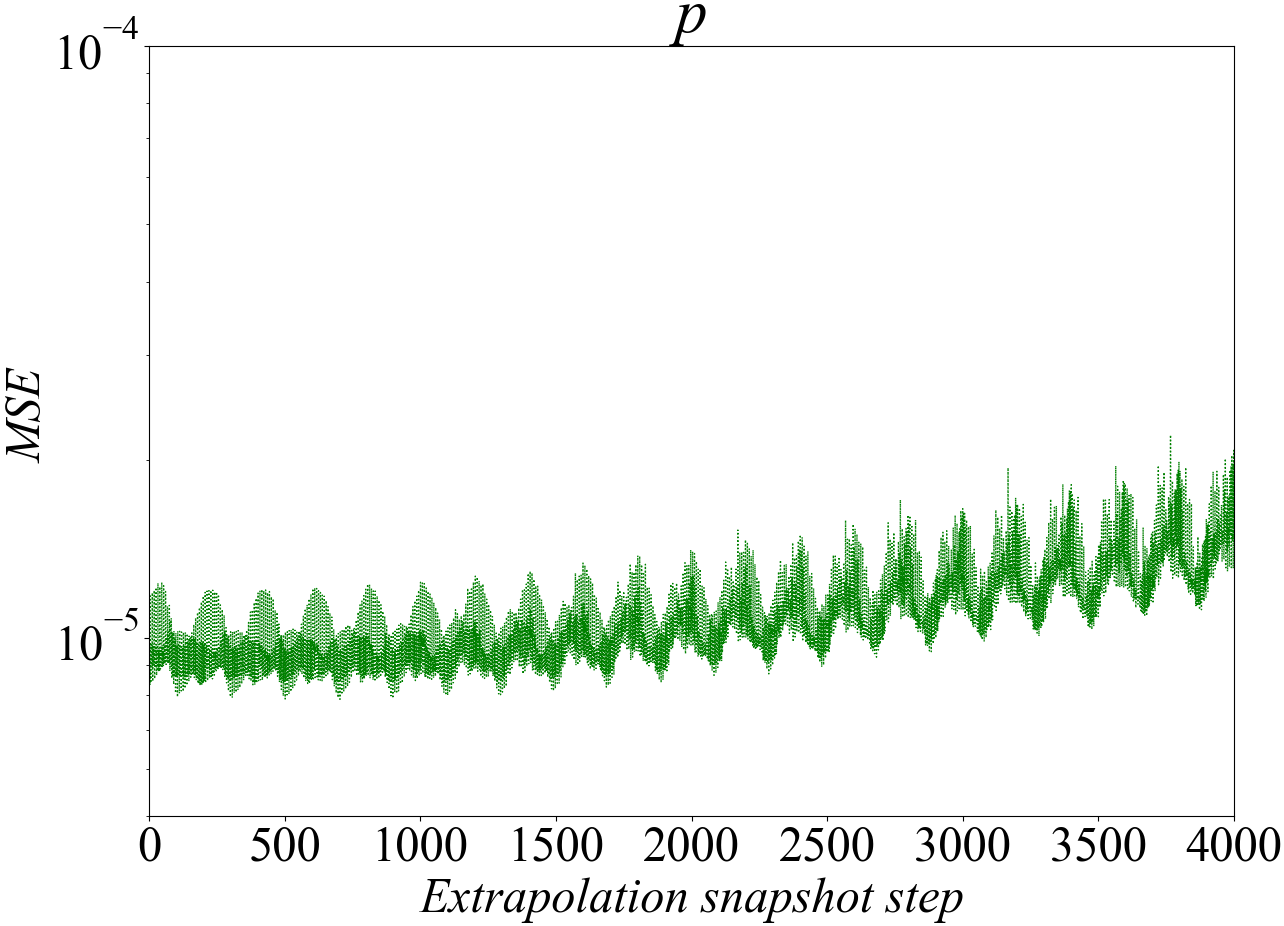}}
 \caption{The evolution of MSEs between the DNS data and the predictions of DeepONet + GSAM for $\Delta t = 0.6$. (a) $u_{MSE}$. (b) $v_{MSE}$. (c) $p_{MSE}$.}
 \label{0.6MSE line of GSAM}
\end{figure*}

The prediction results for the $4000th$ extrapolation snapshot based on the DeepONet + GSAM and the evolution of the MSEs between the DNS data and the predictions of DeepONet + GSAM over 4000 extrapolation snapshots are shown in Fig.~\ref{0.6GSAMcontour compare} and Fig.~\ref{0.6MSE line of GSAM}, respectively.

As $\Delta t$ increases to $0.6$, the prediction error of the DeepONet + GSAM significantly amplifies with an increase in extrapolation step, accompanied by noticeable fluctuations. This observation highlights the influence of snapshot time step on the performance of DeepONet + GSAM. Further improvement is still required in the capability of the neural network to predict flow fields with larger snapshot time step. The noteworthy aspect of this case lies in the fact that the error magnitudes of $u$, $v$ and $p$ remain at $10^{-5}$.

\begin{figure*}[htbp] 
 \centering  
  \subfigure[]{
  \label{0.6UMSE diff DMDmodes}
  \includegraphics[scale=0.47]{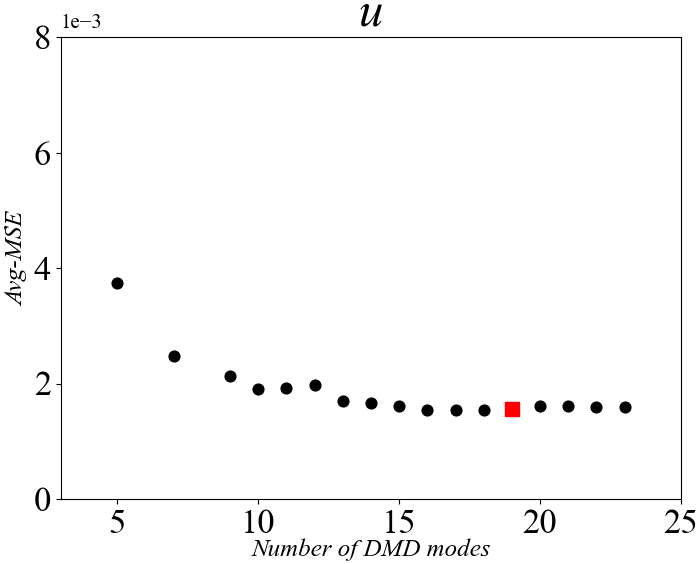}}
  \subfigure[]{
  \label{0.6VMSE diff DMDmodes}
  \includegraphics[scale=0.47]{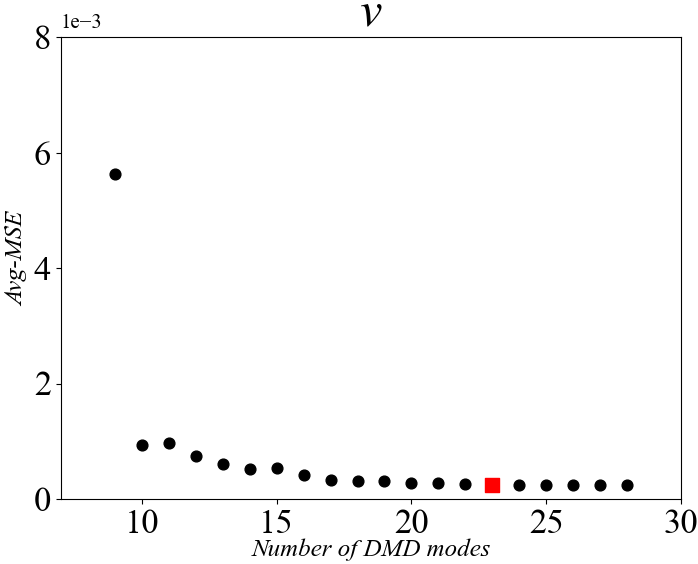}}
  \subfigure[]{
  \label{0.6PMSE diff DMDmodes}
  \includegraphics[scale=0.47]{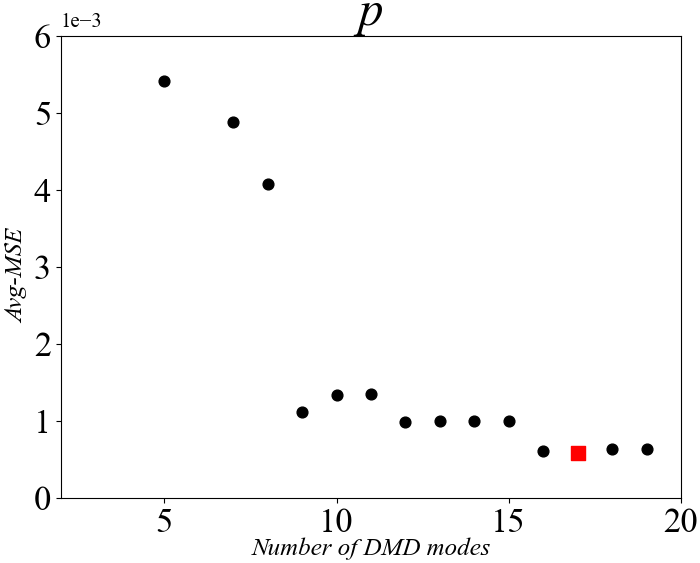}}
 \caption{The MSEs of $u$,$v$ and $p$ between the DNS data and the predictions of DMD for various DMD modes averaged from $3500th$ to $4000th$ extrapolation snapshots for $\Delta t = 0.6$. (a) $u_{MSE}$. (b) $v_{MSE}$. (c) $p_{MSE}$.}
 \label{0.6MSE diff DMDmodes}
\end{figure*}

Subsequently, the prediction results of the DeepONet + GSAM are compared with those of DMD, where 19 modes for $u$, 23 modes for $v$, and 17 modes for $p$ are considered as demonstrated in Fig.~\ref{0.6MSE diff DMDmodes}.

\begin{figure*}[htbp] 
 \centering  
  \subfigure[]{
  \label{0.6UDMD compare GSAM}
  \includegraphics[scale=0.27]{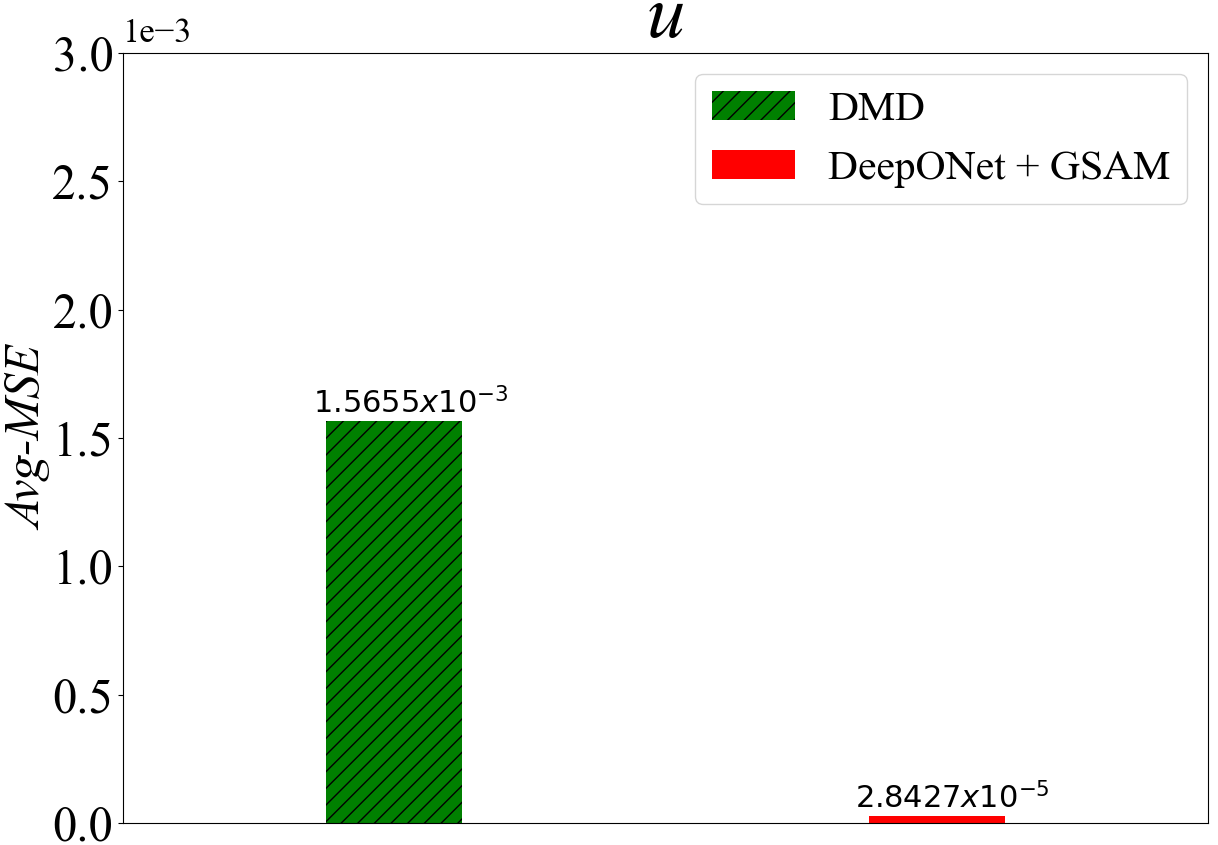}}
  \subfigure[]{
  \label{0.6VDMD compare GSAM}
  \includegraphics[scale=0.27]{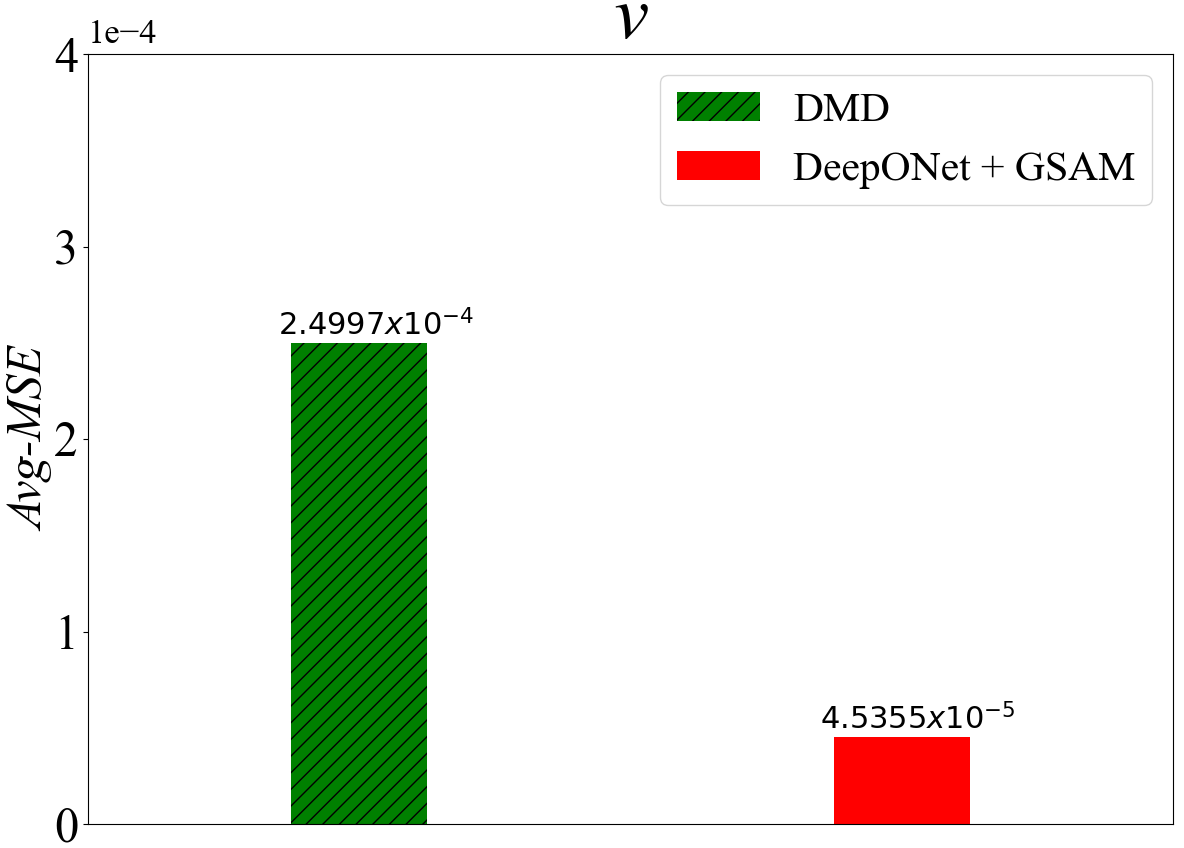}}
  \subfigure[]{
  \label{0.6PDMD compare GSAM}
  \includegraphics[scale=0.27]{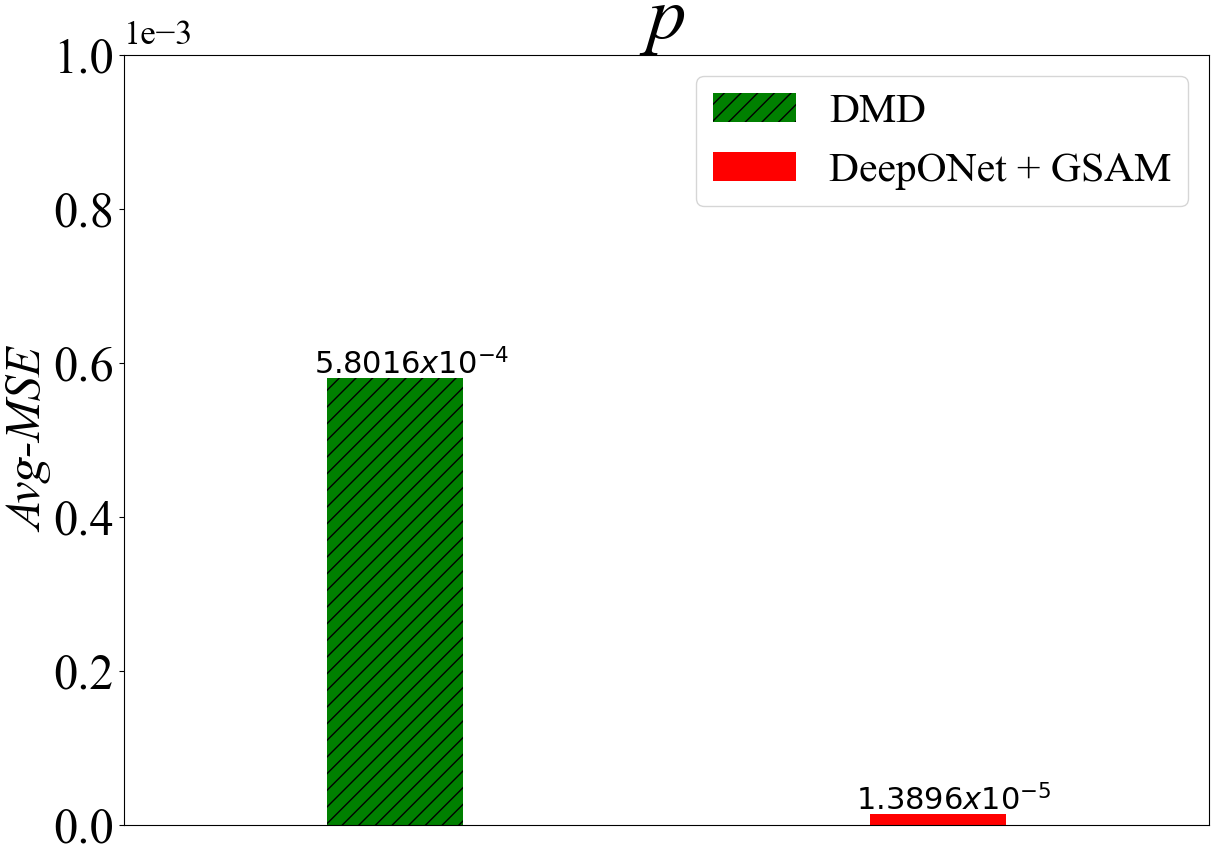}}
 \caption{The MSEs of $u$, $v$ and $p$ between the DNS data and the predictions of DeepONet + GSAM and DMD averaged from $3500th$ to $4000th$ extrapolation snapshots for $\Delta t = 0.6$. (a) $u_{MSE}$. (b) $v_{MSE}$. (c) $p_{MSE}$.}
 \label{0.6DMD compare GSAM}
\end{figure*}

Referring to Fig.~\ref{0.6DMD compare GSAM}, we see that the performance of DeepONet + GSAM remains significantly superior to that of DMD for $\Delta t=0.6$.


\nocite{*}
\bibliography{main}

\begin{thebibliography}{58}%
\makeatletter
\providecommand \@ifxundefined [1]{%
 \@ifx{#1\undefined}
}%
\providecommand \@ifnum [1]{%
 \ifnum #1\expandafter \@firstoftwo
 \else \expandafter \@secondoftwo
 \fi
}%
\providecommand \@ifx [1]{%
 \ifx #1\expandafter \@firstoftwo
 \else \expandafter \@secondoftwo
 \fi
}%
\providecommand \natexlab [1]{#1}%
\providecommand \enquote  [1]{``#1''}%
\providecommand \bibnamefont  [1]{#1}%
\providecommand \bibfnamefont [1]{#1}%
\providecommand \citenamefont [1]{#1}%
\providecommand \href@noop [0]{\@secondoftwo}%
\providecommand \href [0]{\begingroup \@sanitize@url \@href}%
\providecommand \@href[1]{\@@startlink{#1}\@@href}%
\providecommand \@@href[1]{\endgroup#1\@@endlink}%
\providecommand \@sanitize@url [0]{\catcode `\\12\catcode `\$12\catcode `\&12\catcode `\#12\catcode `\^12\catcode `\_12\catcode `\%12\relax}%
\providecommand \@@startlink[1]{}%
\providecommand \@@endlink[0]{}%
\providecommand \url  [0]{\begingroup\@sanitize@url \@url }%
\providecommand \@url [1]{\endgroup\@href {#1}{\urlprefix }}%
\providecommand \urlprefix  [0]{URL }%
\providecommand \Eprint [0]{\href }%
\providecommand \doibase [0]{http://dx.doi.org/}%
\providecommand \selectlanguage [0]{\@gobble}%
\providecommand \bibinfo  [0]{\@secondoftwo}%
\providecommand \bibfield  [0]{\@secondoftwo}%
\providecommand \translation [1]{[#1]}%
\providecommand \BibitemOpen [0]{}%
\providecommand \bibitemStop [0]{}%
\providecommand \bibitemNoStop [0]{.\EOS\space}%
\providecommand \EOS [0]{\spacefactor3000\relax}%
\providecommand \BibitemShut  [1]{\csname bibitem#1\endcsname}%
\let\auto@bib@innerbib\@empty
\bibitem [{\citenamefont {Zhang}\ \emph {et~al.}(2021)\citenamefont {Zhang}, \citenamefont {Xie}, \citenamefont {Ji}, \citenamefont {Zhu},\ and\ \citenamefont {Zheng}}]{zhang2021shapeoptir1}%
  \BibitemOpen
  \bibfield  {author} {\bibinfo {author} {\bibfnamefont {X.}~\bibnamefont {Zhang}}, \bibinfo {author} {\bibfnamefont {F.}~\bibnamefont {Xie}}, \bibinfo {author} {\bibfnamefont {T.}~\bibnamefont {Ji}}, \bibinfo {author} {\bibfnamefont {Z.}~\bibnamefont {Zhu}}, \ and\ \bibinfo {author} {\bibfnamefont {Y.}~\bibnamefont {Zheng}},\ }\bibfield  {title} {\enquote {\bibinfo {title} {Multi-fidelity deep neural network surrogate model for aerodynamic shape optimization},}\ }\href@noop {} {\bibfield  {journal} {\bibinfo  {journal} {Comput. Methods Appl. Mech. Eng.}\ }\textbf {\bibinfo {volume} {373}},\ \bibinfo {pages} {113485} (\bibinfo {year} {2021})}\BibitemShut {NoStop}%
\bibitem [{\citenamefont {Viquerat}\ \emph {et~al.}(2021)\citenamefont {Viquerat}, \citenamefont {Rabault}, \citenamefont {Kuhnle}, \citenamefont {Ghraieb}, \citenamefont {Larcher},\ and\ \citenamefont {Hachem}}]{structuralshapeoptimizationr2}%
  \BibitemOpen
  \bibfield  {author} {\bibinfo {author} {\bibfnamefont {J.}~\bibnamefont {Viquerat}}, \bibinfo {author} {\bibfnamefont {J.}~\bibnamefont {Rabault}}, \bibinfo {author} {\bibfnamefont {A.}~\bibnamefont {Kuhnle}}, \bibinfo {author} {\bibfnamefont {H.}~\bibnamefont {Ghraieb}}, \bibinfo {author} {\bibfnamefont {A.}~\bibnamefont {Larcher}}, \ and\ \bibinfo {author} {\bibfnamefont {E.}~\bibnamefont {Hachem}},\ }\bibfield  {title} {\enquote {\bibinfo {title} {Direct shape optimization through deep reinforcement learning},}\ }\href@noop {} {\bibfield  {journal} {\bibinfo  {journal} {J. Comput. Phys.}\ }\textbf {\bibinfo {volume} {428}},\ \bibinfo {pages} {110080} (\bibinfo {year} {2021})}\BibitemShut {NoStop}%
\bibitem [{\citenamefont {Rabault}\ \emph {et~al.}(2019)\citenamefont {Rabault}, \citenamefont {Kuchta}, \citenamefont {Jensen}, \citenamefont {R{\'e}glade},\ and\ \citenamefont {Cerardi}}]{afcr30}%
  \BibitemOpen
  \bibfield  {author} {\bibinfo {author} {\bibfnamefont {J.}~\bibnamefont {Rabault}}, \bibinfo {author} {\bibfnamefont {M.}~\bibnamefont {Kuchta}}, \bibinfo {author} {\bibfnamefont {A.}~\bibnamefont {Jensen}}, \bibinfo {author} {\bibfnamefont {U.}~\bibnamefont {R{\'e}glade}}, \ and\ \bibinfo {author} {\bibfnamefont {N.}~\bibnamefont {Cerardi}},\ }\bibfield  {title} {\enquote {\bibinfo {title} {Artificial neural networks trained through deep reinforcement learning discover control strategies for active flow control},}\ }\href@noop {} {\bibfield  {journal} {\bibinfo  {journal} {J. Fluid Mech.}\ }\textbf {\bibinfo {volume} {865}},\ \bibinfo {pages} {281--302} (\bibinfo {year} {2019})}\BibitemShut {NoStop}%
\bibitem [{\citenamefont {Zheng}\ \emph {et~al.}(2021)\citenamefont {Zheng}, \citenamefont {Ji}, \citenamefont {Xie}, \citenamefont {Zhang}, \citenamefont {Zheng},\ and\ \citenamefont {Zheng}}]{zheng2021activer31}%
  \BibitemOpen
  \bibfield  {author} {\bibinfo {author} {\bibfnamefont {C.}~\bibnamefont {Zheng}}, \bibinfo {author} {\bibfnamefont {T.}~\bibnamefont {Ji}}, \bibinfo {author} {\bibfnamefont {F.}~\bibnamefont {Xie}}, \bibinfo {author} {\bibfnamefont {X.}~\bibnamefont {Zhang}}, \bibinfo {author} {\bibfnamefont {H.}~\bibnamefont {Zheng}}, \ and\ \bibinfo {author} {\bibfnamefont {Y.}~\bibnamefont {Zheng}},\ }\bibfield  {title} {\enquote {\bibinfo {title} {From active learning to deep reinforcement learning: Intelligent active flow control in suppressing vortex-induced vibration},}\ }\href@noop {} {\bibfield  {journal} {\bibinfo  {journal} {Phys. Fluids}\ }\textbf {\bibinfo {volume} {33}} (\bibinfo {year} {2021})}\BibitemShut {NoStop}%
\bibitem [{\citenamefont {Wang}\ \emph {et~al.}(2023)\citenamefont {Wang}, \citenamefont {Fan}, \citenamefont {Jiang}, \citenamefont {Triantafyllou},\ and\ \citenamefont {Karniadakis}}]{Wang2023}%
  \BibitemOpen
  \bibfield  {author} {\bibinfo {author} {\bibfnamefont {Z.}~\bibnamefont {Wang}}, \bibinfo {author} {\bibfnamefont {D.}~\bibnamefont {Fan}}, \bibinfo {author} {\bibfnamefont {X.}~\bibnamefont {Jiang}}, \bibinfo {author} {\bibfnamefont {M.~S.}\ \bibnamefont {Triantafyllou}}, \ and\ \bibinfo {author} {\bibfnamefont {G.~E.}\ \bibnamefont {Karniadakis}},\ }\bibfield  {title} {\enquote {\bibinfo {title} {Deep reinforcement transfer learning of active control for bluff body flows at high reynolds number},}\ }\href@noop {} {\bibfield  {journal} {\bibinfo  {journal} {J. Fluid Mech.}\ }\textbf {\bibinfo {volume} {973}},\ \bibinfo {pages} {A32} (\bibinfo {year} {2023})}\BibitemShut {NoStop}%
\bibitem [{\citenamefont {Mo}, \citenamefont {Fu},\ and\ \citenamefont {Bian}(2023)}]{mo2023}%
  \BibitemOpen
  \bibfield  {author} {\bibinfo {author} {\bibfnamefont {C.}~\bibnamefont {Mo}}, \bibinfo {author} {\bibfnamefont {Q.}~\bibnamefont {Fu}}, \ and\ \bibinfo {author} {\bibfnamefont {X.}~\bibnamefont {Bian}},\ }\bibfield  {title} {\enquote {\bibinfo {title} {Chemotaxis of an elastic flagellated microrobot},}\ }\href@noop {} {\bibfield  {journal} {\bibinfo  {journal} {Phys. Rev. E}\ }\textbf {\bibinfo {volume} {108}},\ \bibinfo {pages} {044408} (\bibinfo {year} {2023})}\BibitemShut {NoStop}%
\bibitem [{\citenamefont {Ma}\ and\ \citenamefont {Pan}(2021)}]{ROMr40}%
  \BibitemOpen
  \bibfield  {author} {\bibinfo {author} {\bibfnamefont {Z.}~\bibnamefont {Ma}}\ and\ \bibinfo {author} {\bibfnamefont {W.}~\bibnamefont {Pan}},\ }\bibfield  {title} {\enquote {\bibinfo {title} {Data-driven nonintrusive reduced order modeling for dynamical systems with moving boundaries using gaussian process regression},}\ }\href@noop {} {\bibfield  {journal} {\bibinfo  {journal} {Comput. Methods Appl. Mech. Eng.}\ }\textbf {\bibinfo {volume} {373}},\ \bibinfo {pages} {113495} (\bibinfo {year} {2021})}\BibitemShut {NoStop}%
\bibitem [{\citenamefont {Pegolotti}\ \emph {et~al.}(2021)\citenamefont {Pegolotti}, \citenamefont {Pfaller}, \citenamefont {Marsden},\ and\ \citenamefont {Deparis}}]{ROMr41}%
  \BibitemOpen
  \bibfield  {author} {\bibinfo {author} {\bibfnamefont {L.}~\bibnamefont {Pegolotti}}, \bibinfo {author} {\bibfnamefont {M.~R.}\ \bibnamefont {Pfaller}}, \bibinfo {author} {\bibfnamefont {A.~L.}\ \bibnamefont {Marsden}}, \ and\ \bibinfo {author} {\bibfnamefont {S.}~\bibnamefont {Deparis}},\ }\bibfield  {title} {\enquote {\bibinfo {title} {Model order reduction of flow based on a modular geometrical approximation of blood vessels},}\ }\href@noop {} {\bibfield  {journal} {\bibinfo  {journal} {Comput. Methods Appl. Mech. Eng.}\ }\textbf {\bibinfo {volume} {380}},\ \bibinfo {pages} {113762} (\bibinfo {year} {2021})}\BibitemShut {NoStop}%
\bibitem [{\citenamefont {Sirovich}(1987)}]{r5POD}%
  \BibitemOpen
  \bibfield  {author} {\bibinfo {author} {\bibfnamefont {L.}~\bibnamefont {Sirovich}},\ }\bibfield  {title} {\enquote {\bibinfo {title} {Turbulence and the dynamics of coherent structures. i. coherent structures},}\ }\href@noop {} {\bibfield  {journal} {\bibinfo  {journal} {Q. Appl. Math.}\ }\textbf {\bibinfo {volume} {45}},\ \bibinfo {pages} {561--571} (\bibinfo {year} {1987})}\BibitemShut {NoStop}%
\bibitem [{\citenamefont {Schmid}(2010)}]{r6DMD}%
  \BibitemOpen
  \bibfield  {author} {\bibinfo {author} {\bibfnamefont {P.~J.}\ \bibnamefont {Schmid}},\ }\bibfield  {title} {\enquote {\bibinfo {title} {Dynamic mode decomposition of numerical and experimental data},}\ }\href@noop {} {\bibfield  {journal} {\bibinfo  {journal} {J. Fluid Mech.}\ }\textbf {\bibinfo {volume} {656}},\ \bibinfo {pages} {5--28} (\bibinfo {year} {2010})}\BibitemShut {NoStop}%
\bibitem [{\citenamefont {Kutz}(2017)}]{drawbackPODDMDr7}%
  \BibitemOpen
  \bibfield  {author} {\bibinfo {author} {\bibfnamefont {J.~N.}\ \bibnamefont {Kutz}},\ }\bibfield  {title} {\enquote {\bibinfo {title} {Deep learning in fluid dynamics},}\ }\href@noop {} {\bibfield  {journal} {\bibinfo  {journal} {J. Fluid Mech.}\ }\textbf {\bibinfo {volume} {814}},\ \bibinfo {pages} {1--4} (\bibinfo {year} {2017})}\BibitemShut {NoStop}%
\bibitem [{\citenamefont {Mikolov}\ \emph {et~al.}(2013)\citenamefont {Mikolov}, \citenamefont {Chen}, \citenamefont {Corrado},\ and\ \citenamefont {Dean}}]{NLPr80}%
  \BibitemOpen
  \bibfield  {author} {\bibinfo {author} {\bibfnamefont {T.}~\bibnamefont {Mikolov}}, \bibinfo {author} {\bibfnamefont {K.}~\bibnamefont {Chen}}, \bibinfo {author} {\bibfnamefont {G.}~\bibnamefont {Corrado}}, \ and\ \bibinfo {author} {\bibfnamefont {J.}~\bibnamefont {Dean}},\ }\bibfield  {title} {\enquote {\bibinfo {title} {Efficient estimation of word representations in vector space},}\ }\href@noop {} {\bibfield  {journal} {\bibinfo  {journal} {arXiv preprint arXiv:1301.3781}\ } (\bibinfo {year} {2013})}\BibitemShut {NoStop}%
\bibitem [{\citenamefont {Pennington}, \citenamefont {Socher},\ and\ \citenamefont {Manning}(2014)}]{NLPr81}%
  \BibitemOpen
  \bibfield  {author} {\bibinfo {author} {\bibfnamefont {J.}~\bibnamefont {Pennington}}, \bibinfo {author} {\bibfnamefont {R.}~\bibnamefont {Socher}}, \ and\ \bibinfo {author} {\bibfnamefont {C.~D.}\ \bibnamefont {Manning}},\ }\bibfield  {title} {\enquote {\bibinfo {title} {Glove: Global vectors for word representation},}\ }in\ \href@noop {} {\emph {\bibinfo {booktitle} {Proceedings of the 2014 Conference on Empirical Methods in Natural Language Processing}}}\ (\bibinfo {year} {2014})\ pp.\ \bibinfo {pages} {1532--1543}\BibitemShut {NoStop}%
\bibitem [{\citenamefont {He}\ \emph {et~al.}(2016{\natexlab{a}})\citenamefont {He}, \citenamefont {Zhang}, \citenamefont {Ren},\ and\ \citenamefont {Sun}}]{CVr90}%
  \BibitemOpen
  \bibfield  {author} {\bibinfo {author} {\bibfnamefont {K.}~\bibnamefont {He}}, \bibinfo {author} {\bibfnamefont {X.}~\bibnamefont {Zhang}}, \bibinfo {author} {\bibfnamefont {S.}~\bibnamefont {Ren}}, \ and\ \bibinfo {author} {\bibfnamefont {J.}~\bibnamefont {Sun}},\ }\bibfield  {title} {\enquote {\bibinfo {title} {Deep residual learning for image recognition},}\ }in\ \href@noop {} {\emph {\bibinfo {booktitle} {Proceedings of the IEEE Conference on Computer Vision and Pattern Recognition}}}\ (\bibinfo {year} {2016})\ pp.\ \bibinfo {pages} {770--778}\BibitemShut {NoStop}%
\bibitem [{\citenamefont {Simonyan}\ and\ \citenamefont {Zisserman}(2014)}]{CVr91}%
  \BibitemOpen
  \bibfield  {author} {\bibinfo {author} {\bibfnamefont {K.}~\bibnamefont {Simonyan}}\ and\ \bibinfo {author} {\bibfnamefont {A.}~\bibnamefont {Zisserman}},\ }\bibfield  {title} {\enquote {\bibinfo {title} {Very deep convolutional networks for large-scale image recognition},}\ }\href@noop {} {\bibfield  {journal} {\bibinfo  {journal} {arXiv preprint arXiv:1409.1556}\ } (\bibinfo {year} {2014})}\BibitemShut {NoStop}%
\bibitem [{\citenamefont {Raissi}, \citenamefont {Perdikaris},\ and\ \citenamefont {Karniadakis}(2019)}]{2019PINNsr10}%
  \BibitemOpen
  \bibfield  {author} {\bibinfo {author} {\bibfnamefont {M.}~\bibnamefont {Raissi}}, \bibinfo {author} {\bibfnamefont {P.}~\bibnamefont {Perdikaris}}, \ and\ \bibinfo {author} {\bibfnamefont {G.~E.}\ \bibnamefont {Karniadakis}},\ }\bibfield  {title} {\enquote {\bibinfo {title} {Physics-informed neural networks: A deep learning framework for solving forward and inverse problems involving nonlinear partial differential equations},}\ }\href@noop {} {\bibfield  {journal} {\bibinfo  {journal} {J. Comput. Phys.}\ }\textbf {\bibinfo {volume} {378}},\ \bibinfo {pages} {686--707} (\bibinfo {year} {2019})}\BibitemShut {NoStop}%
\bibitem [{\citenamefont {Zhu}\ \emph {et~al.}(2024)\citenamefont {Zhu}, \citenamefont {Kong}, \citenamefont {Deng},\ and\ \citenamefont {Bian}}]{zhu2024physics}%
  \BibitemOpen
  \bibfield  {author} {\bibinfo {author} {\bibfnamefont {Y.}~\bibnamefont {Zhu}}, \bibinfo {author} {\bibfnamefont {W.}~\bibnamefont {Kong}}, \bibinfo {author} {\bibfnamefont {J.}~\bibnamefont {Deng}}, \ and\ \bibinfo {author} {\bibfnamefont {X.}~\bibnamefont {Bian}},\ }\bibfield  {title} {\enquote {\bibinfo {title} {{Physics-informed neural networks for incompressible flows with moving boundaries}},}\ }\href@noop {} {\bibfield  {journal} {\bibinfo  {journal} {Phys. Fluids}\ }\textbf {\bibinfo {volume} {36}} (\bibinfo {year} {2024})}\BibitemShut {NoStop}%
\bibitem [{\citenamefont {Song}\ \emph {et~al.}(2024)\citenamefont {Song}, \citenamefont {Cao}, \citenamefont {Liao},\ and\ \citenamefont {Zhang}}]{song2024vwpinns}%
  \BibitemOpen
  \bibfield  {author} {\bibinfo {author} {\bibfnamefont {J.}~\bibnamefont {Song}}, \bibinfo {author} {\bibfnamefont {W.}~\bibnamefont {Cao}}, \bibinfo {author} {\bibfnamefont {F.}~\bibnamefont {Liao}}, \ and\ \bibinfo {author} {\bibfnamefont {W.}~\bibnamefont {Zhang}},\ }\bibfield  {title} {\enquote {\bibinfo {title} {Vw-pinns: A volume weighting method for pde residuals in physics-informed neural networks},}\ }\href@noop {} {\bibfield  {journal} {\bibinfo  {journal} {arXiv preprint arXiv:2401.06196}\ } (\bibinfo {year} {2024})}\BibitemShut {NoStop}%
\bibitem [{\citenamefont {E}\ and\ \citenamefont {Yu}(2018)}]{deepritzr13}%
  \BibitemOpen
  \bibfield  {author} {\bibinfo {author} {\bibfnamefont {W.}~\bibnamefont {E}}\ and\ \bibinfo {author} {\bibfnamefont {B.}~\bibnamefont {Yu}},\ }\bibfield  {title} {\enquote {\bibinfo {title} {The deep ritz method: a deep learning-based numerical algorithm for solving variational problems},}\ }\href@noop {} {\bibfield  {journal} {\bibinfo  {journal} {Commun. Math. Stat.}\ }\textbf {\bibinfo {volume} {6}},\ \bibinfo {pages} {1--12} (\bibinfo {year} {2018})}\BibitemShut {NoStop}%
\bibitem [{\citenamefont {Sirignano}\ and\ \citenamefont {Spiliopoulos}(2018)}]{Sirignano2018}%
  \BibitemOpen
  \bibfield  {author} {\bibinfo {author} {\bibfnamefont {J.}~\bibnamefont {Sirignano}}\ and\ \bibinfo {author} {\bibfnamefont {K.}~\bibnamefont {Spiliopoulos}},\ }\bibfield  {title} {\enquote {\bibinfo {title} {{DGM: A deep learning algorithm for solving partial differential equations}},}\ }\href@noop {} {\bibfield  {journal} {\bibinfo  {journal} {J. Comput. Phys.}\ }\textbf {\bibinfo {volume} {375}},\ \bibinfo {pages} {1339--1364} (\bibinfo {year} {2018})}\BibitemShut {NoStop}%
\bibitem [{\citenamefont {Karnakov}, \citenamefont {Litvinov},\ and\ \citenamefont {Koumoutsakos}(2022)}]{karnakov2022ODILr161}%
  \BibitemOpen
  \bibfield  {author} {\bibinfo {author} {\bibfnamefont {P.}~\bibnamefont {Karnakov}}, \bibinfo {author} {\bibfnamefont {S.}~\bibnamefont {Litvinov}}, \ and\ \bibinfo {author} {\bibfnamefont {P.}~\bibnamefont {Koumoutsakos}},\ }\bibfield  {title} {\enquote {\bibinfo {title} {Optimizing a discrete loss (odil) to solve forward and inverse problems for partial differential equations using machine learning tools},}\ }\href@noop {} {\bibfield  {journal} {\bibinfo  {journal} {arXiv preprint arXiv:2205.04611}\ } (\bibinfo {year} {2022})}\BibitemShut {NoStop}%
\bibitem [{\citenamefont {Jin}\ \emph {et~al.}(2021)\citenamefont {Jin}, \citenamefont {Cai}, \citenamefont {Li},\ and\ \citenamefont {Karniadakis}}]{jin2021nsfnetsr18}%
  \BibitemOpen
  \bibfield  {author} {\bibinfo {author} {\bibfnamefont {X.}~\bibnamefont {Jin}}, \bibinfo {author} {\bibfnamefont {S.}~\bibnamefont {Cai}}, \bibinfo {author} {\bibfnamefont {H.}~\bibnamefont {Li}}, \ and\ \bibinfo {author} {\bibfnamefont {G.~E.}\ \bibnamefont {Karniadakis}},\ }\bibfield  {title} {\enquote {\bibinfo {title} {Nsfnets (navier-stokes flow nets): Physics-informed neural networks for the incompressible navier-stokes equations},}\ }\href@noop {} {\bibfield  {journal} {\bibinfo  {journal} {J. Comput. Phys.}\ }\textbf {\bibinfo {volume} {426}},\ \bibinfo {pages} {109951} (\bibinfo {year} {2021})}\BibitemShut {NoStop}%
\bibitem [{\citenamefont {Mao}, \citenamefont {Jagtap},\ and\ \citenamefont {Karniadakis}(2020)}]{mao2020pinnsr20}%
  \BibitemOpen
  \bibfield  {author} {\bibinfo {author} {\bibfnamefont {Z.}~\bibnamefont {Mao}}, \bibinfo {author} {\bibfnamefont {A.~D.}\ \bibnamefont {Jagtap}}, \ and\ \bibinfo {author} {\bibfnamefont {G.~E.}\ \bibnamefont {Karniadakis}},\ }\bibfield  {title} {\enquote {\bibinfo {title} {Physics-informed neural networks for high-speed flows},}\ }\href@noop {} {\bibfield  {journal} {\bibinfo  {journal} {Comput. Methods Appl. Mech. Eng.}\ }\textbf {\bibinfo {volume} {360}},\ \bibinfo {pages} {112789} (\bibinfo {year} {2020})}\BibitemShut {NoStop}%
\bibitem [{\citenamefont {Cai}\ \emph {et~al.}(2021)\citenamefont {Cai}, \citenamefont {Wang}, \citenamefont {Fuest}, \citenamefont {Jeon}, \citenamefont {Gray},\ and\ \citenamefont {Karniadakis}}]{cai2021flowr21}%
  \BibitemOpen
  \bibfield  {author} {\bibinfo {author} {\bibfnamefont {S.}~\bibnamefont {Cai}}, \bibinfo {author} {\bibfnamefont {Z.}~\bibnamefont {Wang}}, \bibinfo {author} {\bibfnamefont {F.}~\bibnamefont {Fuest}}, \bibinfo {author} {\bibfnamefont {Y.~J.}\ \bibnamefont {Jeon}}, \bibinfo {author} {\bibfnamefont {C.}~\bibnamefont {Gray}}, \ and\ \bibinfo {author} {\bibfnamefont {G.~E.}\ \bibnamefont {Karniadakis}},\ }\bibfield  {title} {\enquote {\bibinfo {title} {Flow over an espresso cup: inferring 3-d velocity and pressure fields from tomographic background oriented schlieren via physics-informed neural networks},}\ }\href@noop {} {\bibfield  {journal} {\bibinfo  {journal} {J. Fluid Mech.}\ }\textbf {\bibinfo {volume} {915}},\ \bibinfo {pages} {A102} (\bibinfo {year} {2021})}\BibitemShut {NoStop}%
\bibitem [{\citenamefont {Brunton}\ and\ \citenamefont {Kutz}(2018)}]{Brunton2018}%
  \BibitemOpen
  \bibfield  {author} {\bibinfo {author} {\bibfnamefont {S.~L.}\ \bibnamefont {Brunton}}\ and\ \bibinfo {author} {\bibfnamefont {J.~N.}\ \bibnamefont {Kutz}},\ }\href@noop {} {\emph {\bibinfo {title} {{Data-Driven Science and Engineering}}}},\ \bibinfo {edition} {2nd}\ ed.\ (\bibinfo  {publisher} {Cambridge Universtiy Press},\ \bibinfo {year} {2018})\BibitemShut {NoStop}%
\bibitem [{\citenamefont {Hochreiter}\ and\ \citenamefont {Schmidhuber}(1997)}]{LSTM1997r22}%
  \BibitemOpen
  \bibfield  {author} {\bibinfo {author} {\bibfnamefont {S.}~\bibnamefont {Hochreiter}}\ and\ \bibinfo {author} {\bibfnamefont {J.}~\bibnamefont {Schmidhuber}},\ }\bibfield  {title} {\enquote {\bibinfo {title} {Long short-term memory},}\ }\href@noop {} {\bibfield  {journal} {\bibinfo  {journal} {Neural Comput.}\ }\textbf {\bibinfo {volume} {9}},\ \bibinfo {pages} {1735--1780} (\bibinfo {year} {1997})}\BibitemShut {NoStop}%
\bibitem [{\citenamefont {Han}\ \emph {et~al.}(2019)\citenamefont {Han}, \citenamefont {Wang}, \citenamefont {Zhang},\ and\ \citenamefont {Chen}}]{han2019pofr23}%
  \BibitemOpen
  \bibfield  {author} {\bibinfo {author} {\bibfnamefont {R.}~\bibnamefont {Han}}, \bibinfo {author} {\bibfnamefont {Y.}~\bibnamefont {Wang}}, \bibinfo {author} {\bibfnamefont {Y.}~\bibnamefont {Zhang}}, \ and\ \bibinfo {author} {\bibfnamefont {G.}~\bibnamefont {Chen}},\ }\bibfield  {title} {\enquote {\bibinfo {title} {A novel spatial-temporal prediction method for unsteady wake flows based on hybrid deep neural network},}\ }\href@noop {} {\bibfield  {journal} {\bibinfo  {journal} {Phys. Fluids}\ }\textbf {\bibinfo {volume} {31}} (\bibinfo {year} {2019})}\BibitemShut {NoStop}%
\bibitem [{\citenamefont {Nakamura}\ \emph {et~al.}(2021)\citenamefont {Nakamura}, \citenamefont {Fukami}, \citenamefont {Hasegawa}, \citenamefont {Nabae},\ and\ \citenamefont {Fukagata}}]{nakamura2021r26}%
  \BibitemOpen
  \bibfield  {author} {\bibinfo {author} {\bibfnamefont {T.}~\bibnamefont {Nakamura}}, \bibinfo {author} {\bibfnamefont {K.}~\bibnamefont {Fukami}}, \bibinfo {author} {\bibfnamefont {K.}~\bibnamefont {Hasegawa}}, \bibinfo {author} {\bibfnamefont {Y.}~\bibnamefont {Nabae}}, \ and\ \bibinfo {author} {\bibfnamefont {K.}~\bibnamefont {Fukagata}},\ }\bibfield  {title} {\enquote {\bibinfo {title} {Convolutional neural network and long short-term memory based reduced order surrogate for minimal turbulent channel flow},}\ }\href@noop {} {\bibfield  {journal} {\bibinfo  {journal} {Phys. Fluids}\ }\textbf {\bibinfo {volume} {33}} (\bibinfo {year} {2021})}\BibitemShut {NoStop}%
\bibitem [{\citenamefont {Wu}\ \emph {et~al.}(2021)\citenamefont {Wu}, \citenamefont {Gong}, \citenamefont {Pan}, \citenamefont {Qiu}, \citenamefont {Feng},\ and\ \citenamefont {Pain}}]{wu2021pofr27}%
  \BibitemOpen
  \bibfield  {author} {\bibinfo {author} {\bibfnamefont {P.}~\bibnamefont {Wu}}, \bibinfo {author} {\bibfnamefont {S.}~\bibnamefont {Gong}}, \bibinfo {author} {\bibfnamefont {K.}~\bibnamefont {Pan}}, \bibinfo {author} {\bibfnamefont {F.}~\bibnamefont {Qiu}}, \bibinfo {author} {\bibfnamefont {W.}~\bibnamefont {Feng}}, \ and\ \bibinfo {author} {\bibfnamefont {C.}~\bibnamefont {Pain}},\ }\bibfield  {title} {\enquote {\bibinfo {title} {Reduced order model using convolutional auto-encoder with self-attention},}\ }\href@noop {} {\bibfield  {journal} {\bibinfo  {journal} {Phys. Fluids}\ }\textbf {\bibinfo {volume} {33}} (\bibinfo {year} {2021})}\BibitemShut {NoStop}%
\bibitem [{\citenamefont {Zhang}\ \emph {et~al.}(2022)\citenamefont {Zhang}, \citenamefont {Ji}, \citenamefont {Xie}, \citenamefont {Zheng},\ and\ \citenamefont {Zheng}}]{zhang2022CMAMEr28}%
  \BibitemOpen
  \bibfield  {author} {\bibinfo {author} {\bibfnamefont {X.}~\bibnamefont {Zhang}}, \bibinfo {author} {\bibfnamefont {T.}~\bibnamefont {Ji}}, \bibinfo {author} {\bibfnamefont {F.}~\bibnamefont {Xie}}, \bibinfo {author} {\bibfnamefont {H.}~\bibnamefont {Zheng}}, \ and\ \bibinfo {author} {\bibfnamefont {Y.}~\bibnamefont {Zheng}},\ }\bibfield  {title} {\enquote {\bibinfo {title} {Unsteady flow prediction from sparse measurements by compressed sensing reduced order modeling},}\ }\href@noop {} {\bibfield  {journal} {\bibinfo  {journal} {Comput. Methods Appl. Mech. Eng.}\ }\textbf {\bibinfo {volume} {393}},\ \bibinfo {pages} {114800} (\bibinfo {year} {2022})}\BibitemShut {NoStop}%
\bibitem [{\citenamefont {Lu}\ \emph {et~al.}(2021)\citenamefont {Lu}, \citenamefont {Jin}, \citenamefont {Pang}, \citenamefont {Zhang},\ and\ \citenamefont {Karniadakis}}]{lu2021DNOr30}%
  \BibitemOpen
  \bibfield  {author} {\bibinfo {author} {\bibfnamefont {L.}~\bibnamefont {Lu}}, \bibinfo {author} {\bibfnamefont {P.}~\bibnamefont {Jin}}, \bibinfo {author} {\bibfnamefont {G.}~\bibnamefont {Pang}}, \bibinfo {author} {\bibfnamefont {Z.}~\bibnamefont {Zhang}}, \ and\ \bibinfo {author} {\bibfnamefont {G.~E.}\ \bibnamefont {Karniadakis}},\ }\bibfield  {title} {\enquote {\bibinfo {title} {Learning nonlinear operators via deeponet based on the universal approximation theorem of operators},}\ }\href@noop {} {\bibfield  {journal} {\bibinfo  {journal} {Nat. Mach. Intell.}\ }\textbf {\bibinfo {volume} {3}},\ \bibinfo {pages} {218--229} (\bibinfo {year} {2021})}\BibitemShut {NoStop}%
\bibitem [{\citenamefont {Li}\ \emph {et~al.}(2020)\citenamefont {Li}, \citenamefont {Kovachki}, \citenamefont {Azizzadenesheli}, \citenamefont {Liu}, \citenamefont {Bhattacharya}, \citenamefont {Stuart},\ and\ \citenamefont {Anandkumar}}]{li2020FNOr31}%
  \BibitemOpen
  \bibfield  {author} {\bibinfo {author} {\bibfnamefont {Z.}~\bibnamefont {Li}}, \bibinfo {author} {\bibfnamefont {N.}~\bibnamefont {Kovachki}}, \bibinfo {author} {\bibfnamefont {K.}~\bibnamefont {Azizzadenesheli}}, \bibinfo {author} {\bibfnamefont {B.}~\bibnamefont {Liu}}, \bibinfo {author} {\bibfnamefont {K.}~\bibnamefont {Bhattacharya}}, \bibinfo {author} {\bibfnamefont {A.}~\bibnamefont {Stuart}}, \ and\ \bibinfo {author} {\bibfnamefont {A.}~\bibnamefont {Anandkumar}},\ }\bibfield  {title} {\enquote {\bibinfo {title} {Fourier neural operator for parametric partial differential equations},}\ }\href@noop {} {\bibfield  {journal} {\bibinfo  {journal} {arXiv preprint arXiv:2010.08895}\ } (\bibinfo {year} {2020})}\BibitemShut {NoStop}%
\bibitem [{\citenamefont {Li}, \citenamefont {Meidani},\ and\ \citenamefont {Farimani}(2022)}]{li2022transformerr32}%
  \BibitemOpen
  \bibfield  {author} {\bibinfo {author} {\bibfnamefont {Z.}~\bibnamefont {Li}}, \bibinfo {author} {\bibfnamefont {K.}~\bibnamefont {Meidani}}, \ and\ \bibinfo {author} {\bibfnamefont {A.~B.}\ \bibnamefont {Farimani}},\ }\bibfield  {title} {\enquote {\bibinfo {title} {Transformer for partial differential equations' operator learning},}\ }\href@noop {} {\bibfield  {journal} {\bibinfo  {journal} {arXiv preprint arXiv:2205.13671}\ } (\bibinfo {year} {2022})}\BibitemShut {NoStop}%
\bibitem [{\citenamefont {Xu}, \citenamefont {Zhou},\ and\ \citenamefont {Bian}(2024)}]{xu2024pof}%
  \BibitemOpen
  \bibfield  {author} {\bibinfo {author} {\bibfnamefont {B.}~\bibnamefont {Xu}}, \bibinfo {author} {\bibfnamefont {Y.}~\bibnamefont {Zhou}}, \ and\ \bibinfo {author} {\bibfnamefont {X.}~\bibnamefont {Bian}},\ }\bibfield  {title} {\enquote {\bibinfo {title} {Self-supervised learning based on transformer for flow reconstruction and prediction},}\ }\href@noop {} {\bibfield  {journal} {\bibinfo  {journal} {Phys. Fluids}\ }\textbf {\bibinfo {volume} {36}} (\bibinfo {year} {2024})}\BibitemShut {NoStop}%
\bibitem [{\citenamefont {Chen}\ and\ \citenamefont {Chen}(1995)}]{chenandchen1995r33}%
  \BibitemOpen
  \bibfield  {author} {\bibinfo {author} {\bibfnamefont {T.}~\bibnamefont {Chen}}\ and\ \bibinfo {author} {\bibfnamefont {H.}~\bibnamefont {Chen}},\ }\bibfield  {title} {\enquote {\bibinfo {title} {Universal approximation to nonlinear operators by neural networks with arbitrary activation functions and its application to dynamical systems},}\ }\href@noop {} {\bibfield  {journal} {\bibinfo  {journal} {IEEE T. Neural Network}\ }\textbf {\bibinfo {volume} {6}},\ \bibinfo {pages} {911--917} (\bibinfo {year} {1995})}\BibitemShut {NoStop}%
\bibitem [{\citenamefont {Lin}\ \emph {et~al.}(2021)\citenamefont {Lin}, \citenamefont {Maxey}, \citenamefont {Li},\ and\ \citenamefont {Karniadakis}}]{lin2021JFMr35}%
  \BibitemOpen
  \bibfield  {author} {\bibinfo {author} {\bibfnamefont {C.}~\bibnamefont {Lin}}, \bibinfo {author} {\bibfnamefont {M.}~\bibnamefont {Maxey}}, \bibinfo {author} {\bibfnamefont {Z.}~\bibnamefont {Li}}, \ and\ \bibinfo {author} {\bibfnamefont {G.~E.}\ \bibnamefont {Karniadakis}},\ }\bibfield  {title} {\enquote {\bibinfo {title} {A seamless multiscale operator neural network for inferring bubble dynamics},}\ }\href@noop {} {\bibfield  {journal} {\bibinfo  {journal} {J. Fluid Mech.}\ }\textbf {\bibinfo {volume} {929}},\ \bibinfo {pages} {A18} (\bibinfo {year} {2021})}\BibitemShut {NoStop}%
\bibitem [{\citenamefont {Oommen}\ \emph {et~al.}(2022)\citenamefont {Oommen}, \citenamefont {Shukla}, \citenamefont {Goswami}, \citenamefont {Dingreville},\ and\ \citenamefont {Karniadakis}}]{oommen2022operatorr36}%
  \BibitemOpen
  \bibfield  {author} {\bibinfo {author} {\bibfnamefont {V.}~\bibnamefont {Oommen}}, \bibinfo {author} {\bibfnamefont {K.}~\bibnamefont {Shukla}}, \bibinfo {author} {\bibfnamefont {S.}~\bibnamefont {Goswami}}, \bibinfo {author} {\bibfnamefont {R.}~\bibnamefont {Dingreville}}, \ and\ \bibinfo {author} {\bibfnamefont {G.~E.}\ \bibnamefont {Karniadakis}},\ }\bibfield  {title} {\enquote {\bibinfo {title} {Learning two-phase microstructure evolution using neural operators and autoencoder architectures},}\ }\href@noop {} {\bibfield  {journal} {\bibinfo  {journal} {Npj Comput. Mater.}\ }\textbf {\bibinfo {volume} {8}},\ \bibinfo {pages} {190} (\bibinfo {year} {2022})}\BibitemShut {NoStop}%
\bibitem [{\citenamefont {Mao}\ \emph {et~al.}(2021)\citenamefont {Mao}, \citenamefont {Lu}, \citenamefont {Marxen}, \citenamefont {Zaki},\ and\ \citenamefont {Karniadakis}}]{mao2021deepmr37}%
  \BibitemOpen
  \bibfield  {author} {\bibinfo {author} {\bibfnamefont {Z.}~\bibnamefont {Mao}}, \bibinfo {author} {\bibfnamefont {L.}~\bibnamefont {Lu}}, \bibinfo {author} {\bibfnamefont {O.}~\bibnamefont {Marxen}}, \bibinfo {author} {\bibfnamefont {T.~A.}\ \bibnamefont {Zaki}}, \ and\ \bibinfo {author} {\bibfnamefont {G.~E.}\ \bibnamefont {Karniadakis}},\ }\bibfield  {title} {\enquote {\bibinfo {title} {Deepm\&mnet for hypersonics: Predicting the coupled flow and finite-rate chemistry behind a normal shock using neural-network approximation of operators},}\ }\href@noop {} {\bibfield  {journal} {\bibinfo  {journal} {J. Comput. Phys.}\ }\textbf {\bibinfo {volume} {447}},\ \bibinfo {pages} {110698} (\bibinfo {year} {2021})}\BibitemShut {NoStop}%
\bibitem [{\citenamefont {Wang}\ and\ \citenamefont {Perdikaris}(2023)}]{wang2023longphysicDNO}%
  \BibitemOpen
  \bibfield  {author} {\bibinfo {author} {\bibfnamefont {S.}~\bibnamefont {Wang}}\ and\ \bibinfo {author} {\bibfnamefont {P.}~\bibnamefont {Perdikaris}},\ }\bibfield  {title} {\enquote {\bibinfo {title} {Long-time integration of parametric evolution equations with physics-informed deeponets},}\ }\href@noop {} {\bibfield  {journal} {\bibinfo  {journal} {J. Comput. Phys.}\ }\textbf {\bibinfo {volume} {475}},\ \bibinfo {pages} {111855} (\bibinfo {year} {2023})}\BibitemShut {NoStop}%
\bibitem [{\citenamefont {Micha{\l}owska}\ \emph {et~al.}(2023)\citenamefont {Micha{\l}owska}, \citenamefont {Goswami}, \citenamefont {Karniadakis},\ and\ \citenamefont {Riemer-S{\o}rensen}}]{michalowska2023r38}%
  \BibitemOpen
  \bibfield  {author} {\bibinfo {author} {\bibfnamefont {K.}~\bibnamefont {Micha{\l}owska}}, \bibinfo {author} {\bibfnamefont {S.}~\bibnamefont {Goswami}}, \bibinfo {author} {\bibfnamefont {G.~E.}\ \bibnamefont {Karniadakis}}, \ and\ \bibinfo {author} {\bibfnamefont {S.}~\bibnamefont {Riemer-S{\o}rensen}},\ }\bibfield  {title} {\enquote {\bibinfo {title} {Neural operator learning for long-time integration in dynamical systems with recurrent neural networks},}\ }\href@noop {} {\bibfield  {journal} {\bibinfo  {journal} {arXiv preprint arXiv:2303.02243}\ } (\bibinfo {year} {2023})}\BibitemShut {NoStop}%
\bibitem [{\citenamefont {Karniadakis}\ and\ \citenamefont {Sherwin}(2005)}]{GK_CFDbook}%
  \BibitemOpen
  \bibfield  {author} {\bibinfo {author} {\bibfnamefont {G.}~\bibnamefont {Karniadakis}}\ and\ \bibinfo {author} {\bibfnamefont {S.}~\bibnamefont {Sherwin}},\ }\href@noop {} {\emph {\bibinfo {title} {Spectral/hp Element Methods for Computational Fluid Dynamics, 2nd edition}}}\ (\bibinfo  {publisher} {Oxford University Press},\ \bibinfo {address} {Oxford,UK},\ \bibinfo {year} {2005})\BibitemShut {NoStop}%
\bibitem [{\citenamefont {Lu}\ \emph {et~al.}(2022)\citenamefont {Lu}, \citenamefont {Meng}, \citenamefont {Cai}, \citenamefont {Mao}, \citenamefont {Goswami}, \citenamefont {Zhang},\ and\ \citenamefont {Karniadakis}}]{lu2022comprehensiver43}%
  \BibitemOpen
  \bibfield  {author} {\bibinfo {author} {\bibfnamefont {L.}~\bibnamefont {Lu}}, \bibinfo {author} {\bibfnamefont {X.}~\bibnamefont {Meng}}, \bibinfo {author} {\bibfnamefont {S.}~\bibnamefont {Cai}}, \bibinfo {author} {\bibfnamefont {Z.}~\bibnamefont {Mao}}, \bibinfo {author} {\bibfnamefont {S.}~\bibnamefont {Goswami}}, \bibinfo {author} {\bibfnamefont {Z.}~\bibnamefont {Zhang}}, \ and\ \bibinfo {author} {\bibfnamefont {G.~E.}\ \bibnamefont {Karniadakis}},\ }\bibfield  {title} {\enquote {\bibinfo {title} {A comprehensive and fair comparison of two neural operators (with practical extensions) based on fair data},}\ }\href@noop {} {\bibfield  {journal} {\bibinfo  {journal} {Comput. Methods Appl. Mech. Eng.}\ }\textbf {\bibinfo {volume} {393}},\ \bibinfo {pages} {114778} (\bibinfo {year} {2022})}\BibitemShut {NoStop}%
\bibitem [{\citenamefont {Szegedy}\ \emph {et~al.}(2015)\citenamefont {Szegedy}, \citenamefont {Liu}, \citenamefont {Jia}, \citenamefont {Sermanet}, \citenamefont {Reed}, \citenamefont {Anguelov}, \citenamefont {Erhan}, \citenamefont {Vanhoucke},\ and\ \citenamefont {Rabinovich}}]{googlenet2015r42}%
  \BibitemOpen
  \bibfield  {author} {\bibinfo {author} {\bibfnamefont {C.}~\bibnamefont {Szegedy}}, \bibinfo {author} {\bibfnamefont {W.}~\bibnamefont {Liu}}, \bibinfo {author} {\bibfnamefont {Y.}~\bibnamefont {Jia}}, \bibinfo {author} {\bibfnamefont {P.}~\bibnamefont {Sermanet}}, \bibinfo {author} {\bibfnamefont {S.}~\bibnamefont {Reed}}, \bibinfo {author} {\bibfnamefont {D.}~\bibnamefont {Anguelov}}, \bibinfo {author} {\bibfnamefont {D.}~\bibnamefont {Erhan}}, \bibinfo {author} {\bibfnamefont {V.}~\bibnamefont {Vanhoucke}}, \ and\ \bibinfo {author} {\bibfnamefont {A.}~\bibnamefont {Rabinovich}},\ }\bibfield  {title} {\enquote {\bibinfo {title} {Going deeper with convolutions},}\ }in\ \href@noop {} {\emph {\bibinfo {booktitle} {Proceedings of the IEEE Conference on Computer Vision and Pattern Recognition}}}\ (\bibinfo {year} {2015})\ pp.\ \bibinfo {pages} {1--9}\BibitemShut {NoStop}%
\bibitem [{\citenamefont {Wandel}\ \emph {et~al.}(2022)\citenamefont {Wandel}, \citenamefont {Weinmann}, \citenamefont {Neidlin},\ and\ \citenamefont {Klein}}]{wandel2022spliner16}%
  \BibitemOpen
  \bibfield  {author} {\bibinfo {author} {\bibfnamefont {N.}~\bibnamefont {Wandel}}, \bibinfo {author} {\bibfnamefont {M.}~\bibnamefont {Weinmann}}, \bibinfo {author} {\bibfnamefont {M.}~\bibnamefont {Neidlin}}, \ and\ \bibinfo {author} {\bibfnamefont {R.}~\bibnamefont {Klein}},\ }\bibfield  {title} {\enquote {\bibinfo {title} {Spline-pinn: Approaching pdes without data using fast, physics-informed hermite-spline cnns},}\ }in\ \href@noop {} {\emph {\bibinfo {booktitle} {Proceedings of the AAAI Conference on Artificial Intelligence}}},\ Vol.~\bibinfo {volume} {36}\ (\bibinfo {year} {2022})\ pp.\ \bibinfo {pages} {8529--8538}\BibitemShut {NoStop}%
\bibitem [{\citenamefont {Wang}\ \emph {et~al.}(2018)\citenamefont {Wang}, \citenamefont {Girshick}, \citenamefont {Gupta},\ and\ \citenamefont {He}}]{wang2018nonlocalr45}%
  \BibitemOpen
  \bibfield  {author} {\bibinfo {author} {\bibfnamefont {X.}~\bibnamefont {Wang}}, \bibinfo {author} {\bibfnamefont {R.}~\bibnamefont {Girshick}}, \bibinfo {author} {\bibfnamefont {A.}~\bibnamefont {Gupta}}, \ and\ \bibinfo {author} {\bibfnamefont {K.}~\bibnamefont {He}},\ }\bibfield  {title} {\enquote {\bibinfo {title} {Non-local neural networks},}\ }in\ \href@noop {} {\emph {\bibinfo {booktitle} {Proceedings of the IEEE Conference on Computer Vision and Pattern Recognition}}}\ (\bibinfo {year} {2018})\ pp.\ \bibinfo {pages} {7794--7803}\BibitemShut {NoStop}%
\bibitem [{\citenamefont {Buades}, \citenamefont {Coll},\ and\ \citenamefont {Morel}(2005)}]{buades2005nonlocalalgorithmr46}%
  \BibitemOpen
  \bibfield  {author} {\bibinfo {author} {\bibfnamefont {A.}~\bibnamefont {Buades}}, \bibinfo {author} {\bibfnamefont {B.}~\bibnamefont {Coll}}, \ and\ \bibinfo {author} {\bibfnamefont {J.-M.}\ \bibnamefont {Morel}},\ }\bibfield  {title} {\enquote {\bibinfo {title} {A non-local algorithm for image denoising},}\ }in\ \href@noop {} {\emph {\bibinfo {booktitle} {2005 IEEE Computer Society Conference on Computer Vision and Pattern Recognition}}},\ Vol.~\bibinfo {volume} {2}\ (\bibinfo {organization} {Ieee},\ \bibinfo {year} {2005})\ pp.\ \bibinfo {pages} {60--65}\BibitemShut {NoStop}%
\bibitem [{\citenamefont {Vaswani}\ \emph {et~al.}(2017)\citenamefont {Vaswani}, \citenamefont {Shazeer}, \citenamefont {Parmar}, \citenamefont {Uszkoreit}, \citenamefont {Jones}, \citenamefont {Gomez}, \citenamefont {Kaiser},\ and\ \citenamefont {Polosukhin}}]{vaswani2017attentionr47}%
  \BibitemOpen
  \bibfield  {author} {\bibinfo {author} {\bibfnamefont {A.}~\bibnamefont {Vaswani}}, \bibinfo {author} {\bibfnamefont {N.}~\bibnamefont {Shazeer}}, \bibinfo {author} {\bibfnamefont {N.}~\bibnamefont {Parmar}}, \bibinfo {author} {\bibfnamefont {J.}~\bibnamefont {Uszkoreit}}, \bibinfo {author} {\bibfnamefont {L.}~\bibnamefont {Jones}}, \bibinfo {author} {\bibfnamefont {A.~N.}\ \bibnamefont {Gomez}}, \bibinfo {author} {\bibfnamefont {{\L}.}~\bibnamefont {Kaiser}}, \ and\ \bibinfo {author} {\bibfnamefont {I.}~\bibnamefont {Polosukhin}},\ }\bibfield  {title} {\enquote {\bibinfo {title} {Attention is all you need},}\ }\href@noop {} {\bibfield  {journal} {\bibinfo  {journal} {Advances in Neural Information Processing Systems}\ }\textbf {\bibinfo {volume} {30}} (\bibinfo {year} {2017})}\BibitemShut {NoStop}%
\bibitem [{\citenamefont {He}\ \emph {et~al.}(2016{\natexlab{b}})\citenamefont {He}, \citenamefont {Zhang}, \citenamefont {Ren},\ and\ \citenamefont {Sun}}]{he2016residualconnection}%
  \BibitemOpen
  \bibfield  {author} {\bibinfo {author} {\bibfnamefont {K.}~\bibnamefont {He}}, \bibinfo {author} {\bibfnamefont {X.}~\bibnamefont {Zhang}}, \bibinfo {author} {\bibfnamefont {S.}~\bibnamefont {Ren}}, \ and\ \bibinfo {author} {\bibfnamefont {J.}~\bibnamefont {Sun}},\ }\bibfield  {title} {\enquote {\bibinfo {title} {Deep residual learning for image recognition},}\ }in\ \href@noop {} {\emph {\bibinfo {booktitle} {Proceedings of the IEEE Conference on Computer Vision and Pattern Recognition}}}\ (\bibinfo {year} {2016})\ pp.\ \bibinfo {pages} {770--778}\BibitemShut {NoStop}%
\bibitem [{\citenamefont {Park}\ and\ \citenamefont {Kim}(2022)}]{park2022visiontfworkr48}%
  \BibitemOpen
  \bibfield  {author} {\bibinfo {author} {\bibfnamefont {N.}~\bibnamefont {Park}}\ and\ \bibinfo {author} {\bibfnamefont {S.}~\bibnamefont {Kim}},\ }\bibfield  {title} {\enquote {\bibinfo {title} {How do vision transformers work?}}\ }\href@noop {} {\bibfield  {journal} {\bibinfo  {journal} {arXiv preprint arXiv:2202.06709}\ } (\bibinfo {year} {2022})}\BibitemShut {NoStop}%
\bibitem [{\citenamefont {Arnab}\ \emph {et~al.}(2021)\citenamefont {Arnab}, \citenamefont {Dehghani}, \citenamefont {Heigold}, \citenamefont {Sun}, \citenamefont {Lu{\v{c}}i{\'c}},\ and\ \citenamefont {Schmid}}]{visiontransformerr49}%
  \BibitemOpen
  \bibfield  {author} {\bibinfo {author} {\bibfnamefont {A.}~\bibnamefont {Arnab}}, \bibinfo {author} {\bibfnamefont {M.}~\bibnamefont {Dehghani}}, \bibinfo {author} {\bibfnamefont {G.}~\bibnamefont {Heigold}}, \bibinfo {author} {\bibfnamefont {C.}~\bibnamefont {Sun}}, \bibinfo {author} {\bibfnamefont {M.}~\bibnamefont {Lu{\v{c}}i{\'c}}}, \ and\ \bibinfo {author} {\bibfnamefont {C.}~\bibnamefont {Schmid}},\ }\bibfield  {title} {\enquote {\bibinfo {title} {Vivit: A video vision transformer},}\ }in\ \href@noop {} {\emph {\bibinfo {booktitle} {Proceedings of the IEEE/CVF International Conference on Computer Vision}}}\ (\bibinfo {year} {2021})\ pp.\ \bibinfo {pages} {6836--6846}\BibitemShut {NoStop}%
\bibitem [{\citenamefont {Keskar}\ \emph {et~al.}(2016)\citenamefont {Keskar}, \citenamefont {Mudigere}, \citenamefont {Nocedal}, \citenamefont {Smelyanskiy},\ and\ \citenamefont {Tang}}]{keskar2016larger530}%
  \BibitemOpen
  \bibfield  {author} {\bibinfo {author} {\bibfnamefont {N.~S.}\ \bibnamefont {Keskar}}, \bibinfo {author} {\bibfnamefont {D.}~\bibnamefont {Mudigere}}, \bibinfo {author} {\bibfnamefont {J.}~\bibnamefont {Nocedal}}, \bibinfo {author} {\bibfnamefont {M.}~\bibnamefont {Smelyanskiy}}, \ and\ \bibinfo {author} {\bibfnamefont {P.~T.~P.}\ \bibnamefont {Tang}},\ }\bibfield  {title} {\enquote {\bibinfo {title} {On large-batch training for deep learning: Generalization gap and sharp minima},}\ }\href@noop {} {\bibfield  {journal} {\bibinfo  {journal} {arXiv preprint arXiv:1609.04836}\ } (\bibinfo {year} {2016})}\BibitemShut {NoStop}%
\bibitem [{\citenamefont {Dziugaite}\ and\ \citenamefont {Roy}(2017)}]{dziugaite2017computingr531}%
  \BibitemOpen
  \bibfield  {author} {\bibinfo {author} {\bibfnamefont {G.~K.}\ \bibnamefont {Dziugaite}}\ and\ \bibinfo {author} {\bibfnamefont {D.~M.}\ \bibnamefont {Roy}},\ }\bibfield  {title} {\enquote {\bibinfo {title} {Computing nonvacuous generalization bounds for deep (stochastic) neural networks with many more parameters than training data},}\ }\href@noop {} {\bibfield  {journal} {\bibinfo  {journal} {arXiv preprint arXiv:1703.11008}\ } (\bibinfo {year} {2017})}\BibitemShut {NoStop}%
\bibitem [{\citenamefont {Jiang}\ \emph {et~al.}(2019)\citenamefont {Jiang}, \citenamefont {Neyshabur}, \citenamefont {Mobahi}, \citenamefont {Krishnan},\ and\ \citenamefont {Bengio}}]{jiang2019fantasticr532}%
  \BibitemOpen
  \bibfield  {author} {\bibinfo {author} {\bibfnamefont {Y.}~\bibnamefont {Jiang}}, \bibinfo {author} {\bibfnamefont {B.}~\bibnamefont {Neyshabur}}, \bibinfo {author} {\bibfnamefont {H.}~\bibnamefont {Mobahi}}, \bibinfo {author} {\bibfnamefont {D.}~\bibnamefont {Krishnan}}, \ and\ \bibinfo {author} {\bibfnamefont {S.}~\bibnamefont {Bengio}},\ }\bibfield  {title} {\enquote {\bibinfo {title} {Fantastic generalization measures and where to find them},}\ }\href@noop {} {\bibfield  {journal} {\bibinfo  {journal} {arXiv preprint arXiv:1912.02178}\ } (\bibinfo {year} {2019})}\BibitemShut {NoStop}%
\bibitem [{\citenamefont {Foret}\ \emph {et~al.}(2020)\citenamefont {Foret}, \citenamefont {Kleiner}, \citenamefont {Mobahi},\ and\ \citenamefont {Neyshabur}}]{foret2020SAMr50}%
  \BibitemOpen
  \bibfield  {author} {\bibinfo {author} {\bibfnamefont {P.}~\bibnamefont {Foret}}, \bibinfo {author} {\bibfnamefont {A.}~\bibnamefont {Kleiner}}, \bibinfo {author} {\bibfnamefont {H.}~\bibnamefont {Mobahi}}, \ and\ \bibinfo {author} {\bibfnamefont {B.}~\bibnamefont {Neyshabur}},\ }\bibfield  {title} {\enquote {\bibinfo {title} {Sharpness-aware minimization for efficiently improving generalization},}\ }\href@noop {} {\bibfield  {journal} {\bibinfo  {journal} {arXiv preprint arXiv:2010.01412}\ } (\bibinfo {year} {2020})}\BibitemShut {NoStop}%
\bibitem [{\citenamefont {Kingma}\ and\ \citenamefont {Ba}(2014)}]{kingma2014adamr44}%
  \BibitemOpen
  \bibfield  {author} {\bibinfo {author} {\bibfnamefont {D.~P.}\ \bibnamefont {Kingma}}\ and\ \bibinfo {author} {\bibfnamefont {J.}~\bibnamefont {Ba}},\ }\bibfield  {title} {\enquote {\bibinfo {title} {Adam: A method for stochastic optimization},}\ }\href@noop {} {\bibfield  {journal} {\bibinfo  {journal} {arXiv preprint arXiv:1412.6980}\ } (\bibinfo {year} {2014})}\BibitemShut {NoStop}%
\bibitem [{\citenamefont {Zhuang}\ \emph {et~al.}(2022)\citenamefont {Zhuang}, \citenamefont {Gong}, \citenamefont {Yuan}, \citenamefont {Cui}, \citenamefont {Adam}, \citenamefont {Dvornek}, \citenamefont {Tatikonda}, \citenamefont {Duncan},\ and\ \citenamefont {Liu}}]{zhuang2022GSAMr51}%
  \BibitemOpen
  \bibfield  {author} {\bibinfo {author} {\bibfnamefont {J.}~\bibnamefont {Zhuang}}, \bibinfo {author} {\bibfnamefont {B.}~\bibnamefont {Gong}}, \bibinfo {author} {\bibfnamefont {L.}~\bibnamefont {Yuan}}, \bibinfo {author} {\bibfnamefont {Y.}~\bibnamefont {Cui}}, \bibinfo {author} {\bibfnamefont {H.}~\bibnamefont {Adam}}, \bibinfo {author} {\bibfnamefont {N.}~\bibnamefont {Dvornek}}, \bibinfo {author} {\bibfnamefont {S.}~\bibnamefont {Tatikonda}}, \bibinfo {author} {\bibfnamefont {J.}~\bibnamefont {Duncan}}, \ and\ \bibinfo {author} {\bibfnamefont {T.}~\bibnamefont {Liu}},\ }\bibfield  {title} {\enquote {\bibinfo {title} {Surrogate gap minimization improves sharpness-aware training},}\ }\href@noop {} {\bibfield  {journal} {\bibinfo  {journal} {arXiv preprint arXiv:2203.08065}\ } (\bibinfo {year} {2022})}\BibitemShut {NoStop}%
\bibitem [{\citenamefont {Xu}\ \emph {et~al.}(2019)\citenamefont {Xu}, \citenamefont {Zhang}, \citenamefont {Luo}, \citenamefont {Xiao},\ and\ \citenamefont {Ma}}]{xu2019spectralbiasr52}%
  \BibitemOpen
  \bibfield  {author} {\bibinfo {author} {\bibfnamefont {Z.-Q.~J.}\ \bibnamefont {Xu}}, \bibinfo {author} {\bibfnamefont {Y.}~\bibnamefont {Zhang}}, \bibinfo {author} {\bibfnamefont {T.}~\bibnamefont {Luo}}, \bibinfo {author} {\bibfnamefont {Y.}~\bibnamefont {Xiao}}, \ and\ \bibinfo {author} {\bibfnamefont {Z.}~\bibnamefont {Ma}},\ }\bibfield  {title} {\enquote {\bibinfo {title} {Frequency principle: Fourier analysis sheds light on deep neural networks},}\ }\href@noop {} {\bibfield  {journal} {\bibinfo  {journal} {arXiv preprint arXiv:1901.06523}\ } (\bibinfo {year} {2019})}\BibitemShut {NoStop}%
\bibitem [{\citenamefont {Henderson}(1995)}]{henderson1995physics}%
  \BibitemOpen
  \bibfield  {author} {\bibinfo {author} {\bibfnamefont {R.~D.}\ \bibnamefont {Henderson}},\ }\bibfield  {title} {\enquote {\bibinfo {title} {Details of the drag curve near the onset of vortex shedding},}\ }\href@noop {} {\bibfield  {journal} {\bibinfo  {journal} {Phys. Fluids}\ }\textbf {\bibinfo {volume} {7}},\ \bibinfo {pages} {2102--2104} (\bibinfo {year} {1995})}\BibitemShut {NoStop}%
\end{thebibliography}%

\end{document}